\documentclass[longauth,traditabstract]{aa}

\usepackage[switch, displaymath, mathlines]{lineno}
\usepackage{setspace}

\usepackage{amsmath}

\usepackage{graphicx}

\usepackage{txfonts}
\usepackage{natbib}
\bibpunct{(}{)}{;}{a}{}{,}    
\usepackage{units}

\usepackage{longtable}
\usepackage{lscape}

\usepackage{tabularx}

\usepackage[colorlinks=false, final, draft]{hyperref}

\newcommand*{\aligncell}[2]{\multicolumn{1}{#1}{#2}}

\clearpage{}

\newcommand{\hess}{H.E.S.S.}
\newcommand{\hessone}{H.E.S.S.~I}

\newcommand{\hone}{\ion{H}{I}}
\newcommand{\xmm}{\emph{XMM-Newton}}
\newcommand{\chandra}{\emph{Chandra}}
\newcommand{\suzaku}{\emph{Suzaku}}
\newcommand{\fermi}{\emph{Fermi}-LAT}
\newcommand{\gammaray}{$\gamma$-ray}
\newcommand{\gammarays}{$\gamma$-rays}

\newcommand{\hgpsurl}{\url{https://www.mpi-hd.mpg.de/hfm/HESS/hgps}}
\newcommand{\urlAstropy}{\url{http://www.astropy.org}}
\newcommand{\urlSherpa}{\url{http://cxc.cfa.harvard.edu/sherpa}}
\newcommand{\urlGammapy}{\url{https://github.com/gammapy/gammapy}}
\newcommand{\urlGammacat}{\url{https://github.com/gammapy/gamma-cat}}
\newcommand{\urlSnrcat}{\url{http://www.physics.umanitoba.ca/snr/SNRcat}}
\newcommand{\urlSimbad}{\url{http://simbad.u-strasbg.fr/simbad}}
\newcommand{\urlAtnf}{\url{http://www.atnf.csiro.au/research/pulsar/psrcat}}

\newcommand{\hgpsregion}{$-114\degr < l < 75\degr$ and $-5\degr < b < +5\degr$}
\newcommand{\hgpsGlat}{$|b|\leqslant3\degr$}
\newcommand{\hgpsMeanPSF}{$0.08\degr$}
\newcommand{\hgpsSensitivity}{$\lesssim 1.5$\%}
\newcommand{\hgpsObsTimeApprox}{$2700\,\mathrm{h}$}
\newcommand{\hgpsAssumedSpecIndex}{2.3}

\newcommand{\hgpsComponentCountTotal}{98}

\newcommand{\hgpsSourceCountAnalysed}{64 } 
\newcommand{\hgpsSourceCountCutout}{14 } 
\newcommand{\hgpsSourceCountMissingStr}{four} 
\newcommand{\hgpsSourceCountTotal}{78 }
\newcommand{\hgpsSourceCountNew}{16 }
\newcommand{\hgpsSourceCountReAnalysed}{48} 
\newcommand{\hgpsSourceCountCrossChecked}{45} 

\newcommand{\InHGPSFGLSourceCount}{352 }     
\newcommand{\InHGPSFHLSourceCount}{44 }      
\newcommand{\InHGPSSNRSourceCount}{211}      
\newcommand{\InHGPSPWNSourceCount}{29 }       
\newcommand{\InHGPSCOMPSourceCount}{42 }      
\newcommand{\InHGPSPSRSourceCount}{222 }      

\newcommand{\HGPSFGLAssociationCount}{64 }     
\newcommand{\HGPSFHLAssociationCount}{31 }     
\newcommand{\HGPSSNRAssociationCount}{24 }     
\newcommand{\HGPSPWNAssociationCount}{16 }     
\newcommand{\HGPSCOMPAssociationCount}{21 }    
\newcommand{\HGPSPSRAssociationCount}{47 }     
\newcommand{\HGPSEXTAssociationCount}{20 }     

\newcommand{\HGPSTotalAssociationCount}{223 }

\newcommand{\oneFGLAssocCount}{40 }     
\newcommand{\oneFHLAssocCount}{29 }     
\newcommand{\oneSNRAssocCount}{21 }     
\newcommand{\onePWNAssocCount}{16 }     
\newcommand{\oneCOMPAssocCount}{20 }    
\newcommand{\onePSRAssocCount}{42 }     

\newcommand{\FHLnonAssociated}{13 }   

\newcommand{\hgpsSourceCountID}{31 }
\newcommand{\hgpsSourceCountPWN}{12 }
\newcommand{\hgpsSourceCountSNR}{8 }
\newcommand{\hgpsSourceCountCOMP}{8 }
\newcommand{\hgpsSourceCountBin}{3 }
\newcommand{\hgpsSourceCountNotAssoc}{11 }
\newcommand{\hgpsSourceCountAssoc}{67 }

\newcommand{\hgpsPointLikeSourceCount}{17 } 

\newcommand{\hgpsRSpecIsNotChanged}{34 } 
\newcommand{\hgpsRSpecIsExtended}{21 }   
\newcommand{\hgpsRSpecIsReduced}{9 }     

\newcommand{\lima}{Li~\&~Ma}

\newcommand{\SingleSourceCaption}[2]{
\caption[New source image: #1]{
VHE \gammaray\ image: #1.  See Fig.~\ref{fig:HESS_J1119m614} for a general
description.  #2
}
}

\newcommand{\wstaturl}{\url{https://heasarc.gsfc.nasa.gov/xanadu/xspec/manual/XSappendixStatistics.html}}
\newcommand{\specstackurl}{\url{http://cxc.harvard.edu/ciao/download/doc/combine.pdf}}

\newcommand{\TS}{TS}
\newcommand{\TSinFormula}{\mathrm{TS}}

\newcommand{\runDuration}{28~min}

\clearpage{}

\usepackage{footmisc}
\makeatletter
\renewcommand*{\fnsymbol}[1]{\ifcase#1\or*\or$\dagger$\or$\ddagger$\or**\or$\dagger\dagger$\or$\ddagger\ddagger$ \fi}
\makeatother

\begin{document}

\title{The \hess{} Galactic plane survey}

\author{\tiny H.E.S.S. Collaboration
\and H.~Abdalla \inst{1}
\and A.~Abramowski \inst{2}
\and F.~Aharonian \inst{3,4,5}
\and F.~Ait~Benkhali \inst{3}
\and E.O.~Ang\"uner \inst{21}
\and M.~Arakawa \inst{43}
\and M.~Arrieta \inst{15}
\and P.~Aubert \inst{24}
\and M.~Backes \inst{8}
% ...
}

\institute{
Centre for Space Research, North-West University, Potchefstroom 2520, South Africa \and
Universit\"at Hamburg, Institut f\"ur Experimentalphysik, Luruper Chaussee 149, D 22761 Hamburg, Germany \and
Max-Planck-Institut f\"ur Kernphysik, P.O. Box 103980, D 69029 Heidelberg, Germany \and
Dublin Institute for Advanced Studies, 31 Fitzwilliam Place, Dublin 2, Ireland \and
National Academy of Sciences of the Republic of Armenia,  Marshall Baghramian Avenue, 24, 0019 Yerevan, Republic of Armenia  \and
Yerevan Physics Institute, 2 Alikhanian Brothers St., 375036 Yerevan, Armenia \and
Institut f\"ur Physik, Humboldt-Universit\"at zu Berlin, Newtonstr. 15, D 12489 Berlin, Germany \and
University of Namibia, Department of Physics, Private Bag 13301, Windhoek, Namibia \and
GRAPPA, Anton Pannekoek Institute for Astronomy, University of Amsterdam,  Science Park 904, 1098 XH Amsterdam, The Netherlands \and
Department of Physics and Electrical Engineering, Linnaeus University,  351 95 V\"axj\"o, Sweden \and
Institut f\"ur Theoretische Physik, Lehrstuhl IV: Weltraum und Astrophysik, Ruhr-Universit\"at Bochum, D 44780 Bochum, Germany \and
GRAPPA, Anton Pannekoek Institute for Astronomy and Institute of High-Energy Physics, University of Amsterdam,  Science Park 904, 1098 XH Amsterdam, The Netherlands \and
Institut f\"ur Astro- und Teilchenphysik, Leopold-Franzens-Universit\"at Innsbruck, A-6020 Innsbruck, Austria \and
School of Physical Sciences, University of Adelaide, Adelaide 5005, Australia \and
LUTH, Observatoire de Paris, PSL Research University, CNRS, Universit\'e Paris Diderot, 5 Place Jules Janssen, 92190 Meudon, France \and
Sorbonne Universit\'es, UPMC Universit\'e Paris 06, Universit\'e Paris Diderot, Sorbonne Paris Cit\'e, CNRS, Laboratoire de Physique Nucl\'eaire et de Hautes Energies (LPNHE), 4 place Jussieu, F-75252, Paris Cedex 5, France \and
Laboratoire Univers et Particules de Montpellier, Universit\'e Montpellier, CNRS/IN2P3,  CC 72, Place Eug\`ene Bataillon, F-34095 Montpellier Cedex 5, France \and
IRFU, CEA, Universit\'e Paris-Saclay, F-91191 Gif-sur-Yvette, France \and
Astronomical Observatory, The University of Warsaw, Al. Ujazdowskie 4, 00-478 Warsaw, Poland \and
Aix Marseille Universit\'e, CNRS/IN2P3, CPPM, Marseille, France \and
Instytut Fizyki J\c{a}drowej PAN, ul. Radzikowskiego 152, 31-342 Krak{\'o}w, Poland \and
Funded by EU FP7 Marie Curie, grant agreement No. PIEF-GA-2012-332350,  \and
School of Physics, University of the Witwatersrand, 1 Jan Smuts Avenue, Braamfontein, Johannesburg, 2050 South Africa \and
Laboratoire d'Annecy-le-Vieux de Physique des Particules, Universit\'{e} Savoie Mont-Blanc, CNRS/IN2P3, F-74941 Annecy-le-Vieux, France \and
Landessternwarte, Universit\"at Heidelberg, K\"onigstuhl, D 69117 Heidelberg, Germany \and
Universit\'e Bordeaux, CNRS/IN2P3, Centre d'\'Etudes Nucl\'eaires de Bordeaux Gradignan, 33175 Gradignan, France \and
Oskar Klein Centre, Department of Physics, Stockholm University, Albanova University Center, SE-10691 Stockholm, Sweden \and
Wallenberg Academy Fellow,  \and
Institut f\"ur Astronomie und Astrophysik, Universit\"at T\"ubingen, Sand 1, D 72076 T\"ubingen, Germany \and
Laboratoire Leprince-Ringuet, Ecole Polytechnique, CNRS/IN2P3, F-91128 Palaiseau, France \and
APC, AstroParticule et Cosmologie, Universit\'{e} Paris Diderot, CNRS/IN2P3, CEA/Irfu, Observatoire de Paris, Sorbonne Paris Cit\'{e}, 10, rue Alice Domon et L\'{e}onie Duquet, 75205 Paris Cedex 13, France \and
Univ. Grenoble Alpes, CNRS, IPAG, F-38000 Grenoble, France \and
Department of Physics and Astronomy, The University of Leicester, University Road, Leicester, LE1 7RH, United Kingdom \and
Nicolaus Copernicus Astronomical Center, Polish Academy of Sciences, ul. Bartycka 18, 00-716 Warsaw, Poland \and
Institut f\"ur Physik und Astronomie, Universit\"at Potsdam,  Karl-Liebknecht-Strasse 24/25, D 14476 Potsdam, Germany \and
Friedrich-Alexander-Universit\"at Erlangen-N\"urnberg, Erlangen Centre for Astroparticle Physics, Erwin-Rommel-Str. 1, D 91058 Erlangen, Germany \and
DESY, D-15738 Zeuthen, Germany \and
Obserwatorium Astronomiczne, Uniwersytet Jagiello{\'n}ski, ul. Orla 171, 30-244 Krak{\'o}w, Poland \and
Centre for Astronomy, Faculty of Physics, Astronomy and Informatics, Nicolaus Copernicus University,  Grudziadzka 5, 87-100 Torun, Poland \and
Department of Physics, University of the Free State,  PO Box 339, Bloemfontein 9300, South Africa \and
Heisenberg Fellow (DFG), ITA Universit\"at Heidelberg, Germany  \and
GRAPPA, Institute of High-Energy Physics, University of Amsterdam,  Science Park 904, 1098 XH Amsterdam, The Netherlands \and
Department of Physics, Rikkyo University, 3-34-1 Nishi-Ikebukuro, Toshima-ku, Tokyo 171-8501, Japan
% ...
}

\offprints{H.E.S.S.~collaboration,
\protect\\\email{\href{mailto:contact.hess@hess-experiment.eu}{contact.hess@hess-experiment.eu}};
\protect\\\protect\footnotemark[1] Corresponding authors
}

\date{\today}

\clearpage{}

\abstract{
We present the results of the most comprehensive survey of the Galactic plane in
very high-energy (VHE) \gammarays, including a public release of Galactic sky
maps, a catalog of VHE sources, and the discovery of \hgpsSourceCountNew new
sources of VHE \gammarays. The High Energy Spectroscopic System (\hess)\
Galactic plane survey (HGPS) was a decade-long observation program carried out
by the \hessone\ array of Cherenkov telescopes in Namibia from 2004 to 2013. The
observations amount to nearly \hgpsObsTimeApprox\ of quality-selected data,
covering the Galactic plane at longitudes from $\ell=250\degr$ to $65\degr$ and
latitudes \hgpsGlat. In addition to the unprecedented spatial coverage, the HGPS
also features a relatively high angular resolution (\hgpsMeanPSF $\approx 5$
arcmin mean point spread function 68\% containment radius), sensitivity
(\hgpsSensitivity\ Crab flux for point-like sources), and energy range (0.2 to
100~TeV). We constructed a catalog of VHE \gammaray\ sources from the HGPS data
set with a systematic procedure for both source detection and characterization
of morphology and spectrum. We present this likelihood-based method in detail,
including the introduction of a model component to account for unresolved,
large-scale emission along the Galactic plane.  In total, the resulting HGPS
catalog contains \hgpsSourceCountTotal VHE sources, of which
\hgpsSourceCountCutout are not reanalyzed here, for example, due to their
complex morphology, namely shell-like sources and the Galactic center region.
Where possible, we provide a firm identification of the VHE source or plausible
associations with sources in other astronomical catalogs.  We also studied the
characteristics of the VHE sources with source parameter distributions. The
\hgpsSourceCountNew new sources were previously unknown or unpublished, and we
individually discuss their identifications or possible associations.  We firmly
identified \hgpsSourceCountID  sources as pulsar wind nebulae (PWNe), supernova
remnants (SNRs), composite SNRs, or gamma-ray binaries. Among the 47 sources not
yet identified, most of them (36) have possible associations with cataloged
objects, notably PWNe and energetic pulsars that could power VHE PWNe.
}
\clearpage{}


\maketitle

\makeatletter
\renewcommand*{\fnsymbol}[1]{\ifcase#1\@arabic{#1}\fi}
\makeatother

\newpage
\twocolumn

\onehalfspacing

\section{Introduction}
\label{sec:introduction}

In this paper, we present the results from the High Energy Spectroscopic System
(\hess)\ Galactic plane survey (HGPS), the deepest and most comprehensive survey
of the inner Milky Way Galaxy undertaken so far in very high-energy (VHE; $0.1
\la E \la 100$~TeV) \gammarays. Results include numerous sky images (maps) and a
new source catalog that is the successor of two previous HGPS releases. The
first release \citep{2005Sci...307.1938A} was based on $\sim$140~h of
observations with the imaging atmospheric Cherenkov telescope (IACT) array
\hess\ and contained eight previously unknown sources of VHE \gammarays. In the
second release \citep{ref:gps2006}, we used 230~h of data, covering
$\ell=330\degr$ to $30\degr$ in Galactic longitude and $|b|\leq{}3\degr$ in
latitude. In total, we detected 22 sources of \gammarays\ in that data set.
Since then, the HGPS data set enlarged by more than one order of magnitude in
observation time, now comprising roughly \hgpsObsTimeApprox\ of high-quality
data recorded in the years 2004 -- 2013. The spatial coverage is also
significantly larger, now encompassing the region from $\ell=250\degr$ to
$65\degr$ in longitude. \hess\ provided periodic updates on this progress by
publishing new unidentified sources~\citep{ref_gps_unids2008} and through
conference proceedings \citep{2008ICRC....2..579H, Chaves08_GPS, ref:icrc09,
ref:icrc11, Deil12_GPS, Carrigan13a_GPS, Carrigan13b_GPS}.

\begin{figure*}
\includegraphics[width=\textwidth]{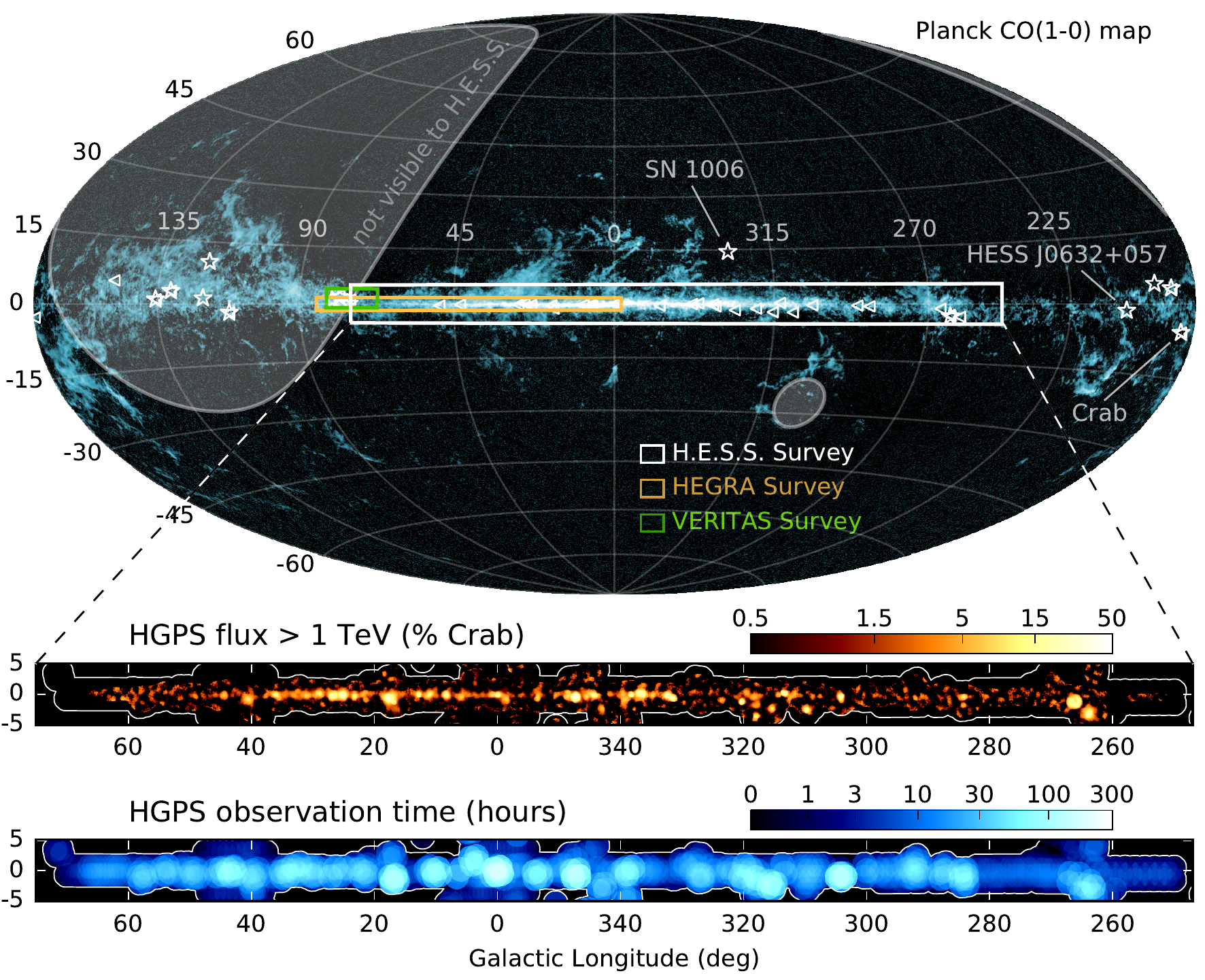}
\caption[HGPS region, flux, exposure illustration in all-sky context]{
Illustration of HGPS region superimposed an all-sky image of
\emph{Planck} CO(1-0) data \citep{Planck15} in Galactic coordinates and
Hammer-Aitoff projection. For comparison, we overlay the HEGRA Galactic plane
survey \citep{ref:hegrasurvey} and VERITAS Cygnus survey \citep{Weinstein:2009}
footprints. Triangles denote the \fermi\ 2FHL \gammaray\ sources \citep{2FHL}
identified as Galactic, and stars indicate the 15 Galactic VHE \gammaray\ sources
outside the HGPS region. \hess\ has detected three of these, which are labeled
SN~1006 \citep{2010A&A...516A..62A}, the Crab Nebula \citep{2006A&A...457..899A,
2014A&A...562L...4H}, and  HESS~J0632$+$057 \citep{2007A&A...469L...1A,
2014ApJ...780..168A}. The gray shaded regions denote the  part of the sky that
cannot be observed from the \hess\ site at reasonable zenith angles (less than
60$\degr$). The lower panels show the HGPS \gammaray\ flux above 1~TeV for
regions where the sensitivity is better than 10\%~ Crab (correlation radius
$R_{\mathrm{c}} = 0.4\degr$; see Sect.~\ref{sec:maps}) and observation time,
both also in Galactic coordinates. The white contours in the lower panels
delineate the boundaries of the survey region; the HGPS has little or no
exposure beyond Galactic latitudes of $|b|\leq{}3\degr$ at most locations along
the Galactic plane.
}
\label{fig:hgps_region_exposure_illustration}
\end{figure*}

Compared to the first HGPS releases over a decade ago, the deeper exposure over
a much larger sky area of the Galaxy, combined with improved \gammaray\
reconstruction, analysis, and modeling techniques, now results in a new catalog
containing \hgpsSourceCountTotal VHE \gammaray\ sources.
Figure~\ref{fig:hgps_region_exposure_illustration} illustrates the HGPS region
and compares this region to the structure of the Galaxy, represented by an
all-sky \emph{Planck} CO(1-0) map, and the smaller regions of previous surveys
performed by the IACT arrays HEGRA \citep[High-Energy-Gamma-Ray
Astronomy,][]{ref:hegrasurvey} and VERITAS \citep[Very Energetic Radiation
Imaging Telescope Array System,][]{Weinstein:2009}. Even though the HGPS covers
only a few percent of the entire sky, this region contains the vast majority of
the known Galactic \fermi\ 2FHL \gammaray\ sources \citep{2FHL}\footnote{In this
paper, we compare the HGPS with the \fermi\ 2FHL catalog, but not with 3FHL
\citep{2017arXiv170200664T} or the HAWC 2HWC catalog
\citep{2017ApJ...843...40A}, which were not published at the time this paper was
written and which already contain comparisons with Galactic \hess\ sources.}.
The figure also shows the measured integral VHE \gammaray\ flux and the HGPS
observation times. As can be seen from the map of observation times
(Fig.~\ref{fig:hgps_region_exposure_illustration}, lower panel), the HGPS data
set is not homogeneous. Nonetheless, the HGPS features on average a point-source
sensitivity better than 1.5\%~Crab\footnote{Throughout this paper, and as is
generally the case in VHE \gammaray\ astronomy, we use the Crab Nebula flux as a
standard candle reference: 1~Crab unit is defined here as
$\Phi\left(>1\mathrm{TeV}\right) = 2.26 \cdot 10^{-11}$ cm$^{-2}$ s$^{-1}$
\citep{ref:hesscrab}.} in the core survey region within 60\degr\ in longitude of
the Galactic center~(see Fig.~\ref{fig:hgps_sensitivity}, lower panel).

In this paper, we aim to present the entire data set of the HGPS in a way that is
accessible and useful for the whole astronomical community.  We have made the
maps of VHE \gammaray\ significance, flux, upper limits, and sensitivity
available online\footnote{\hgpsurl} for the first time in \emph{FITS} format
\citep{Pence:2010}. We developed a semi-automatic analysis pipeline to construct
a catalog by detecting and modeling discrete sources of VHE \gammaray\ emission
present in these survey maps. We applied a standardized methodology to the
characterization of the \gammaray\ sources to measure their
morphological and spectral properties. The goal was to perform a robust analysis
of sources in the survey region with as little manual intervention as possible.
With such a generic approach, the catalog pipeline is not optimal for the few
very bright and extended sources with complex (non-Gaussian) morphology. For
these sources, dedicated analyses are more appropriate, and in all cases, they
have already been performed and published elsewhere. We therefore exclude these
sources, which are listed in Table~\ref{tab:hgps_external_sources} below, from the
pipeline analysis but include the results from the dedicated analysis in the
HGPS catalog for completeness.

We have structured the present paper as follows: we describe the \hess\
telescope array, the data set, and the analysis techniques in
Sect.~\ref{sec:dataset}. We provide the maps of the VHE \gammaray\ sky in
various representations and details of their production in Sect.~\ref{sec:maps}.
Section~\ref{sec:cc} explains how the HGPS catalog of \gammaray\ sources was
constructed, then Sect.~\ref{sec:results} presents and discusses the results,
including source associations and identifications with other astronomical
objects. Section~\ref{sec:conclusions_outlook} concludes the main paper with a
summary of the HGPS and its results. In Sect.~\ref{sec:online}, we describe the
supplementary online material (maps and catalog in electronic form), including
caveats concerning measurements derived from the maps and catalog.

\section{Data set}
\label{sec:dataset}

\subsection{The High Energy Stereoscopic System (\hess)}
\label{sec:dataset:hess}

\hess\ is an array of five IACTs located at an altitude of 1800~m above sea
level in the Khomas highland of Namibia. It detects Cherenkov light emitted by
charged particles in an electromagnetic extensive air shower (EAS) initiated
when a primary photon (\gammaray) of sufficient energy enters Earth's
atmosphere. This array consists of four smaller telescopes, built and operated in the
first phase of the experiment (\hess\ \textit{Phase I}) and a fifth much larger
telescope, which was added to the center of the array in 2012 to launch the
second phase (\hess\ \textit{Phase II}) of the experiment.

\hess\ accumulated the data presented here exclusively with the \hess\ array
during its first phase. These four \hess~\textit{Phase I} telescopes have
tessellated mirrors with a total area of 107~m$^2$ and cameras consisting of 960
photomultipliers. The energy threshold of the four-telescope array is roughly
200~GeV at zenith and increases with increasing zenith angle. We can reconstruct
the arrival direction and energy of the primary photon with accuracies of
$\sim$\hgpsMeanPSF\ and $\sim$15\%, respectively.  Because of its comparatively
large field of view (FoV), 5\degr\ in diameter, the \hess~\textit{Phase I} array
is well suited for survey operations. The relative acceptance for \gammarays\ is
roughly uniform for the innermost 2\degr\ of the FoV and gradually drops toward
the edges to 40\% of the peak value at 4\degr\ diameter \citep{ref:hesscrab}.

\subsection{Observations, quality selection, and survey region}
\label{sec:dataset:selection}

The HGPS data set covers the period from January 2004 to January 2013. \hess\
acquired this data set by pointing the IACT array to a given position in the sky
for a nominal duration of \runDuration\ (referred to as an observation run
hereafter). We considered all runs with zenith angles up to 65\degr\ and
observation positions centered in the Galactic coordinate range $\ell =
244.5\degr$ to $77.5\degr$ and $|b|<7.0\degr$. To reduce systematic effects
arising from imperfect instrument or atmospheric conditions, we carefully
selected good-quality runs as close as possible to the nominal description of
the instrument used in the Monte Carlo (MC) simulations
\citep[see][]{ref:hesscrab}. For example, the IACT cameras suffer from
occasional hardware problems affecting individual or groups of camera pixels, so
we did not use observation runs with significant pixel problems. In addition, we
only used those runs with at least three operational telescopes.

Furthermore, despite the very good weather conditions at the \hess\ site, both
nightly and seasonal variations of the atmospheric transparency occur and
require monitoring. Layers of dust or haze in the atmosphere effectively act as
a filter of the Cherenkov light created in an EAS, thereby raising the energy
threshold for triggering the IACTs. Since we calculated the instrument response
tables describing the performance of the instrument (e.g., the effective areas)
with MC simulations, deviations from the atmospheric conditions assumed in the
simulations lead to systematic uncertainties in the determination of energy
thresholds, reconstructed energies, and \gammaray\ fluxes. To account for this,
we applied a further quality cut using only observations where the Cherenkov
transparency coefficient~$T$~\citep{2014APh....54...25H}, which characterizes
the atmospheric conditions, falls within the range $0.8<T<1.2$ (for clear skies,
$T=1$).

After applying the aforementioned data quality selection cuts, 6239~observation
runs remain, $\sim$77\% of which are runs with four telescopes operational. The
total observation time is 2864~h, corresponding to a total livetime of 2673~h
(6.7\% average dead time). The third panel of
Fig.~\ref{fig:hgps_region_exposure_illustration} is a map of the observation
time over the survey region, clearly showing a non-uniform exposure. This is a
result of the HGPS observation strategy, summarized as follows:

\begin{itemize}

\item Dedicated survey observations, taken with a typical spacing between
pointings of $0.7\degr$ in longitude and in different latitude bands located
between $b=-1.8\degr$ and $b=1\degr$. In addition, for the longitude bands
$\ell = 355\degr$ to $5\degr$ and $\ell = 38\degr$ to $48\degr$, we extended
the survey observations in latitude, adding observation pointings from
$b=-3.5\degr$ to $b=3.5\degr$ to explore the possibility of high-latitude
emission.

\item Deeper follow-up observations of source candidates (``hot spots'') seen in
previous survey observations.

\item Exploratory and follow-up observations of astrophysical objects located
inside the survey region that were promising candidates for emitting VHE
\gammarays.

\item Observations to extend the HGPS spatial coverage and fill-up
observations to achieve a more uniform sensitivity across the Galactic plane.

\end{itemize}

Combining all of these observations, we achieved a more uniform, minimum
2\%~Crab flux sensitivity in the region between $\ell=283\degr$ to $58\degr$ and
$b=-0.3\degr\,\pm\,0.7\degr$ (see the sensitivity map in
Fig.~\ref{fig:hgps_sensitivity}).

\subsection{Event reconstruction and selection}
\label{sec:dataset:events}

We first converted the camera pixel data to amplitudes measured in units of
photoelectrons (p.e.), identifying the non-operational pixels for a given
observation following the procedures described by \citet{ref:hesscalib}.  We
then applied standard \hess\ techniques for the analysis of the camera images:
image cleaning, Hillas moment analysis, and the stereo reconstruction of the
direction of the primary photon, described by \citet{ref:hesscrab}. To suppress
the hadronic background and select photon candidate events, we used a
multivariate machine learning technique using boosted decision trees based on
EAS and image shape parameters \citep{ref:tmva}. For the generation of the
survey maps (Sect.~\ref{sec:maps}), we applied the hard cuts configuration
whereas for the extraction of source spectra (Sect.~\ref{sec:results}) we used
the standard cuts. The most important distinguishing cut is a minimum of
160~p.e. for hard cuts and 60~p.e. for standard cuts, but there are other
differences. See \citet{ref:tmva} for further information; specifically, we used
the $\zeta$ analysis cuts listed in Table~2(a) for the HGPS.

We cross-checked the results presented in this paper with an alternative
calibration, reconstruction, and gamma-hadron separation method based on a
semi-analytical description of the EAS development \citep{2009APh....32..231D}
with hard cuts of 120~p.e. for maps and standard cuts of 60~p.e. for spectra.

For the energy reconstruction of the primary photons, we compared the image
amplitudes in the cameras to the mean amplitudes found in MC simulations of the
array \citep{ref:simulations}. Those simulations, which were analyzed with the
same chain as the real data for the sake of consistency, include the detailed
optical and electronic response of the instrument. The range of optical
efficiencies encountered in the HGPS data set is large; efficiencies start at
100\% of the nominal value and drop to almost 50\% for some telescopes prior to
the mirror refurbishments conducted in 2009--2011. Therefore, we produced
several sets of MC simulations, each with optical efficiencies of the four
telescopes corresponding to their states at suitably chosen times: at the start
of \hess\ operations; at the point when efficiencies had dropped to $\sim$70\%,
before the first mirror refurbishment campaign; and after the mirror
refurbishment of each telescope. We then chose the set of simulations most
closely matching the state of the system at a given time.  Finally, we corrected
the remaining difference between simulated and actual optical efficiencies using
a calibration technique based on the intensity of ring-shaped images from
individual muons producing Cherenkov radiation above a telescope
\citep{ref:muonsbolz, ref:muonsleroy}.

\section{HGPS sky maps}
\label{sec:maps}

In this section, we describe the methods used to produce the HGPS sky maps. We
used the sky maps as the basis for subsequent construction of the HGPS source
catalog; this catalog is also a data product that we release to the community
along with this work.

We first computed sky maps for each individual observation run. We then summed
these maps over all observations. We chose to use a Cartesian projection in
Galactic coordinates, covering the region from $\ell=70\degr$ to $250\degr$ and
$b=\pm5\degr$, and we set the pixel size to $0.02\degr$~/~pixel.

In Sect.~\ref{sec:events_map}, we describe the production of the map containing
the detected events (events map). In Sect.~\ref{sec:background_estimation}, we
describe the map of expected background events (acceptance map,
Sect.~\ref{sec:acceptance_map}), the estimation of a refined background map by
introducing exclusion regions (Sect.~\ref{sec:exclusion_regions}), and the usage
of the adaptive ring background method (Sect.~\ref{sec:adaptiveringmethod}). We
then continue in Sect.~\ref{sec:significancemaps} by describing the computation
of the significance map, and, in Sect.~\ref{sec:highlevelmaps}, the exposure map
(Sect.~\ref{sec:exposure_map}), which is used to derive quantities such as flux
(Sect.~\ref{sec:fluxmaps}), flux error and upper limits
(Sect.~\ref{sec:errfluxmap}), and sensitivities (Sect.~\ref{sec:sensmaps}).

\subsection{Events map}
\label{sec:events_map}

The events map consists of the reconstructed positions of the primary \gammaray\
photons from all events in the sky. To avoid systematic effects near the edge of
the FoV in each observation run, we only include events for which the direction
of the primary photon is reconstructed within $2\degr$ of the center of the FoV.
This choice results in an effective analysis FoV of $4\degr$ diameter.

At the lowest energies, the energy reconstruction is biased by EASs with upward
fluctuations in the amount of detected Cherenkov light; downward fluctuations do
not trigger the cameras. In order to derive reliable flux maps (see
Sect.~\ref{sec:fluxmaps}), we only kept events with an energy reconstructed
above a defined safe energy threshold. We chose the level of this safe threshold
such that, for each run, the energy bias as determined by MC simulations is
below 10\% across the entire FoV. This conservative approach (together with the
use of hard analysis cuts defined in Sect.~\ref{sec:dataset:events}) leads to
energy threshold values ranging from $\sim$400~GeV, where the array observed
close to zenith, up to 2~TeV at 65$\degr$ from zenith.
Figure~\ref{fig:hgps_energy_threshold_profiles} plots the variation of the safe
energy threshold with Galactic longitude, showing the energy threshold for each
observation together with the minimum value for each longitude. The variations
observed are mainly due to the zenith angle dependency, and regions of different
Galactic longitude generally are observable at different zenith angles.

\subsection{Background estimation}
\label{sec:background_estimation}

Events passing the event reconstruction and selection procedure are considered
\gammaray\ candidate events. Since these events are still dominantly from EASs
induced by \gammaray-like cosmic rays and electrons or positrons, we estimated the
amount of remaining background events on a statistical basis using a ring
model~\citep{ref:bgmodeling} as detailed further below. For each test position,
we counted the photon candidates found in a suitable ring-shaped region around
that position in the same FoV. This yields an estimate of the background level
after proper normalization and after excluding regions with actual \gammaray\
emission from the background estimate.

\subsubsection{Acceptance map}
\label{sec:acceptance_map}

The acceptance map represents the number of expected events from cosmic-ray
backgrounds estimated from runs of sky regions at similar zenith angles but
without VHE \gammaray\ sources. As for the events map (see
Sect.~\ref{sec:events_map}), we computed the acceptance map for energies above
the safe energy threshold. To account for the differences in optical efficiency
and observation time between these runs and those under analysis, we normalized
the acceptance map such that, outside the exclusion regions (see
Sect.~\ref{sec:exclusion_regions}), the number of expected counts matches the
number of measured counts. The acceptance maps are used to derive the
normalization coefficient between the region of interest and the background
region (see Sect.~\ref{sec:significancemaps}).

\begin{figure*}
\includegraphics[width=\textwidth]{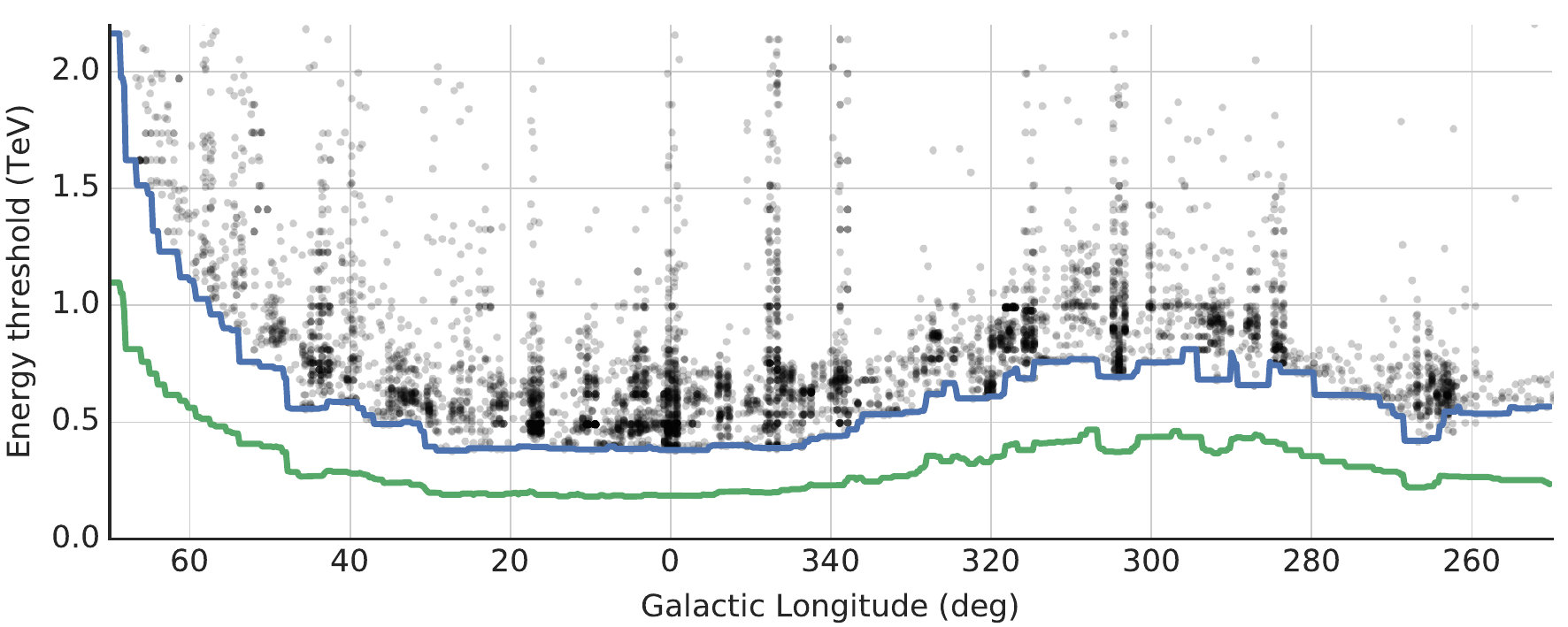}
\caption[HGPS survey energy threshold]{
HGPS minimum safe energy threshold as a function of Galactic longitude for a
latitude of $b=0\degr$. The blue curve shows the minimum threshold for hard cuts
(used for maps), and the green curve indicates standard cuts (used for spectra).
The black dots represent the safe threshold for each observation run obtained
for the hard cuts configuration. The few black dots below the blue line
correspond to runs at Galactic latitude $|b| > 2\degr$.
}
\label{fig:hgps_energy_threshold_profiles}
\end{figure*}

\subsubsection{Exclusion regions}
\label{sec:exclusion_regions}

The background estimation method described above only works if regions with VHE
\gammaray\ emission are excluded from the background estimation region. We
defined exclusion regions automatically using an iterative algorithm to avoid
potential observer bias and to treat the entire data set in a uniform way. The
procedure starts with the significance maps (see
Sect.~\ref{sec:significancemaps}) produced for the two standard correlation
radii $R_{\mathrm{c}} = 0.1\degr$ and $0.2\degr$. These radii define the
circular region over which a quantity (e.g., \gammaray\ excess) is integrated.
The procedure identifies regions above $5\sigma$ and expands them by excluding
an additional $0.3\degr$ beyond the $5\sigma$ contour. This procedure is
conservative; it minimizes the amount of surrounding signal that could
potentially contaminate the background estimation. A first estimation of the
exclusion regions is then included in the significance map production and a  new
set of exclusion regions is derived. We iterated this procedure until stable
regions are obtained, which typically occurs after three iterations. The
resulting regions are shown in Fig.~\ref{fig:catalog:rois} below.

\subsubsection{Adaptive ring method}
\label{sec:adaptiveringmethod}

\begin{figure}
\resizebox{\hsize}{!}{\includegraphics{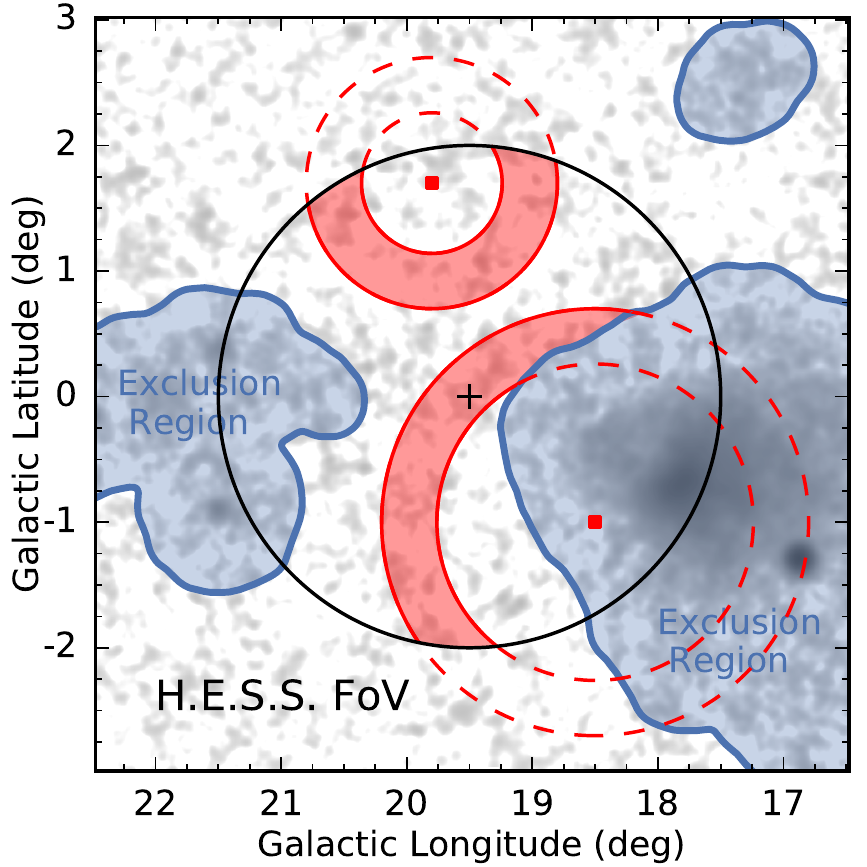}}
\caption[Illustration of adaptive ring background estimation for maps]{
Illustration of the adaptive ring method for background estimation for a single
observation (see Sect.~\ref{sec:adaptiveringmethod}). The HGPS significance
image is shown in inverse grayscale and exclusion regions as blue contours. The
analysis FoV for one observation is shown as a black circle with 2\degr\ radius
and a black cross at the observation pointing position. The red rings illustrate
the regions in which the background is estimated for two positions in the FoV
(illustrated as red squares). Only regions in the ring inside the FoV and
outside exclusion regions are used for background estimation. For the position
in the lower right, the ring was adaptively enlarged to ensure an adequate
background estimate (see text).
}
\label{fig:hgps_map_background_estimation}
\end{figure}

In the HGPS, often exclusion regions cover a significant fraction of the FoV;
therefore, we could not use the standard ring background method
\citep{ref:bgmodeling}. For example, using a typical outer ring radius of
$\sim$0.8\degr\ would lead to numerous holes in the sky maps at positions where
the entire ring would be contained inside an exclusion region (i.e., where no
background estimation was possible). A much larger outer radius (e.g.,
$\sim$1.5\degr) would be necessary to prevent these holes but would lead to
unnecessarily large uncertainties in the background estimation in regions
without, or with small, exclusion regions where smaller ring radii are feasible.

To address the limitations of the standard method, we do not use a static ring
geometry but rather adaptively change the inner and outer ring radii, as
illustrated in Fig.~\ref{fig:hgps_map_background_estimation}, depending on the
exclusion regions present in a given FoV. For a given test position within a
FoV, we begin with a minimum inner ring radius of $0.7\degr$ and constant ring
thickness $0.44\degr$ and enlarge the inner radius if a large portion of the
ring area overlaps with exclusion regions. We do this until the acceptance
integrated in the ring (but outside exclusion regions) is more than four times the
acceptance integrated at the test position. A maximum outer radius of $1.7\degr$
avoids large uncertainties in the acceptance toward the edge of the FoV.

\subsection{Significance maps}
\label{sec:significancemaps}

We produced significance maps to determine the exclusion regions (see
Sect.~\ref{sec:exclusion_regions}).  For each grid position $(\ell,b)$ in a
significance map, we counted the number of photon candidates $N_\mathrm{ON}$ in
the circular ON region, defined a priori by the correlation radius
$R_{\mathrm{c}}$.  We determined the background level by counting the number of
photon candidates $N_\mathrm{OFF}$ in the ring centered at $(\ell,b)$.  The
background normalization factor is $\alpha \equiv \xi_\mathrm{ON} /
\xi_\mathrm{OFF}$, where $\xi_\mathrm{ON}$ is the integral of the acceptance map
within $R_{\mathrm{c}}$ and $\xi_\mathrm{OFF}$ is the integral of the acceptance
map within the ring. The number of excess events $N_{\gamma}$ within
$R_{\mathrm{c}}$ is then
\begin{equation}
\label{eq:excess}
N_{\gamma}=N_\mathrm{ON}-\alpha{}N_\mathrm{OFF}.
\end{equation}
We computed the significance of this \gammaray\ excess according to
Eq.~17 of \cite{ref:lima} without correcting further for trials.

\subsection{High-level maps}
\label{sec:highlevelmaps}

We can derive additional high-level maps based on the measurement of
$N_{\gamma}$ within a given $R_{\mathrm{c}}$ and the instrument response
functions. In this work, we computed flux, flux error, sensitivity,  and upper
limit maps, starting from the formula
\begin{equation}
\label{eq:flux}
F = \frac{N_{\gamma}}{N_{\mathrm{exp}}}
\,\int_{E_1}^{E_2}\phi_\mathrm{ref}(E)\,\mathrm{d}E,
\end{equation}
where $F$ is the integral flux computed between the energies $E_1$ and $E_2$,
$N_{\gamma}$ is the measured excess, and $N_{\mathrm{exp}}$ is the total
predicted number of excess events, also called exposure (see
Sect.~\ref{sec:exposure_map}).

\subsubsection{Exposure maps}
\label{sec:exposure_map}

The exposure $N_{\mathrm{exp}}$ in Eq.~\ref{eq:flux} is given by
\begin{equation}
\label{eq:expcounts}
N_{\mathrm{exp}} \equiv \mathcal{E} = \sum_\mathrm{R \in runs} T_\mathrm{R}\,\int_{E_\mathrm{min}}^\infty
\phi_\mathrm{ref}(E_\mathrm{r})\,A_\mathrm{eff}(E_\mathrm{r},q_\mathrm{R})\,\mathrm{d}E_\mathrm{r}.
\end{equation}
Here, $E_\mathrm{r}$ is the reconstructed energy, $T_\mathrm{R}$ is the
observation livetime, $q_\mathrm{R}$ symbolizes the observation parameters for a
specific run (zenith, off-axis, and azimuth angle; pattern of telescopes
participating in the run; and optical efficiencies); $A_\mathrm{eff}$ is the
effective area obtained from MC simulations, which is assumed constant during a
\runDuration\ run; and $E_\mathrm{min}$ is the safe threshold energy appropriate
for the observation (as described in Sect.~\ref{sec:events_map}). We computed
the quantity $N_{\mathrm{exp}}$ for each position in the sky to create the
expected $\gamma$-ray count map, also referred to as the exposure map
$\mathcal{E}$ in the following. The function $\phi_{\mathrm{ref}}(E)$ is the
reference differential photon number $\gamma$-ray source flux, assumed to be
following a power law (PL) with a predefined spectral index, i.e.,
\begin{equation}
\label{eq:refflux}
\phi_\mathrm{ref}(E)=\phi_0\,(E/E_0)^{-\Gamma}.
\end{equation}

\subsubsection{Flux maps}
\label{sec:fluxmaps}

In Eq.~\ref{eq:flux}, the flux value $F$ is completely determined by the scaling
factor $N_{\gamma}/N_{\mathrm{exp}}$ once the spectral shape is fixed. We chose
to use $E_1=1\,\mathrm{TeV}$ and $E_2=\infty$. We stress that $E_1$ is not the
threshold energy used in the analysis, but the energy above which the integral
flux is given. In Eq.~\ref{eq:refflux}, one can choose the flux normalization
$\phi_0$ arbitrarily, since it cancels out in the computation of the flux.  We
also chose the spectral index $\Gamma = \hgpsAssumedSpecIndex$ in the released
maps to be compatible with the average index of known Galactic VHE \gammaray\
sources. To test the impact of this latter assumption, we performed tests that
show that, on average, flux variations are less than $5\%$ if the assumed
spectral index is varied by $\pm0.2$ (our systematic uncertainty of the spectral
index).

The released flux maps contain values of integral flux above $1$~TeV, calculated
according to Eq.~\ref{eq:flux}, in units of cm$^{-2}$~s$^{-1}$.  This should be
interpreted as the flux of a potential source, assuming a spectrum
$\phi_\mathrm{ref}(E)$, that is centered on a given pixel position in the map
and fully enclosed within $R_{\mathrm{c}}$.

Figures~\ref{fig:hgps_region_exposure_illustration} and \ref{fig:fluxmap} show
two example flux maps computed with $R_{\mathrm{c}} = 0.4\degr$ and $0.1\degr$,
respectively. The maps contain nonzero values only in regions in which the
sensitivity is better than 2.5\% Crab to prevent very large (positive and
negative) values due to statistical fluctuations in low-exposure regions.

\subsubsection{Flux error and upper limit maps}
\label{sec:errfluxmap}

Statistical uncertainties on the flux were computed by replacing $N_{\gamma}$ in
Eq.~\ref{eq:flux} by $N_{\gamma}^\mathrm{\pm1\sigma}$, which are the upper and
lower boundaries of the measured excess for a 68\% confidence level. Those
errors were computed with a Poisson likelihood method described in
\citet{ref:rolke}, using the same $N_\mathrm{ON}$ and $N_\mathrm{OFF}$
integrated within the circle of radius $R_{\mathrm{c}}$ used when computing the
excess maps. The values reported in the flux-error maps are the average of the
upper and lower error bars.

Similarly, an upper-limit map can be calculated by replacing $N_{\gamma}$ in
Eq.~\ref{eq:flux} by $N_{\gamma}^\mathrm{UL}$, that is, the upper limit on the
excess found for a predefined confidence level of 95\%; we used the same profile
likelihood method as for the error bar.

\subsubsection{Sensitivity maps}
\label{sec:sensmaps}

\begin{figure*}
\includegraphics[width=\textwidth]{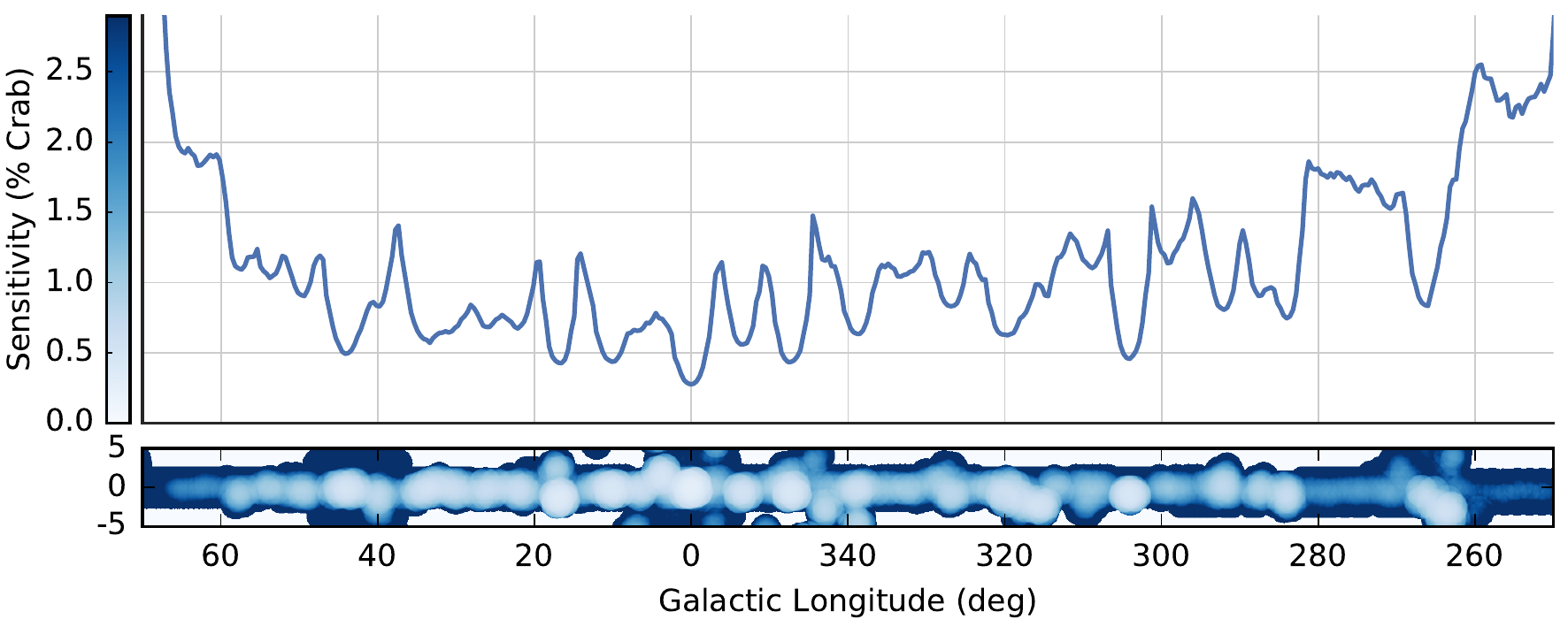}
\caption[Sensitivity image and longitude profile] {
HGPS point-source sensitivity map and profile along the Galactic plane at a
Galactic latitude $b = 0\degr$. The sensitivity is given in \%~Crab, for a
correlation radius $R_{\mathrm{c}} = 0.1\degr$, assuming a spectral index
$\Gamma = \hgpsAssumedSpecIndex$. This sensitivity is computed under
the isolated point source assumption and is thus better than the actual
sensitivity achieved for the HGPS source catalog (see
Sect.~\ref{sec:cc:discussion}).
}
\label{fig:hgps_sensitivity}
\end{figure*}

The sensitivity is defined as the minimal flux needed for a source with the
assumed spectrum and fully contained within the correlation circle
$R_{\mathrm{c}}$ to be detected above the background at $5\sigma$ statistical
significance. Alternatively this can be thought of as a measure of
$\hat{N_{\gamma}}$, the number of photons needed to reach such a significance
level above the background determined by $N_\mathrm{OFF}$ and $\alpha$. To
compute the sensitivity map, $N_{\gamma}$ in Eq.~\ref{eq:flux} is replaced by
$\hat{N_{\gamma}}$, which is determined by numerically solving Eq.~17 of
\citet{ref:lima} for $N_\mathrm{ON}$ (related to $\hat{N_{\gamma}}$ by
Eq.~\ref{eq:excess} above). We note that possible background systematics are not
taken into account in this computation.

The point-source sensitivity level reached by \hess\ at all points in the HGPS
data set is depicted in Fig.~\ref{fig:hgps_sensitivity}, where a projection of
the sensitivity map along Galactic longitude at a Galactic latitude of
$b=0\degr$ is also shown. It is typically at the level of 1\% to 2\% Crab. The
deepest observations were obtained around interesting objects for which
additional pointed observations were performed. Examples include the Galactic
center region (around $\ell=0\degr$, where the best sensitivity of $\sim 0.3$\%
Crab is reached), the Vela region ($\ell=266\degr$), the regions around
\object{HESS J1825$-$137} and \object{LS 5039} ($\ell=17\degr$), or around
\object{HESS J1303$-$631} and \object{PSR B1259$-$63} ($\ell=304\degr$).

Similarly, the sensitivity values along Galactic latitude for two values of
longitude are shown in Fig.~\ref{fig:hgps_sources_glat}. For most of the
surveyed region, the sensitivity decreases rapidly above $|b|>2\degr$ due to the
finite FoV of the \hess\ array and the observation pattern taken, except for a
few regions, such as at $\ell=0\degr$ where high latitude observations were
performed (see Sect.~\ref{sec:dataset}). The best sensitivity is obtained around
$b=-0.3\degr$, reflecting the \hess\ observation strategy; the latitude
distribution of the sources peaks in this region.

We note that the sensitivity shown in Fig.~\ref{fig:hgps_sensitivity} does not
correspond to the completeness of the HGPS source catalog. One major effect is
that the HGPS sensitivity is dependent on source size; it is less sensitive for
larger sources, as shown in Fig.~\ref{fig:hgps_sources_flux_extension} and
discussed at the end of Sect.~\ref{sec:results:distributions}. Other effects
that reduce the effective sensitivity or completeness limit of HGPS are
the detection threshold, which corresponds to $\sim 5.5\sigma$; the large-scale
emission model; and source confusion, as discussed in the following
Sect.~\ref{sec:cc}

\section{HGPS source catalog}
\label{sec:cc}

\subsection{Introduction and overview}
\label{sec:cc:introduction}

The HGPS source catalog construction procedure intends to improve upon previous
\hess\ survey publications both in sensitivity and homogeneity of the analysis
performed. The previous iteration, the second \hess\ survey paper of 2006
\citep{ref:gps2006}, used a 230 h data set with inhomogeneous exposure that was
limited to the innermost region of the Galaxy. This survey detected a total of
14 sources by locating peaks in significance maps on three different spatial
scales: $0.1\degr$, $0.22\degr$, and $0.4\degr$. It then modeled the sources by
fitting two-dimensional symmetric Gaussian morphological models to determine the
position, size and flux of each source, using a Poissonian maximum-likelihood
method.

\begin{figure*}
\includegraphics[width=\textwidth]{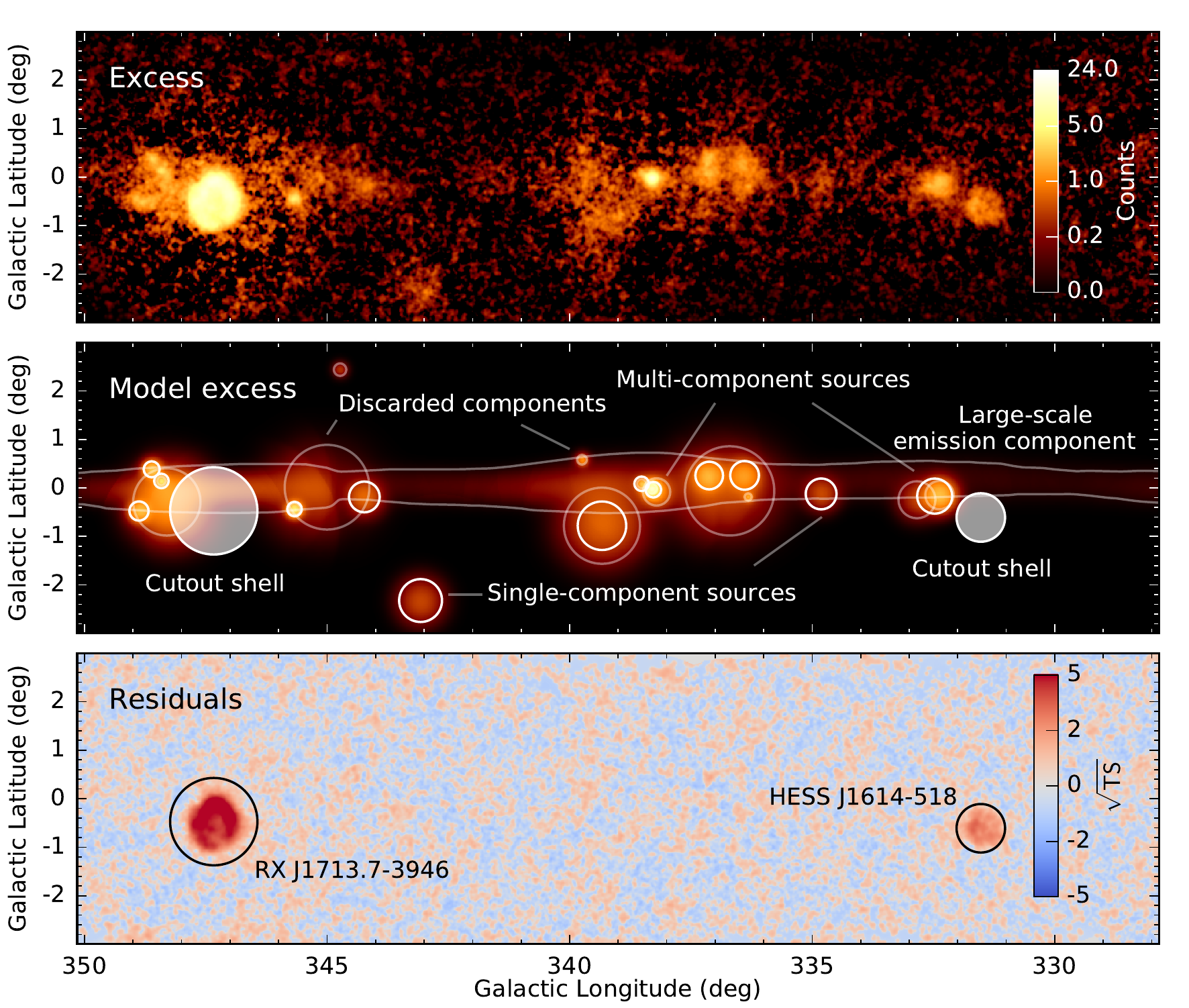}
\caption
[Source catalog model construction illustration]{
Illustration of the catalog model construction in the region of
350\degr~to~328\degr\ in Galactic longitude. The upper panel shows the
\gammaray\ excess counts smoothed by the PSF, the middle panel the PSF-convolved
and smoothed excess model, and the lower panel the significance map of the
residuals for a point-like source hypothesis (given in $\mathrm{sign}
(\mathrm{Flux}) \sqrt{\TSinFormula}$). The middle panel shows examples of the
steps taken in the excess map modeling part of the source catalog procedure (see
Sect.~\ref{sec:cc} for details). It starts by cutting out shell-type supernova
remnants (SNRs; RX~J1713.7$-$3946 and the SNR candidate HESS~J1614$-$518 in this
region) and by assuming a fixed large-scale emission component. Then a
multi-Gaussian model was fitted with the significant components shown in the
middle panel as thin transparent circles. Some of these were discarded and are
not part of the emission attributed to HGPS catalog sources. White circles show
examples of single-component as well as multicomponent sources. For a complete
overview of all analysis regions (ROIs) and excluded sources, see
Fig.~\ref{fig:catalog:rois}.
}
\label{fig:hgps_catalog_model}
\end{figure*}

Since 2006, \hess\ has increased its exposure tenfold and enlarged the survey
region more than twofold, while also improving the homogeneity of the exposure.
As illustrated in the upper panel of Fig.~\ref{fig:hgps_catalog_model}, the data
now show many regions of complex emission, for example, overlapping emission of
varying sizes and multiple sources with clearly non-Gaussian morphologies. Apart
from discrete emission, the Galactic plane also exhibits significant emission on
large spatial scales \citep{2014PhRvD..90l2007A}. For these reasons, we needed
to develop a more complex analysis procedure to construct a more realistic model
of the \gammaray\ emission in the entire survey region. Based on this model, we
compiled the HGPS source catalog.

We first introduce the maximum-likelihood method used for fitting the emission
properties (Sect.~\ref{sec:cc:ml}).  Next, we describe the \hess\ point spread
function (PSF; Sect.~\ref{sec:cc:maps:psf}) and the \TS\ maps
(Sect.~\ref{sec:maps:ts}), which are two important elements in the analysis and
catalog construction. The procedure is then as follows:

\begin{enumerate}

\item Cut out the Galactic center (GC) region and shell-type supernova remnants
from the data set because of their complex morphologies
(Sect.~\ref{sec:cc:cutout_sources}).

\item Model the large-scale emission in the Galactic plane globally
(Sect.~\ref{sec:cc:large-scale-emission}).

\item Split the HGPS region into manageable regions of interest (ROIs)
(Sect.~\ref{sec:cc:maps:roi}).

\item Model the emission in each ROI as a superposition of components with
Gaussian morphologies (Sect.~\ref{sec:cc:components}).

\item Merge Gaussian components into astrophysical VHE \gammaray\ sources
(Sect.~\ref{sec:cc:component_classification}).

\item Determine the total flux, position, and size of each \gammaray\ source
(Sect.~\ref{sec:cc:source_characterization}).

\item Measure the spectrum of each source (Sect.~\ref{sec:cc:spectra}).

\item Associate the HGPS sources with previously published \hess\ sources and
multiwavelength (MWL) catalogs of possible counterparts
(Sect.~\ref{sec:results:assoc_id}).

\end{enumerate}

\subsection{Poisson maximum-likelihood morphology fitting}
\label{sec:cc:ml}

To detect and characterize sources and to model the large-scale emission
in the Galactic plane, we used a spatially-binned likelihood analysis based on
the following generic model:
\begin{equation}
\label{eq:generic_model}
N_{\mathrm{Pred}} =
N_{\mathrm{Bkg}} + \mathrm{PSF} \ast \left( \mathcal{E} \cdot S \right)
,\end{equation}
where $N_{\mathrm{Pred}}$ represents the predicted number of counts,
$N_{\mathrm{Bkg}}$ the background model created with the adaptive ring method
(described in Sect.~\ref{sec:adaptiveringmethod}), $\mathcal{E}$ the exposure
map (see Eq.~\ref{eq:expcounts} in Sect.~\ref{sec:fluxmaps}), and $S$ a
two-dimensional parametric morphology model that we fit to the data.
Additionally, we took into account the angular resolution of \hess\ by
convolving the flux model with a model of the PSF of the instrument.

Assuming Poisson statistics per bin, the maximum-likelihood fit then minimizes
the \textit{Cash} statistic \citep{1979ApJ...228..939C},
\begin{equation}
\mathcal{C} = 2 \sum_{i} \left(M_i - D_i \log M_{i} \right),
\label{eq:cash_statistic}
\end{equation}
where the sum is taken over all bins $i$, and $M_i$ (model) represents the
expected number of counts according to Eq.~\ref{eq:generic_model} and $D_i$
(data) the actual measured counts per bin.

To determine the statistical significance of a best-fit source model compared to
the background-only model, we use a likelihood ratio test with test statistic
\TS. This is defined by the likelihood ratio or equivalently as the difference in
$\mathcal{C}$ between both hypotheses,
\begin{equation}
\TSinFormula = \mathcal{C}_0 - \mathcal{C}_S,
\label{eq:ts_definition}
\end{equation}
where $\mathcal{C}_0$ corresponds to the value of the \textit{Cash} statistic of
the background-only hypothesis and $\mathcal{C}_S$ the best-fit model that
includes the source emission.

For a large number of counts, according to Wilks' theorem \citep{Wilks38}, \TS\
is asymptotically distributed as $\chi^2_N$, where $N$ is the number of free
parameters defining the flux model. In this limit, the statistical significance
corresponds approximately to $\mathrm{sign}(\mathrm{Flux}) \cdot
\sqrt{|\TSinFormula|}$, where the sign of the best-fit flux is needed to allow
for negative significance values in regions where the number of counts is
smaller than the background estimate (e.g.,~due to a statistical downward
fluctuation).

We performed the modeling and fitting described above in
Eqs.~\ref{eq:generic_model}, \ref{eq:cash_statistic}, and \ref{eq:ts_definition}
in pixel coordinates using the HGPS maps in Cartesian projection. Spatial
distortion of flux models are negligible as a result of the projection from the
celestial sphere because the HGPS observations only cover a latitude range of
\hgpsGlat. We implemented the analysis in Python using Astropy version~1.3
\citep{2013AandA...558A..33A}, Sherpa version~4.8 \citep{Freeman:2001}, and
Gammapy version~0.6 \citep{Donath2015, 2017arXiv170901751D}.

\subsection{Point spread function}
\label{sec:cc:maps:psf}

For HGPS, the PSF was computed for a given sky position assuming a power-law
point source with a spectral index of \hgpsAssumedSpecIndex\ (average index of
known VHE \gammaray\ sources) and assuming rotational symmetry of the PSF. Since
the  \hess\ PSF varies with \gammaray\ energy and observing parameters such as
the number of participating telescopes, zenith angle, and offset angle in the
field of view, an effective PSF corresponding to the HGPS survey counts maps was
computed by applying the same cuts (especially safe energy threshold) and
exposure weighting the PSF of contributing runs (i.e., within the FoV of
2\degr). The per-run PSF was computed by interpolating PSFs with similar
observation parameters, using precomputed lookups from MC EAS simulations. All
computations were carried out using two-dimensional histograms with axes
$\theta^2$, where $\theta$ is the offset between the MC source position and the
reconstructed event position, and $\log(E_r)$, where $E_r$ is the reconstructed
event energy; at the very end, the integration over energy was performed,
resulting in a one-dimensional histogram with axis $\theta^2$, which was fitted
by a triple-exponential analytical function to obtain a smooth distribution,
\begin{equation}
\frac{\mathrm{d}P}{\mathrm{d}\theta^2}(\theta^2) =
\sum_{i=1}^3 A_i \exp\left(-\frac{\theta^2}{2\sigma_i^2}\right),
\end{equation}
where $P$ is the event probability, and $A_i$ and $\sigma_i$ are the weights and
widths of the corresponding components, respectively. This ad hoc model
corresponds to a triple-Gaussian, two-dimensional, PSF model when projected onto
a sky map.

For the HGPS catalog, the 68\% containment radius of the PSF model adopted is
typically $\theta\sim$\hgpsMeanPSF\ and varies by approximately $\pm 20\%$ at
the locations of the HGPS sources. For observations with large FoV offsets, the
68\% containment increases by almost a factor of two to $\theta\sim$0.15\degr,
which is mostly relevant for high Galactic latitude sources at the edge of the
HGPS survey region. The HGPS PSF has a 95\% containment radius of
$\theta\sim$0.2\degr and approximately varies by $\pm 20\%$ at the locations of
the HGPS sources. The PSF at large FoV offsets (corresponding to high-GLAT
regions in the survey map) is more tail heavy; there the 95\% to 68\%
containment radius ratio increases from $\sim$2.5 up to 4.
Section~\ref{sec:cc:extension_ul} discusses systematic uncertainties related to
the PSF model in connection with upper limits on source sizes.

\subsection{Test statistics maps}
\label{sec:maps:ts}

In addition to the standard \lima\ significance maps described in
Sect.~\ref{sec:significancemaps}, we also used \TS\ maps in the analysis. The
\TS\ denotes the likelihood ratio of the assumed source hypothesis versus the
null hypothesis (i.e.,\ background only) for every position (pixel) in the map.
We computed these maps assuming various spatial templates: a point-like source
morphology (i.e., PSF only), and PSF-convolved Gaussian morphologies with widths
$0.05\degr$, $0.10\degr$, and $0.20\degr$.  During the computation of each map,
at the center of each map pixel, we performed a single-parameter likelihood fit
of the amplitude of the template, according to Eq.~\ref{eq:generic_model}. We
then filled the map with the \TS\ value defined in Eq.~\ref{eq:ts_definition}.

We used the resulting \TS\ maps primarily to compute residual maps and residual
distributions. The main advantage over standard \lima\ significance maps is that
source morphology and PSF information can be taken into account. Additionally,
this paper uses \TS\ maps when presenting sky maps because they contain uniform
statistical noise everywhere in the map.  In contrast, flux or excess maps that
are smoothed with the same spatial templates still show increased noise in
regions of low exposure. We implemented the \TS\ map algorithm available in
Gammapy; see also \citet{Stewart:2009} for a more detailed description of \TS\
maps.

\subsection{Sources not reanalyzed}
\label{sec:cc:cutout_sources}

\hess\ observations have revealed many sources with complex morphology, e.g.,
\object{RX J0852.0$-$4622} (also known as \object{Vela Junior}), which has a
very pronounced shell-like structure~\citep{HESS:VelaJnr}, or the Galactic
center region, which has multiple point-sources embedded in a very elongated
ridge-like emission \citep{HESS:Arc}. Dedicated studies model such regions of
emission using complex parametric models, for example, model templates based on
molecular data, shell-like models, asymmetric Gaussian models, and combinations
thereof. It is challenging to systematically model the emission across the
entire Galactic plane using these more complex models, which tend to yield
unstable or non-converging fit results because of the large number of free and
often poorly constrained parameters. This can be especially problematic in ROIs
with multiple, complex overlapping sources.

Given the difficulties with modeling complex source morphologies, we decided to
restrict the HGPS analyses to a symmetrical Gaussian model assumption and
exclude all firmly identified shell-like sources and the very complex GC region
from reanalysis. A complete list of the ten excluded (or cut-out) sources in the
HGPS region is given in Table~\ref{tab:hgps_external_sources}. The table also
contains four sources that were not significant in the current HGPS analysis but
were found to be significant in other dedicated, published analyses; these cases
are discussed in detail in Sect.~\ref{sec:results:previously:missing}. We refer
to these \hgpsSourceCountCutout sources in total listed in
Table~\ref{tab:hgps_external_sources} as ``EXTERN'' HGPS sources and have
included these sources in the HGPS source catalog because we wanted to give a
complete list of sources in the HGPS region. We also have these sources included
in the various distributions, histograms, and other plots exploring the global
properties of the HGPS sources in Sect.~\ref{sec:results:distributions}. The
morphological and spectral parameters for those sources were adapted from the
most recent \hess\ publication (listed in
Table~\ref{tab:hgps_external_sources}).\footnote{We note that the values in the
HGPS catalog for EXTERN sources do not fully reflect the results of the original
publication. Specifically, in some cases the information is incomplete (e.g.,
when certain measurements were not given in the paper) or not fully accurate
(e.g., when the published measurements do not fully agree with the definition of
measurements in this paper, or when parameter errors are different due to error
inaccuracies in the error propagation when converting to HGPS measures.)}

\begin{table*}
\caption
[Sources in the HGPS catalog with parameters taken from previous publications]{
Fourteen EXTERN sources in the HGPS catalog, i.e., VHE sources in the HGPS
region previously detected by \hess\ that were not reanalyzed in this paper. For
each source, we list the reason why it was not reanalyzed and give the reference
that was used to fill the parameters in the HGPS source catalog. See
Sect.~\ref{sec:cc:cutout_sources} and for sources not significant in the HGPS
analysis also Sect.~\ref{sec:results:previously:missing}.
}
\label{tab:hgps_external_sources}
\centering
\begin{tabular}{lllll}
\hline\hline
Source name      & Common name     & Reason for not reanalyzing & Reference \\
\hline
\object{HESS J0852$-$463} & Vela Junior       & Shell morphology        & \cite{HESS:VelaJnr} \\
\object{HESS J1442$-$624} & \object{RCW 86}   & Shell morphology        & \cite{2016arXiv160104461H} \\
\object{HESS J1534$-$571} & G323.7$-$1.0      & Shell morphology        & \cite{HESS:Shells} \\
\object{HESS J1614$-$518} & ---               & Shell morphology        & \cite{HESS:Shells} \\
\object{HESS J1713$-$397} & \object{RX J1713.7$-$3946} & Shell morphology        & \cite{HESS:RXJ1713} \\
\object{HESS J1731$-$347} & G353.6$-$0.7      & Shell morphology        & \cite{2011AandA...531A..81H} \\
\object{HESS J1912$+$101} & ---               & Shell morphology        & \cite{HESS:Shells} \\
\object{HESS J1745$-$290} & Galactic~center   & Galactic center region  & \cite{GCPevatron} \\
\object{HESS J1746$-$285} & Arc~source        & Galactic center region  & \cite{HESS:Arc} \\
\object{HESS J1747$-$281} & G0.9$+$0.1        & Galactic center region  & \cite{Aharonian:2005d} \\
\hline
\object{HESS J1718$-$374} & G349.7$+$0.2      & Not significant in HGPS & \cite{2015AandA...574A.100H} \\
\object{HESS J1741$-$302} & ---               & Not significant in HGPS & \cite{HESS:1741} \\
\object{HESS J1801$-$233} & \object{W 28}     & Not significant in HGPS & \cite{Aharonian:2008f}\\
\object{HESS J1911$+$090} & \object{W 49B}    & Not significant in HGPS & \cite{HESS:W49} \\

\hline
\end{tabular}
\end{table*}

\subsection{Large-scale emission model}
\label{sec:cc:large-scale-emission}

We previously demonstrated that there exists VHE \gammaray\ emission that is
large scale and diffuse along the Galactic plane \citep{2014PhRvD..90l2007A}.
In that paper, we constructed a mask to exclude the regions of the plane where
significant emission was detected. The latitude profile of excess \gammarays\
outside this mask clearly showed the presence of significant large-scale
\gammaray\ emission. We do not extend the analysis of this diffuse emission any
further here. Whether the emission originates from interactions of diffuse
cosmic rays in the interstellar medium or from faint, unresolved \gammaray\
sources (or a combination thereof) is not investigated. Instead, we take a
pragmatic approach and model the large-scale emission present in the HGPS
empirically as described in the following.

The presence of a large-scale component of \gammaray\ emission along the
Galactic plane complicates the extraction of the Gaussian \gammaray\ source
components. This large-scale emission can mimic the presence of spurious
degree-scale sources in some regions of the plane and it also tends to broaden
the Gaussian components that describe otherwise well-defined sources. It is
therefore necessary to model the large-scale \gammaray\ emission to measure the
flux and morphology of the HGPS sources more accurately.

\begin{figure*}
\includegraphics[width=\textwidth]{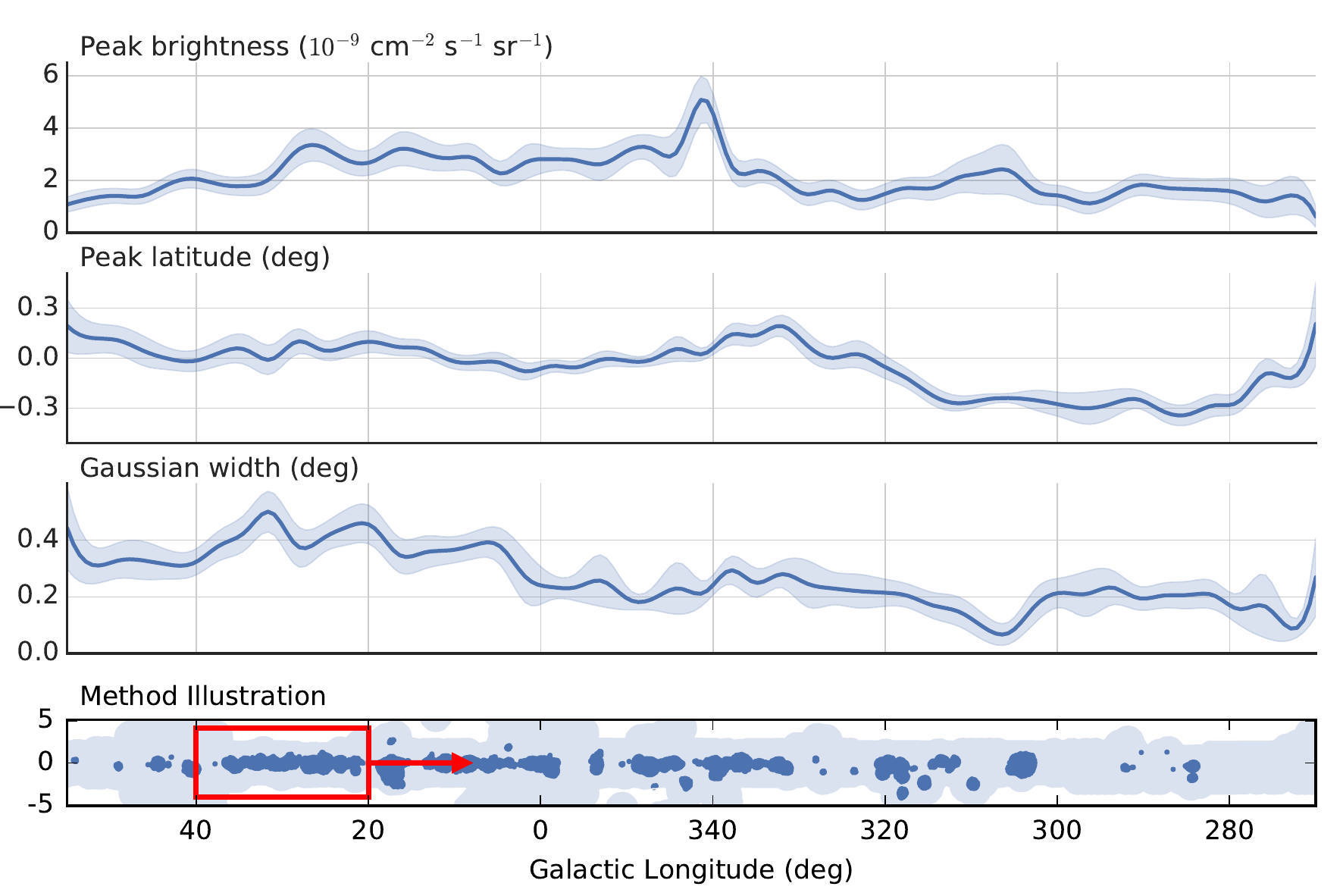}
\caption[Large-scale emission model]{
Distribution of the fit large-scale emission model parameters with Galactic
longitude. The first panel gives the peak brightness of the large-scale emission
model in units of $10^{-9}$~cm$^{-2}$~s$^{-1}$~sr$^{-1}$ ($\approx
1.3$\%~Crab~deg$^{-2}$). The second panel shows the peak position of the
Gaussian along the Galactic latitude axis in degrees and the third panel shows
the width ($\sigma$) of the Gaussian in degrees. The solid lines are the result
of fitting each set of parameters every $2.5\degr$ in longitude and
interpolating. The light blue bands show the 1$\sigma$ error region obtained
from the covariance matrix of the likelihood function. The lower panel
illustrates the $20\degr$ wide sliding-window method (red rectangle) that was
used to determine the large-scale emission model in areas (shown in light blue)
where the HGPS sensitivity is better than 2.5\%~Crab but outside exclusion
regions (shown in dark blue); this is explained in further detail in the main
text.
}
\label{fig:hgps_diffuse_model}
\end{figure*}

To do so, we built an empirical surface brightness model of the large-scale
emission (see Fig.~\ref{fig:hgps_diffuse_model}), where the latitude profile is
Gaussian and defined by three parameters: the peak position in latitude, the
width, and amplitude of the Gaussian. We estimated the parameters using a
maximum-likelihood fit in regions where no significant emission is measurable on
small scales, i.e.,\ outside the exclusion regions defined for the ring
background model, taking exposure into account. Regardless of the physical
origin of the large-scale emission, it is likely to be structured along the
plane and not constant.

To estimate the variable parameters of the model, we fit the Gaussian parameters
in rectangular regions of width $20\degr$ in longitude and height $6\degr$ in
latitude.  We excluded all pixels inside the standard exclusion regions used to
produce the background maps (see Sect.~\ref{sec:background_estimation}). The
Gaussian parameters were dependent on the size of both the exclusion regions and
rectangular regions.  We found that the typical variations were $\sim$25\%. To
obtain a smooth sampling of the variations, we followed a sliding-window
approach, distributing the centers of the rectangular regions every $2.5\degr$
in longitude and interpolating between these points.

The maximum-likelihood fit compares the description of the data between the
cosmic-ray (CR) background only and the CR background plus the model.  We used
the likelihood ratio test to estimate the significance of adding the large-scale
component in each 20-deg-wide window, finding it to be larger than $3\sigma$ (TS
difference of 9) over most of the HGPS region.
Figure~\ref{fig:hgps_diffuse_model} shows the resulting best-fit Gaussian
parameters together with the associated uncertainty intervals estimated from the
likelihood error function.  After this fit, we froze the parameters of the model
for use in the \gammaray\ source detection and morphology fitting procedure.

While the approach presented here provides an estimate of the large-scale
emission present in the HGPS maps, it does not comprise a measurement of the
total Galactic diffuse emission (see discussion in
Sect.~\ref{sec:results:large}).

\subsection{Regions of interest}
\label{sec:cc:maps:roi}

To search for sources, we divided the whole HGPS region into smaller overlapping
ROIs. This was necessary to limit both the number of simultaneously fit
parameters and the number of pixels involved in the fit.

We manually applied the following criteria to define the ROIs:

\begin{enumerate}

\item[(a)] All significant emission (above $5\sigma$) in the HGPS region should
be contained in at least one ROI.

\item[(b)] No significant emission should be present close to the edges of an
ROI.

\item[(c)] The width of each ROI should not exceed $\sim$10\degr\ in longitude
to limit the number of sources involved in the fit.

\item[(d)] ROIs should cover the full HGPS latitude range from $-5\degr$ to
$5\degr$.

\end{enumerate}

In cases in which criterion~(b) could not be fulfilled, we excluded the
corresponding emission from the ROI and assigned it to a different, overlapping
ROI. Figure~\ref{fig:catalog:rois} illustrates the boundaries of the 18 ROIs
defined with these criteria. Some of the ROIs show regions without any exposure;
these regions were masked out and ignored in the subsequent likelihood fit.

\subsection{Multi-Gaussian source emission model}
\label{sec:cc:components}

After excluding shell-type supernova remnants (SNRs) and the GC region from
reanalysis and adding a model for large-scale emission to the background, we
modeled all remaining emission as a superposition of Gaussian components. We
took the following model as a basis:
\begin{equation}
N_{\mathrm{Pred}} = N_{\mathrm{Bkg}} + \mathrm{PSF} \ast
\left( \mathcal{E} \cdot \sum_{i} S_{\mathrm{Gauss},i} \right) +
\mathcal{E} \cdot S_{\mathrm{LS}},
\label{eq:expected_signal}
\end{equation}
where $N_{\mathrm{Pred}}$ corresponds to the predicted number of counts,
$N_{\mathrm{Bkg}}$ to the number of counts from the background model,
$S_{\mathrm{LS}}$ the contribution of the large-scale emission model, $\sum_{i}
S_{\mathrm{Gauss},i}$ the sum of the Gaussian components, and $\mathcal{E}$ the
exposure as defined in Eq.~\ref{eq:expcounts}.

For a given set of model parameters, we integrated the surface brightness
distribution $S$ over each spatial bin, multiplied it by the exposure
$\mathcal{E}$, and convolved it with the PSF to obtain the predicted number of
counts per pixel. For every ROI, we took the PSF at the position of the
brightest emission and assumed it to be constant within the ROI.

For the Gaussian components, we chose the following parametrization:
\begin{equation}
S_{\mathrm{Gauss}}(r| \phi, \sigma) =
\phi \frac{1}{2\pi\sigma^2}\exp\left(-\frac{r^2}{2\sigma^2}\right),
\label{eq:gauss}
\end{equation}
where $S_{\mathrm{Gauss}}$ is the surface brightness, $\phi$ the total
spatially integrated flux, and $\sigma$ the width of the Gaussian component. The
offset $r~=~\sqrt{(\ell - \ell_{0})^2 + (b - b_{0})^2}$ is defined with respect
to the position $(\ell_{0}, b_{0})$ of the component measured in Galactic
coordinates.

We conducted the manual fitting process following a step-by-step procedure.
Starting with one Gaussian component per ROI, we added Gaussian components
successively and refit all of the parameters simultaneously until no significant
residuals were left. In each step, we varied the starting parameters of the fit
to avoid convergence toward a local minimum. The significance of the tested
component was estimated from
\begin{equation}
\TSinFormula =
\mathcal{C}(\mbox{with component}) -
\mathcal{C}(\mbox{best solution without component})
\label{eq:detection:alternative_threshold}
.\end{equation}
We considered the component to be statistically significant and kept it in the
model when the TS value exceeded a threshold of $\TSinFormula=30$. The
probability of having one false detection in the HGPS survey from statistical
background fluctuations is small ($p=0.03$). This number was determined by
simulating 100 HGPS survey counts maps as Poisson-fluctuated background model
maps, followed by a multi-Gaussian peak finding method, resulting in three peaks
with $\TSinFormula\ge30$. However, we note that this assessment of expected
false detections lies on the assumption that the hadronic background as well as
the large-scale and source gamma-ray emission model are perfect. In HGPS, as in any
other Galactic plane survey with complex emission features, this is not the
case. Several components with $\TSinFormula\ge30$ are not confirmed by the
cross-check analysis (see Sect.~\ref{sec:cc:component_classification}).

The definition of \TS\ above differs slightly from the definition given in
Eq.~\ref{eq:ts_definition}. For a single, isolated component, both values are
identical. However, if a second, overlapping component exists, some of the
emission of the first source is modeled by the second source, reducing the
significance of the first. We therefore estimated the significance of a
component from the TS difference in the total model of the ROI and not from the
TS difference compared to the background-only model.

\begin{figure}
\resizebox{\hsize}{!}{\includegraphics{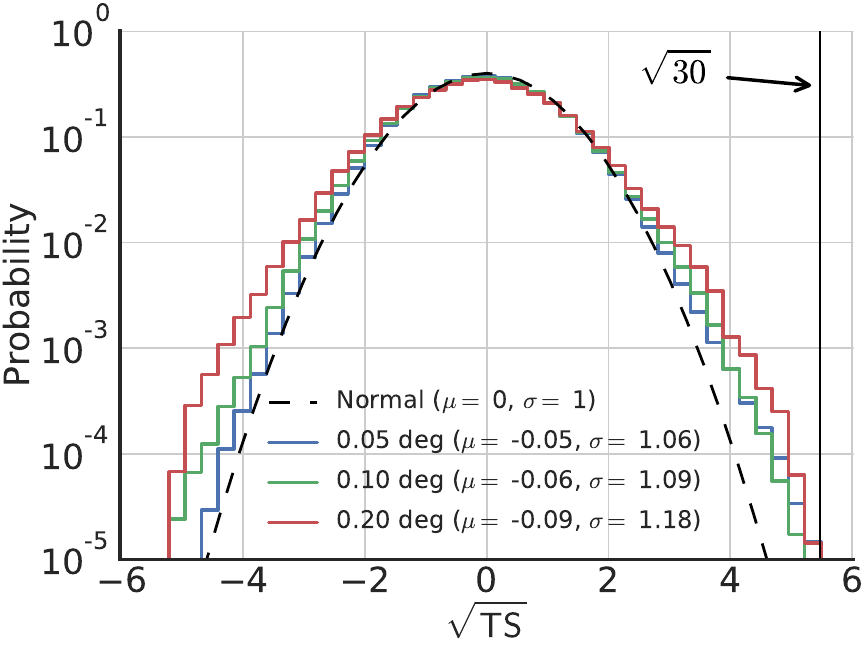}}
\caption[Residual significance distribution]{
Residual significance distribution after taking the HGPS emission model into
account (see Fig.~\ref{fig:hgps_catalog_model}, middle panel). The significance
was computed using a Gaussian source morphology of size $\sigma = 0.05\degr$,
0.10\degr\, and 0.20\degr. A vertical line at $\sqrt{\TSinFormula} = \sqrt{30}$
is shown, corresponding to the detection threshold for the HGPS multi-Gaussian
modeling. The sky region corresponding to this distribution includes pixels
inside exclusion regions, except for the Galactic center and shell-type SNRs,
which were not modeled for the HGPS (see Table~\ref{tab:hgps_external_sources},
lower panel of Fig.~\ref{fig:hgps_catalog_model} and
Fig.~\ref{fig:catalog:rois}).
}
\label{fig:hgps_residual_significance_distribution_ts}
\end{figure}

Applied to real data, we found a total of \hgpsComponentCountTotal\ significant
Gaussian components using this procedure and \TS\ threshold.
Figure~\ref{fig:hgps_residual_significance_distribution_ts} depicts the residual
$\sqrt{\TSinFormula}$ distributions over the entire HGPS region. These
distributions demonstrate that there is approximate agreement with a normal
Gaussian distribution; in particular, we find no features above the
$\sqrt{\TSinFormula} = \sqrt{30}$ detection threshold. Inherent imperfections in
the background, large-scale emission models and source emission models lead to a
slight broadening of the distributions with respect to a normal distribution, as
expected.

For reference, the \hgpsComponentCountTotal\ Gaussian components have been
assigned identifiers in the format \verb=HGPSC NNN=, where \verb=NNN= is a
three-digit number (counting starts at 1), sorted by right ascension (which is
right to left in the survey maps). The complete list of components is provided
in the electronic catalog table (see Table~\ref{tab:hgps_component_columns}).

\subsection{Component selection, merging, and classification}
\label{sec:cc:component_classification}

We repeated the entire modeling procedure described in the previous section with
a second set of maps produced with an independent analysis framework (see
Sect.~\ref{sec:dataset:events}). Five of the \hgpsComponentCountTotal\ HGPS
components were not significant in the cross-check analysis and were therefore
discarded (see Fig.~\ref{fig:hgps_catalog_model} and
Table~\ref{tab:hgps_component_columns}). Those components we labeled with
\verb=Discarded Small= in the column \verb=Component_Class= of the FITS table.

We observed two other side effects of the modeling procedure. Firstly, very
bright VHE sources, even some with center-filled morphologies such as Vela~X,
decomposed into several Gaussian components, modeling various morphological
details of the source. Figure~\ref{fig:hgps_catalog_model} illustrates this
effect: there are two multicomponent sources shown. Therefore in cases where
overlapping components were not clearly resolved into separate emission peaks,
we merged them into a single source in the HGPS catalog. In total, we found 15
such multicomponent sources: ten consisting of two Gaussian components and five
consisting of three Gaussian components. It would be intriguing to analyze the
complex morphology of these multicomponent sources in greater detail, but this
kind of analysis is beyond the scope of this survey paper. We labeled components
that are part of a multicomponent source as \verb=Source Multi=. We used the
label \verb=Source Single,= respectively, if there is only one component
modeling the source.

The second side effect was that some of the Gaussian components appeared to have
very large sizes coupled with very low surface brightness. We interpret these
components as artifacts of the modeling procedure, which picks up additional
diffuse \gammaray\ emission that is not covered by our simple large-scale
emission model (Sect.~\ref{sec:cc:large-scale-emission}). For example, as shown
in Fig.~\ref{fig:hgps_catalog_model}, the emission around $\ell \sim 345\degr$
initially comprised three model components: two components that clearly
converged on the two discrete emission peaks visible in the excess map and one
very large and faint component that appeared to be modeling large-scale emission
along the Galactic plane in between the two and not clearly related to either of
the two peaks. In total, we found ten such large-scale components (see
Table~\ref{tab:hgps_component_columns}), which we discarded and did not include
in the final HGPS source catalog as they are likely low-brightness diffuse
emission. We labeled this class of components as \verb=Discarded Large= in the
component list.

\subsection{Source characterization}
\label{sec:cc:source_characterization}

\subsubsection{Position, size, and flux}
\label{sec:cc:merged_source_parameters}

For HGPS sources that consist of several components, we determined the final
catalog parameters of the sources as follows:

\paragraph*{Flux\\}

The total flux is the sum of the fluxes of the individual components
\begin{equation}
F_{\mathrm{Source}} = \sum_i F_i
\label{eq:source_flux}
.\end{equation}

\paragraph*{Position\\}

We calculated the position by weighting the individual component positions with
the respective fluxes. The final $\ell_{\mathrm{Source}}$ and
$b_{\mathrm{Source}}$ coordinates of the source are written as%
\begin{linenomath}
\begin{align}
\ell_{\mathrm{Source}} = \frac{1}{F_{\mathrm{Source}}}\sum_i \ell_iF_i
&& \mathrm{and} &&
b_{\mathrm{Source}} = \frac{1}{F_{\mathrm{Source}}}\sum_i b_iF_i.
\end{align}
\end{linenomath}

\paragraph*{Size\\}

We obtained the size in $\ell$ and $b$ directions from the second moment of the
sum of the components as follows:
\begin{linenomath}
\begin{align}
\sigma_{\ell, \mathrm{Source}}^2 =&
\frac{1}{F_{\mathrm{Source}}}
\sum_i F_i \cdot (\sigma_i^2 + \ell_i^2) - \ell_{\mathrm{Source}}^2
\\
\sigma_{b, \mathrm{Source}}^2 =&
\frac{1}{F_{\mathrm{Source}}}
\sum_i F_i \cdot (\sigma_i^2 + b_i^2) - b_{\mathrm{Source,}}^2
\end{align}
\end{linenomath}
where additionally we defined the average circular size as
\begin{equation}
\sigma_{\mathrm{Source}} = \sqrt{
\sigma_{\ell, \mathrm{Source}}\,
\sigma_{b, \mathrm{Source}}
}
\label{eq:source_size}
.\end{equation}
We computed the uncertainties of the parameters using Gaussian error
propagation, taking the full covariance matrix estimate from the fit into
account.

\subsubsection{Size upper limits}
\label{sec:cc:extension_ul}

In the morphology fit, we did not take into account uncertainties in the PSF
model. However, studies using \hess\ data \citep[e.g.,][]{Stycz16} have revealed
a systematic bias on the size of point-like extragalactic sources on the order
of $\sigma_{\mathrm{syst}} = 0.03\degr$, so we have adopted this number as the
systematic uncertainty of the PSF.

Given a measured source extension $\sigma_{\mathrm{Source}}$ and corresponding
uncertainty $\Delta\sigma_{\mathrm{Source}}$, we used the following criterion to
claim a significant extension beyond the PSF:
\begin{equation}
\sigma_{\mathrm{Source}} - 2\Delta \sigma_{\mathrm{Source}}
> \sigma _{\mathrm{syst}},
\end{equation}
i.e.,\ if the extension of a source is $2\Delta \sigma_{\mathrm{Source}}$ beyond
the systematic minimum $\sigma_{\mathrm{syst}}$. If this criterion is not met,
we consider the source to be compatible with being point-like and define an
upper limit on the source size as follows:
\begin{equation}
\sigma_{\mathrm{UL}} =
\max(\sigma_{\mathrm{syst}}, \sigma_{\mathrm{Source}}
+ 2\Delta \sigma_{\mathrm{Source}}).
\label{eq:extension_ul}
\end{equation}

\subsubsection{Localization}
\label{sec:cc:localisation}

The HGPS source location error is characterized by error circles with radius
$R_{\alpha}$ at confidence levels $\alpha = 0.68$ and $\alpha = 0.95$, computed
as
\begin{equation}
R_{\alpha} = f_{\alpha} \times
\sqrt{\Delta \ell_{\mathrm{stat}}^2 + \Delta \ell_{\mathrm{syst}}^2 +
\Delta b_{\mathrm{stat}}^2 + \Delta b_{\mathrm{syst}}^2}.
\label{eq:position_error}
\end{equation}

The values $\Delta \ell_{\mathrm{stat}}$ and $\Delta b_{\mathrm{stat}}$ are the
statistical errors on Galactic longitude $\ell$ and latitude $b$, respectively,
from the morphology fit. For the \hess\ systematic position error, a value of
$\Delta \ell_{\mathrm{syst}} = \Delta b_{\mathrm{syst}} = 20\arcsec =
0.0056\degr$ per axis was assumed, following the method and value in
\citep{2010MNRAS.402.1877A}.

Assuming a Gaussian probability distribution, the factor $f_{\alpha}$ is chosen
as $f_{\alpha} = \sqrt{-2\log(1-\alpha)}$ for a given confidence level $\alpha$
\citep[see Eq.~1 in][]{Abdo:2009e}.

\subsubsection{Source naming}
\label{sec:cc:identifier}

The \hgpsSourceCountTotal\ HGPS catalog sources have been assigned source names
in the format \verb=HESS JHHMM=$\pm$\verb=DDd=, where \verb=HHMM= and
$\pm$\verb=DDd= are the source coordinates in right ascension and declination,
respectively. For new sources, the source name is based on the source location
reported in this paper. For sources that had been assigned names in previous
\hess\ publications or conference presentations, the existing name was kept for
the HGPS catalog, even if the position in the HGPS analysis would have led to a
different name.  Similarly, the source candidates (or hotspots, see
Sect.~\ref{sec:sourcecandidates}) have been assigned names in the format
\verb=HOTS JHHMM=$\pm$\verb=DDd=.

\subsection{Source spectra}
\label{sec:cc:spectra}

After detection and subsequent morphological analysis of the sources, we
measured a spectrum for each of the sources using an aperture photometry method.
In this method we sum the ON counts within an aperture defined as a circular
region centered on the best-fit position of each source. We fit a spectral model
within that aperture using an ON-OFF likelihood method \citep{Piron:2001}, where
the OFF background is estimated using reflected regions defined on a run-by-run
basis \citep{1994APh.....2..137F, ref:bgmodeling}. Based on the morphology
model, we then corrected the measured flux for containment and contamination
from other nearby sources. For the spectral analysis, we applied standard cuts,
resulting in energy thresholds in the range 0.2--0.5~TeV, lower than the
thresholds achieved using hard cuts in the detection and morphology steps.
Figure~\ref{fig:hgps_energy_threshold_profiles} shows the variation of the
threshold with longitude. In the following sections, we describe the spectral
analysis process in more detail.

\subsubsection{Circular apertures and reflected region background estimate}
\label{sec:cc:spectra:reg}

The optimal choice for the size for the spectral extraction region is a balance
between including a large percentage of flux from the source and limiting the
contamination of the measurement by hadronic background events, large-scale
emission, and other nearby sources. Following these requirements, we chose the
aperture radius $R_{\mathrm{spec}}$ as follows:

\begin{itemize}

\item $R_{\mathrm{spec}}=R_{\mathrm{70}}$ for \hgpsRSpecIsNotChanged medium-size
sources, where $R_{\mathrm{70}}$ is the 70\% containment radius measured on the
PSF-convolved excess model image (\texttt{R70} in the catalog),

\item minimum $R_{\mathrm{spec}}=0.15\degr$ for \hgpsRSpecIsExtended small
($\texttt{R70} < 0.15\degr$) sources,

\item maximum $R_{\mathrm{spec}}=0.5\degr$ for \hgpsRSpecIsReduced very large
($\texttt{R70} > 0.5\degr$) sources.

\end{itemize}

A minimal aperture radius of $0.15\degr$ was imposed to make the measurement of
the source spectrum more robust against systematic uncertainties of the PSF and
the source morphology assumption.

The aperture radius was limited to a maximum radius of $R_{\mathrm{spec}} =
0.50\degr$ to limit the fraction of observations that cannot be used for the
spectrum measurement because no background estimate could be obtained.

\begin{figure}
\resizebox{\hsize}{!}{\includegraphics{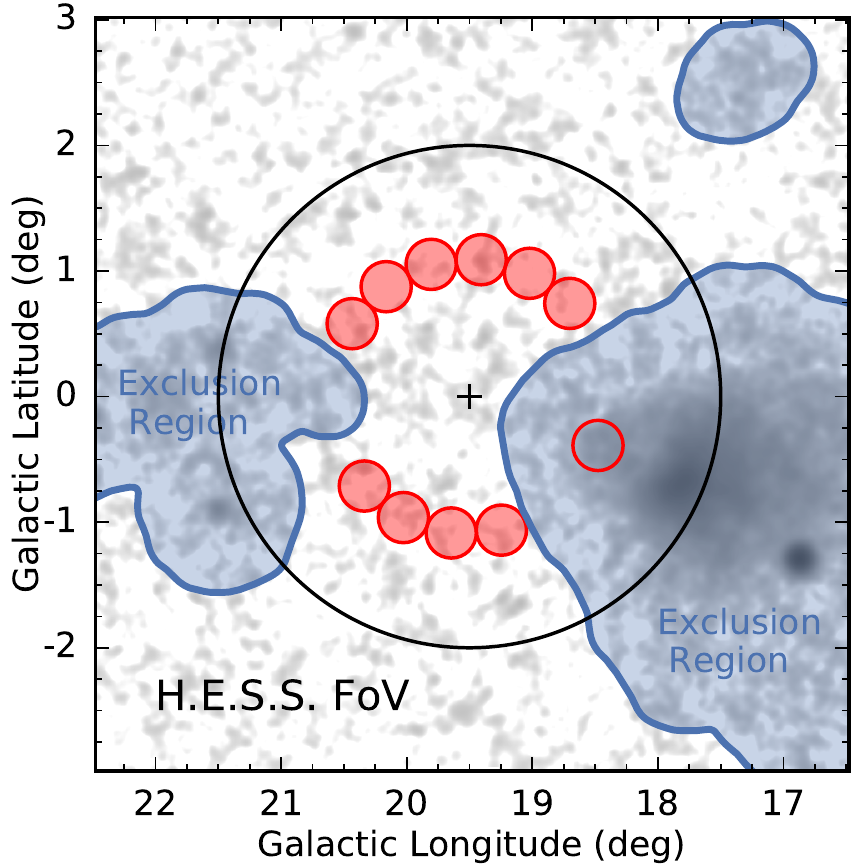}}
\caption[Illustration of reflected region background estimation for spectra] {
Illustration of reflected region background estimation for spectra
(Sect.~\ref{sec:cc:spectra:reg}). The HGPS significance image is shown in
inverse grayscale and exclusion regions as blue contours. The analysis FoV for
one observation is shown as a black circle with $2\degr$ radius and a black
cross at the observation pointing position. The non-filled red circle
illustrates the ON region for spectral analysis; the filled red circles indicate
the OFF regions.
}
\label{fig:hgps_spectrum_background_estimation}
\end{figure}

As illustrated in Fig.~\ref{fig:hgps_spectrum_background_estimation}, the
background is estimated using the reflected region method
\citep{1994APh.....2..137F, ref:bgmodeling}. For every spectral extraction
region (ON region), corresponding OFF regions with the same shape and offset to
the pointing position are chosen outside exclusion regions.

The method works well for small, isolated \gammaray\ sources such as active
galactic nuclei (AGNs) or the Crab nebula, where typically $\sim$10 OFF regions
are found in every observation. This results in a well-constrained background,
and all the exposure can be used for the spectral measurement. Because of the
high density of sources in the Galactic plane, large areas of emission are
excluded and only few reflected regions can be found.  This effectively results
in a loss of exposure for the spectrum measurement compared to the map
measurement. For the HGPS analysis this is a large problem because of the very
extensive exclusion regions used: 64\% of the livetime is lost for spectral
analysis compared to the total available livetime that is used in the map-based
analysis. For each source, see the \texttt{Livetime\_Spec} and \texttt{Livetime}
information in the source catalog. In cases where the loss of exposure is very
high, the background cannot be well constrained, which consequently results in
spectral parameters that are not well constrained. The following sources are
affected by this issue:

\begin{itemize}

\item Sources located in or near large exclusion regions (see
Fig.~\ref{fig:catalog:rois}). An area of width $\sim$2\degr\ is often excluded
along the Galactic plane, and this covers a significant portion of the analysis
FoV, which has a diameter of $4\degr$.
{}
\item Sources with large ON regions.

\item Sources observed with too small or too large offsets because they are
located close to other sources that were covered with dedicated observations.

\end{itemize}

\subsubsection{Flux containment and contamination correction}
\label{sec:cc:flux_correction}

By construction and because of additional effects such as PSF leakage or source
morphologies featuring tails, the spectral extraction region does not contain
the full flux of the source. Additionally, the large-scale emission model and
other nearby Gaussian components bias the flux measurement within the spectral
region. Based on this emission model, we separate the contributions from the
different components and derive a correction factor for the spectral flux
measurement.

The total flux in the spectral measurement region is
\begin{equation}
\label{eq:flux_total}
F_{\mathrm{Total}}^{\mathrm{ON}} =
F_{\mathrm{Source}}^{\mathrm{ON}}
+ F_{\mathrm{LS}}^{\mathrm{ON}}
+ F_{\mathrm{Other}}^{\mathrm{ON}},
\end{equation}
where $F_{\mathrm{Source}}^{\mathrm{ON}}$ is the contribution from the source
itself, $F_{\mathrm{LS}}^{\mathrm{ON}}$ is the contribution from the large-scale
emission model, and $F_{\mathrm{Other}}^{\mathrm{ON}}$ is the contribution from
nearby sources and other, discarded Gaussian emission components.

Assuming $F_{\mathrm{Source}}$ is the flux measurement from the morphology fit,
we define the correction factor as
\begin{equation}
\label{eq:correction_factor_1}
C_{\mathrm{Correction}} = F_{\mathrm{Source}} / F_{\mathrm{Total}}^{\mathrm{ON}}.
\end{equation}

To summarize the contributions from the large-scale emission model and other
sources in close (angular) proximity, we define a quantity called
contamination. This quantity measures the fraction of flux within the
spectral region that does not originate from the source itself and is written as
\begin{equation}
\label{eq:contamination}
C_{\mathrm{Contamination}} = \frac{F_{\mathrm{LS}}^{\mathrm{ON}}
  + F_{\mathrm{Other}}^{\mathrm{ON}}}{F_{\mathrm{Total}}^{\mathrm{ON}}}
.\end{equation}
Additionally, we define the containment of a source as the ratio
between the flux of the source within the spectral measurement region
$F_{\mathrm{Source}}^{\mathrm{ON}}$ (taking the morphology model into account)
and the total flux obtained from the morphology fit $F_{\mathrm{Source}}$ as follows:
\begin{equation}
\label{eq:containment}
C_{\mathrm{Containment}} = F_{\mathrm{Source}}^{\mathrm{ON}} / F_{\mathrm{Source}}
.\end{equation}
The HGPS catalog provides all the quantities mentioned in this section,
and all aperture-photometry based flux measurements in the HGPS catalog (see
Table~\ref{tab:hgps_sources_columns}) are corrected by the factor given in
Eq.~\ref{eq:correction_factor_1} (see Sect.~\ref{sec:cc:spectra:model} and
~\ref{sec:cc:spectra:points}).

We note that this region-based spectral analysis method with a single integral
flux correction factor assumes energy-independent source morphology. The spectra
obtained for sources with energy-dependent morphology does not correspond to the
correct total emission spectra of the sources. Currently energy-dependent
morphology has been clearly established for two sources (HESS J1303$-$631 and
HESS J1825$-$137), and there are hints of energy-dependent morphology for a few
more. Furthermore, using an integral flux correction factor is not fully correct
because the HGPS PSF is somewhat dependent on energy (smaller PSF at higher
energies). The resulting inaccuracy on HGPS spectral results is small, because
we have chosen a minimal spectral aperture radius of 0.15\degr, which contains
most of the emission for point sources at all energies. Generally, spectra for
sources with large correction factors are likely to be less accurate, because
the morphology models used to compute the correction are only approximations.

\subsubsection{Spectral model fit}
\label{sec:cc:spectra:model}

We performed the spectral fits on the stacked\footnote{Observation stacking was
performed as described here: \specstackurl} observations, using the ON-OFF
Poisson likelihood function, referred to as the $W$ statistic (WSTAT) in
XSPEC\footnote{See \wstaturl\ or Appendix~A of \citet{Piron:2001}.}. For each
observation, we applied a safe energy threshold (see Sect.~\ref{sec:events_map})
cut at low energies, and the maximum energy was chosen at the highest event
energy in the stacked counts spectrum for the on region (resulting in a maximum
energy of 30~TeV to 90~TeV). Energy dispersion was not taken into account via a
matrix, but in an approximate way in which the effective area is computed in
such a way that it results in fully correct spectral results for power-law
spectra with spectral index 2, and, given the good energy resolution of \hess,
only small errors are made for other spectral shapes \citep{Hoppe:2008c}.

To describe the spectral shape of the VHE \gammaray\ emission, we fit a PL
model to the data, i.e.,
\begin{equation}
\phi(E) =
\frac{{\rm d}N}{{\rm d}E} =
\phi_0 \left(\frac{E}{E_0}\right)^{-\Gamma},
\label{eqn:pl}
\end{equation}
where $\phi_0$ is the differential flux at a reference (pivot) energy
$E_0$ and $\Gamma$ is the spectral index. In addition, we also fit an
exponential cutoff power-law (ECPL) model,
\begin{equation}
\phi(E) = \phi_0 \left(\frac{E}{E_0}\right)^{-\Gamma} \exp(-\lambda E)
\label{eqn:ecpl}
,\end{equation}
which additionally contains the inverse cutoff energy $\lambda = 1 /
E_{\mathrm{cutoff}}$ as a third, free parameter. The reference (pivot) energy
$E_0$ is not a free parameter in either model; we compute this parameter on a
source-by-source basis to minimize the correlation between the other fit
parameters.

We computed integral fluxes as
\begin{equation}
F(E_1, E_2) = \int_{E_1}^{E_2}\phi(E)\,\mathrm{d}E,
\label{eq:spec_int_flux}
\end{equation}
usually for the energy band above 1~TeV, with integral flux errors computed
using Gaussian error propagation. We computed energy fluxes for a given energy
band as
\begin{equation}
G(E_1, E_2) = \int_{E_1}^{E_2}E\,\phi(E)\,\mathrm{d}E.
\end{equation}

The source catalog provides the PL fit results (see
Table~\ref{tab:hgps_sources_columns} for a description of columns) for every
source and the ECPL parameters where the ECPL model is more likely
($\TSinFormula = W_{\mathrm{ECPL}} - W_{\mathrm{PL}} > 9$). All
aperture-photometry based flux measurements are corrected by the factor given in
Eq.~\ref{eq:correction_factor_1}.

\subsubsection{Flux points}
\label{sec:cc:spectra:points}

Flux points are estimates of the differential flux $\phi$ at a given set of
reference energies $E_{ref}$. To compute flux points for the HGPS catalog, we
chose a method similar to that used for the \fermi\ catalogs \citep[see, e.g.,
Sect.~5.3 in][]{3FGL}. For every source we selected a total number of six bins
$(E_1, E_2)$ in reconstructed energy, logarithmically spaced between the safe
energy threshold and a maximum energy of 50~TeV. The reference energy for the
flux point estimation was set to the logarithmic bin center $E_{ref} = \sqrt{E_1
E_2}$. The differential flux $\phi$ was computed via a one-parameter likelihood
fit (same method as described in Sect.~\ref{sec:cc:spectra:model}), under the
assumption of the global best-fit PL and using only the data within the bin of
reconstructed energy $(E_1, E_2)$. An 1$\sigma$ asymmetric error on $\phi$ was
computed from the likelihood profile, and for spectral points of small
significance ($TS<1$), in addition an upper limit on $\phi$ was computed at 95\%
confidence level. All spectral point measurements in the HGPS catalog are
corrected by the factor given in Eq.~\ref{eq:correction_factor_1}.

\subsection{Method discussion}
\label{sec:cc:discussion}

The sensitivity profile and map shown in Fig.~\ref{fig:hgps_sensitivity}  were
computed assuming a point-like source morphology and using the \lima\
significance estimation. The likelihood fit method including the large-scale
emission model component used for the catalog production fundamentally differs
from that. We qualitatively discuss below the most important differences and
their influence on the effective sensitivity with which the catalog was
produced.

In Sect.~\ref{sec:sensmaps}, the sensitivity was defined as the minimum required
flux for a source to be detected with a certain level of confidence. Assuming
the source is extended, which applies to  most of the Galactic sources found by
\hess, the total flux of the source is distributed over a larger area on the
sky. Given a fixed background level, the signal-to-noise ratio is decreased and
the sensitivity scales with the size of the source as
\begin{equation}
F_{\mathrm{min}}(\sigma_{\mathrm{source}}) \propto
\sqrt{\sigma_{\mathrm{source}}^2 + \sigma_{\mathrm{PSF}}^2},
\label{eq:sensitivity_extended}
\end{equation}
where $\sigma_{\mathrm{source}}$ is the size of the source and
$\sigma_{\mathrm{PSF}}$ the size of the PSF \citep{ref:hintonhofmann}. It is
constant for sources smaller than the PSF and increases linearly with source
size for sources much larger than the PSF.

For low surface brightness sources close to the Galactic plane, high levels of
contamination (defined as in Eq.~\ref{eq:contamination}) from the large-scale
emission model were observed. This effectively reduces the sensitivity close to
the Galactic plane and even caused a few previously detected \hess\ sources to
fall below the detection threshold (see also
Sect.~\ref{sec:results:previously:missing}) chosen for the HGPS analysis.  For
sources far from the Galactic plane, however, the influence of the large-scale
emission can be neglected.

Systematic and statistical background uncertainties, which are neglected in this
analysis, bias the sensitivity for large, extended sources. Neglecting
background fluctuations in the likelihood fit can lead to an overestimation of
the significance of large sources, which can lead to unreliable detections of
large emission components. In addition, the adaptive ring method
(Sect.~\ref{sec:adaptiveringmethod}), which has a minimal inner ring radius of
$0.7\degr$, does not provide a reliable background estimate for those large
emission components.

Systematic uncertainties of various origins affect the spectral parameters of
the sources. In addition to the transparency of the atmosphere, calibration, and
event reconstruction (see Sect.~\ref{sec:dataset}), the analysis method itself
can introduce uncertainties. In particular, the background and large-scale
emission emission model, and the source extraction and measurement method
(multi-Gaussian morphology and aperture photometry) influence the flux and
spectral index measurement. We estimate the relative systematic uncertainties of
the flux extracted from the maps (Sect.~\ref{sec:maps}) and from the spectrum
(Sect.~\ref{sec:cc:spectra}) to be 30\%; for the spectral index
(Sect.~\ref{sec:cc:spectra}) we estimate an absolute systematic uncertainty of
0.2. This estimate is based on the scatter seen in the cross-check analysis and
other analyses (e.g., a source catalog extracted without a large-scale emission
model component). For individual difficult sources (poor containment, large
contamination, complex, and marginally significant morphology features), larger
systematics may arise (see Sects.~\ref{sec:results:previously} and
\ref{sec:results:xcheck}). We note that the systematic uncertainties quoted here
are the same as in the previous HGPS publication~\citep{ref:gps2006}, and, as
expected for a population of extended sources in the Galactic plane, these
values are slightly larger than the systematic uncertainties previously
estimated for isolated point-like sources such as the Crab nebula
\citep{ref:hesscrab}.

\begin{figure}
\resizebox{\hsize}{!}{\includegraphics{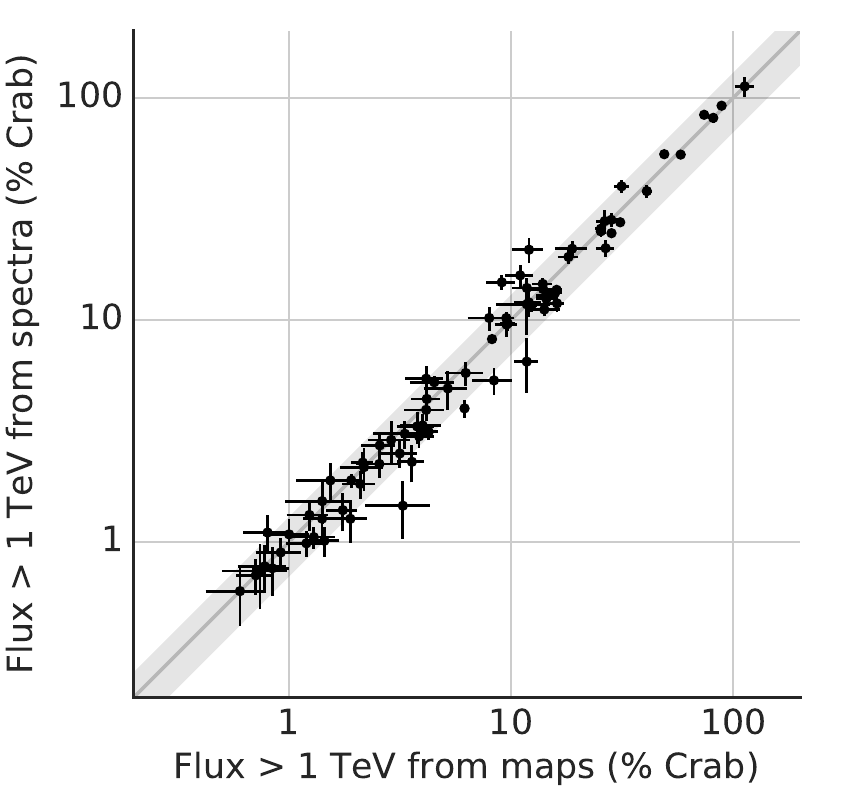}}
\caption[Comparison of source flux measurements]{
Comparison of integral source flux measurements above 1~TeV as calculated with
two different methods. The flux estimate from maps is the total source flux
according to the morphology model fit, assuming a spectral index of $\Gamma =
2.3$ (the \texttt{Flux\_Map} column in the catalog).  The flux estimate from
spectra is computed from the total source best-fit spectral model extracted
using aperture photometry and aperture correction (the
\texttt{Flux\_Spec\_Int\_1TeV} column in the catalog). The gray band in the
background illustrates a systematic uncertainty of 30\% on the flux values.
}
\label{fig:hgps_sources_spectrum_map_flux_comparison}
\end{figure}

A comparison of the two methods presented in this paper for calculating HGPS
source integral flux ($E > 1$~TeV) was performed as a diagnostic test (see
scatter plot in Fig.~\ref{fig:hgps_sources_spectrum_map_flux_comparison}). The
flux show on the x-axis is the total source flux estimated from the source
morphology fit on the maps (given by Eqn.~\ref{eq:source_flux}), assuming a
power-law spectrum with index $\Gamma = 2.3$. The flux estimate on the y-axis
was obtained from a spectral analysis (see Eqn.~\ref{eq:spec_int_flux} in
Sect.~\ref{sec:cc:spectra}), using a PL or ECPL spectral model assuption
(best-fit model) and an aperture photometry method that includes a containment
and contamination correction according to the HGPS multi-Gaussian plus
large-scale emission model. One can see that the two flux estimates agree very
well for most sources within the statistical errors and the 30\% flux systematic
uncertainty that we quote above. There are exceptions however, which can to a
large degree be attributed to differences in the underlying morphology and
spectral model assumptions of the two flux estimators. We note that when
comparing either of these HGPS source flux estimates against the cross-check,
the level of agreement is similar, but not quite as good (see
Sect.~\ref{sec:results:xcheck} for a discussion of individual cases). When
comparing against previous publications, the scatter is even larger (flux
differences up to a factor of 2 in a few cases), which can in many cases be
understood to be the result of differences in morphology model (of the source
itself,  of nearby overlapping sources, or the large-scale emission model) or
the spectral extraction region; most previous publications did not apply
containment or contamination corrections.

\section{Results and discussion}
\label{sec:results}

This section presents the results and a discussion of the HGPS based on the
data set (Sect.~\ref{sec:dataset}), maps (Sect.~\ref{sec:maps}), and catalog
(Sect.~\ref{sec:cc}).

\subsection{Source associations and firm identifications}
\label{sec:results:assoc_id}

Determining the physical nature of a VHE \gammaray\ source often requires
detailed spectral and morphological characterization of the VHE emission and
availability of complementary  MWL information.  Finding a
likely counterpart of a point-like VHE source is generally easy thanks to the
limited region of the sky to investigate.  For an extended source, such as the vast
majority of the HGPS sources, the procedure is often much more involved because
of multiple spatial associations, unless the VHE morphology is similar to that
observed at other wavelengths (e.g.,~for a large shell-type SNR).

We therefore make a distinction between source associations and firm
identifications of sources. The former is a list of astronomical objects,
extracted from catalogs of plausible counterparts, which are are found to be
spatially coincident with the HGPS source. When particularly solid evidence
exists that connects one of these associated objects to the VHE emission, such
as variability or shell-type morphology, we consider the HGPS source to be
firmly identified.

In Sect.~\ref{sec:association_procedure} we first describe the systematic
association procedure, followed by the discussion of the results of this search
for plausible counterparts in Sect.~\ref{sec:association_results}. Finally we
present the list of firmly identified HGPS sources in
Sect.~\ref{sec:identifications}.

\subsubsection{Source association procedure}
\label{sec:association_procedure}

Our objective is to associate each HGPS source with plausible counterparts found
among nearby objects in the most relevant counterpart catalogs (that is catalogs
of objects already identified as VHE emitters such as SNRs and pulsar wind
nebulae and high energy $\gamma$-ray sources, see
Table~\ref{tab:hgps_associations_catalogs}). We search for these counterparts in
a region larger than the nominal HGPS source size (its 68\% containment radius);
we associate all the objects whose cataloged positions are at an angular
distance smaller than the source spectral extraction radius,
\texttt{$R_{\mathrm{SPEC}}$} (Sect.~\ref{sec:cc:spectra:reg}; see also
Fig.~\ref{fig:hgps_survey_mwl_1} ff.). This spatial criterion is motivated by
the fact that often the origin of the relativistic particles is significantly
offset from the VHE centroid, for example, when VHE pulsar wind nebulae (PWNe)
are offset from energetic pulsars or extended well beyond their X-ray PWNe
counterparts. We expect this procedure to be affected by source confusion
(multiple associations that are difficult to disentangle), especially for larger
VHE sources.

\begin{table*}
\caption[Automatic source association results]{
Results of the automatic association procedure for each catalog used (see main
text for details and selections applied). The second column lists the numbers of
objects in the HGPS survey region for each catalog. The third column gives the
total number of associations found. The last column gives the number of HGPS
sources having at least one associated object of a given category. The
difference between the two last columns is only large for 3FGL because 3FGL is
the only counterpart catalog for which the source density is so high that many
HGPS sources are associated with multiple 3FGL sources. Out of the
\hgpsSourceCountTotal HGPS sources, only \hgpsSourceCountNotAssoc are left
without any association.
}
\label{tab:hgps_associations_catalogs}
\centering
\begin{tabular}{lrrr}
\hline\hline
Type                &  Number of objects         &  Total number of             & Number of HGPS sources\\
                    &  in HGPS region            &  associations                & with at least 1 association\\
\hline
2FHL sources        &     \InHGPSFHLSourceCount  &     \HGPSFHLAssociationCount & \oneFHLAssocCount     \\
3FGL sources        &     \InHGPSFGLSourceCount  &     \HGPSFGLAssociationCount & \oneFGLAssocCount     \\
\hline
Supernova remnants  &     \InHGPSSNRSourceCount  &    \HGPSSNRAssociationCount  & \oneSNRAssocCount     \\
Pulsar wind nebulae &     \InHGPSPWNSourceCount  &    \HGPSPWNAssociationCount  & \onePWNAssocCount     \\
Composite remnants  &    \InHGPSCOMPSourceCount  &    \HGPSCOMPAssociationCount & \oneCOMPAssocCount    \\
\hline
Energetic pulsars   &     \InHGPSPSRSourceCount  &     \HGPSPSRAssociationCount &  \onePSRAssocCount    \\
\hline
Extra associations  &      --                    &     \HGPSEXTAssociationCount &  --                   \\
\hline
\end{tabular}
\end{table*}

This criterion is a compromise between the number of spurious associations and
the number of missed associations. A spurious association would be one with a
counterpart that is physically unrelated to the HGPS source (e.g.,~a chance
spatial coincidence in the same region of the sky). A missed association would
be a real counterpart that is not selected by the procedure (e.g.,~a pulsar
significantly offset from a VHE source could be missed even though it is known
to generate a PWN).  As a consequence of this spatial criterion, larger sources
naturally has a larger number of associated objects.  The criterion is intended
to be loose (inclusive) to minimize missed associations at the expense of
including potentially spurious associations.  Nonetheless, this procedure has
certain limitations, for example,~difficulties in associating VHE emission with
an SNR if the emission was produced in offset molecular clouds illuminated by
cosmic rays that escaped from the SNR.

In the following paragraphs, we briefly describe the catalogs used for the
automatic association procedure applied to search for counterparts.  We also
describe a list of additional objects that have been associated with HGPS
sources in previous publications but are not present in the counterpart search
catalogs. We note that some of these catalogs contain a single, specific type of
object (e.g., supernova remnants), whereas other catalogs contain multiple types
of physical objects because they are the result of broad surveys at energies
relevant to the HGPS (e.g., the \fermi\ catalogs).

\paragraph{High-energy \gammaray\ sources\\}

We searched for associated high-energy (HE) \gammaray\ sources in the \fermi\
2FHL source catalog \citep{2FHL} and the full 3FGL catalog \citep{3FGL}. The
2FHL catalog covers the 50~GeV to $\sim$2~TeV energy range, and the 3FGL catalog
covers the 0.1--300~GeV range. They contain \InHGPSFHLSourceCount and
\InHGPSFGLSourceCount sources in the HGPS region, respectively. We expect the
\fermi\ catalogs to contain a significant number of HGPS sources. In the case of
2FHL, this is due to its energy range, which partially overlaps that of the
HGPS, and its sensitivity, which reaches $\sim$3-4\% Crab in the HGPS region
\citep{2FHL}. But even without such overlaps, we expect to find many \fermi\
associations, since many objects emit \gammarays\ following a spectrum that
extends from the HE to the VHE range. Even for noncontinuous spectra we expect
to find numerous associations, for example, when a pulsar emits GeV emission
detected by \fermi\ and its wind nebula emits TeV emission detected by \hess

\paragraph{Supernova remnants and pulsar wind nebulae\\}

Supernova remnants and PWNe are among the most common particle accelerators in
the Galaxy and are well-known VHE \gammaray\ emitters. Nonetheless, it is often
challenging to establish associations between SNRs and VHE sources. For example,
only specific regions of an SNR shell could be emitting or neighboring molecular
clouds could be illuminated by multi-TeV particles that escaped the shock front
of the SNR. Pulsar wind nebulae evolve as their pulsar ages and the available
rotational energy (spin-down power) decreases. Since the X-ray synchrotron
radiation from PWNe arises from higher energy electrons than the IC radiation in
the VHE gamma-ray band, and the cooling time of the electrons decreases with
their energy ($t_c = E/(dE/dt)$, for radiative losses $t_c \propto 1/E$) we
expect PWNe to shine longer in VHE gamma rays. Furthermore, a decreasing
magnetic field with age can limit the emission time in radio and X-rays without
affecting the VHE emission. As a result, some old PWNe should be undetectable
outside the VHE \gammaray\ domain (see, e.g., \cite{1997MNRAS.291..162A,
deJager09, HESS:PWNPOP}). For such old PWNe only the detection of a middle-aged
energetic pulsar in the vicinity of a VHE source can provide evidence toward the
true nature of the VHE emission.

To search for SNR and PWN associations, we take the most complete catalog of
SNRs and PWNe to date into account, SNRcat\footnote{\urlSnrcat, accessed
Oct~10,~2015} \citep{SNRcat}. The SNRcat is a census of Galactic supernova
remnants and their high-energy observations. It is based on the radio Catalogue
of Galactic Supernova Remnants \citep{2014BASI...42...47G} but additionally
includes other types of remnants in an effort to be as complete and up-to-date
as possible. In particular, it contains plerionic objects, PWNe with no observed
shell. The possible presence of a PWN is usually assessed based on the presence
of diffuse, nonthermal emission in radio, X-rays, or even \gammarays. Several of
these cataloged objects have been classified by SNRcat as candidate PWNe solely
because of the presence of VHE emission in the vicinity of an energetic pulsar.
We removed those objects from the catalog used in our association procedure to
avoid cases in which we might misleadingly self-associate.

For the association procedure, we split the SNRcat objects into three subsets
based on their apparent type. The first subset consists of objects that have no
evidence of nebular emission and mostly belong to the shell or filled-center
types in SNRcat; this subset contains \InHGPSSNRSourceCount\ objects within the
HGPS region. The second subset consists of objects that are listed in SNRcat as
PWNe (or PWNe candidates) showing no evidence for shell-like emission; this
subset contains \InHGPSPWNSourceCount objects within the HGPS region. The third
subset consists of objects showing evidence of both shell and nebular emission,
which we refer to as composite objects; this subset contains
\InHGPSCOMPSourceCount objects within the HGPS region. For a further discussion
of a potential PWN nature of these objects see the population study presented in
\cite{HESS:PWNPOP}.

\paragraph{Energetic pulsars\\}

We selected energetic pulsars from version 1.54 of the ATNF catalog of radio
pulsars \citep{Manchester:2005}. We excluded millisecond pulsars because they
are not expected to power VHE PWNe and applied a cut on the spin-down energy
flux $\dot{E} / d^2 > 10^{33}$~erg~s$^{-1}$~kpc$^{-2}$ on the remaining pulsars.
In addition, to take into account energetic pulsars of unknown distance, we
included all objects with a spin-down luminosity $\dot{E} >
10^{34}$~erg~s$^{-1}$, resulting in a total of \InHGPSPSRSourceCount pulsars
used in the association procedure. We did not take into account pulsars that do
not have a measured $\dot{E}$. It is important to note that pulsars represent
indirect associations: the associated pulsars are not directly emitting the
unpulsed VHE \gammaray\ emission found in the HGPS, but rather indicate that
they could be powering a PWN that directly emits such emission.

\subsubsection{Association results and discussion}
\label{sec:association_results}

\paragraph{HE \gammaray\ sources\\}

Of the \InHGPSFGLSourceCount 3FGL sources present in the HGPS region, we find
\HGPSFGLAssociationCount to be associated with an HGPS source.  As expected, we
also find a large portion of the \InHGPSFHLSourceCount 2FHL sources in the HGPS
region to be associated with HGPS sources: only \FHLnonAssociated of these have
no HGPS counterpart. One of these sources is notably coincident with the VHE
source candidate \object{HOTS J1111$-$611} (Sect.~\ref{sec:HOTS_J1111m611}).
Many of the other 2FHL sources lacking an HGPS association tend to be located in
low-sensitivity parts of the HGPS region. Only four 2FHL sources in parts of the
HGPS with good sensitivity show no significant VHE emission in the HGPS:
\object{Puppis A} \citep{2015AA...575A..81H}, \object{2FHL J0826.1$-$4500},
\object{$\eta$ Carinae}, and the composite \object{SNR G326.3$-$1.8}
\citep{2013ApJ...768...61T}.

\paragraph{Supernova remnants\\}

We find 24 of the 78 HGPS sources to be associated with shell-like SNRs. Given
the large number of such objects in the HGPS region (\InHGPSSNRSourceCount) and
given their sizes, the number of chance coincidences is non-negligible. This is
to be expected since we have not tried to specifically match SNR and HGPS source
positions and sizes as in \citet{Fermi_SNR_cat}. Nonetheless, as discussed
below, we find six known shells in the HGPS to be firmly identified and two more
to be VHE shell candidates based on their specific morphologies
\citep{HESS:Shells}. We study the population of known SNRs in the HGPS further
in a companion paper \citep{HESS:SNRUL}.

\paragraph{Pulsar wind nebulae and composites\\}

We find 37 of the SNRcat objects (in the HGPS region) containing a PWN or PWN
candidate to be associated with an HGPS source. Conversely, we find more than
40\% of HGPS sources to have at least one associated object in the PWN or
composite classes. This supports the notion that systems containing PWNe are
prolific VHE emitters. As discussed below, we are able to firmly identify about
half of these associations using additional observational evidence such as
similar MWL morphology or energy-dependent \gammaray\ morphology.

\paragraph{Pulsars\\}

We find 47 of all the HGPS sources to be associated with an energetic pulsar.
This suggests that the population of HGPS sources contains numerous PWNe.
However, we selected a relatively low threshold $\dot{E}$ in our association
criteria to minimize missed associations. We quantitatively study such selection
effects in a companion paper \citep{HESS:PWNPOP} that provides a detailed look
at the physical characteristics of firmly identified PWNe and a list of
candidate PWN identifications based on various expected characteristics.

\paragraph{Extra associations\\}

For completeness, in addition to the associations obtained through the
catalog-based, automatic procedure, we add a list of \HGPSEXTAssociationCount
extra associated objects that are plausible counterparts for some HGPS sources
and are not covered by the limited set of catalogs we use. Previous publications
had proposed most of these associations, often thanks to dedicated MWL
observations of VHE source regions (e.g.,~the X-ray source \object{XMMU J183245$-$0921539}
and \object{HESS J1832$-$093}). We propose other associations in this work for some of
the new sources (Sect.~\ref{sec:results:new}). We also include the original
identifiers of VHE sources discovered first by other instruments (e.g.,
\object{VER J1930$+$188}, which corresponds to \object{HESS J1930$+$188}).
Table~\ref{tab:hgps_associations} includes all of these extra associations,
labeled ``EXTRA''.

\paragraph{Sources without physical associations\\}

Eleven HGPS sources do not have any associations with known physical objects,
although some are associated with HE \gammaray\ sources. We list and discuss
these briefly here (the new VHE sources are discussed in
Sect.~\ref{sec:results:new}):

\begin{enumerate}

\item \object{HESS J1457$-$593} is one of the new sources detected in the HGPS
analysis. Although the automatic association procedure does not find any
counterparts, the VHE \gammaray\ emission may originate in a molecular cloud
illuminated by CRs that escaped from the nearby but offset \object{SNR
G318.2$+$0.1}. This scenario is briefly described in
Sect.~\ref{sec:HESS_J1457m593}.

\item \object{HESS J1503$-$582} is also a new HGPS source and does not have any
compelling association except for the HE \gammaray\ sources \object{3FGL
J1503.5$-$5801} and \object{2FHL J1505.1$-$5808}, neither of which is of a
firmly identified nature. We describe this enigmatic source in
Sect.~\ref{sec:HESS_J1503m582}.

\item \object{HESS J1626$-$490} has only one association, with the HE \gammaray\
source \object{3FGL J1626.2$-$4911}. A dedicated \xmm\ observation did not
reveal any compelling X-ray counterpart either \citep{2011A&A...526A..82E}.

\item \object{HESS J1702$-$420} is near the point-like source \object{2FHL
J1703.4$-$4145}. The elongation of the VHE \gammaray\ emission prevented the
automated procedure from making the association, but a connection between the
objects seems plausible. The small size \object{SNR G344.7$-$0.1} (about
$8^{\prime}$ in diameter) is also in the vicinity and in good positional
coincidence with the (point-like) 2FHL source.

\item \object{HESS J1708$-$410} has no compelling association, even though this
source was the target of dedicated X-ray observations to look for associated
emission \citep{2009ApJ...707.1717V}. Given the brightness and relatively steep
spectrum of this VHE source ($\Gamma = 2.57 \pm 0.09$), the absence of a
counterpart at lower \gammaray\ energies in the \fermi\ catalogs is surprising
and suggests the emission peaks in the hundreds of GeV range.

\item \object{HESS J1729$-$345} is north of the nearby SNR HESS~J1731$-$347
\citep{2011AandA...531A..81H}. An investigation into a potential connection
between the two suggests the VHE emission from the former could be from a
molecular cloud illuminated by hadronic particles that escaped from the SNR
\citep{Cui2016}.

\item HESS~J1741$-$302 is the subject of a dedicated companion paper
\citep{HESS:1741} discussing potential PWNe and SNR-related association
scenarios, among others. These aspects are therefore not discussed here.

\item \object{HESS J1745$-$303} is close to, but offset from, \object{SNR
G359.1$-$0.5}. \suzaku\ observations have revealed neutral iron line emission in
the region, suggesting the presence of molecular matter and making this object
another possible case of a CR-illuminated cloud \citep{2009ApJ...691.1854B}.  We
find this object also to be associated with the HE \gammaray\ sources
\object{2FHL J1745.1$-$3035} and \object{3FGL J1745.1$-$3011}.

\item \object{HESS J1828$-$099} is a new HGPS source described in
Sect.~\ref{sec:HESS_J1828m099}.

\item \object{HESS J1832$-$085} is also a new HGPS source, described in
Sect.~\ref{sec:HESS_J1832m085}.

\item \object{HESS J1858$+$020} has an association with the HE \gammaray\ source
\object{3FGL J1857.9$+$0210} and is close to, but offset from, \object{SNR
G35.6$-$0.4}. A dedicated study \citep{2014A&A...561A..56P} did not find any
compelling X-ray counterpart, although multiple possible scenarios were
investigated, including CR-illuminated molecular clouds.

\end{enumerate}

\subsubsection{Firmly identified HGPS sources}
\label{sec:identifications}

In this section, we go one step further and treat those HGPS sources for which
the physical origin of the VHE \gammaray\ emission has been firmly identified.
Whereas the association criteria were principally based on positional evidence
(angular offset), we also perform a census of the additional evidence that is
available to reinforce spatial associations and arrive at firm identifications.
The supplementary observables we consider are correlated MWL variability,
matching MWL morphology, and energy-dependent \gammaray\ morphology
\citep{ref:hintonhofmann}. Table~\ref{tab:hgps_identified_sources} summarizes
the results, along with the respective references for the additional evidence.
Among the \hgpsSourceCountTotal sources in the HGPS region, we determine
\hgpsSourceCountID to be firmly identified.

\begin{table*}
\caption
[Firmly-identified HGPS sources]{
Table of \hgpsSourceCountID firmly-identified objects among the HGPS sources.
The object classes are \gammaray\ binary, shell-type supernova remnant (SNR),
pulsar wind nebula (PWN), and composite SNR (in cases where it is not possible
to distinguish between the shell and interior nebula). The evidence used to
identify the VHE \gammaray\ emission include position, morphology, variability,
and energy-dependent morphology (ED Morph.).
}
\label{tab:hgps_identified_sources}
\centering
\begin{tabular}{lllll}
\hline\hline
Source name       & Identified object    & Class      & Evidence    & Reference \\
\hline
\object{HESS J1018$-$589} A & \object{1FGL J1018.6$-$5856} & Binary     & Variability & \cite{J1018} \\
\object{HESS J1302$-$638}  & PSR~B1259$-$63       & Binary     & Variability & \cite{2005AandA...442....1A} \\
\object{HESS J1826$-$148}  & LS~5039              & Binary     & Variability & \cite{Aharonian:2006a} \\

\hline
HESS~J0852$-$463  & Vela~Junior          & SNR        & Morphology  & \cite{Aharonian:2005b} \\
HESS~J1442$-$624  & RCW~86               & SNR        & Morphology  & \cite{2016arXiv160104461H} \\
HESS~J1534$-$571  & G323.7$-$1.0         & SNR        & Morphology  & \cite{HESS:Shells} \\
HESS~J1713$-$397  & RX~J1713.7$-$3946    & SNR        & Morphology  & \cite{2004Natur.432...75A} \\
HESS~J1718$-$374  & G349.7$+$0.2         & SNR        & Position    & \cite{2015AandA...574A.100H} \\
HESS~J1731$-$347  & G353.6$-$0.7         & SNR        & Morphology  & \cite{2011AandA...531A..81H} \\
HESS~J1801$-$233  & W~28                 & SNR        & Position    & \cite{Aharonian:2008f} \\  
HESS~J1911$+$090  & W~49B                & SNR        & Position    & \cite{HESS:W49} \\

\hline
\object{HESS J0835$-$455}  & \object{Vela X}               & PWN        & Morphology  & \cite{Aharonian:2006d} \\
HESS~J1303$-$631  & G304.10$-$0.24       & PWN        & ED Morph.   & \cite{2012AandA...548A..46H} \\
\object{HESS J1356$-$645}  & G309.92$-$2.51         & PWN        & Position    & \cite{2011AandA...533A.103H} \\ 
\object{HESS J1418$-$609}  & G313.32$+$0.13         & PWN        & Position    & \cite{Aharonian:2006b} \\ 
\object{HESS J1420$-$607}  & G313.54$+$0.23         & PWN        & Position    & \cite{Aharonian:2006b} \\ 
\object{HESS J1514$-$591}  & \object{MSH 15$-$52}          & PWN        & Morphology  & \cite{Aharonian:2005c} \\
\object{HESS J1554$-$550}  & G327.15$-$1.04       & PWN        & Morphology  & Section~\ref{sec:HESS_J1554m550} \\  
HESS~J1747$-$281  & G0.87$+$0.08           & PWN        & Morphology  & \cite{Aharonian:2005d} \\  
\object{HESS J1818$-$154}  & G15.4$+$0.1            & PWN        & Morphology  & \cite{HESS2014_J1818} \\ 
HESS~J1825$-$137  & G18.00$-$0.69        & PWN        & ED Morph.   & \cite{Aharonian:2006g} \\
\object{HESS J1837$-$069}  & G25.24$-$0.19          & PWN        & Morphology  & \cite{2008AIPC.1085..320M} \\ 
\object{HESS J1849$-$000}  & G32.64$+$0.53          & PWN        & Position    & Section~\ref{sec:HESS_J1849m000} \\

\hline
\object{HESS J1119$-$614}  & G292.2$-$0.5         & Composite  & Position    & Section~\ref{sec:HESS_J1119m614} \\ %
\object{HESS J1640$-$465}  & G338.3$-$0.0         & Composite  & Position    & \cite{2014MNRAS.439.2828A}, \cite{2014ApJ...788..155G} \\
\object{HESS J1714$-$385}  & \object{CTB 37A}              & Composite  & Position    & \cite{2008AA...490..685A} \\
\object{HESS J1813$-$178}  & G12.8$-$0.0          & Composite  & Position    & \cite{2007AA...470..249F}, \cite{2009ApJ...700L.158G} \\  
\object{HESS J1833$-$105}  & G21.5$-$0.9          & Composite  & Position    & Section~\ref{sec:HESS_J1833m105} \\
\object{HESS J1834$-$087}  & \object{W 41}                 & Composite  & Morphology  & \cite{2015AA...574A..27H} \\
\object{HESS J1846$-$029}  & G29.7$-$0.3          & Composite  & Position    & Section~\ref{sec:HESS_J1846m029} \\
HESS~J1930$+$188  & G54.1$+$0.3            & Composite  & Position    & \cite{2010ApJ...719L..69A}, Sect.~\ref{sec:results:previously} \\
\hline
\end{tabular}
\end{table*}

Firm identifications rely on different forms of evidence that vary depending on
the source class. The VHE \gammaray\ emission from compact binary systems is
always point-like and should exhibit variability that is also seen at lower
energies. In contrast, the VHE emission from shell-type SNRs is extended
(provided the SNR is sufficiently large and close) and nonvariable, but can be
identified based on the specific shell morphology and correlated morphology at
lower energies.

Composite SNRs have both a shell and an interior PWN detected at lower energies
and can be more complex to identify correctly. If the angular size of the shell
emission is larger than the size of the VHE emission, we can identify the VHE
emission as coming from the PWN filling the SNR. This is the case, for example,
for HESS~J1747$-$281 (PWN in \object{SNR G0.9$+$0.1}) and HESS~J1554$-$550 (PWN
in \object{SNR G327.1$-$1.1}). In other cases, we are only able to identify the
HGPS source with the composite SNR as a whole, i.e., we are confident that the
VHE emission originates in the composite object but cannot disentangle whether
it comes predominantly from the PWN or the shell (usually due to PSF
limitations).

More evolved stellar remnant systems are difficult to identify firmly.  We can
make a firm PWN identification when there is a PWN of comparable size and
compatible position detected at lower energies.  This is the case, for example,
for HESS~J1420$-$607 (\object{PWN G313.54$+$0.23}) and HESS~J1356$-$645
(\object{PWN G309.92$-$2.51}).  In the absence of any clear PWN, or when its
size at lower energies is much smaller than the VHE source, we have to rely on
other evidence. The clearest such evidence is the detection of energy-dependent
morphology, expected in PWNe because of the cooling of energetic electrons as
they are transported away from the pulsar. At higher energies, the extent of the
emission shrinks and its barycenter moves closer to the pulsar. This is the case
for two sources thus far, HESS~J1303$-$631 (\object{PWN G304.10$-$0.24}) and
HESS~J1825$-$137 (\object{PWN G18.00$-$0.69}). In the absence of such evidence,
the identification of a VHE source as a PWN remains tentative when the only
evidence is an energetic pulsar in the vicinity. Candidate PWN identifications
are evaluated in detail in a companion paper \citep{HESS:PWNPOP}.

\begin{figure}
\resizebox{\hsize}{!}{\includegraphics{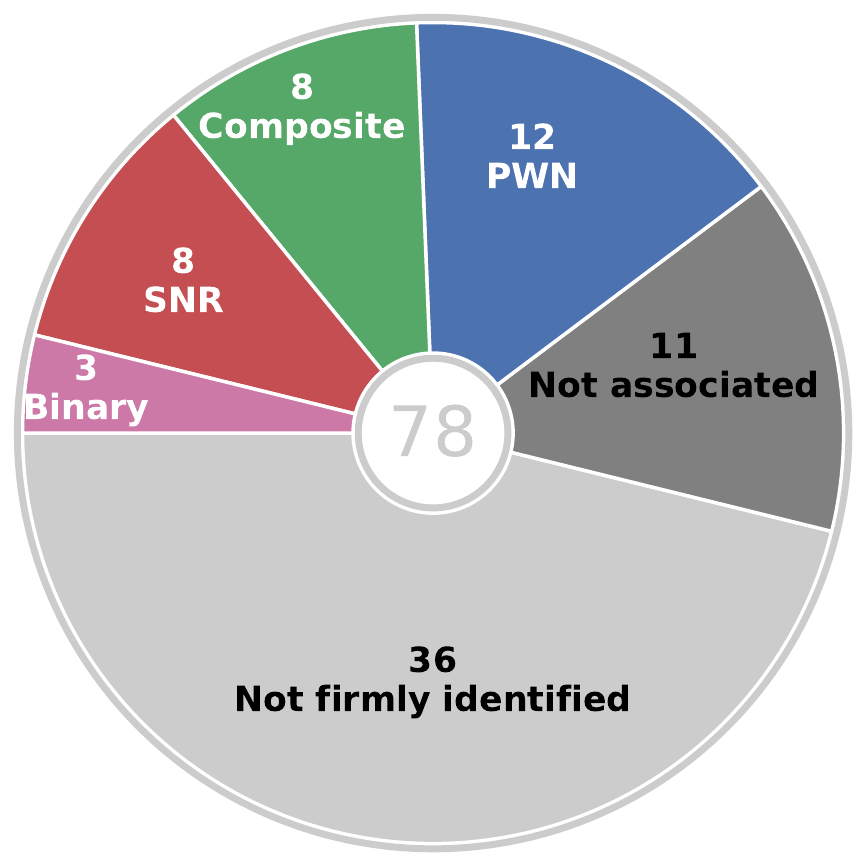}}
\caption[Source identification summary pie chart]{
Source identification summary pie chart. See
Table~\ref{tab:hgps_identified_sources} and Sect.~\ref{sec:identifications}.
}
\label{fig:hgps_source_id}
\end{figure}

A large percentage (39\%) of the \hgpsSourceCountID firmly identified sources
are PWNe. The next largest source classes identified are SNR shells (26\%) and
composite SNRs (26\%). Finally, \gammaray\ binary systems are also identified in
the HGPS. It is not yet possible to identify firmly more than half of the total
\hgpsSourceCountTotal HGPS sources with the conservative criteria we adopted,
although the vast majority have one or more promising spatial associations that
could prove to be real identifications following more in-depth studies beyond
the scope of this work. We do not find any physical associations for 11 of the
VHE sources in the HGPS, although for some of these, potentially related
emission is seen in HE \gammarays, and for others, offset counterparts are
present but simply not found by the automated association procedure adopted (see
previous section). Figure~\ref{fig:hgps_source_id} summarizes these
identifications.

We note that one source in HGPS, \object{HESS J1943$+$213}, is likely an extragalactic
object. It has no measured extension and a radio counterpart that many recent
studies tend to classify as a BL-Lac object \citep{2014A&A...571A..41P,
2016ApJ...822..117S, 2016ApJ...823L..26A}. However, its VHE flux has not
revealed any variability so far, which is unusual for such an object
\citep{2016arXiv161005799S}.

\subsection{Large-scale emission}
\label{sec:results:large}

In Sect.~\ref{sec:cc:large-scale-emission}, we introduced an empirical spatial
model to account for the large-scale VHE \gammaray\ emission we observed along
the Galactic plane to detect and characterize accurately the discrete VHE
\gammaray\ sources. This model provides an estimate of the spatial distribution
of the large-scale VHE emission discovered by \citet{2014PhRvD..90l2007A}. We
find that the fit amplitude, latitudinal width, and position of this model,
shown on Fig.~\ref{fig:hgps_diffuse_model}, are consistent with the latitude
profile of that previous work. The width is also comparable to the HGPS source
latitude distribution (Fig.~\ref{fig:hgps_sources_glat}, ff.) but smaller than
that of molecular gas traced by CO emission \citep{Dame01}.

Owing to the observational constraints and analysis used, the large-scale
emission model cannot be considered a measurement of the total Galactic diffuse
emission. The large-scale emission model provides an estimate of the diffuse
emission present in the HGPS maps. Its parameter values depend on the map
construction technique, in particular the exclusion region mask used in the
analysis (Sect.~\ref{sec:exclusion_regions}), i.e., changes in the mask can
alter the parameters of the model. For instance, the peak observed at $\ell \sim
340\degr$ in Fig.~\ref{fig:hgps_diffuse_model} is due to the presence of
low-level emission that is just below the threshold to be covered by the
exclusion mask we use for the HGPS. While a significant percentage of the
large-scale emission is expected to be truly interstellar diffuse emission, it
is very likely that emission from discrete but unresolved sources contributes as
well. Finally, some features in the HGPS large-scale emission model are likely
artifacts of errors in the estimation of the background model of gamma-like
cosmic ray EAS events (see Sect.~\ref{sec:background_estimation}); these events
are the dominating model component in the HGPS counts maps, thus small relative
errors in that background model can lead to significant changes in the excess
model of the HGPS sources, but even more so the HGPS large-scale emission model.

\subsection{Source parameter distributions}
\label{sec:results:distributions}

In the following section we study the global properties of the VHE \gammaray\
sources in the HGPS catalog. We compare certain key source parameters against
each other and briefly discuss the implications in the context of the Galactic
VHE source population, survey sensitivity, and firmly identified MWL source
classes.


\begin{figure*}[!ht]
\includegraphics[width=\textwidth]{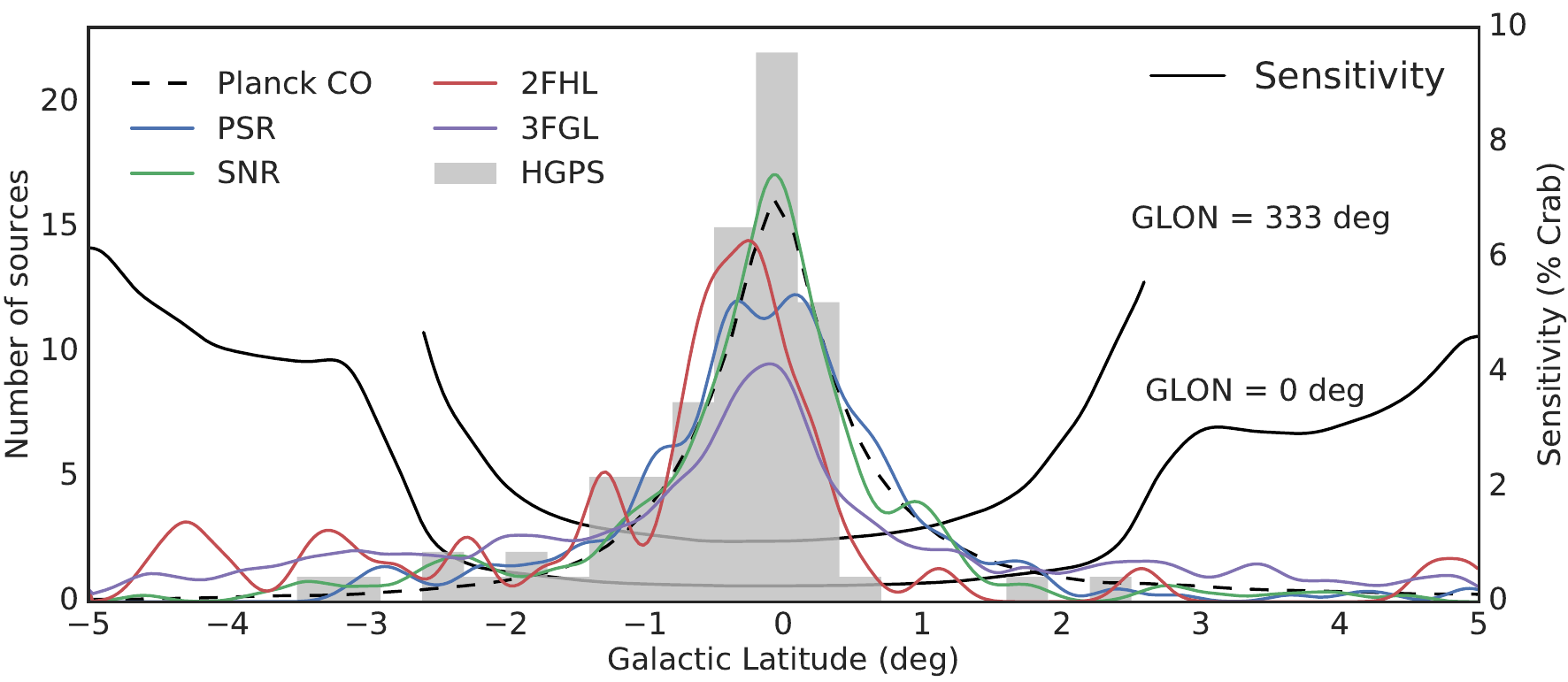}
\caption[Source Galactic latitude distribution]{
Galactic latitude distribution of the HGPS sources (gray histogram). The bin
size of this histogram is $0.3\degr$. The HGPS point source sensitivity is shown
(in units of \%~Crab) at two different longitudes of $0\degr$ and $333\degr$.
For comparison, the pulsar (PSR), supernova remnant (SNR), 3FGL and 2FHL source
distributions in the HGPS longitude range are shown as overlaid curves, smoothed
with Gaussians of width $0.15\degr$. The dashed line shows \emph{Planck}
measurements of CO(1-0) line emission as an estimate for matter density in the
Galaxy and similarly smoothed. All curves are normalized to the area of the
histogram.
}
\label{fig:hgps_sources_glat}
\end{figure*}

The latitude distribution of the \hgpsSourceCountTotal HGPS sources is shown in
Fig.~\ref{fig:hgps_sources_glat}. The distribution has a mean of $b =
-0.41\degr$ and a width of $0.87\degr$. For visual comparison, the latitude
distributions of the main classes of associated counterparts
(Sect.~\ref{sec:results:assoc_id}) --- SNRs, energetic pulsars, 3FGL sources,
and 2FHL sources --- are shown in this figure. Also shown for reference is an
estimate of the matter density profile as traced by \emph{Planck} measurements
of CO(1-0) line emission \citep{Planck15}. It should be kept in mind throughout
this section that the HGPS sensitivity is not uniform as a function of longitude
or latitude (Sect.~\ref{sec:sensmaps}).

The HGPS latitude distribution of sources correlates well with both potential
counterparts and tracers of matter density.  The distribution is somewhat skewed
toward negative latitudes even though the HGPS sensitivity has a relatively wide
and flat coverage in latitude. In Fig.~\ref{fig:hgps_sources_glat}, the
sensitivity is illustrated by two curves showing regions of relatively good
sensitivity (e.g., at $\ell = 0\degr$) and relatively poor sensitivity (e.g., at
$\ell = 333\degr$). These curves demonstrate that the HGPS sensitivity coverage
in latitude is, in general, much wider than the HGPS source distribution.
Although there are local exceptions at some longitudes, the latitude coverage is
generally flat in the range $-2.0\degr < b < 1.5\degr$, at various locations
even in $-2.5\degr < b < 2.5\degr$.  However, the counterpart catalogs are known
to suffer from various selection biases and the Galactic disk itself is known to
not be perfectly symmetric as observed across the spectrum.

In addition, one might still argue that, given the narrow range of latitudes
observed with respect to surveys at other wavelengths, the HGPS sources may not
be representative of the underlying distribution of VHE \gammaray\ sources.
However, in light of the counterpart distributions, in particular the 2FHL
sources, it can be reasonably assumed that the limited latitude coverage only
has a weak effect on the observed source population distribution.

\begin{figure*}
\includegraphics[width=\textwidth]{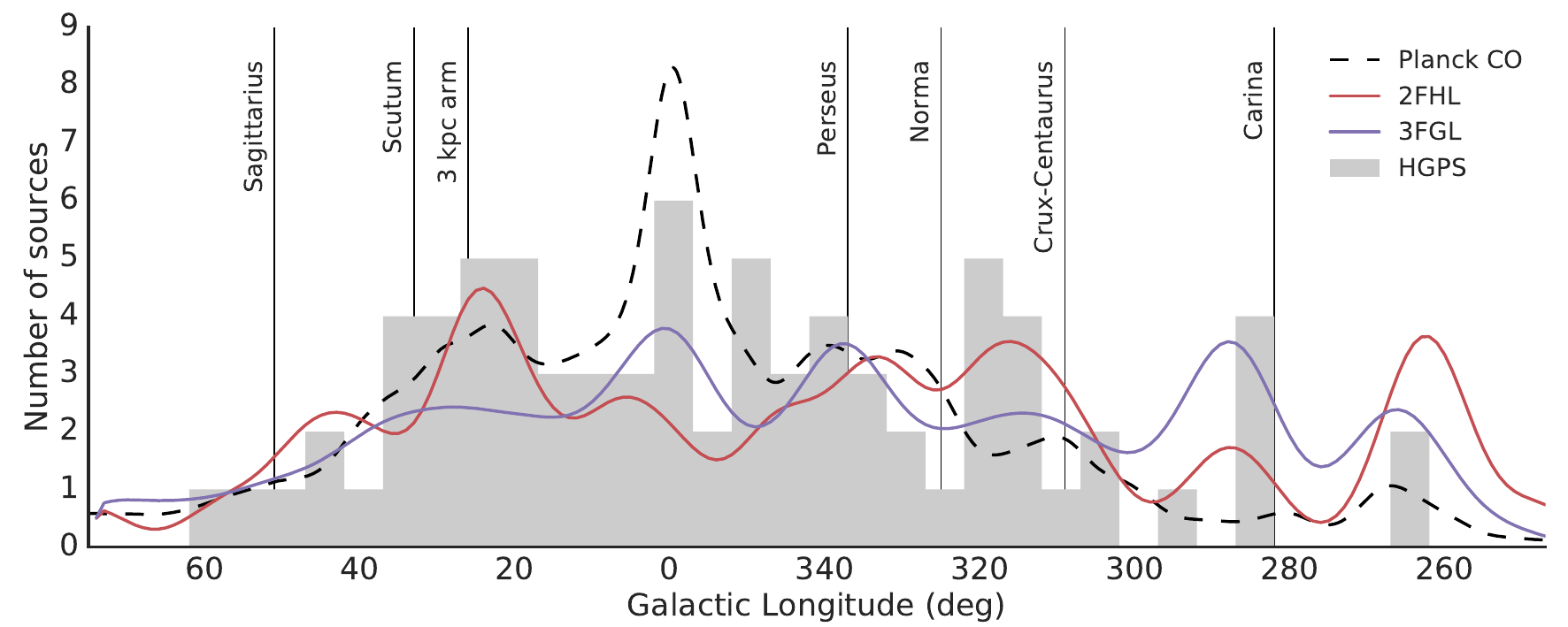}
\caption[Source Galactic longitude distribution]{
Galactic longitude distribution of the HGPS sources (gray histogram).
The bin size of this histogram is $5\degr$. For comparison, the 3FGL and 2FHL
source distributions (smoothed with a Gaussian of width $5\degr$) and the
\emph{Planck} measurements of CO(1-0) line emission as an estimate for matter
density in the Galaxy (smoothed with a Gaussian of width $2.5\degr$) are shown
for the range in Galactic latitude $b \le 5\degr$ and normalized to the area of
the histogram. Spiral arm tangent locations shown are from
\cite{2014ApJS..215....1V}.
}
\label{fig:hgps_sources_glon}
\end{figure*}

\begin{figure*}
\includegraphics[width=\textwidth]{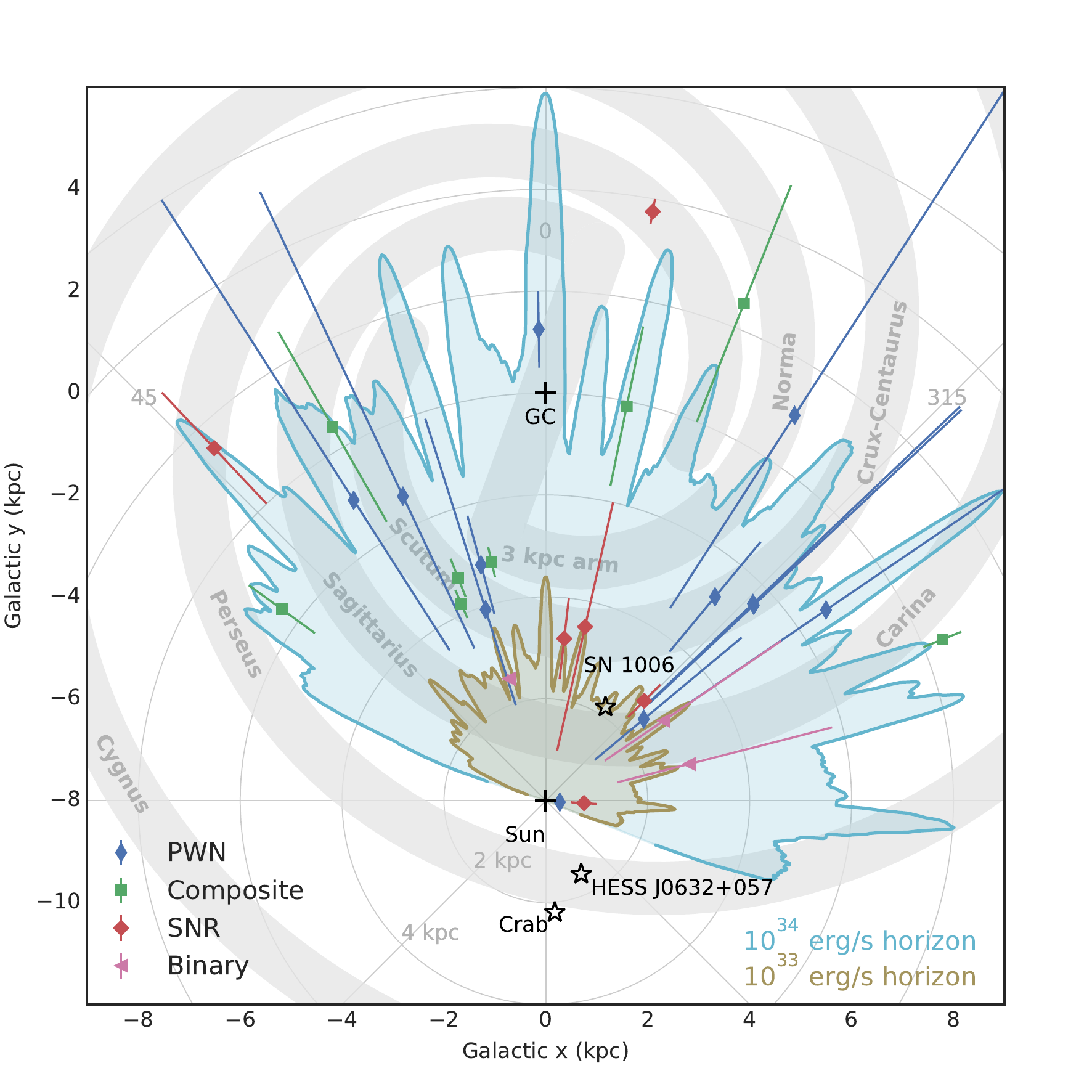}
\caption[Location of Galactic \hess\ sources in the Galaxy]{
Illustration of the location of identified \hess\ sources in the Galaxy with
respect to HGPS completeness (sensitivity limits). This is a face-on view; the
spiral arms \citep{2014ApJS..215....1V} are schematically drawn as gray bars.
The HGPS horizons for source luminosities of $10^{33}$ and $10^{34}$~erg/s (for
a putative 5$\sigma$ detection of a point-like source, same as
Fig.~\ref{fig:hgps_sensitivity}) are depicted by light blue and light brown
lines (and shaded regions therein), respectively. The source distances are from
SNRcat \citep{SNRcat} and ATNF pulsar catalog \citep{Manchester:2005}.  When no
distance uncertainties were available, we applied a generic uncertainty of
factor two on the distance. The three labeled sources are the Galactic
\gammaray\ sources outside the HGPS region detected by \hess
}
\label{fig:hgps_face_on_milky_way}
\end{figure*}

The longitude distribution of the \hgpsSourceCountTotal HGPS sources is shown in
Fig.~\ref{fig:hgps_sources_glon}, together with the molecular interstellar
matter column density profile as traced by CO(1-0) line emission (same as in the
previous figure). The latter, measured by \emph{Planck} \citep{Planck15}, has a
uniform exposure (sensitivity) over the sky, unlike the HGPS, adding caveats to
potential detailed correlations seen in this figure. We can nevertheless
robustly conclude that there is a very general correlation in longitude between
the number of HGPS sources and the molecular matter column density and that the
HGPS sources are mostly found in the inner $\sim$60\degr\ of the Galaxy.
Additionally, the spiral arm tangents as traced by CO
\citep{2014ApJS..215....1V} are shown in Fig.~\ref{fig:hgps_sources_glon}. An
increased number of sources could be expected in the directions of the near
spiral arm tangents (see Fig.~\ref{fig:hgps_face_on_milky_way}).  In the
longitude distribution, a slight excess of sources in the direction of
\textit{Scutum} and between \textit{Norma} and \textit{Crux-Centaurus} can be
observed. However, because of the limited sample size of 1--6 sources per bin, no
significant increased source density in the direction of spiral arm tangents can
be observed.

For comparison, we also added distributions for the Fermi-LAT
catalogs 3FGL and 2FHL to Fig.~\ref{fig:hgps_sources_glon}. While
Fermi-LAT has a roughly uniform exposure, their sensitivity in the HGPS region
is reduced in the inner Galaxy where diffuse emission is brighter, and also the
source extraction is very different from the HGPS approach, so that a direct
comparison is not possible. Finally we have chosen not to show the SNR and
pulsar distributions in the Galactic longitude distribution at all because the
coverage of those catalogs is not uniform.

\begin{figure*}[ht!]
\centering
\includegraphics[width=18cm]{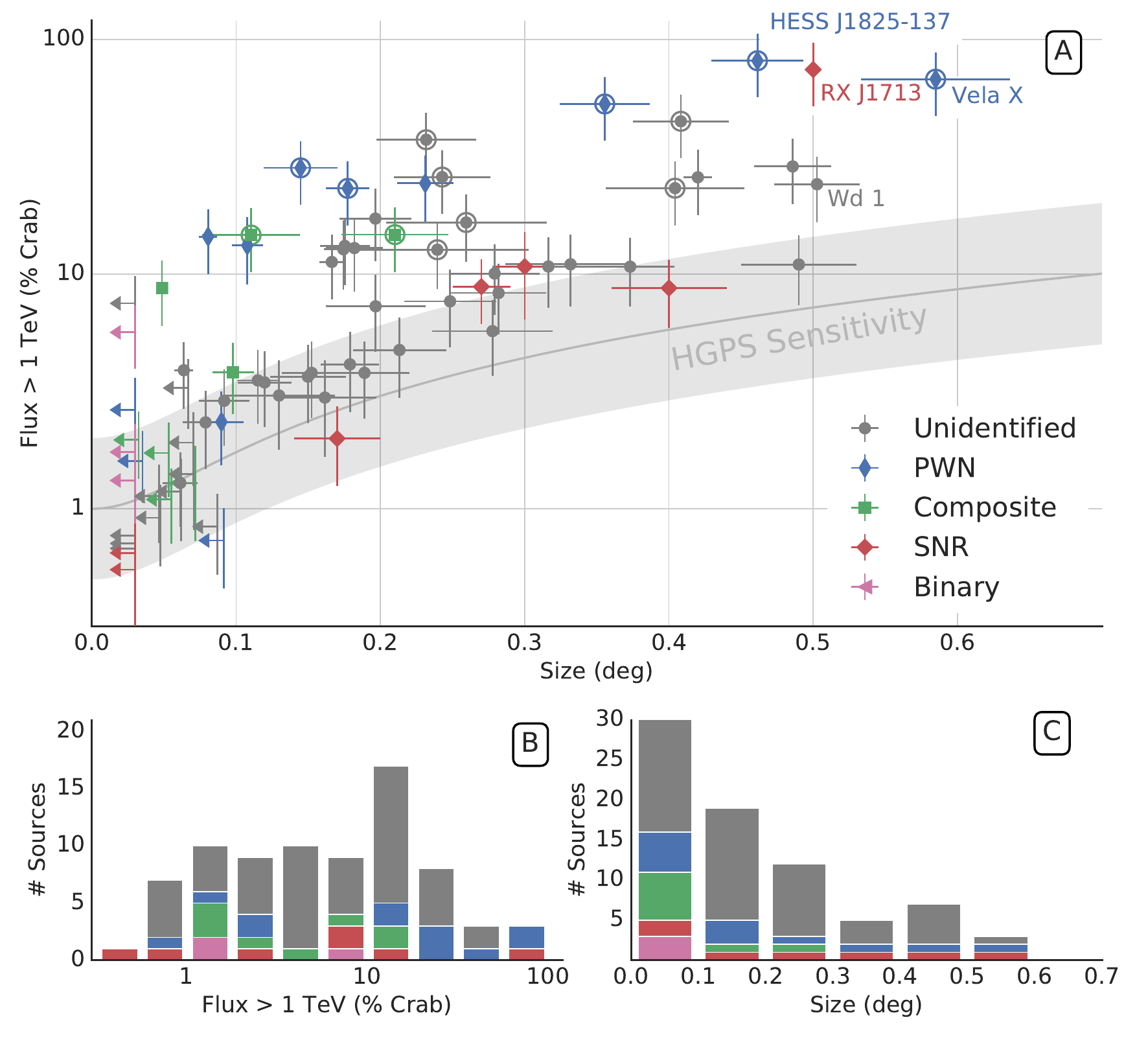}
\caption[Source flux versus size scatter plot]{
\textbf{Panel A:} Integral source flux ($E > 1$~TeV) vs. source size scatter
plot with colors representing the different classes of firmly identified
sources. For HGPS sources modeled as single Gaussians, the size is its width
($\sigma$). For sources modeled as multiple Gaussians (indicated with a circle
around the marker), the size is the RMS of the two-dimensional intensity
distribution (see Eq.~\ref{eq:source_size}). For sources with shell-like
morphology (SNRs), the size is the outer shell radius. To improve the
visibility of the plot, we do not show the SNR Vela~Junior (HESS~J0852$-$463)
at a size of $1\degr$ and a flux of 103\%~Crab. We illustrate the approximate
sensitivity limit of the HGPS, as defined in Eq.~\ref{eq:sensitivity_extended},
with an assumed point-source sensitivity of 1\%~Crab and an uncertainty band
with a factor $\pm$2 to represent the sensitivity variations in the survey
region (see caveats in main text).
\textbf{Panel B:} Distribution of the integral fluxes
($E > 1$~TeV)of the HGPS sources; colors are shown as in panel~A.
\textbf{Panel C:} Distribution of the HGPS source sizes; colors shown as in
panel~A. The first bin contains 30 sources, of which 17 are compatible with
point-like sources according to Eq.~\ref{eq:extension_ul}. As in panel~A, we
omit Vela~Junior, at a size of $1\degr$.
}
\label{fig:hgps_sources_flux_extension}
\end{figure*}

We compare the HGPS source integral fluxes ($E > 1$~TeV) to source sizes in
panel A of Fig.~\ref{fig:hgps_sources_flux_extension} and show the distributions
of fluxes and sizes separately in panels~B and C, respectively. In the
flux--size figure, we plot the approximate flux sensitivity limit of the HGPS as
a function of source size. One can see that the sensitivity worsens as the
source size increases, as expressed by Eq.~\ref{eq:sensitivity_extended}. The
HGPS sources indeed generally follow this trend. From
Fig.~\ref{fig:hgps_sources_flux_extension}, we therefore conclude that the HGPS
can be considered complete down to $\sim$10\%~Crab for sources $< 0.7\degr$. For
smaller sources ($< 0.1\degr$), the HGPS achieves completeness at a few \%~Crab
(see also Fig.~\ref{fig:hgps_sensitivity}).

We show the distribution of HGPS source integral fluxes ($E > 1$~TeV), which are
calculated assuming a spectral index of $\Gamma = 2.3$, in panel B of
Fig.~\ref{fig:hgps_sources_flux_extension}. At higher fluxes, we naturally
expect the number of sources to decrease. At the lowest fluxes, we also expect
the number to be small, because we reached the sensitivity limit of the HGPS.

As can be seen in panel C of Fig.~\ref{fig:hgps_sources_flux_extension} and
despite the modest \hess\ PSF (\hgpsMeanPSF), the majority of sources are not
compatible with being point-like but rather found to be significantly extended
and as large as $1\degr$. Owing to the methods used for background subtraction
(see Sect.~\ref{sec:adaptiveringmethod}), the HGPS is not sensitive to sources
with larger sizes.

The firmly identified HGPS sources (Sect.~\ref{sec:results:assoc_id}) are
highlighted in Fig.~\ref{fig:hgps_sources_flux_extension}. It can be seen that
all identified binary systems appear as point-like sources in the HGPS, as
expected. The PWNe appear to have various angular sizes, in agreement with the
diversity observed in the VHE PWN population \citep{HESS:PWNPOP}.  Most
identified SNRs are extended, likely owing to selection bias (smaller SNRs are
difficult to identify, e.g., through shell-like morphology) and the \hess\ PSF.
The identified composite SNRs, on the other hand, are typically smaller, owing
to the difficulty in disentagling VHE emission from the SNR shell and interior
PWN, similarly related to the \hess\ PSF. In any case, it does not seem possible
to identify the nature of the many unidentified sources solely on the basis of
their sizes or a flux--size comparison.

\begin{figure}[!t]
\resizebox{\hsize}{!}{\includegraphics{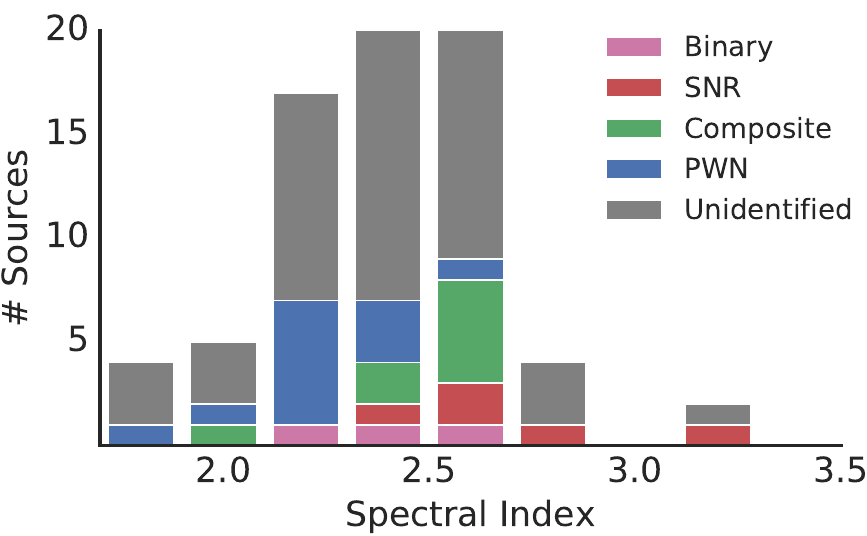}}
\caption[Catalog: Spectral index distribution]{
Distribution of the HGPS source power-law (PL) spectral indices. For
consistency, the PL model spectral index is used for all sources, even those
for which an exponential cutoff power law (ECPL) fits better. Taking statistical and
systematic uncertainties into account, all indices are compatible within
$2\sigma$ with the mean $\Gamma = 2.4 \pm 0.3$ of the distribution.
}
\label{fig:sources_index}
\end{figure}

Figure~\ref{fig:sources_index} shows the distribution of the HGPS source
power-law (PL) spectral indices $\Gamma$. For consistency, the PL model spectral
index is used for all sources, even those for which an exponential cutoff power
law (ECPL) fits better. The index distribution has a mean $\Gamma = 2.4 \pm
0.3$.  This is compatible with the index ($\Gamma = 2.3$) adopted in the
production of the HGPS flux maps (Sect.~\ref{sec:fluxmaps}) and the HGPS PSF
computation (Sect.~\ref{sec:cc:maps:psf}). We note that individual source indices
have typical statistical uncertainties of order $\pm 0.2$ and a similar
systematic uncertainty; HGPS data are often not sufficient to precisely
constrain the index because the energy range covered with good statistical
precision is typically only about one decade ($1 \la E \la 10$~TeV). Finally,
the figure also shows how the firmly identified HGPS sources are distributed in
index, showing no strong tendency with respect to source class.

\begin{figure}[!th]
\resizebox{\hsize}{!}{\includegraphics{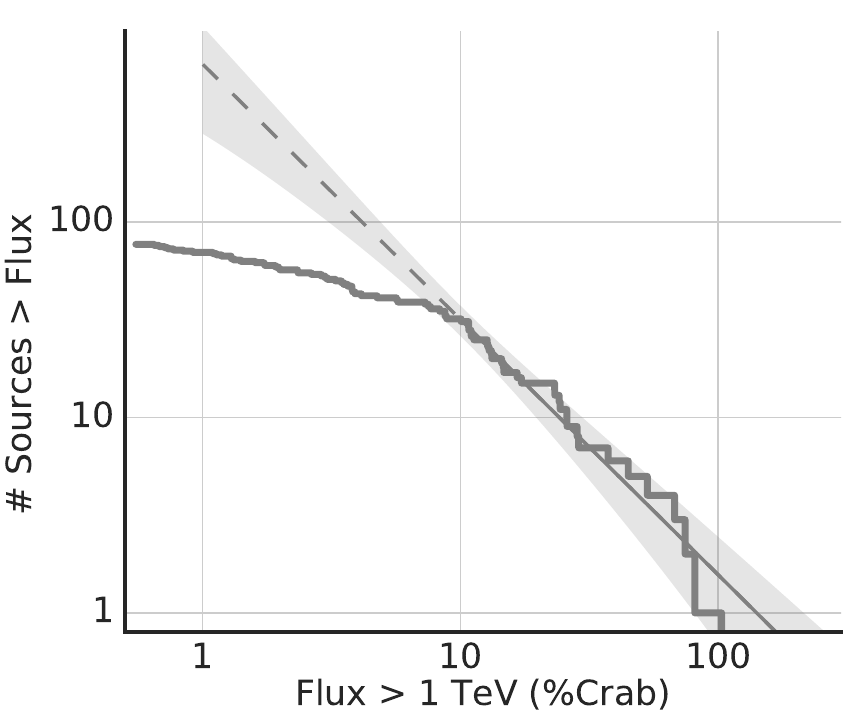}}
\caption[Source log N -- log S distributions]{
Cumulative $\log N(>S)$ -- $\log S$ distribution for the HGPS sources, showing
the number of sources $N$ above given flux thresholds $S$ (integral flux above
1~TeV in \%~Crab). The line and error band show the result of an unbinned
PL fit above a flux threshold of 10\%~Crab; the dashed line in the
1-10\%~Crab flux range illustrates the extension of the PL to
fluxes below 10\%~Crab (for comparison, not fitted in that range).
}
\label{fig:hgps_sources_log_n_log_s}
\end{figure}

We show the cumulative $\log N(>S)$ -- $\log S$ distribution of HGPS source
integral fluxes ($E > 1$~TeV, obtained from the maps) in
Fig.~\ref{fig:hgps_sources_log_n_log_s}. The \hgpsSourceCountTotal HGPS sources
span a range in flux from 0.6\% Crab to 103\% Crab; 32 sources are above 10\%
Crab. We performed an unbinned likelihood fit of a PL model to the  $\log N$ --
$\log S$ distribution (also shown in Fig.~\ref{fig:hgps_sources_log_n_log_s}),
using only the range $S > 10$\% Crab where we consider the HGPS survey mostly
complete. The best-fit value of the PL slope is $-1.3~\pm~0.2$ (for the
cumulative distribution), and the amplitude corresponds to $32\pm 5$ sources
above 10\%~Crab. This slope is consistent with Galactic models in which
equal-luminosity sources are homogeneously distributed in a thin disk, which
predict a slope of $-1.0$.\footnote{The flux $S$ of a source scales with the
distance $d$ like $S \propto L / d^2$, where $L$ is the intrinsic luminosity of
the source. For a thin disk, we have $N(>S) \propto d^2 \propto L / S$, which
corresponds to a slope of $-1.0$ in the cumulative $\log N$ -- $\log S$
distribution.}

The only robust statement that can be inferred from the $\log N$ -- $\log S$
distribution of HGPS sources is that it provides a lower limit on the true $\log
N$ -- $\log S$ distribution; that is, there are at least, for example, 70
sources above 1\%~Crab. If one assumes that $\log N$ -- $\log S$ distributions
are always concave (which most ``reasonable'' spatial distributions and source
luminosity functions encountered in the literature are), then the extrapolation
of the PL fit shown in Fig.~\ref{fig:hgps_sources_log_n_log_s} sets an upper
limit of $\sim 600$ sources above 1\%~Crab, with a statistical error of a factor
of 2.

More detailed analyses of the $\log N$ -- $\log S$ distribution or of the
flux-size distribution are possible in principle but in practice do not yield
robust results because of the limited number of sources and the large
uncertainties concerning the effective sensitivity achieved. We emphasize that
the catalog creation procedure is complex (special treatment of known shell-type
sources, large-scale emission model component, 15 discarded and several merged
components; see Sect.~\ref{sec:cc:component_classification}), with the net
effect that the sensitivities shown in Fig.~\ref{fig:hgps_sensitivity} and
panel~A of Fig.~\ref{fig:hgps_sources_flux_extension} are not reliably achieved,
because those sensitivity estimates assume isolated sources, there is no
underlying large-scale emission or source confusion, and there is a detection
threshold of $5\sigma$, whereas the component detection threshold of $TS=30$
corresponds to $\sim$~$5.5\sigma$.

A representation of the Galaxy seen face-on is depicted in
Fig.~\ref{fig:hgps_face_on_milky_way} to visualize how much of the Galaxy the
HGPS has been able to probe at different sensitivity levels. Two limits are
shown, illustrating the sensitivity detection limit (horizon) of the HGPS for
potential point-like sources with presumed luminosity of $10^{33}$ and
$10^{34}$~erg/s. Given the achieved sensitivity in the Galactic plane, it is
clear that \hess\ has only probed a small fraction of the Galaxy -- just up to a
median distance of $7.3$~kpc for bright ($10^{34}$~erg/s) point-like sources
(and less for extended sources). Furthermore, this illustrative look at survey
completeness strengthens the hypothesis that the large-scale emission described
in Sect.~\ref{sec:cc:large-scale-emission} could be partly explained by a
population of unresolved sources, presumed to be distant.

\subsection{Comparison with previous VHE publications}
\label{sec:results:previously}

In total, we reanalyzed \hgpsSourceCountReAnalysed\ VHE \gammaray\ sources that
have been the subject of past \hess\ publications. In this section we present a
systematic comparison of the present HGPS results, with the latest published
results, as summarized in \textit{gamma-cat}\footnote{\urlGammacat, accessed
July~24,~2017}, the open TeV source catalog.

We associated HGPS sources with previous analyses simply by the name of the
source, which was unique except for three cases: \object{HESS J1800$-$240},
\object{HESS J1746$-$308}, and HESS~J1930$+$188, which we discuss in detail in
Sect.~\ref{sec:results:previously:changed}. We excluded these sources from the
systematic comparison in the first place.

To further identify the cases for which we obtained significantly different
results from previously published analyses, we compared the position, size,
spectral index, and flux of the remaining uniquely associated sources, taking
statistical and systematic errors of the measurements into account. For each of
these parameters, we estimated the total uncertainty $\sigma_{\mathrm{tot}}$ as
the 1$\sigma$ statistical and systematic uncertainties added in quadrature. We
estimated this quantity for both the HGPS-derived source parameters and
previously published \hess\ values.

The systematic uncertainties on position and size are given in
Sect.~\ref{sec:cc:localisation} and Sect.~\ref{sec:cc:extension_ul},
respectively. Additionally, we assumed a systematic uncertainty
$\Delta\Gamma_{\mathrm{syst}} = 0.2$ on the spectral index and 30\% on the flux
of the source, in agreement with previous estimates \citep{ref:hesscrab}.  We
then defined the criterion for significant outliers as
\begin{equation}
\label{eq:morphoutliercriterion}
\Delta_{\mathrm{HGPS-\hess}} >
2 \sqrt{\sigma_{\mathrm{tot, HGPS}}^2 + \sigma_{\mathrm{tot,\hess}}^2}
,\end{equation}
where $\Delta_{\mathrm{HGPS-\hess}}$ is the difference between the corresponding
parameter values. When comparing the position we chose the angular separation as
comparison parameter. We note that for many sources, the data sample used here
is significantly different from that used in the publication, hence the
correlation of statistical errors is usually not too large.

We first discuss the general level of agreement between the current and previous
analyses (excluding the outliers) in Sect.~\ref{sec:results:previously:overview}
and later discuss the outliers of the comparison individually in
Sect.~\ref{sec:results:previously:changed}.

\subsubsection{Agreement with previous publications}
\label{sec:results:previously:overview}

For the vast majority of sources, we find that there is good agreement between
the HGPS-derived position, morphology, and spectrum within the statistical and
systematic uncertainties.

\paragraph*{Position\\}

We found the position of 43 (out of \hgpsSourceCountCrossChecked) sources to be
compatible with the previously published value, according to
Eq.~\ref{eq:morphoutliercriterion}. For point-like sources we found an average
shift of $0.02\pm0.01$~deg, while for extended sources the value was
$0.06\pm0.05$~deg. Both values agree well with the expected scatter considering
the statistical and systematic uncertainties on the measurements. As an
additional check, we also verified that the positions of the identified
\gammaray\ binaries (known point sources) HESS~J1826$-$148 and HESS~J1302$-$638
are in good agreement (within 40$\arcsec$) with the reference positions of the
corresponding objects LS~5039 and PSR~B1259$-$63 as listed in
SIMBAD\footnote{\urlSimbad}.

\paragraph*{Size\\}

Comparing the sizes of the extended sources we found 30 (out of 35) sources to
be compatible with the previously published value. The average size difference
for the extended sources was on the order of $\sim18$~\%, the distribution of
values having a width of $\sim40$~\%. This indicates that with the current
analysis we measured slightly larger sizes of the sources on average, but the
distribution is dominated by a large scatter. We expect the scatter to result
mainly from differences in the analysis procedure. Previous analyses mainly
fitted single Gaussian morphologies, while in this analysis we allowed for
multiple Gaussian components. Further differences are the addition of the
large-scale emission model and the systematic modeling of emission from
neighboring sources.

Previous publications found seven sources to be compatible with a point-like
source. In the current analysis we found all these sources to be compatible with
a point-like source again. Additionally, we identified the following three cases
that are compatible with a point-like source according to
Eq.~\ref{eq:extension_ul}, which were previously found to be extended:

\begin{enumerate}

\item For \object{HESS J1427$-$608} we measured a size of $0.048\pm0.009$\degr,
compared to $0.063\pm0.010$\degr\ in \cite{ref_gps_unids2008}. This source is a
borderline case that just meets our criterion for a point-like source.

\item For HESS~J1714$-$385 we found a size of $0.034\pm0.011$\degr\ compared to
$0.067\pm0.017$\degr\ in \cite{2008AA...490..685A}.  With the current analysis,
a smaller size was found because underlying emission was modeled by separate
emission components (see Fig.~\ref{fig:hgps_catalog_model}).

\item We now measure the size of \object{HESS J1808$-$204} to be
$0.058\pm0.014$\degr\ (consistent with point-like, in the definition of
Eq.~\ref{eq:extension_ul}), compared to the previously measured size
$0.095\pm0.015$\degr\ (extended) \citep{HESS:1808}. This discrepancy is due to
the HGPS's  inclusion of a large-scale emission component that now models
\gammaray\ excess previously accounted for in the source component itself.

\end{enumerate}

\paragraph*{Flux\\}

We found the flux of 42 (out of \hgpsSourceCountCrossChecked) sources to be
compatible with the previous published value, according to
Eq.~\ref{eq:morphoutliercriterion}.

The average difference in flux for extended sources was $3$~\% with a width of
$43$~\% for the distribution of values. While the average value is compatible
with previous analyses, we still found a large scatter (albeit compatible to the
systematic and statistical errors) of the distribution.

A fair comparison between flux values obtained with the current method and
earlier analyses proved to be difficult again because of fundamental differences
between the methods used. In previous publications, aperture photometry was
mostly used, while in this analysis the main flux measurement was based on a
model fit, taking the PSF and morphology of the source and large-scale emission
into account. Flux estimate differences with these two methods are shown in
Fig.~\ref{fig:hgps_sources_spectrum_map_flux_comparison} (both measures from the
HGPS analysis, not with respect to previous publications). Many of the
differences in spectra and fluxes measured in the HGPS analysis and previous
publications are the result of changes in the spectral extraction region
(position and size).

\paragraph*{Spectral index\\}

For all sources we found the spectral power-law indices to be compatible with
the previously published values. The mean difference in spectral index was
$0.04$ with a width of $0.23$ for the distribution. This is well compatible with
the expected scatter taking statistical and systematic uncertainties of the
measured spectral indices into account.

\subsubsection{Differences with previous publications}
\label{sec:results:previously:changed}

In the following paragraphs, we list and discuss the outliers as identified by
Eq.\ref{eq:morphoutliercriterion}.

\paragraph*{HESS~J0835$-$455\\}

This source (\object{Vela X}) exhibits complex morphology, and the HGPS analysis
best models the VHE emission as a superposition of three Gaussian components
with an average size $0.58\degr \pm 0.052\degr$.  This value is somewhat larger
than the value published first in \citet{Aharonian:2006d}, where it was modeled
as a single asymmetric Gaussian of size $0.48\degr \pm 0.03\degr \times
0.36\degr \pm 0.03\degr$. However, a more recent \hess\ publication
\citep{2012A&A...548A..38A} studied the complex emission more thoroughly.  It
fit profiles of the emission along two perpendicular axes, the main one aligned
with the primary orientation of the emission.  Along the major axis, the study
measured a Gaussian size $0.52\degr \pm 0.02\degr$, and along the minor axis,
two Gaussians (sizes $0.12\degr \pm 0.02\degr$ and $0.60\degr \pm 0.04\degr$)
were required to best fit the emission. The HGPS model of the emission from
HESS~J0835$-$455 is thus largely compatible with the most recent dedicated study
of the VHE emission, and the apparent discrepancy is simply a result of
comparing two different multi-component models with our general outlier
criterion (Eq.~\ref{eq:morphoutliercriterion}).

\paragraph*{HESS~J1646$-$458\\}

\object{HESS J1646$-$458} is a complex emission region located in the vicinity
of the stellar cluster \object{Westerlund 1}. Its morphology suggests it
consists of multiple sources. \citet{2012A&A...537A.114A} separated the emission
into at least two distinct features (with radii $0.35\degr$ and $0.25\degr$,
respectively) as well as some structured extended emission, distributed over the
signal region of 2.2\degr\ diameter, and even extending beyond. A flux above
1~TeV in the signal region of $7.6 \pm 1.3 \pm 1.5\times
10^{-12}$~cm$^{-2}$~s$^{-1}$ was derived, and a spectral index of $2.19 \pm 0.08
\pm 0.20$. An ON-OFF background estimation technique was used to cope with the
large source size.

In the HGPS analysis, this complex emission is modeled by a single Gaussian
component of 0.5\degr\ size shifted by 0.47\degr\ from the center of the region
used in \citet{2012A&A...537A.114A}, with a lower flux above 1~TeV of $5.48 \pm
0.46 \times 10^{-12}$~cm$^{-2}$~s$^{-1}$, and steeper index of $2.54 \pm 0.13$.
Given the complex morpology and the large scale of the spectral extraction
region used in \citet{2012A&A...537A.114A}, significant differences in source
parameters are to be expected; in the HGPS analysis part of the flux is absorbed
in the large-scale diffuse background.

\paragraph*{HESS~J1708$-$410\\}

The flux above 1 TeV of HESS~J1708$-$410 is found to be smaller in the HGPS
analysis than in \citet{ref_gps_unids2008}. While the size of the source is
similar in both cases, the different approaches used in the HGPS analysis lead
to different integration radii used to derive the source spectrum. The HGPS
analysis uses an integration radius about two times smaller than in the
dedicated analysis, which explains the apparent discrepancy.

\paragraph*{HESS~J1729$-$345\\}

For HESS~J1729$-$345, the HGPS analysis finds a flux above 1 TeV larger than in
\citet{2011AandA...531A..81H}. Because of the HGPS morphology modeling of the
source and its procedure to define the integration radius, the spectrum of this
source is derived in a region with a radius about two times larger than in the
dedicated publication, accounting for the observed difference.

\paragraph*{HESS~J1745$-$303\\}

HESS~J1745$-$303 was studied in \citet{2008A&A...483..509A} with 80~h of data.
Its morphology is complex and three subregions, called A, B, and C, were
discussed. In the HGPS analysis, with more than 160~h on the region, two
distinct sources are detected: HESS~J1745$-$303 and HESS~J1746$-$308.  The
former encloses the hotspots A and C and a fraction of region B. A second
source is now detected at $b = -1.11\degr$ latitude. This source contains part
of hotspot B and emission at large latitudes that was not significant before,
likely due to the additional livetime obtained since 2008.  It is fainter and
its spectrum is very steep but poorly constrained.  There is also a third
extended ($\sigma \sim 0.5\degr$) Gaussian component in the region. It is
currently considered to be a diffuse component. The association of the two
sources and the extended component is unclear and the exact morphology of the
VHE emission in the region will require dedicated studies.

\paragraph*{HESS~J1800$-$240\\}

In \citet{Aharonian:2008f} the emission in the region of W~28 was found to be
split into two components: HESS~J1801$-$233 (addressed below), which is not
significant in the HGPS analysis and is coincident with the W~28 SNR itself, and
a complex region HESS~J1800$-$240 offset by $0.5\degr$ to the south. The latter
was previously found to be resolved into three hotspots dubbed HESS~J1800$-$240
A, B, and C \citep{Aharonian:2008f}. Since sources HESS~J1800$-$240 A and B are
spatially coincident with molecular clouds, \citet{Aharonian:2008f} suggested
that they were produced by CRs that had escaped the SNR and had illuminated
ambient gas clouds, making this system an archetype of CR escape from evolved
SNRs \citep[see, e.g.,][]{Aharonian1996, 2015SSRv..188..187S,
2015arXiv151002102G}.

In the HGPS analysis, however, only one source is redetected, HESS~J1800$-$240,
as one large Gaussian component centered on the hotspot B. The separation into
several components does not result in a high enough \TS\ to separate it into
several significant sources in the analysis shown here.

\paragraph*{HESS~J1825$-$137\\}

HESS~J1825$-$137 is a large PWN with a bright core surrounded by extended,
asymmetric emission. The HGPS analysis finds it has a size of $0.46\degr \pm
0.03\degr$, using three Gaussian components to model the VHE entire \gammaray\
emission. This is significantly larger than the $0.24\degr \pm 0.02\degr$
obtained with a single symmetric Gaussian model or the $0.23\degr \pm 0.02\degr
\times 0.26\degr \pm 0.02\degr$ with a single asymmetric Gaussian in
\citet{Aharonian:2006g}. These models were stated to have rather poor $\chi^2$
goodness-of-fit values. The more complex approach taken for the morphology
modeling in the HGPS improves the description of the \gammaray\ emission from
this PWN and accounts for the differences with respect to previous, simpler
modeling.

\paragraph*{HESS~J1837$-$069\\}

The HGPS analysis finds HESS~J1837$-$069 to have a size of $0.36\degr \pm
0.03\degr$ based on modeling the VHE \gammaray\ emission as three Gaussian
components.  This is larger than the size previously derived using a single
asymmetric Gaussian \citep{ref:gps2006}, i.e., $0.12\degr$ by $0.05\degr$; and
using a single Gaussian \citep{2008AIPC.1085..320M}, i.e., $0.22\degr$. The more
complex modeling of the HGPS, which also takes into account more of the extended
nebular emission from this identified PWN, explains the apparent discrepancy.
Consequently, we used a larger region (twice the radius compared to
\citet{ref:gps2006}) to derive the spectrum, leading to an integral flux above
1~TeV that is larger by a factor of $\sim$3 than in the dedicated publication.

\paragraph*{HESS~J1857$+$026\\}

The size of the source \object{HESS J1857$+$026} is significantly larger in this
analysis than previously published in \citet{ref_gps_unids2008}.  In the latter,
the source is fit with an asymmetric Gaussian ($0.11\degr \pm 0.08\degr \times
0.08\degr \pm 0.03\degr$), whereas the HGPS analysis best models the source with
two Gaussian components for an approximate size of $0.26\degr \pm 0.06\degr$.
The difference in size is explained by the multicomponent approach of the HGPS
that better takes into account the larger scale emission underneath the central
bright core.

\paragraph*{\object{HESS J1908$+$063}\\}

The position and size published in \citet{2009A&A...499..723A} are significantly
different from those obtained in the HGPS analysis. The position is offset by
$0.17\degr$ and the size is found to be $0.48\degr \pm 0.04\degr$, which is
$0.14\degr$ larger. We note that the size we find is consistent with that
measured by the VERITAS Collaboration~\citep{2014ApJ...787..166A}, even though
the positions differ by $0.3\degr$. A plausible cause for these discrepancies is
that this is a large source likely composed of multiple components, where
results are expected to be sensitive to the morphology assumptions and to
details in background modeling techniques, in particular, if those tend to
absorb large-scale features.

\paragraph*{HESS~J1923$+$141\\}

The VHE \gammaray\ source \object{HESS J1923$+$141} \citep[preliminary \hess\
results published in][]{W51:HESS_ICRC} is spatially coincident with the
\object{W 51} region, studied in detail with the MAGIC IACT \citep{MAGIC:W51}.
The HGPS results are generally compatible with those from MAGIC. However, the
latter shows evidence for a \gammaray\ source composed of two components above
1~TeV, which cannot yet be confirmed by \hess\ One component is coincident with
the interaction region between \object{W 51C} and \object{W 51B}, while the
other is coincident with the potential PWN \object{CXOU J192318.5$+$140305}
\citep{Koo2005}, suggesting that HESS~J1923$+$141 may be a composite of VHE
emission of different astrophysical origins.

\paragraph*{HESS~J1930$+$188\\}
\label{sec:HESS_J1930p188}

The VHE \gammaray\ source, discovered with VERITAS \citep[with the identifier
VER~J1930$+$188,][]{2010ApJ...719L..69A}, is coincident with the composite
\object{SNR G54.1$+$0.3} and the pulsar \object{PSR J1930$+$1852}. We report on
the \hess\ observations of this source for the first time here. The HGPS source
is found to have a slightly displaced position from the pulsar and the VERITAS
best fit (by $0.04\degr$).  Despite the agreement with the VERITAS spectral
index, the integral flux above 1 TeV found in our analysis is $\sim$40\% lower
than their published flux. We note, however, that the apparent discrepancy with
VERITAS is not confirmed by our cross-check analysis, which yields a flux for
this source that is larger by more than the nominal 30\% systematic flux
uncertainty, and is in agreement with the VERITAS measurement.

\subsubsection{Sources not redetected}
\label{sec:results:previously:missing}

In total, there are \hgpsSourceCountMissingStr\ previously published VHE
\gammaray\ sources that are not redetected with the current HGPS analysis. All
of these are rather faint sources which, for the HGPS analysis, yield
significances close to the HGPS detection threshold of $TS=30$. We consider
these as real sources of \gammarays; the nondetection in the HGPS is primarily a
result of differences between the HGPS analysis and specific analysis methods.
We found that some of the most relevant differences are

\begin{enumerate}

\item event reconstruction and \gammaray-hadron separation cuts that are less
sensitive compared to more specialized methods that have been used in individual
source analyses;

\item higher energy threshold in the HGPS analysis, in conjunction with a soft
spectrum of the tested source;

\item use of the $2\degr$ FoV offset cut (see Sect.~\ref{sec:events_map}), which
is tighter than the value used in many previous \hess\ publications ($2.5\degr$
or even $3\degr$).

\end{enumerate}

In addition, the use of a large-scale emission model and the modeling of nearby
extended components and overlapping sources modifies the measured flux and hence
the significance of a source compared to previous analyses, where larger scale
background features were accounted for in different ways (e.g., partly absorbed
in the ring background). Given these differences, it is not surprising that few
faint sources fail the HGPS detection criteria.

In the following paragraphs, we describe the individual cases in more detail.
For completeness, we added all missing sources to the final HGPS catalog;
the source parameters were taken from the corresponding publication (see also
Table~\ref{tab:hgps_external_sources}).

\paragraph*{HESS~J1718$-$374 and HESS~J1911$+$090\\}

The VHE \gammaray\ sources HESS~J1718$-$374 and HESS~J1911$+$090
(Figs.~\ref{fig:hgps_survey_mwl_2} and \ref{fig:hgps_survey_mwl_1}) were
previously detected toward the \object{SNR G349.7$+$0.2} and W~49B /
SNR~G43.3$-$0.2, respectively. Both are thought to result from interactions
with molecular clouds and exhibit correspondingly steep (soft) spectra, which
have PL indices $\Gamma = 2.80\pm0.27$ \citep{2015AandA...574A.100H} and
$3.14\pm0.24$ \citep{HESS:W49}, respectively. The energy threshold of the
analyses is therefore key to detecting these sources. As described in
Sect.~\ref{sec:maps}, the maps that serve as a starting point for the source
catalog have been produced using the hard cuts configuration and a conservative
safe energy threshold, explaining the lack of detection of these sources in the
HGPS analysis.

\paragraph*{HESS~J1741$-$302\\}

The unidentified source HESS~J1741$-$302 is located on the Galactic plane ($b =
0.05\degr$) and $\sim$1.7\degr\ away from the Galactic center. With an integral
flux of $\sim$1\% Crab above 1~TeV it is one of the faintest \hess\ sources
detected so far \citep{HESS:1741}. Because of the addition of the large-scale
emission model in the HGPS analysis, HESS~J1741$-$302 does not reach the HGPS
$TS=30$ detection threshold.

\paragraph*{HESS~J1801$-$233\\}

HESS~J1801$-$233 is part of the HESS~J1800$-$240 and HESS~J1801$-$233 source
complex discussed above, characterizing emission features of the SNR~W~28 region
\citep{Aharonian:2008f}. The emission was found to be split into two components:
HESS~J1801$-$233, which is coincident with the northeastern boundary of W~28
where the shockwave is interacting with a molecular cloud, and a complex region
HESS~J1800$-$240 offset by $0.5\degr$ to the south. HESS~J1801$-$233 does not
reach the $TS=30$ threshold and is therefore not found to be significant in the
HGPS analysis. We note that the \gammaray\ emission from W~28 is bright in the
GeV range and is clearly detected above 50~GeV \citep{2FHL}.  It has a  steep
spectral index of $2.7\pm 0.3$ at VHE \citep{Aharonian:2008f}. It is therefore
not detected here because of our higher analysis energy threshold (about
400\,GeV at a longitude of $7\degr$, see
Fig.~\ref{fig:hgps_energy_threshold_profiles}) and because of the inclusion of
the large-scale emission model in our analysis, which reduces the significance
of such a faint source. Furthermore, we reiterate that HESS~J1800$-$240 is
detected in the HGPS as one large Gaussian source, see
Sect.~\ref{sec:results:previously:changed}, rather than three individual
hotspots as in \citet{Aharonian:2008f}. This potentially also contributes to a
reduction of the significance of this previously established source
HESS~J1801$-$233.

\subsection{Comparison with the cross-check analysis}
\label{sec:results:xcheck}

For most sources, the spectral fit results reported in this catalog agree with
those obtained from the independent cross-check analysis (see
Sect.~\ref{sec:cc:discussion}). For the following sources, however, larger
differences, exceeding the systematic errors, are observed. Several factors
could explain these differences, such as the lower energy threshold in the
cross-check analysis, the differences in the morphology models, or the fact that
the cross-check spectrum analysis is run for the positions and sizes obtained
with the main analysis.

\begin{itemize}

\item HESS~J1503$-$582 (see Sect.~\ref{sec:HESS_J1503m582}): While the spectral
indices are compatible, the derived integral flux above 1 TeV is about two times
higher in the main analysis than in the cross-check analysis.

\item HESS~J1646$-$458 (Westerlund 1): the cross-check analysis gives a spectrum
that is about two times brighter around 1~TeV, with a curvature or cutoff
leading to similar fluxes as the main analysis at low and high energies. We
would like to stress again that the HGPS analysis for this source is not very
reliable, because the source size is similar to the \hess\ field of view and a
more careful individual study and background estimation is needed, as explained
in Sect.~\ref{sec:results:previously:changed} which points out the differences
with the previously published measurement.

\item \object{HESS J1718$-$385}: For both the main and cross-check analyses, the
preferred spectral model is a power law with an exponential cutoff. The cutoff
energies are compatible and the spectra are in agreement above $\sim$ 3 TeV.
However, below this energy, some discrepancy is observed as the main analysis
spectral fit yields a spectral index that is harder than in the cross-check
analysis, resulting in an integral flux above 1 TeV about two times lower in the
main analysis than in the cross-check analysis.

\item HESS~J1729$-$345: While the derived spectral indices are compatible, the
integral flux above 1 TeV is about two times higher in the cross-check analysis
than in the main analysis.

\item HESS~J1746$-$308: The large spectral index derived from the main analysis
could not be confirmed by the cross-check analysis. The differential flux values
at 1 TeV are compatible, but the discrepancy in the obtained spectral indices
leads to an integral flux above 1 TeV about two times higher in the cross-check
analysis than in the main analysis.

\item \object{HESS J1852$-$000} (see Sect.~\ref{sec:HESS_J1852m000}): The derived
spectral indices are compatible, but the integral flux above 1 TeV is about two
times higher in the cross-check analysis than in the main analysis.

\end{itemize}

Spectral model results for these six sources should therefore be treated with
caution.

\subsection{New VHE sources}
\label{sec:results:new}

\begin{figure*}
\includegraphics[width=\textwidth]{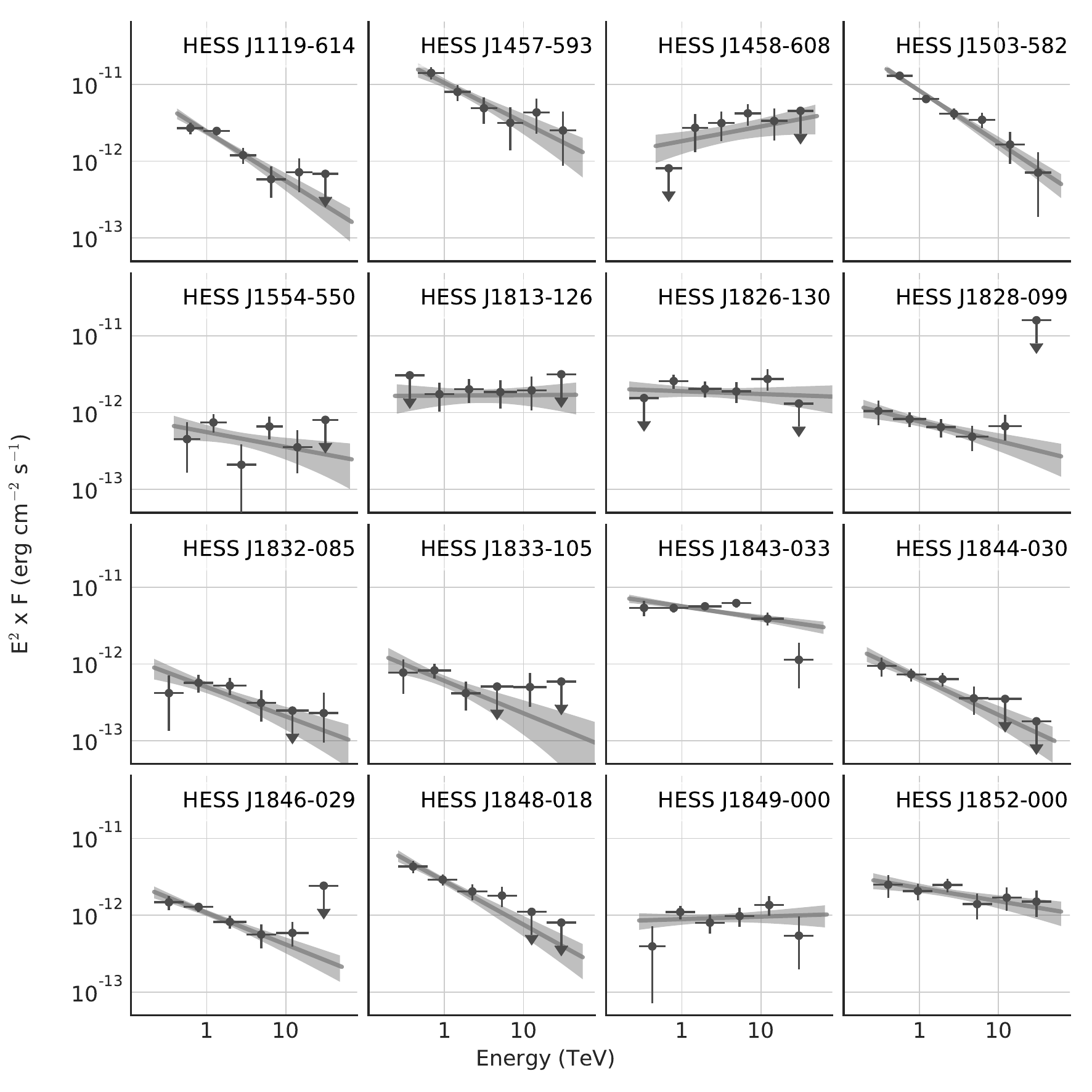}
\caption[Spectra of the new sources]{
Fitted power-law spectral models with uncertainty bands and flux points for new
sources.
}
\label{fig:hgps_spectra_new_sources}
\end{figure*}

During the construction of the HGPS catalog, statistically significant VHE
\gammaray\ emission was detected from \hgpsSourceCountNew sources which were not
previously known or for which only preliminary detections had been published
(e.g. in conference proceedings). All of these new sources are confirmed by the
cross-check analysis --- we do not expect any of these new sources to be a false
detection (see Sect.~\ref{sec:cc:components} and
\ref{sec:cc:component_classification}) The morphological and spectral properties
of these new, confirmed VHE sources are provided in
Tables~\ref{tab:hgps_catalog}~and~\ref{tab:hgps_spectra}, their spectra are
shown in Fig.~\ref{fig:hgps_spectra_new_sources}. Each new source is also
briefly described in the following sections, in the context of its MWL
environment and possible origin of the VHE \gammarays.

\subsubsection{HESS~J1119$-$614}
\label{sec:HESS_J1119m614}

\begin{figure}
\resizebox{\hsize}{!}{\includegraphics{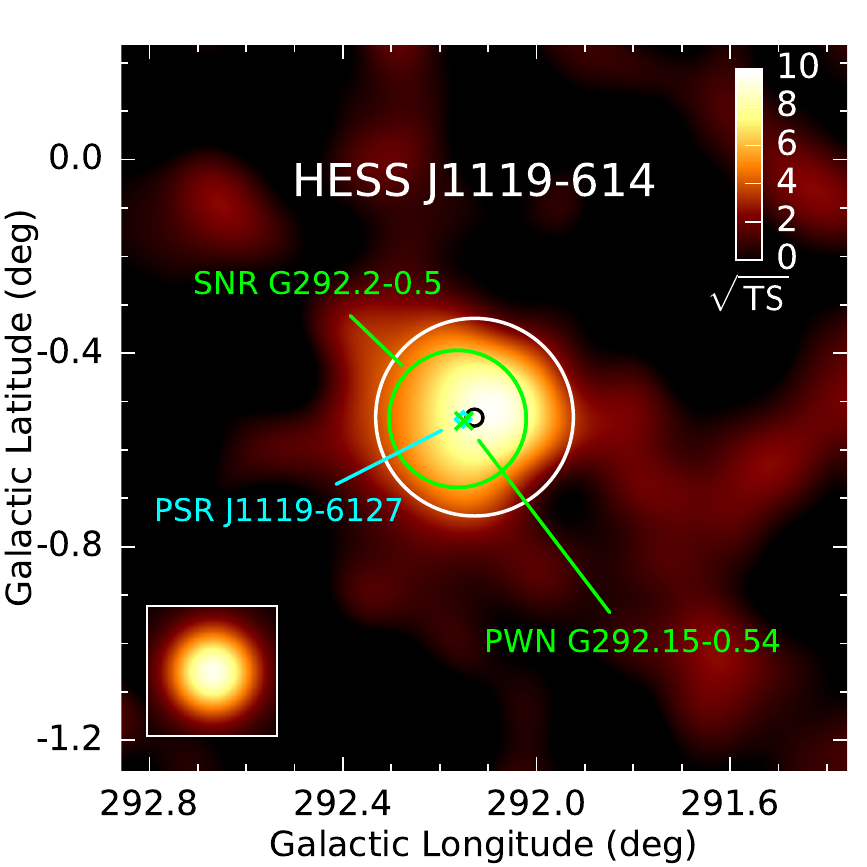}}
\caption[New source image: HESS~J1119$-$614]{
Significance ($\approx\sqrt{TS}$) of the VHE \gammaray\ excess, centered on the
new source HESS~J1119$-$614, with the \hess\ The PSF for this data set shown
inset. The black circle at the center indicates the 68\% uncertainty in the
best-fit centroid of the VHE emission. The white circle represents the 70\%
containment region of the emission (\texttt{R\_SPEC}, used also for spectral
measurement). The approximate size of the radio shell of \object{SNR
G292.2$-$0.5} is shown as a green circle and the \object{PWN G292.15$-$0.54} as
a green marker. The position of the pulsar \object{PSR J1119$-$6127} is denoted
by a cyan diamond. The FoV is $1.5\degr \times 1.5\degr$.
}
\label{fig:HESS_J1119m614}
\end{figure}

We confirm the discovery of VHE \gammaray\ emission from HESS~J1119$-$614
(Fig.~\ref{fig:HESS_J1119m614}) and identify it as the composite SNR
G292.2$-$0.5. We base the firm identification on the basis of spatial
coincidence with the SNR and its associated PWN G292.15$-$0.54 and highly
magnetized pulsar PSR~J1119$-$6127. \hess\ previously published
\citep{Djannati-Atai09} preliminary source properties that are compatible with
the HGPS results.

A compact (size $6\arcsec \times 15\arcsec$), nonthermal PWN has been detected
in X-rays \citep{Gonzalez03,Safi-Harb08} and is considered a candidate PWN in HE
\gammarays\ \citep{Acero13}.  It is powered by the energetic pulsar PSR
J1119$-$6127, with spin-down luminosity $\dot{E} = 2.3 \times
10^{36}$~erg~s$^{-1}$ and distance $d = 8.4 \pm 0.4$~kpc \citep{Caswell04}. The
pulsar has been detected in radio \citep{Camilo00} and HE \gammarays\
\citep[][as \object{3FGL J1119.1$-$6127} in the latter]{Parent11,3FGL} and is
characterized by a relatively high surface B-field ($4.1 \times 10^{13}$~G).
Despite it being a rotation-powered pulsar, it has recently joined the other
high-B pulsar \object{PSR J1846$-$0258} in revealing a magnetar-like behavior
\citep{2016ApJ...829L..25G, 2016ATel.9378....1Y, 2016ATel.9282....1A}. It is
further notable for being among the handful of pulsars for which braking indices
have been measured, in this case $n = 2.684 \pm 0.002$ \citep{Weltevrede11}, as
opposed to simply assuming $n = 3$, giving a more precise characteristic age
$\tau_{\mathrm{c}} = \frac{P}{(n-1)\dot{P}} = 1.9$~kyr, where $P$ and $\dot{P}$
are the currently measured period and period derivative, respectively.

Considering the luminosity of HESS~J1119$-$614, $L_{\gamma}(\mathrm{1-10~TeV}) =
2.4\times10^{34} (d/8.4\mathrm{kpc})^2 $~erg~s$^{-1}$, the apparent efficiency
of converting the pulsar's rotational energy to \gammarays,
$\epsilon_{\mathrm{1-10~TeV}} \equiv L_{\gamma} / \dot{E} = 1.1\%$, is
compatible with the efficiencies ($\la 10\%$) of other VHE sources that have
been identified as PWNe \citep{Kargaltsev13}. The offset of the VHE emission
from this young pulsar, where the X-ray PWN is located, is not statistically
significant with respect to the uncertainty on the best-fit VHE centroid ($\pm
0.02\degr$).

The age of SNR~G292.2$-$0.5 is in the range 4.2$-$7.1~kyr \citep{Kumar12}. This
can be reconciled with the characteristic age of the pulsar if the braking
index~$n$ was much smaller than the current value until recently. This
assumption is reasonable in light of recent evidence for erratic radio timing
behavior from the pulsar \citep{Weltevrede11}. The X-ray emission from the SNR
is predominantly thermal and has an additional hard, nonthermal, X-ray
component. This nonthermal emission is likely from the PWN, although an origin
in the SNR reverse shock could not be ruled out \citep{Kumar12}.

The X-ray spectral measurements suggest the SNR is generally expanding in a
low-density medium, appearing to disfavor a hadronic origin for the VHE
\gammarays\ \citep{Drury94}. However, there is also evidence for localized,
high-density regions near the eastern SNR shell, including dark clouds and CO
features \citep{Kumar12}. We cannot confirm the claim by \cite{Kumar12}, based
on preliminary \hess\ results \citep{Djannati-Atai09}, that no VHE emission is
detected from the eastern SNR shell, as it is well within the VHE emission
region in the HGPS analysis.

In conclusion, while the identification with the composite SNR and PWN system is
firm, it is not yet clear whether the VHE emission originates in the SNR shock,
either leptonically, from the shell itself, or hadronically, from interactions
with ambient media; the PWN; or some combination thereof.

\subsubsection{HESS~J1457$-$593}
\label{sec:HESS_J1457m593}

\begin{figure}
\resizebox{\hsize}{!}{\includegraphics{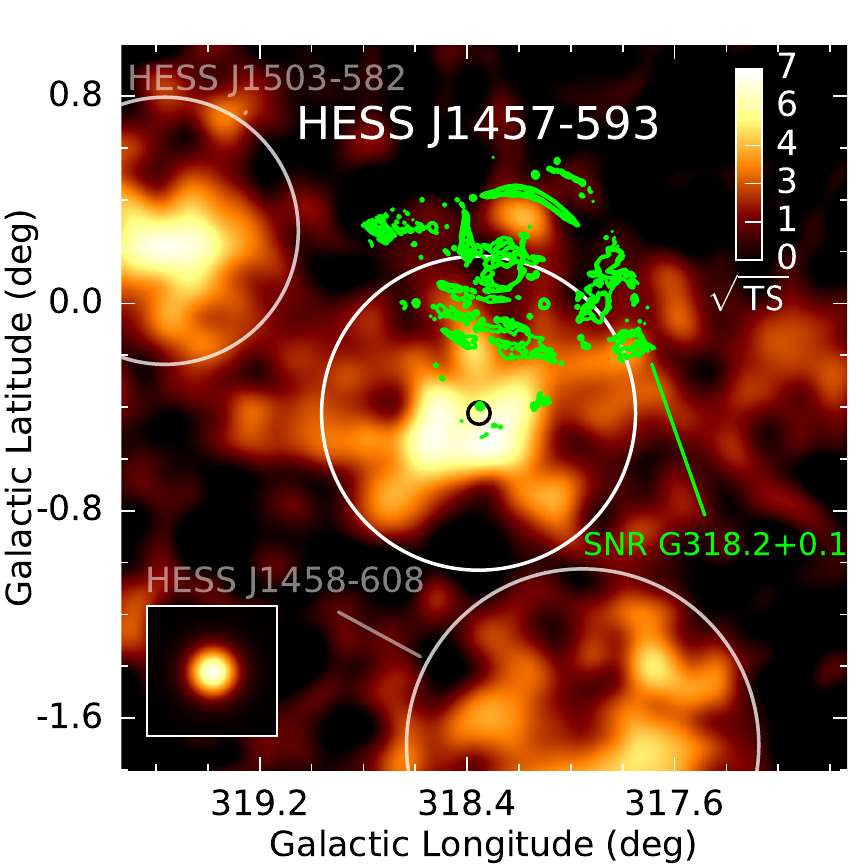}}
\SingleSourceCaption{HESS~J1457$-$593}{
Additionally, the SNR G318.2$+$0.1 is shown by plotting its 843-MHz radio
intensity \citep{Whiteoak96} with contours at 4, 8, and 12~mJy/beam. The FoV is
$2.8\degr \times 2.8\degr$.
}
\label{fig:HESS_J1457m593}
\end{figure}

VHE \gammaray\ emission from the new source HESS~J1457$-$593
(Fig.~\ref{fig:HESS_J1457m593}) is associated with the \object{SNR
G318.2$+$0.1}, on the basis of a spatial coincidence with a shell-type SNR and
lack of other potential MWL counterparts. Preliminary \hess\ morphological
properties were initially published by \cite{Hofverberg10}. The HGPS source
position is compatible with the preliminary position; however, the size of the
source in the catalog is different because of a difference in the assumed
morphological model. Previously, the source was modeled as an  asymmetric
Gaussian ($0.31\degr \pm 0.07\degr$ by $0.17\degr \pm 0.05\degr$) whereas the
HGPS source is modeled, like all HGPS sources, as a symmetric Gaussian
($0.33\degr \pm 0.04\degr$). Nonetheless, the spatial overlap between
HESS~J1457$-$593 and the southern part of the SNR shell still holds.

G318.2+0.1 is observed as a relatively large ($40^{\prime} \times 35^{\prime}$)
shell in radio \citep[e.g.,][]{Whiteoak96}, which is characterized by two
arc-like, nonthermal filaments in the northwest and southeast (SE) that together
form the shell.  The VHE emission is much larger than the SNR shell, and the VHE
centroid is significantly offset ($\sim$0.4\degr) from the SNR center, although
it is partially coincident with the SE rim of the shell. Furthermore, there is
evidence in $^{12}$CO \citep{Dame01} of a giant molecular cloud (GMC) at $(\ell,
b) \approx (318.4\degr, -0.5\degr)$ coincident with both the VHE emission and
the SE rim; this GMC is $1.8\degr \times 1.1\degr$ (average physical size 80 pc)
in size and has mass $\sim$3$\times 10^5$~M$_{\odot}$ and density
$\sim$40~cm$^{-3}$, assuming the near solution of the kinematic distance $3.5
\pm 0.2$~kpc \citep{Hofverberg10}.  Little is known about G318.2$+$0.1 itself,
but assuming it is at the same distance as the GMC and further assuming a
Sedov-Taylor model for the SNR evolution, its physical diameter would be
$\sim$40~pc and its age $\sim$8~kyr. These data suggest a plausible SNR and
molecular cloud interaction scenario \citep[e.g.,][]{Gabici07}, where particles
are accelerated in the shell, escape, and interact with a nearby but offset MC,
producing \gammarays\ via hadronic p-p collisions.

An X-ray study of the SNR with \emph{BeppoSAX} and \emph{ROSAT} did not find
evidence for shell-like, nonthermal emission, nor thermal X-ray emission that
should trace the interaction between the SNR and ISM \citep{Bocchino01}.
However, several hard X-ray sources were found, suggestive of at least localized
nonthermal electron acceleration. Additional MWL observations and spectral
modeling are required to further investigate the scenario responsible for the
production of VHE \gammarays.

\subsubsection{HESS~J1458$-$608}
\label{sec:HESS_J1458m608}

\begin{figure}
\resizebox{\hsize}{!}{\includegraphics{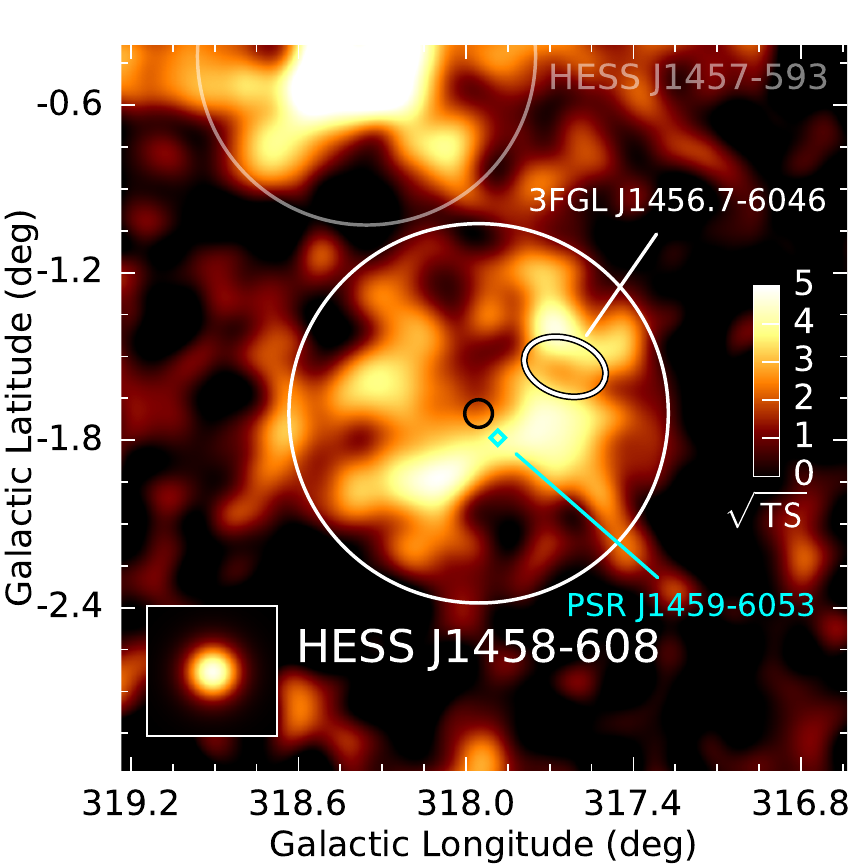}}
\SingleSourceCaption{HESS~J1458$-$608}{
Additionally, the ellipse represents the 95\% uncertainty in the position of the
HE \gammaray\ point source 3FGL~J1456.7$-$6046, and the cyan diamond indicates the
position of the pulsar. The FoV is $2.6\degr \times 2.6\degr$.
}
\label{fig:HESS_J1458m608}
\end{figure}

VHE \gammaray\ emission from the new source \object{HESS J1458$-$608}
(Fig.~\ref{fig:HESS_J1458m608}) is associated with the pulsar \object{PSR
J1459$-$6053} and can likely be identified as a heretofore undetected PWN, on
the basis of a spatial coincidence with an energetic pulsar and the absence of
other plausible MWL counterparts. Preliminary VHE morphological and spectral
properties were first announced by \cite{delosReyes12}. The updated
morphological properties from the HGPS catalog differ from those preliminary
ones, which had underestimated the extent of the large, complex emission region
($0.37\degr \pm 0.03\degr$ vs. $0.17\degr \pm 0.07\degr$; both morphological
models 2D symmetric Gaussian), likely due to the irregular shape of the
emission. Previously there was a hint for additional structure, possibly a
second source hidden in the tail of a dominant source, but this remains
statistically insignificant in the HGPS analysis with respect to a single-source
Gaussian morphology. Also of note, the best-fit centroid of the VHE emission is
now located closer to the \gammaray\ pulsar ($0.11\degr$ versus $0.16\degr$
offset), bolstering the scenario in which the VHE emission is interpreted as a
PWN powered by the pulsar. As expected for such changes in morphological
properties, the HGPS spectral results also differ from the previously derived
preliminary values.

The pulsar PSR~J1459$-$6053 (also \object{3FGL J1459.4$-$6053}) is a relatively
old ($\tau_{\mathrm{c}} = 65$ kyr) but still very energetic HE $\gamma$-ray
pulsar with a spin-down luminosity $9.1 \times 10^{35}$ erg s$^{-1}$  and
unknown distance ($d < 22$ kpc) \citep{Abdo13}. As noted above, it is offset
$0.11\degr$ from the VHE centroid, which is consistent with offsets observed in
other PSR and VHE PWN systems \citep[e.g.,][]{Kargaltsev13}. The putative PWN
has not been detected in X-rays potentially because of the age of the system
\citep{Ray11} or HE $\gamma$-rays \citep{Acero13}.

The new VHE spectrum ($E > 0.46$ TeV) is consistent with the 31-316 GeV \fermi\
upper limits. However, the conclusion, made by \cite{Acero13}, that the peak of
the PWN's inverse Compton emission is located in this energy range has to be
revised as the peak can now only be inferred to be at higher energies.

Apart from the HE \gammaray\ pulsar, there is a second HE source (\object{3FGL
J1456.7$-$6046}) in the FoV. However, it is unclear if it is related to the PSR and
PWN scenario, since it exhibits a highly curved, log-parabolic spectrum typical
of blazars and a TS that fluctuates strongly with the choice of diffuse model or
analysis method \citep{3FGL}.

\subsubsection{HESS~J1503$-$582}
\label{sec:HESS_J1503m582}

\begin{figure}
\resizebox{\hsize}{!}{\includegraphics{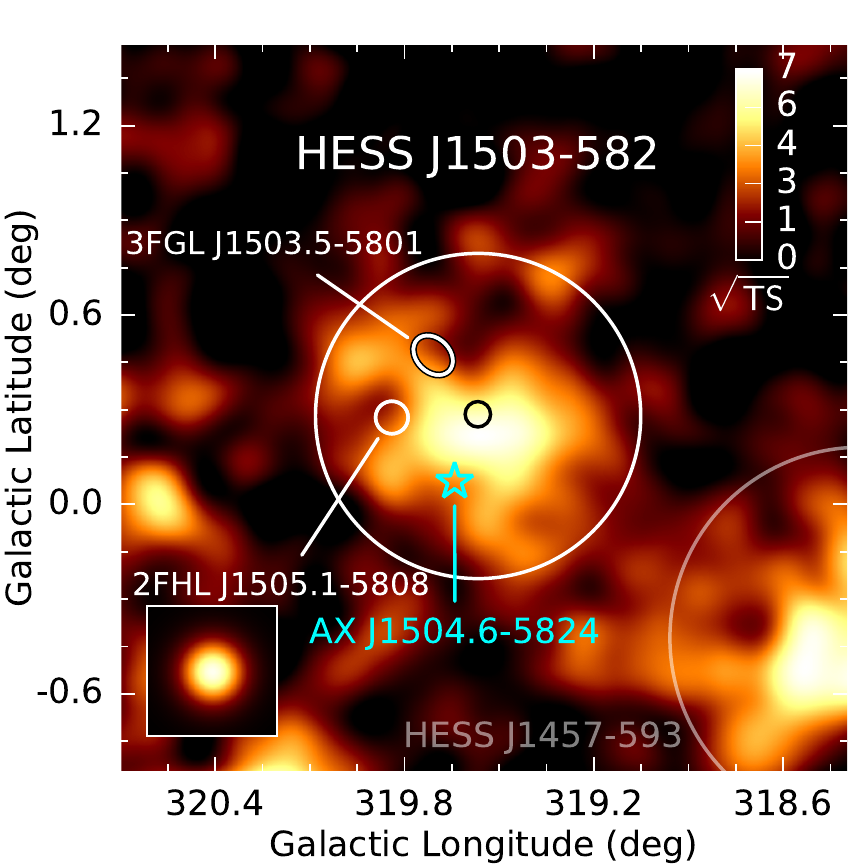}}
\SingleSourceCaption{HESS~J1503$-$582}{
Additionally, the ellipse represents the 95\% uncertainty in the position of the
HE \gammaray\ point source 3FGL~J1503.5$-$5801; the circle represents the 68\%
uncertainty in the position of the HE ($E > 50$~GeV) \gammaray\ point source
2FHL~J1505.1$-$5808; and the star represents the location of the X-ray point
source. The FoV is $2.3\degr \times 2.3\degr$.
}
\label{fig:HESS_J1503m582}
\end{figure}

HESS~J1503$-$582 (Fig.~\ref{fig:HESS_J1503m582}) is a new source for which the
origin of the VHE \gammaray\ emission is unidentified.  \hess\ earlier announced
preliminary morphological and spectral properties for this source
\citep{Renaud08}, which are now superseded by those in this paper. The VHE
emission appears to be one of the softest ($\Gamma = 2.68 \pm
0.08_{\mathrm{stat}}$) in the HGPS, although both its morphological and spectral
properties are affected by systematic uncertainties larger than nominal (see,
e.g., Sect.~\ref{sec:cc:discussion} and Sect.~\ref{sec:results:xcheck}).

A point-like HE ($E > 50$~GeV) \gammaray\ source, 2FHL~J1505.1$-$5808
\citep{2FHL}, is spatially coincident with the VHE emission region.  A
comparison of the VHE and HE ($E > 50$~GeV) spectra suggests that it may be a
PWN \citep{2FHL}, although no PWN or energetic pulsar has been detected so far.
Another, different, point-like HE ($E > 100$~MeV) \gammaray\ source,
3FGL~J1503.5$-$5801 \citep{3FGL}, is also within the VHE region. Its nature is
unknown, but its log-parabolic spectrum suggests it may not be directly related
to HESS~J1503$-$582.

Faint X-ray emission \citep[\object{AX J1504.6$-$5824},][]{Sugizaki01} is
present toward the edge of the VHE emission.  Nominally cataloged as a
cataclysmic variable, its X-ray properties are not well known owing to the low
\emph{ASCA} sensitivity. Analysis of more sensitive data from other X-ray
telescopes is needed to investigate the possibility it may be a PWN; this is the
case despite a lack of an energetic pulsar in the vicinity, but bearing in mind
the unknown nature of the nearby 3FGL source.

A relatively comprehensive search of MWL archives \citep{Renaud08} led to the
investigation of an atypical scenario where the VHE emission could be linked
with a forbidden velocity wing (FVW): faint, characteristic 21-cm \hone\ line
emission structures seen as deviations from the canonical Galactic rotation
curve \citep{Kang07}. The hypothesis is that this FVW, \object{FVW 319.8$+$0.3},
may be related to an older SNR in its radiative phase, as was the case for two
other FVWs \citep{Koo06,Kang12}. Although the SNR would no longer have
sufficient shock velocity to accelerate particles responsible for producing VHE
\gammarays\ \citep{Ptuskin05}, it could nevertheless be indicative of increased
or more recent activity in the region (stellar winds and/or supernova
explosions). A large \hone\ shell, the result of such activity, is nearby
\citep[see][\object{GSH 319$-$01$+$13}]{McClure-Griffiths02}; however, its
centroid is substantially offset by more than 1$\degr$ from HESS~J1503$-$582 and
its extent considerably larger than the VHE emission region, so it seems
unlikely to be related.

On the other hand, VERITAS also searched for VHE emission from an FVW, one which
does show clear shell-type emission in \hone\ (\object{FVW 190.2$+$1.1}).
Despite observations that reached a sensitivity better than 1\%~Crab, VERITAS
did not detect any significant VHE emission \citep{Holder09}. Furthermore, there
is no definitive identification of VHE emission from young stellar clusters,
with the possible exception of the superbubble \object{30 Dor C} in the LMC
\citep{HESS15_LMC}. Therefore, this FVW scenario remains speculative.

\subsubsection{HESS~J1554$-$550}
\label{sec:HESS_J1554m550}

\begin{figure}
\resizebox{\hsize}{!}{\includegraphics{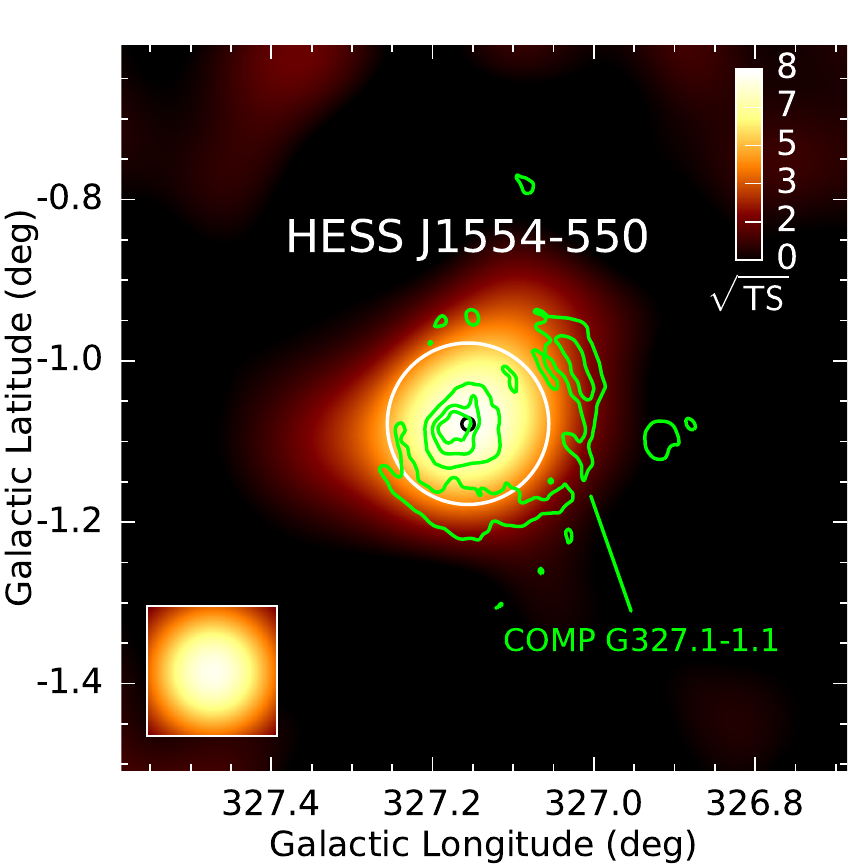}}
\SingleSourceCaption{HESS~J1554$-$550}{
Additionally, the composite SNR is shown by plotting its 843-MHz radio intensity
\citep{Whiteoak96} with contours at 1, 8, and 15~mJy/beam. The FoV is $0.9\degr
\times 0.9\degr$.
}
\label{fig:HESS_J1554m550}
\end{figure}

VHE \gammaray\ emission from the new source HESS~J1554$-$550
(Fig.~\ref{fig:HESS_J1554m550}) is firmly identified with the \object{PWN
G327.15$-$1.04} within the composite \object{SNR G327.1$-$1.1}, on the basis of
both a spatial coincidence with the PWN and the size of the VHE emission region,
which can be constrained to less than $0.035\degr$. Preliminary \hess\
morphological and spectral properties of the VHE source were first published by
\citet{Acero:2012} and are compatible with the HGPS results. However, while
previously the source size was given as $0.03\degr \pm 0.01\degr$, the more
conservative HGPS analysis procedure used here (Sect.~\ref{sec:cc:extension_ul})
finds the source to be compatible with being point-like, and there is a limit on
the size that is nonetheless compatible.

The VHE size limit rules out significant emission from the outer shell of the
SNR and is compatible with the compact, nonthermal PWN, which is observed in
both radio and X-rays \citep{Temim09,Temim15} but not HE \gammarays\
\citep{Acero13,2FHL}. Furthermore, the VHE centroid is compatible with the peak
of the radio emission from the PWN and the tail of X-ray PWN. Although pulsed
emission from the putative pulsar at the heart of the composite SNR has not been
detected in radio, X-ray, nor HE \gammaray\ bands, the X-ray data provide
evidence for the existence of a powerful pulsar, that has an estimated $\dot{E} =
3.1 \times 10^{36}$~erg~s$^{-1}$ \citep{Temim15}. The distance to the SNR is not
well determined, but has been estimated to be roughly 9~kpc \citep{Sun99}.
Assuming this distance, the VHE luminosity of HESS~J1554$-$550 is
$L_{\gamma}(1-10~\mathrm{TeV})=1.0\times10^{34}(d/9~\mathrm{kpc})^2$~erg~s$^{-1}$
and the apparent efficiency $\epsilon_{\mathrm{1-10 TeV}} \equiv L_{\gamma} /
\dot{E} = 0.3\%$, which is compatible with the efficiencies ($\la 10\%$) of
other VHE sources that have been identified as PWNe \citep{Kargaltsev13}.

\subsubsection{\object{HESS J1813$-$126}}
\label{sec:HESS_J1813m126}

\begin{figure}
\resizebox{\hsize}{!}{\includegraphics{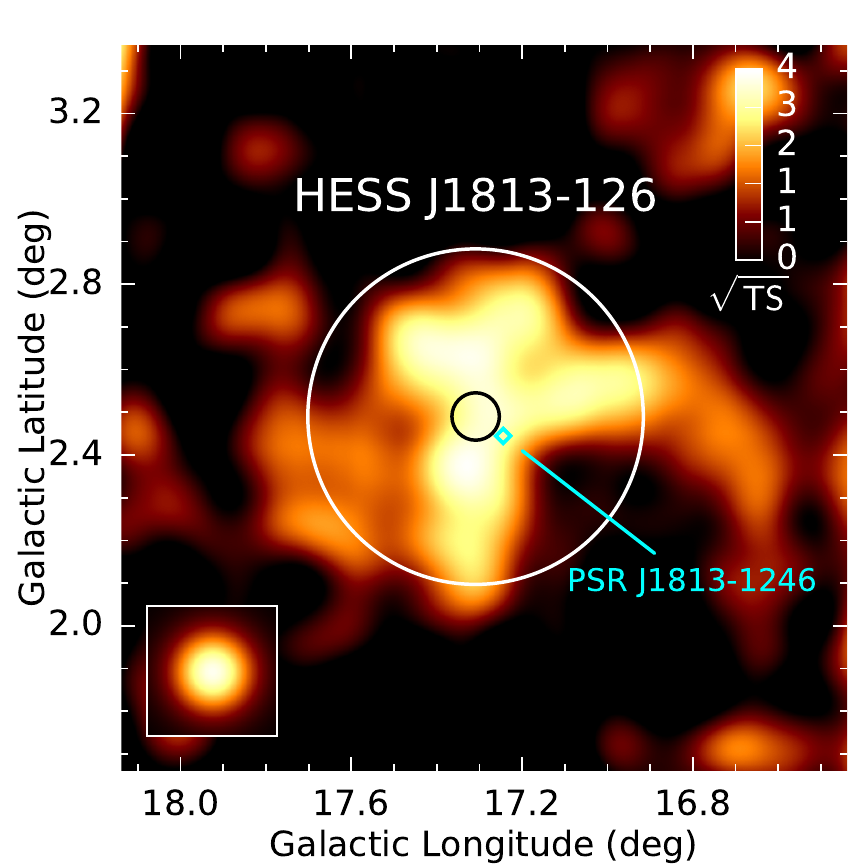}}
\SingleSourceCaption{HESS~J1813$-$126}{
Additionally, the position of the pulsar is denoted by a cyan diamond. The FoV is
$1.7\degr \times 1.7\degr$.
}
\label{fig:HESS_J1813m126}
\end{figure}

The HGPS catalog analysis has revealed an intriguing new source of VHE
\gammarays\ (Fig.~\ref{fig:HESS_J1813m126}) not previously detected, one of the
few off-plane VHE sources ($b = 2.5\degr$).  The only plausible MWL counterpart
associated with this emission is the energetic pulsar \object{PSR J1813$-$1246}
\citep{Abdo09}, marginally coincident with the VHE best-fit centroid.  This
suggests the VHE emission originates in a PWN powered by the pulsar, which has a
spin-down luminosity $\dot{E} = 6.3 \times 10^{36}$~erg~s$^{-1}$ and
characteristic age $\tau_{\mathrm{c}} = 43$~kyr.  The pulsar is one of the
brightest \gammaray\ pulsars \citep[\object{3FGL J1813.4$-$1246},][]{3FGL} and
the second-most energetic, radio-quiet pulsar. This pulsar also been found to
exhibit strong X-ray pulsations, and its distance has been recently constrained
to $d > 2.5$~kpc \citep{Marelli14}.  This implies a lower limit on the VHE
luminosity $L_{\gamma}(1-10~\mathrm{TeV}) > 2.9 \times 10^{33}$~erg~s$^{-1}$ and
a corresponding limit on the apparent efficiency $\epsilon_{\mathrm{1-10 TeV}} >
0.05\%$.

In other energy bands, no off-pulse emission (e.g. emission from the putative
PWN) is detected in HE \gammarays\ (0.1--100~GeV) based on the analysis of five
years of \fermi\ data \citep{Marelli14}, dismissing earlier hints for a GeV PWN
\citep{Ackermann11,Abdo13}, likely owing to the larger data set and improved models
for diffuse emission used in the new analysis. In X-rays ($0.3-10$~keV), despite
relatively deep \xmm\ (130~ks) and \chandra\ (50~ks) observations, no PWN is
detected beyond $1-1.5\arcsec$ from the pulsar \citep{Marelli14}.  This is very
unusual for a pulsar this energetic; that is the derived upper limits in X-rays are
only marginally compatible with known relations between PWN and pulsar
luminosities \citep{Kargaltsev08} and between PSR luminosity and distance to the
PWN termination shock \citep{Gaensler06}.

Therefore, HESS~J1813$-$126 appears to be a rare case of a relic PWN
\citep{deJager09} currently detected exclusively in the VHE domain. Observations
in the hard X-ray domain with \emph{NuSTAR} would be useful to investigate the
hint of a signal seen at 30--520~keV with \emph{INTEGRAL} \citep{Marelli14} and
to determine if there is an unpulsed, nebular component visible at those energies.
Regardless, further work modeling the MWL spectral energy distribution is
necessary to fully investigate this intriguing system.

\subsubsection{HESS~J1826$-$130}
\label{sec:HESS_J1826m130}

\begin{figure}
\resizebox{\hsize}{!}{\includegraphics{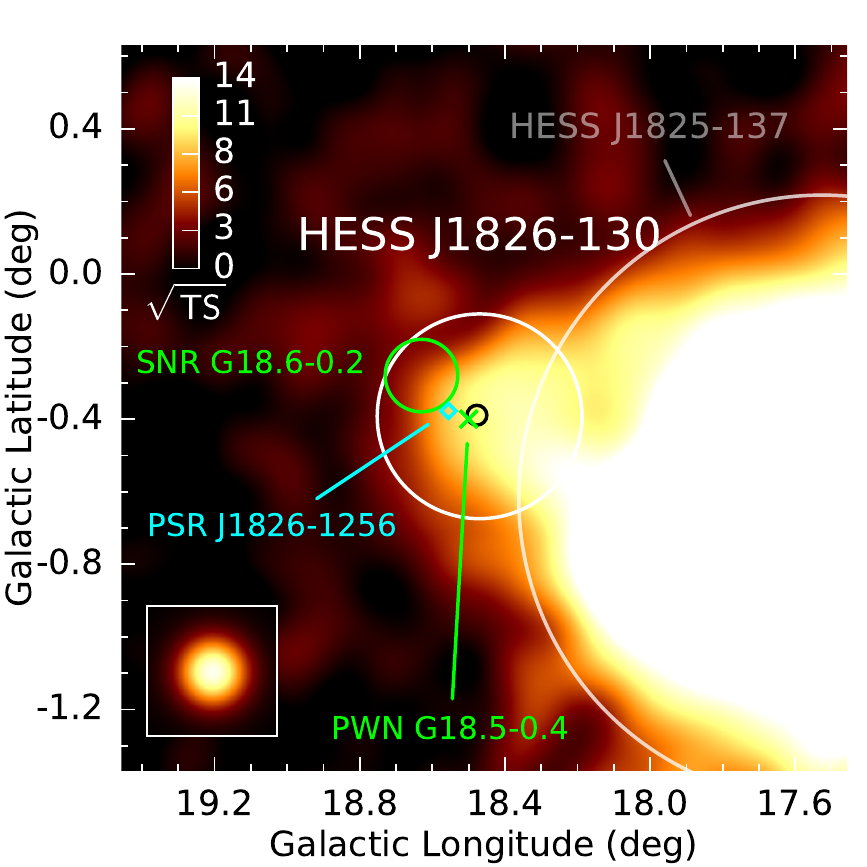}}
\SingleSourceCaption{HESS~J1826$-$130}{
Additionally, the approximate centroid of the PWN is marked by a green cross,
and the pulsar position is marked by a cyan diamond. The FoV is $2.0\degr \times
2.0\degr$.
}
\label{fig:HESS_J1826m130}
\end{figure}

The HGPS catalog analysis reveals a distinct new source of VHE \gammarays,
\object{HESS J1826$-$130} (Fig.~\ref{fig:HESS_J1826m130}), in what was previously
considered extended emission from the nearby PWN HESS~J1825$-$137
\citep{Aharonian:2006g}. Because of the very close proximity to its bright neighbor,
the spectral measurement is highly contaminated (41\%).
\citet{2017arXiv170107002A} reported preliminary findings for this new source.

HESS~J1826$-$130 is associated with the ``Eel'' PWN\footnote{\citet{Roberts07},
based on visual inspection of the VHE images, first suggested that the VHE
emission is separate from the PWN HESS~J1825$-$137 and associated it with the
Eel.} (\object{PWN G18.5$-$0.4}), an elongated, nonthermal, X-ray source
observed with \chandra\ \citep{Roberts07}, and the energetic pulsar \object{PSR
J1826$-$1256} \citep{Abdo09}, on the basis of a spatial coincidence. The
best-fit VHE centroid is compatible with the Eel, while the pulsar is somewhat
offset ($0.09\degr$) from the centroid but well within the VHE emission region
(size $0.15\degr \pm 0.02\degr$). The pulsar is now notable for being one of the
brightest radio-quiet \gammaray\ pulsars \cite[\object{3FGL
J1826.1$-$1256};][]{3FGL}. The distance of the pulsar is unfortunately not
known, which precludes conclusions on the energetics, but its position, $\dot{E} =
3.6 \times 10^{36}$~erg~s$^{-1}$, and $\tau_{\mathrm{c}} = 14$~kyr suggest it is
probably powering the Eel. The PWN is not detected in HE \gammarays\
\citep{Ackermann11,2FHL}. Finally, we note that dense molecular gas was also
found overlapping HESS~J1826$-$130 at a distance matching that of the dispersion
measure of the pulsar \citep{2016MNRAS.458.2813V}, suggesting a possible
hadronic origin for this VHE source.

The \object{SNR G18.6$-$0.2} \citep{Brogan06} is also coincident with the VHE
emission region, although it is significantly smaller in size ($0.1\degr$
diameter). Very little is known about this SNR, except that a partial shell-type
morphology has been observed so far only in radio and IR and that its distance
is estimated to be 4.0--5.2~kpc \citep{Johanson09}.

A firm identification of the VHE source as a PWN is not possible at this time,
in part resulting from the unknown distance to the Eel PWN and PSR system and the
poorly studied SNR.  We are currently preparing more advanced VHE spectral
analysis methods that can account for contamination in crowded FoVs. These
methods will enable more accurate modeling of the SED.

\subsubsection{HESS~J1828$-$099}
\label{sec:HESS_J1828m099}

\begin{figure}
\resizebox{\hsize}{!}{\includegraphics{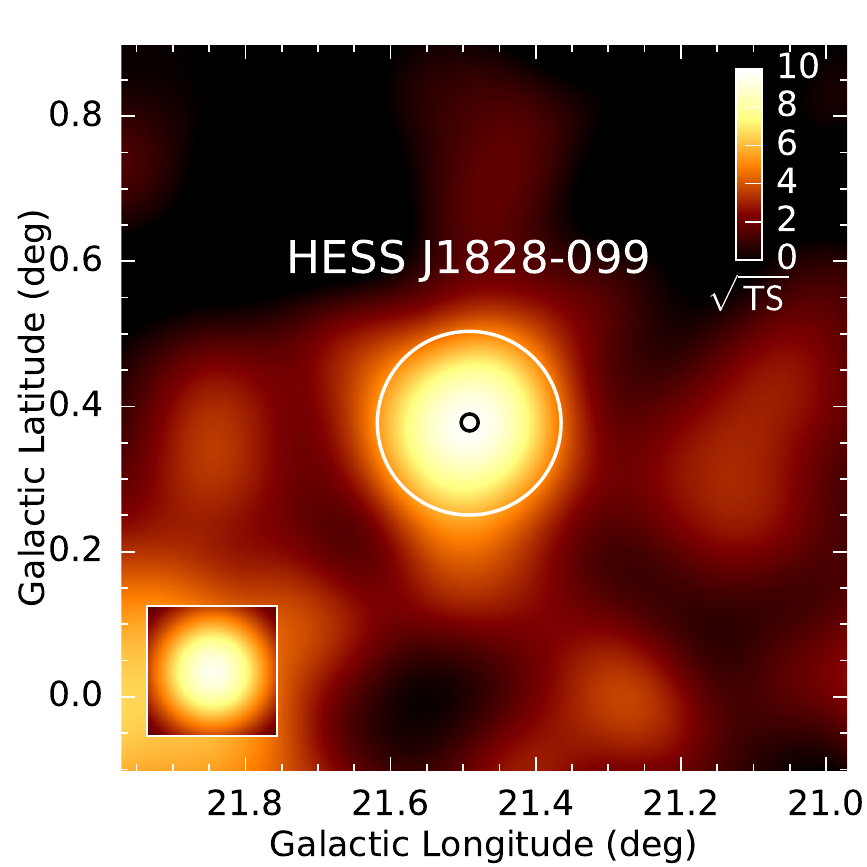}}
\SingleSourceCaption{HESS~J1828$-$099}{
The FoV is $1.0\degr \times 1.0\degr$.
}
\label{fig:HESS_J1828m099}
\end{figure}

HESS~J1828$-$099 is a new source of VHE \gammarays\
(Fig.~\ref{fig:HESS_J1828m099}), which is unique because it appears to be
completely dark at lower energies with no apparent associations (see
Table~\ref{tab:hgps_associations}). It is also notable for being one of the
{\hgpsPointLikeSourceCount point-like sources in the HGPS catalog with a size
(Gaussian std. dev.) less than $0.07\degr$. The detection of a spatially
coincident HE \gammaray\ source has been claimed \citep{Neronov10} but not
confirmed with the latest, significantly larger \fermi\ data sets, i.e. there is
no respective source neither in the 3FGL catalog \citep{3FGL} nor 2FHL catalog
\citep[$E > 50$~GeV;][]{2FHL}. Deeper follow-up observations in especially the
radio and X-ray bands are strongly encouraged to probe nonthermal emission.

\subsubsection{HESS~J1832$-$085}
\label{sec:HESS_J1832m085}

\begin{figure}
\resizebox{\hsize}{!}{\includegraphics{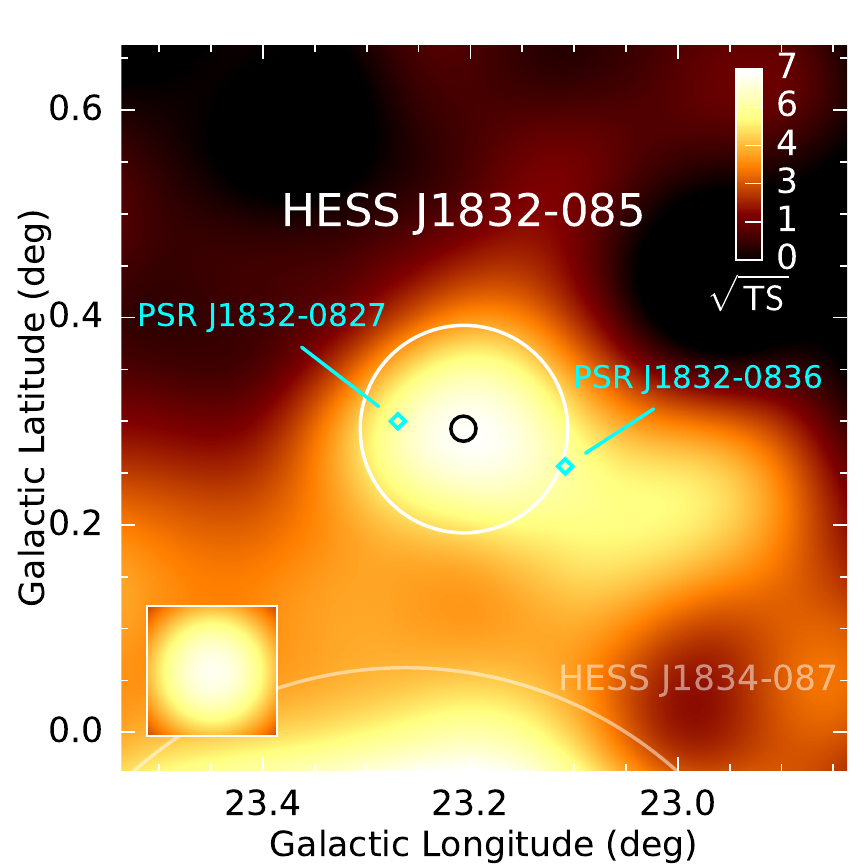}}
\SingleSourceCaption{HESS~J1832$-$085}{
Additionally, the position of the two pulsars are denoted by cyan diamonds. The
FoV is $0.7\degr \times 0.7\degr$.
}
\label{fig:HESS_J1832m085}
\end{figure}

HESS~J1832$-$085 (Fig.~\ref{fig:HESS_J1832m085}) is an unidentified source of
VHE \gammarays. It is notable for its point-like morphology, which is measured
to be less than $0.05\degr$ in extension, and its scarcity of promising MWL
counterparts.

An interesting object that is spatially coincident with HESS~J1832$-$085 is the
pulsar \object{PSR~J1832$-$0827} \citep{Clifton1986}, which has so far only been
detected in radio wavelengths. The pulsar is likely at a distance of
$\approx$4.9~kpc \citep{CordesLazio:2002}, in agreement with other estimates in
the range 4.4--6.1~kpc \citep{Frail91} and has a spin-down
luminosity\footnote{This pulsar was not selected by the standardized HGPS
association procedure (Sect.~\ref{sec:results:assoc_id}) as a possible
counterpart because its luminosity is just below the $\dot{E} >
10^{34}$~erg~s$^{-1}$ threshold.} $\dot{E} = 9.3 \times 10^{33}$~erg~s$^{-1}$
\citep{Hobbs2004}.  It is one of the few pulsars with a measured braking index,
$n = 2.5 \pm 0.9$ \citep{Johnston1999}, providing a characteristic age
$\tau_{\mathrm{c}} \approx 200$~kyr. Another very intriguing object in the FoV
is the energetic millisecond pulsar \object{PSR~J1832$-$0836}, which has a 2.7
ms period \citep{Burgay2013}.  It has a spin-down luminosity $\dot{E} = 1.7
\times 10^{34}$~erg~s$^{-1}$, a very large characteristic age (typical of
millisecond pulsars) $\tau_{\mathrm{c}} = 5 \times 10^9$~kyr, and distance
1.1~kpc \citep{CordesLazio:2002}.

There are no known PWNe associated with these two pulsars nor close to
HESS~J1832$-$085. If either or both of these pulsars are powering VHE PWNe, a
relatively large conversion efficiency of $\epsilon_{\mathrm{1-10~TeV}} \sim
23\%$ would be required for PSR~J1832$-$0827, and a more reasonable
$\epsilon_{\mathrm{1-10~TeV}} \sim 0.6\%$ for PSR~J1832$-$0836. The older ages
are at odds with the inferred small sizes of the VHE PWNe, constrained to be
less than $\approx 4~(d~/~4.9~\mathrm{kpc})$~pc and $\approx
1~(d~/~1.1~\mathrm{kpc})$~pc, respectively. These circumstances, plus the
borderline low spin-down luminosity of PSR~J1832$-$0827, combine to disfavor a
PSR and PWN scenario as the origin of the VHE emission in light of the known VHE
PWN population \citep{HESS:PWNPOP}. The millisecond pulsar scenario is even more
uncertain. That pulsar is slightly more energetic and much closer, but thus far
millisecond pulsars, with ages of billions of years, are not known to produce
PWNe that emit detectable levels of \gammarays\ at TeV energies. Therefore, the
origin of the emission from this new, enigmatic, VHE \gammaray\ source is still
very much unclear.

\subsubsection{HESS~J1833$-$105}
\label{sec:HESS_J1833m105}

\begin{figure}
\resizebox{\hsize}{!}{\includegraphics{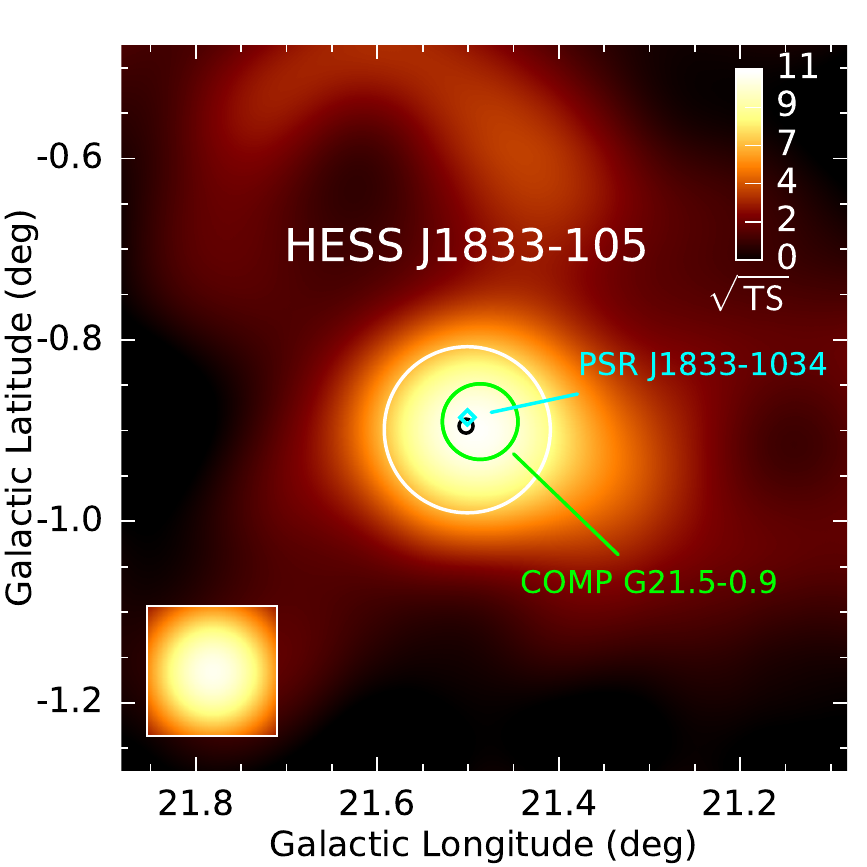}}
\SingleSourceCaption{HESS~J1833$-$105}{
Additionally, the composite SNR is shown by plotting a green circle
approximating the radio shell, and the pulsar position is denoted by a cyan
diamond. The FoV is $0.8\degr \times 0.8\degr$.
}
\label{fig:HESS_J1833m105}
\end{figure}

VHE \gammaray\ emission from the new source \object{HESS~J1833$-$105}
(Fig.~\ref{fig:HESS_J1833m105}) can now be firmly identified with the composite
\object{SNR G21.5$-$0.9} \citep{Wilson76}, which contains a Crab-like PWN.
Preliminary \hess\ source properties were previously shown
\citep{Djannati-Atai:2008a} and are compatible with the HGPS results, although
at the time it was not yet possible to disentangle the possible contributions to
the VHE emission from the PWN and SNR shell.

The new identification is supported by a positional coincidence between the VHE
emission centroid and the PWN center, but most importantly by the lack of
extension of the VHE emission region; this region is constrained to be less than
0.03\degr, which is our systematic limit for source sizes. This implies that we cannot
claim significant VHE emission from the forward shock of the spherical, faint
SNR shell at a radius of 0.038\degr\ \citep{Bocchino05}.

The PWN has also been detected in X-rays \citep{Safi-Harb01,Bocchino05} and IR
\citep{Zajczyk12} although not in HE \gammarays\ \citep{Acero13}, and its
distance has been estimated to be $d \approx 4.8$ kpc \citep{Tian08}.  It is
powered by the very energetic \object{PSR J1833$-$1034}, currently the fifth
most energetic pulsar known in the Galaxy, and has a spin-down luminosity $\dot{E}
\approx 3.4 \times 10^{37}$ erg s$^{-1}$. The pulsar has been detected in radio
\citep{Gupta05,Camilo06} and HE $\gamma$-rays \citep[as \object{3FGL
J1833.5$-$1033};][]{3FGL}. The age of the system has been argued
to be $870^{+200}_{-150}$~yr~\citep{2008MNRAS.386.1411B}, which is significantly less
than the $\tau_{\mathrm{c}} = 4.9$~kyr of the pulsar.

Considering the luminosity of \object{HESS~J1833$-$105}, $L_{\gamma}
(1-10~\mathrm{TeV}) = 2.6 \times 10^{33}$ ($d$ / 4.9 kpc)$^2$ erg s$^{-1}$, the
apparent efficiency converting the rotational energy of the pulsar to $\gamma$-rays,
$\epsilon_{\mathrm{1-10 TeV}} \equiv L_{\gamma} / \dot{E} = 0.08\%$, is
compatible with the efficiencies ($\la 10\%$) of other VHE sources that have
been identified as PWNe \citep{Kargaltsev13}.

The HGPS results confirm predictions that the PWN would emit VHE \gammarays\
at the level of a few percent of the Crab Nebula and exhibit a relatively
hard spectrum \citep{deJager95}.

\subsubsection{HESS~J1843$-$033}
\label{sec:HESS_J1843m033}

\begin{figure}
\resizebox{\hsize}{!}{\includegraphics{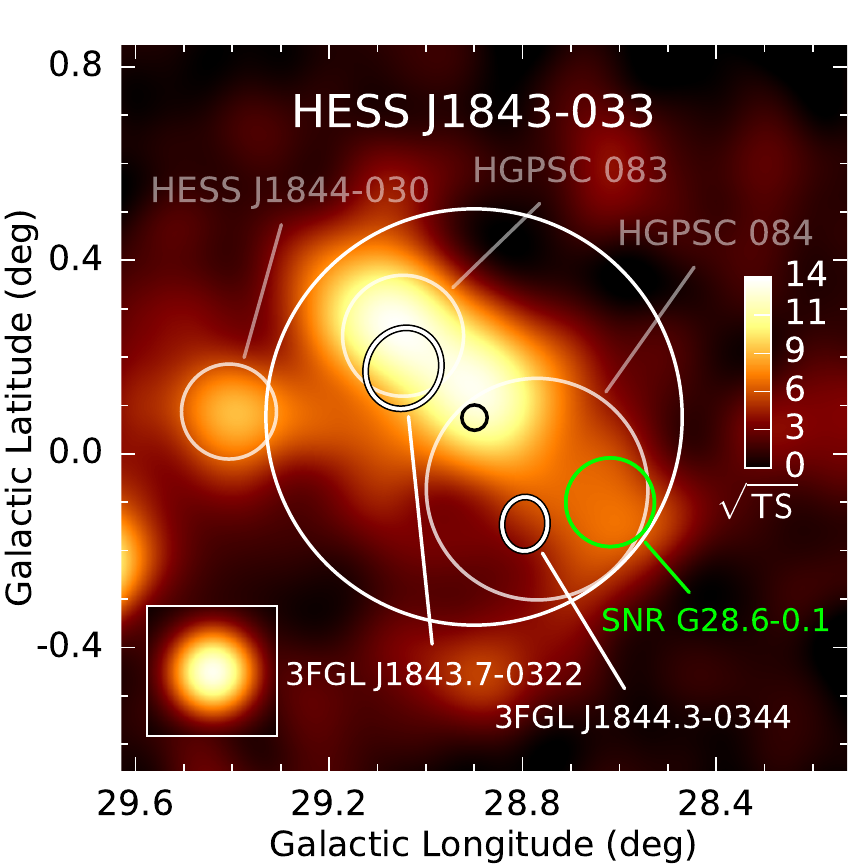}}
\SingleSourceCaption{HESS~J1843$-$033}{
Additionally, the SNR is shown by plotting a green circle approximating the
radio shell. The 3FGL ellipses represents the 95\% uncertainty in the position
of the HE \gammaray\ point sources. The FoV is $1.4\degr \times 1.4\degr$.
}
\label{fig:HESS_J1843m033}
\end{figure}

An extended region of VHE emission, called \object{HESS~J1843$-$033}, was first
published by \citet{2008ICRC....2..579H}. This emission is resolved by the HGPS
catalog analysis into three components that were merged into two distinct
sources: HESS~J1843$-$033 and HESS~J1844$-$030.

HESS~J1843$-$033 consists of two merged offset components (HGPSC~83 and
HGPSC~84) and is therefore highly structured (see
Fig.~\ref{fig:HESS_J1843m033}). The image of the source shows two peaks
separated by $\sim$0.2\degr. The first Gaussian component is clearly associated
with the upper peak. The second Gaussian component is larger and offset with
respect to the lower peak. This is due to more diffuse, low-brightness emission
around $(\ell, b) = (28.6\degr, -0.1\degr)$, suggesting the presence of another
currently unresolved source that shifts the position of the second component.
HESS~J1843$-$033 is therefore most probably a complex region with overlapping
sources that were merged in the HGPS analysis.

Two GeV sources, \object{3FGL~J1843.7$-$0322} and \object{3FGL~J1844.3$-$0344},
are found within the R80 extension of the source. The former is found in the
main region of emission but does not seem to correlate well with any of the two
main peaks. The latter \fermi\ source is located in the low-brightness region
around $(\ell, b) = (28.6\degr, -0.1\degr)$.

No compelling radio counterpart was found in the VLA Galactic Plane Survey
\citep{2006AJ....132.1158S}. Dedicated X-ray observations show the presence of a
faint absorbed extended source with a nonthermal spectrum that is coincident
with the HGPSC~83 component. No compelling counterpart for the second component
has been found. We note however that the nearby radio \object{SNR G28.6$-$0.1}
\citep{Helfand89} is filled with nonthermal X-rays \citep{Ueno03}. If this
emission is due to synchrotron X-rays produced by energetic electrons, IC
emission at VHE is likely contributing to the low-brightness emission that is
visible around the SNR position in Fig.~\ref{fig:HESS_J1843m033}.

\subsubsection{HESS~J1844$-$030}
\label{sec:HESS_J1844m030}

\begin{figure}
\resizebox{\hsize}{!}{\includegraphics{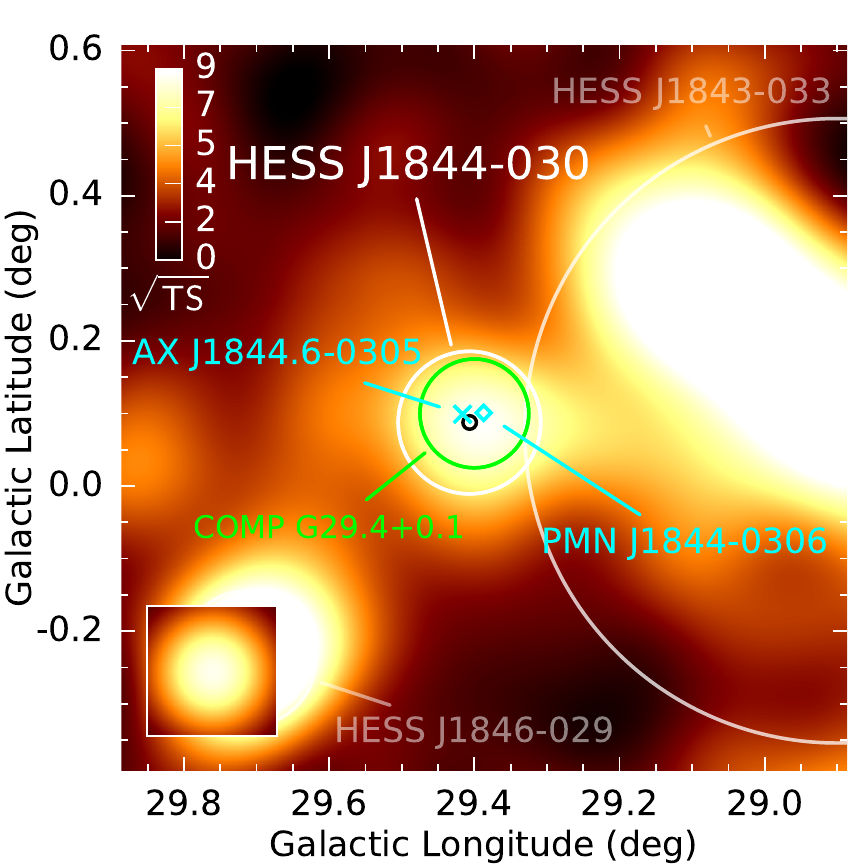}}
\SingleSourceCaption{HESS~J1844$-$030}{
Additionally, the composite SNR is shown by plotting a green circle
approximating the radio shell. The position of the X-ray source is denoted by a
cross, while the position of PMN~J1844$-$0306 is indicated by a diamond. The FoV
is $1.0\degr\ \times 1.0\degr$.
}
\label{fig:HESS_J1844m030}
\end{figure}

HESS~J1844$-$030 is a faint VHE \gammaray\ source that compatible with being
point-like and located in the vicinity of the complex region of
HESS~J1843$-$033. It is positionally coincident with a number of distinct
objects, most notably the radio source \object{PMN J1844$-$0306} (cyan diamond
in Fig.~\ref{fig:HESS_J1844m030}). The nature of the latter is ambiguous. Its
elongated, jet-like morphology is very reminiscent of a radio galaxy, which is
supported by 6 cm VLA observations revealing polarization along the structure
\citep{Helfand89}. This elongated radio feature is surrounded by a partial ring
visible in the 21 cm VLA continuum image. The object is therefore classified as
a SNR candidate in the MAGPIS catalog \citep{Helfand06}, G29.37$+$0.10. It is
also coincident with the X-ray source \object{AX~J1844.7$-$0305}
\citep{Vasisht00,Sugizaki01}.

The association of the jet radio feature and the SNR candidate is unclear.
Although rare, SNRs with jets are plausible, for example PWN structures such as
MSH~15$-$52 \citep{2002ApJ...569..878G} or the \object{SS433} / \object{W 50}
microquasar SNR system with its radio jets and lobes
\citep{1998AJ....116.1842D,HESS:SS433}. The jet structure could also be a
background radio galaxy aligned by chance with a faint radio shell. We note,
however, thanks to \hone\ absorption, \citet{Johanson09} place the source at a
distance between 5 and $\sim$15~kpc. Interestingly, a heavily absorbed X-ray
PWN, dubbed G29.4$+$0.1, is present in SNRcat and overlaps with a part of
PMN~J1844$-$0306. Further MWL observations will be necessary to assess the
nature of the system and the origin of the VHE emission.

\subsubsection{HESS~J1846$-$029}
\label{sec:HESS_J1846m029}

\begin{figure}
\resizebox{\hsize}{!}{\includegraphics{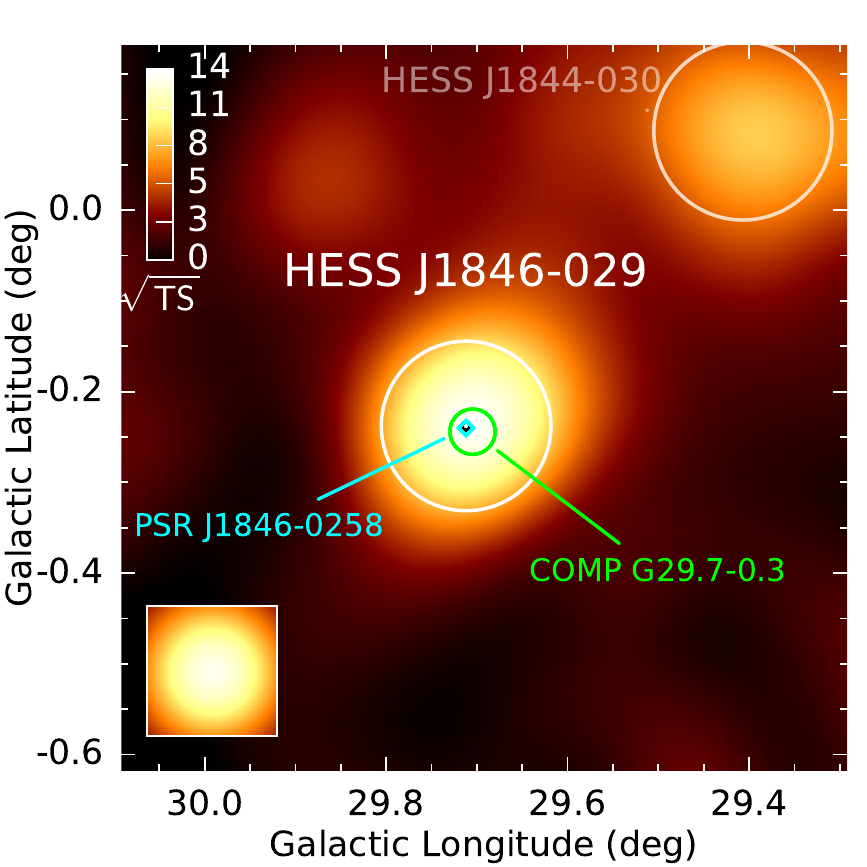}}
\SingleSourceCaption{HESS~J1846$-$029}{
Additionally, the composite SNR is shown by plotting a green circle
approximating the radio shell. The position of the pulsar is indicated by a cyan
diamond. The FoV is $0.8\degr\ \times 0.8\degr$.
}
\label{fig:HESS_J1846m029}
\end{figure}

VHE \gammaray\ emission from the new source HESS~J1846$-$029 (see
Fig.~\ref{fig:HESS_J1846m029}) is spatially coincident with G29.7$-$0.3 (also
known as \object{Kes~75}), one of the youngest composite SNRs in the Galaxy,
which contains the nebula of PSR~J1846$-$0258. Preliminary results were
presented in \cite{Djannati-Atai:2008a} and are compatible with those obtained
in the HGPS analysis.

PSR~J1846$-$0258 is a young, high magnetic-field pulsar. This source has a
rotation period of 324~ms and a spin-down power of $8.3 \times
10^{36}$~erg~s$^{-1}$.  It is among the youngest pulsars in the Galaxy with a
characteristic age of only 723~years \citep{Livingstone06}. It has experienced a
strong increase in its pulsed flux in June 2006 associated with spectral
\citep{Kumar08} and timing \citep{Gavriil08} changes in a similar manner to
magnetars. The result of the search for variations in the VHE source flux at
various timescales was negative \citep{2008AIPC.1085..316T}.

A nebula of 20\arcsec\ in radius surrounds the pulsar in radio and X-ray
wavelengths and \textit{Chandra} high-resolution observations have revealed a
jet and torus \citep{Ng08}. A 3\arcmin\ diameter asymmetric radio shell
surrounds the PSR and PWN system. It consists mainly of two lobes to the south
of the pulsar.  These lobes are emitting X-rays from heated swept-up
interstellar matter and ejecta \citep{Morton07}. Infrared measurements suggest
that the shock is in a region of typical density of 60~cm$^{-3}$
\citep{Temim12}. \cite{Su09} found a bubble in the molecular matter in good
coincidence with the SNR. They proposed that this structure is the wind blown
bubble of the SNR progenitor.

The extension of the VHE emission from HESS~J1846$-$029 is compatible with that
of a point-like source. The upper limit on the size is $0.03\degr$, that is,
comparable with the SNR shell size. The position of this object is compatible
with the position of PSR~J1846$-$0258, within localization uncertainties.
Therefore, we are not able to distinguish between emission from the shell and
emission from the PWN in this composite object.

Assuming a distance of 6 kpc \citep{2008A&A...480L..25L}, which yields a
luminosity of $L_{\gamma} (1-10~\mathrm{TeV}) = 6.9 \times 10^{33}$ ($d$ / 6
kpc)$^2$ erg s$^{-1}$, the apparent conversion efficiency of the rotational
energy of the pulsar to $\gamma$-rays is $\epsilon_{\mathrm{1-10 TeV}} \equiv
L_{\gamma} / \dot{E} = 0.08\%$.  The VHE emission is therefore completely
consistent with an origin in the PWN \citep[see also,
e.g.,][]{2011ApJ...741...40T,2014JHEAp...1...31T}. Yet, given the uncertainties
on extension it is not possible to exclude a contribution from \gammarays\
produced by particles accelerated at the SNR shock, in particular from
collisions of hadrons with ambient and swept-up matter at the shock, or even a
contribution of escaping particles with the molecular shell revealed by
\citet{Su09}.

\subsubsection{HESS~J1848$-$018}
\label{sec:HESS_J1848m018}

\begin{figure}
\resizebox{\hsize}{!}{\includegraphics{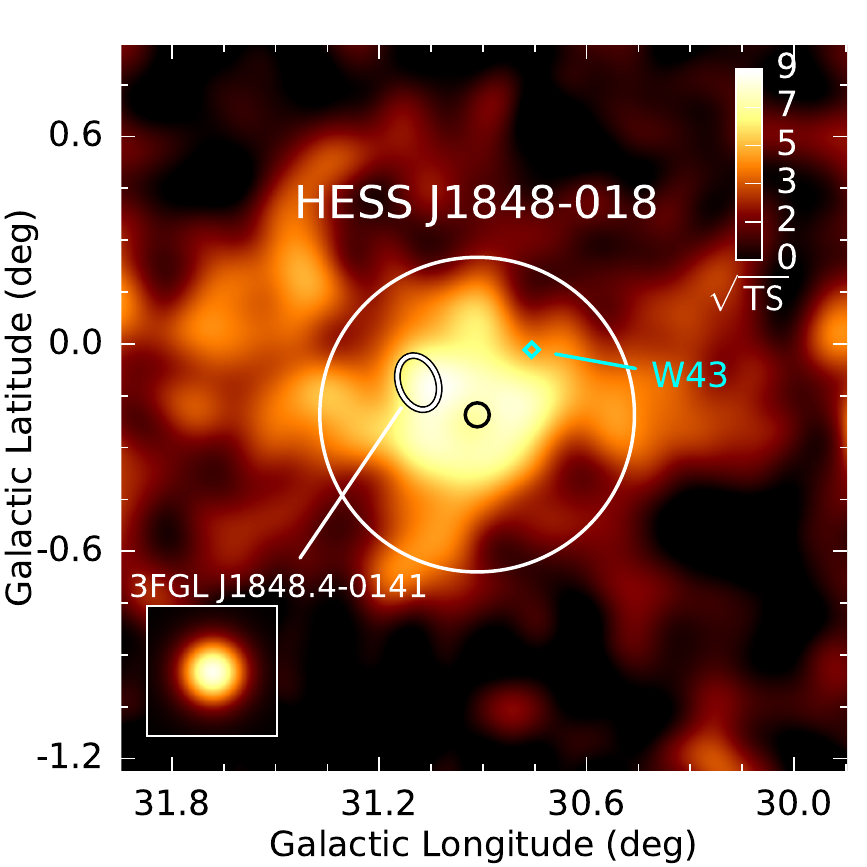}}
\SingleSourceCaption{HESS~J1848$-$018}{
Additionally, the 3FGL ellipse represents the 95\% uncertainty in the position
of the HE \gammaray\ point source. The central stellar cluster of the
mini-starburst W~43 is denoted by a diamond. The FoV is $2.1\degr\ \times
2.1\degr$.
}
\label{fig:HESS_J1848m018}
\end{figure}

For the new source \object{HESS~J1848$-$018} (Fig.~\ref{fig:HESS_J1848m018})
preliminary \hess\ source properties were previously shown \citep{Chaves:2008}.
These properties are compatible with the HGPS results except for the source size
and flux; these were overestimated because the earlier analysis did not include
a model for the diffuse emission (see Sect.~\ref{sec:cc:large-scale-emission}),
which is particularly bright in this region.

The origin of the VHE \gammaray\ emission of \object{HESS~J1848$-$018} is not
yet firmly identified. No SNR or energetic pulsar is currently detected in the
proximity, although we have associated the VHE source with \object{3FGL
J1848.4$-$0141} \citep{3FGL}. This unidentified HE \gammaray\ point source is
significantly offset from the VHE \gammaray\ centroid (by $\sim$0.2\degr) but
well within the VHE emission region. Studies attempting to relate the HE with
the (preliminary) VHE morphology and spectra remained inconclusive
\citep{Tam10,Acero13}. A potential PSR and PWN scenario cannot be confirmed due
to the lack of a detected pulsar (at any wavelength), although the HE spectrum
does exhibit curvature typical of pulsars \citep{3FGL}. Furthermore, there is no
known PWN nearby, although one study has shown marginal statistical evidence for
an extension of the HE source \citep{Lemoine-Goumard11}, which is expected if
the HE emission is from a PWN or the combination of a pulsar and PWN.

An extensive search for other MWL counterparts found the VHE \gammaray\ emission
to be in the direction of the massive star-forming region \object{W~43}, a very
active mini-starburst located at a distance of $6.2 \pm 0.6$~kpc
\citep{Russeil03}. It is one of the closest and most luminous star-forming
regions in the Galaxy \citep{Motte03}, hosting a giant \ion{H}{II} region
(\object{G30.8$-$0.2}), a giant molecular cloud, and the Wolf-Rayet binary star
system \object{WR~121a} in the central stellar cluster together with O-type
stars. The massive stars in the dense central cluster exhibit strong stellar
winds with extreme mass loss rates, in particular the WN7-subtype
\object{WR~121a} \citep{Blum99}.

This unique MWL environment is of interest because the central cluster of W~43
could be the site of efficient particle acceleration in various plausible
hadronic scenarios involving the high-velocity (up to 2000~km~s$^{-1}$) stellar
winds \citep[e.g.,][]{Reimer06,Romero10}.  Furthermore, the very large amount of
molecular gas present in W~43 \citep[$\sim$7$ \times
10^6$~M$_{\odot}$;][]{Nguyen11} provides a natural target for accelerated cosmic
rays (regardless of their potential acceleration site), which would lead to
\gammaray\ production via hadronic p-p collisions \citep[e.g.,][]{Aharonian91}.

It is not yet possible to confirm the W~43 hadronic scenario for the origin of
the VHE emission, in part because of the very complex morphologies present and
the challenges in correlating features observed in radio and infrared
observations at arcsecond scales with the $\sim$5$\arcmin$ resolution in VHE.
The VHE centroid, in particular, is significantly offset from the central
cluster by $\sim$0.2\degr, although the extended VHE emission is generally
coincident with the W~43 complex. This scenario remains under investigation,
especially in light of the recent detection of the superbubble 30~Dor~C in the
LMC \citep{HESS15_LMC}, which suggests that particle acceleration occurring in
the collective winds of massive stars can indeed produce VHE emission.

\subsubsection{HESS~J1849$-$000}
\label{sec:HESS_J1849m000}

\begin{figure}
\resizebox{\hsize}{!}{\includegraphics{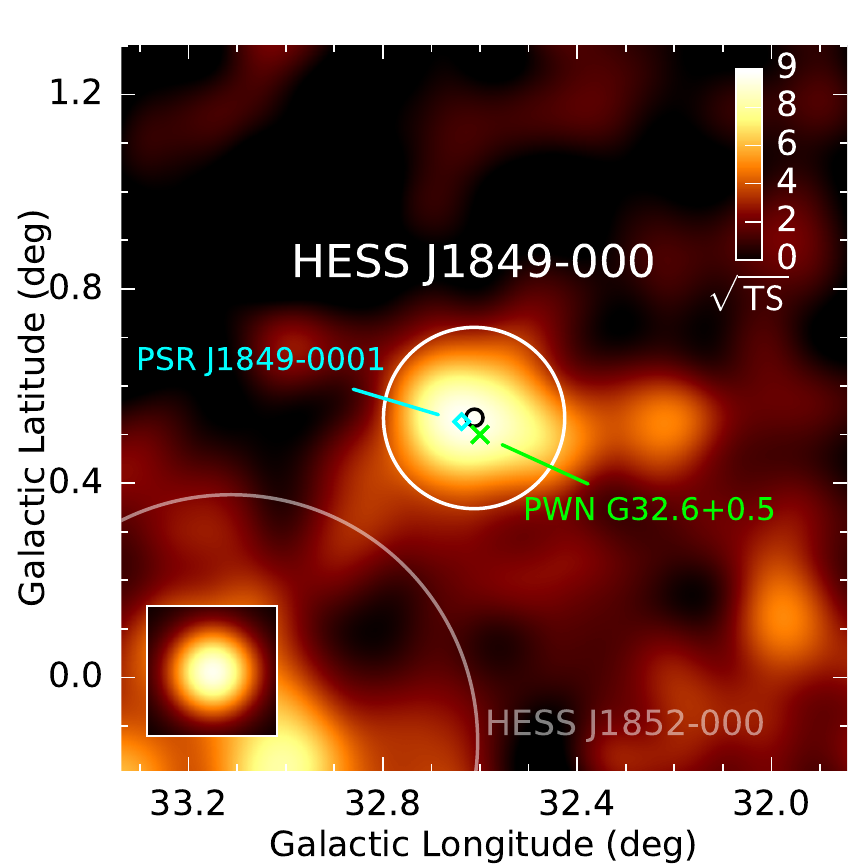}}
\SingleSourceCaption{HESS~J1849$-$000}{
Additionally, the position of the pulsar is indicated by a cyan diamond and the PWN
by a green cross. The FoV is $1.5\degr\ \times 1.5\degr$.
}
\label{fig:HESS_J1849m000}
\end{figure}

The faint, slightly extended source \object{HESS~J1849$-$000} (see
Fig.~\ref{fig:HESS_J1849m000}) was first reported by
\citet{2008AIPC.1085..312T}. It was found to be spatially coincident with the
hard X-ray source \object{IGR~J18490$-$0000} \citep{2012AandA...545A..27K}.
\xmm\ observations revealed a nonthermal, point-like, X-ray source surrounded by
a nebula, making this object a solid PWN candidate. Follow-up observations of
the hard X-ray source with \emph{RXTE} have confirmed this hypothesis with the
discovery of a 38.5~ms periodicity of the X-ray signal
\citep{2011ApJ...729L..16G}. The associated pulsar, \object{PSR~J1849$-$0001},
was found to have a spin-down luminosity $\dot{E} = 9.8 \times
10^{36}$~erg~s$^{-1}$ and a characteristic age $\tau_{\mathrm{c}} = 42.9$~kyr.

The HGPS analysis confirms the existence of a source coincident with
PSR~J1849$-$0001. The best-fit position of HESS~J1849$-$000 is located less than
$0.03\degr$ from the X-ray pulsar position (cyan diamond on
Fig.~\ref{fig:HESS_J1849m000}), well within statistical uncertainties in both
source localizations. The best-fit size of the VHE emission is $0.09\degr$,
which is about a factor of two larger than that of the extended X-ray component
\citep{2011ApJ...729L..16G,2015MNRAS.449.3827K}.

The source has an energy flux $\sim$2.1$\times 10^{-12}$~erg~cm$^{-2}$~s$^{-1}$,
in the range 1--10~TeV, a factor of $\sim$2 above the X-ray nebula energy flux
in the range 2--10~keV \citep{2015MNRAS.449.3827K}. This confirms the likely
nature of HESS~J1849$-$000 as a PWN in transition between a young,
synchrotron-dominated phase and an evolved, IC-dominated phase.

\subsubsection{HESS~J1852$-$000}
\label{sec:HESS_J1852m000}

\begin{figure}
\resizebox{\hsize}{!}{\includegraphics{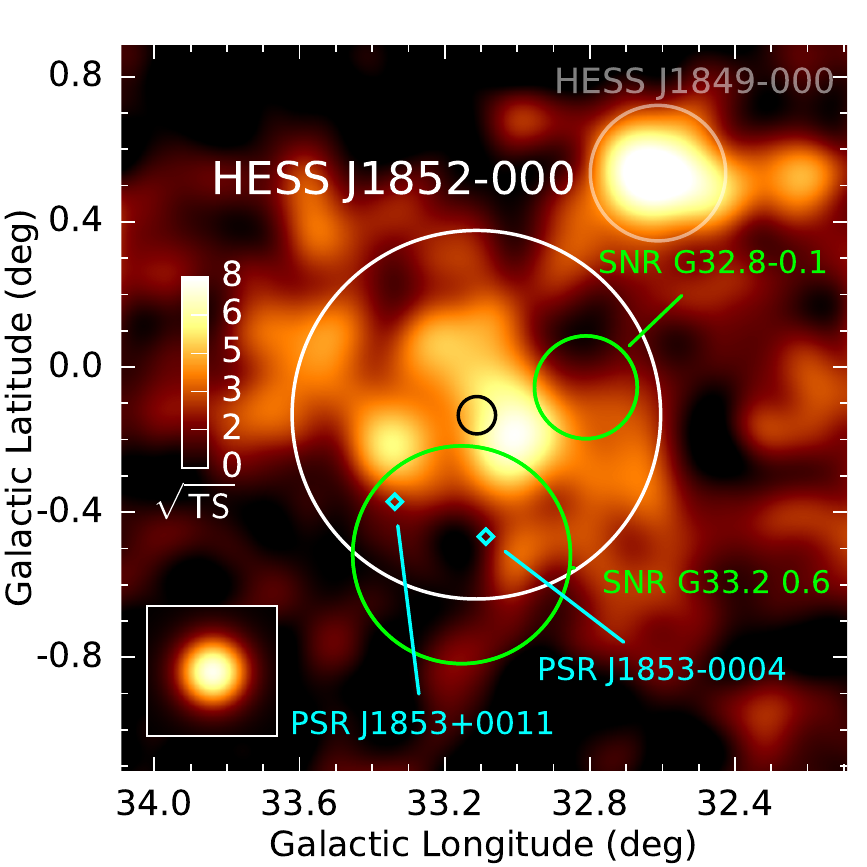}}
\SingleSourceCaption{HESS~J1852$-$000}{
Additionally, the position of the pulsars are indicated by cyan diamonds, and the
SNR is shown by plotting a green circle approximating the radio shell. The FoV
is $2.0\degr\ \times 2.0\degr$.
}
\label{fig:HESS_J1852m000}
\end{figure}

The new source of VHE \gammaray\ emission \object{HESS~J1852$-$000}
(Fig.~\ref{fig:HESS_J1852m000}) is currently unidentified due to multiple source
counterpart confusion. It is spatially associated with the partial shell-type
\object{SNR G32.8$-$0.1} \citep[also known as \object{Kes 78};][]{Kesteven68,
Velusamy74}, the incomplete shell-type \object{SNR G33.2$-$0.6} \citep{Reich82},
and two energetic pulsars, \object{PSR J1853$-$0004} and \object{PSR
J1853$+$0011} \citep{Hobbs04}. Preliminary \hess\ source properties were
previously shown \citep{Kosack11} and are compatible with the HGPS results. As
mentioned in Sect.~\ref{sec:results:xcheck}, the spectral properties of
HESS~J1852$-$000 are affected by systematic uncertainties larger than nominal.

The VHE emission is located along the eastern edge of SNR Kes~78 but extends
well beyond the SNR. The SNR itself is characterized by an elongated and partial
nonthermal shell seen in radio and X-rays \citep{Zhou11,Bamba16}. It is
interacting with adjacent molecular clouds, evidenced by the detection of a
shock-excited OH(1720~MHz) maser on the shell \citep{Koralesky98} and studies of
the CO molecular environment \citep{Zhou11}. The distance of the SNR is
estimated to be $\sim$5~kpc \citep{Koralesky98,Zhou11}, although $\sim$8.8~kpc
has also been suggested \citep[e.g.,][]{Xu09}. A hadronic origin of the VHE
emission has been briefly discussed \citep{Kosack11}, involving escaped cosmic
rays from Kes~78 \cite[e.g.,][]{Aharonian91,Gabici07}.  However, the scenario
remains unconfirmed in the absence of a more detailed study of the gas
environment and its potential correlation with the complex VHE morphology.

The presence of two radio pulsars, PSR~J1853$-$0004 and PSR~J1853$+$0011, within
the VHE emission region also suggests that the VHE \gammarays\ could originate
in one of the PWN or could even be a result of superimposed emission from two
PWNe. Although there are currently no known PWNe  at other energies, the
pulsars' spin-down luminosities $\dot{E} = 2.1 \times 10^{35}$~erg~s$^{-1}$ and
$2.1 \times 10^{34}$~erg~s$^{-1}$, respectively, and distances $d = 6.6$~kpc and
7.5~kpc, are reasonable in the context of other pulsars thought to be powering
VHE PWNe \citep{HESS:PWNPOP}. The pulsars have so far only been detected in
radio, although PSR~J1853$-$0004 has been associated with the HE \gammaray\
source \object{3FGL~J1853.2$+$0006}, which is itself a source whose existence
and properties are currently uncertain \citep[subject to analysis Flags 3 and 4
in][]{3FGL}.

In conclusion, it is not yet clear whether the VHE emission originates from a
hadronic SNR and molecular cloud interaction, previously undetected PWNe
associated with one or both of the spatially coincident pulsars, or some other
yet unknown source.

\subsubsection{Source candidates}
\label{sec:sourcecandidates}

Three VHE \gammaray\ source candidates (hotspots) were found above the $TS = 30$
detection threshold in one HGPS analysis (primary or cross-check), but these
candidates had $TS < 30$ in the other analysis. These should be considered
unconfirmed, or candidate, VHE sources to be confirmed by deeper VHE
observations.

\paragraph{HOTS~J1111$-$611\\}
\label{sec:HOTS_J1111m611}

The VHE emission from the source candidate HOTS~J1111$-$611 has a significance
of $TS = 22$ (cross-check $TS = 41$). It is located at ($\ell$,~$b$)~=~
($291.18\degr \pm 0.03\degr$, $-0.54\degr \pm 0.03\degr$), has a measured
integral flux $F(E > 1~\mathrm{TeV}) = 3.8 \times 10^{-13}$ cm$^{-2}$~s$^{-1}$,
and a size $0.09\degr \pm 0.03\degr$. It is located near ($\sim$0.1\degr) the
very energetic $\tau_{\mathrm{c}} = 32.7$~kyr pulsar \object{PSR J1112$-$6103}
\citep{Manchester:2001} emitting in radio and HE \gammarays\ \citep[\object{3FGL
J1111.9$-$6058},][]{Abdo13}. The pulsar has a high spin-down luminosity
$\dot{E}=4.5\times 10^{36}$~erg~s$^{-1}$ and a distance of 12.2~kpc
\citep{CordesLazio:2002}.  Moreover, a significant HE \gammaray\ source
\citep[\object{2FHL J1112.1$-$6101e},][]{2FHL} above 50~GeV has been reported at
$0.04\degr$ from the pulsar, which makes this HE source likely to be a PWN.  The
characteristics of this pulsar, the apparent efficiency $\epsilon_{E >
1~\mathrm{TeV}} \sim 1\%$, and the presence of a HE component in its vicinity
suggests that it could plausibly power a VHE PWN.

\paragraph*{HESS~J1831$-$098\\}
\label{sec:HESS_J1831m098}

The source candidate \object{HESS J1831$-$098} is found to have $TS = 59$ in the
main HGPS analysis but only $TS = 17$ in the cross-check analysis and is
therefore considered a source candidate. HESS~J1831$-$098 is located in a
complex region with nearby diffuse components, which might explain the
discrepancy observed for this source candidate. Preliminary VHE morphological
and spectral properties on HESS~J1831$-$098 were announced by
\citet{2011ICRC....7..244S}.  This source candidate is coincident with the
energetic pulsar \object{PSR J1831$-$0952}, which exhibits a spin-down
luminosity of $1.1 \times 10^{36}$ erg~s$^{-1}$ and a characteristic age of
128~kyr. According to \citet{2011ICRC....7..244S}, a
$\epsilon_{1-20~\mathrm{TeV}} \sim 1\%$ conversion efficiency from rotational
energy to $\gamma$-rays would be required to power a PWN; this is similar to
values observed in other VHE PWN.

\paragraph{HOTS~J1907$+$091\\}
\label{sec:HOTS_J1907p091}

The VHE emission from source candidate \object{HOTS J1907$+$091} has a
significance of only $TS = 18$ (cross-check $TS = 43$). It is located at
($\ell$,~$b$)~=~ ($42.88\degr \pm 0.08\degr$, $0.69\degr \pm 0.08\degr$), has a
measured integral flux $F(E > 1~\mathrm{TeV}) = 4.3 \times
10^{-13}$~cm$^{-2}$~s$^{-1}$, and an extension of $0.17\degr \pm 0.04\degr$. Two
potential counterparts are found to be spatially coincident with this source
candidate: the magnetar \object{SGR 1900$+$14} \citep{Mazets:1979} and the
\object{SNR G42.8$+$0.6} \citep{Fuerst:1987}. The former has an age
$\tau_{\mathrm{c}} = 0.90$~kyr and a spin-down luminosity $\dot{E} = 2.6 \times
10^{34}$~erg~s$^{-1}$. It is assumed to be at a distance $12.5 \pm 1.7$~kpc
\citep{Davies:2009} based on an association of the magnetar with a massive star
cluster \citep{Wachter:2008}. SGR~1900$+$14 has similar properties to those of
another magnetar, \object{SGR~1806$-$20}, that is associated with the VHE source
HESS~J1808$-$204 \citep{HESS:1808}. It underwent a major burst of soft
\gammaray\ emission in 1998 \citep{Hurley:1999,Frail99} and, similar to
SGR~1806$-$20, it might also be emitting VHE \gammarays.  Little is known about
SNR G42.8$+$0.6. The centroid of the VHE emission is marginally coincident with
the magnetar, while the bulk of the emission overlaps the northeastern half of
the SNR shell.

\section{Summary and conclusions}
\label{sec:conclusions_outlook}


The \hess\ Collaboration has completed its Galactic plane survey, which is an
observation and analysis program that spanned over a decade. This paper presents
the final results of the survey. The four-telescope \hess\ Phase I array was
used for the observations, which features a 5\degr\ FoV that is well suited to
scanning large regions of the sky like the Galactic plane. The Phase I array has
a typical sensitivity to point-like \gammaray\ sources of 1\% Crab Nebula
integral flux ($E > 1$~TeV) in less than 25~h.

The \hess\ Collaboration added a fifth, larger telescope to the array in 2012
(\hess\ Phase II) to extend its sensitivity to lower energies as well as its
ability to rapidly reposition to observe transient phenomena.  However, it also
features a smaller FoV than the four Phase I telescopes, making it much less
suited for scanning large regions. In addition, the HGPS had improved the
uniformity of its exposure and achieved a target sensitivity of 2\% Crab flux in
the inner Galaxy. Primarily for these reasons, as well as the diminishing gains
stemming from source significance scaling approximately as the square root of
livetime, the \hess\ Collaboration in 2013 decided not to continue the HGPS
observation program.

First early results from the HGPS were published in 2005
\citep{2005Sci...307.1938A}. The observations at the time amounted to just 120~h
yet led to the detection of ten VHE \gammaray\ sources in the inner Galaxy,
eight of which were not previously known. Further results followed in 2006
\citep{ref:gps2006}, using 230~h of data and discovering additional four
\gammaray\ sources.  Since then, we provided the community with periodic
updates, which had steadily increasing exposure and additional source
discoveries, by releasing new unidentified sources~\citep{ref_gps_unids2008} and
via published conference proceedings \citep{2008ICRC....2..579H, Chaves08_GPS,
ref:icrc09, ref:icrc11, Deil12_GPS, Carrigan13a_GPS, Carrigan13b_GPS}.

The HGPS data set (Sect.~\ref{sec:dataset}) is now over a factor of ten larger
than in 2006, comprising 2673~h of observations accumulated over the period 2004
to 2013. These data come from a variety of observations: observations from the
initially published surveys, targeted observations of known sources, follow-up
observations of newly discovered source candidates, observations to extend the
HGPS spatial coverage, and fill-up observations to achieve a more uniform
sensitivity across the Galactic plane (Fig.~\ref{fig:hgps_sensitivity}). The
energy threshold of the HGPS varies with the longitude observed but is typically
lower than 0.8~TeV for detections and maps (0.5~TeV for spectral analyses) and
as low as 0.4~TeV (0.2~TeV) in many regions, especially the innermost Galaxy
(Fig.~\ref{fig:hgps_energy_threshold_profiles}).

Compared to the previous publication, the HGPS was also expanded to cover a much
wider range of both longitude and latitude
(Fig.~\ref{fig:hgps_region_exposure_illustration}).  In the first Galactic
quadrant, the HGPS now extends in longitude from the Galactic center to nearly
$\ell = 65\degr$, the northern limit of visibility from the southern-hemisphere
\hess\ site.  In the fourth Galactic quadrant, the HGPS coverage is continuous
and even extends beyond Vela to $\ell = 250\degr$ in the third quadrant.  In
latitude, the coverage varies but is generally $b = \pm 3\degr$ and as large as
$b = \pm 5\degr$ in some regions to explore areas of particular interest off the
plane. The point-source sensitivity is better than 2\%~Crab along the Galactic
plane ($b = 0\degr$) over most of the longitudes covered by the HGPS
(Figs.~\ref{fig:hgps_sensitivity},~\ref{fig:hgps_sources_glat}).  However, the
flux sensitivity varies significantly owing to the mix of observations
comprising the HGPS. It is better than 1\%~Crab in numerous regions but at a
more modest level of 2--10\%~Crab off-plane. The HGPS achieves the best
sensitivity at the Galactic center, reaching 0.3\% of the Crab flux.

To ensure robust results, the HGPS relies on results that agree between two
independent software frameworks (chains; Sect.~\ref{sec:dataset:events}) used to
calibrate raw Cherenkov data as well as reconstruct and analyze the \gammaray\
images and spectra. The primary software chain used the Hillas method for event
reconstruction, and an event classification method using boosted decision trees.
The secondary (cross-check) chain uses an alternative event reconstruction and
classification based on EAS models, returning results that are in very good
agreement globally although there are some variations on a source-by-source
basis (discussed in Sect.~\ref{sec:results:previously}). Monte Carlo simulations
provide the instrument response functions that describe the performance of the
instrument. The mean angular resolution of \hess\ (68\% containment radius of
the PSF) is $\sim$\hgpsMeanPSF\ and varies by approximately 10\% across the
survey region.

We have generated a number of sky maps (images; Sect.~\ref{sec:maps}), which are
public data products\footnote{\hgpsurl} and also form the basis for the HGPS
source catalog construction. To accumulate sufficient signal and search for
\gammaray\ emission of different sizes, we generated three different sets of
maps with events  spatially correlated over radii of $0.1\degr$ (point-like),
$0.2\degr$ and $0.4\degr$, respectively. To subtract background from hadronic
CRs passing \gammaray\ selections in the FoV, we developed an adaptive version
of the classic ring background method that is more flexible and can compensate
for large exclusion regions that minimize signal contaminating background
regions (Sect.~\ref{sec:adaptiveringmethod}). For each point in a sky map, we
calculated the \gammaray\ excess, statistical significance of the \gammaray\
excess, and \gammaray\ flux (or flux upper limit). The main map products
released (Sect.~\ref{sec:online:maps}) are those of significance
(Sect.~\ref{sec:significancemaps}), flux (Fig.~\ref{fig:fluxmap}) and upper
limits.  Auxiliary maps include flux errors and sensitivity
(Fig.~\ref{fig:hgps_sensitivity}).

To detect and characterize the VHE \gammaray\ sources, we developed a
semi-automatic analysis pipeline to construct a source catalog (see
Sect.~\ref{sec:cc}). To disentangle individual \gammaray\ sources in complex
regions of overlapping emission, we implemented morphological modeling based on
two-dimensional maximum-likelihood estimation. We fit the \gammaray\ excess by
two-dimensional symmetric Gaussian components, keeping components with $TS>30$.
To arrive at the HGPS catalog, Gaussian components that that did not correspond
to a clear emission peak in the main and cross-check analysis were rejected.
Some components that strongly overlapped were merged into a single source for
which position, extension, and flux were characterized by the moments of the
multi-Gaussian emission model. In this process, it was necessary to model the
underlying large-scale \gammaray\ emission along the Galactic plane to improve
the modeling of the discrete sources (Sect.~\ref{sec:cc:large-scale-emission}).
We chose to use an empirical model derived with a sliding window method,
$20\degr$ wide in longitude and Gaussian in latitude, whose Gaussian center,
amplitude, and width were fit to the excess outside exclusion regions.  We
calculated source spectra using the reflected region background method when
possible, fitting PL spectral models and determining the best-fit normalization
and spectral index. Flux information is also available from the aforementioned
maps albeit assuming a spectral index of $\Gamma = \hgpsAssumedSpecIndex$.

The HGPS source catalog includes \hgpsSourceCountTotal sources of VHE
\gammarays.  Of these, \hgpsSourceCountAnalysed were detected with the HGPS
pipeline analysis. For completeness, the catalog includes an additional
\hgpsSourceCountCutout~\hess\ sources from regions excluded from the HGPS
pipeline, for example, because of their complexity, such as the Galactic center
region and sources with shell-like morphologies. \hess\ has previously published
the discovery of most of the HGPS sources, although in many cases the available
observation time used for the HGPS analysis is considerably larger.  Of the
total \hgpsSourceCountTotal~sources, \hgpsSourceCountNew are new discoveries
published here for the first time. Five of these new sources are firmly
identified objects: HESS~J1554$-$550 and HESS~J1849$-$000 are PWNe
(Sect.~\ref{sec:HESS_J1554m550},~\ref{sec:HESS_J1849m000}), and
HESS~J1119$-$614, HESS~J1833$-$105, and HESS~J1846$-$029 are composite SNRs
(Sect.~\ref{sec:HESS_J1119m614},~\ref{sec:HESS_J1833m105},~\ref{sec:HESS_J1846m029}).
Three more of the new sources are spatially coincident with HE \gammaray\
pulsars, recently discovered in \fermi\ data, and are thus plausible PWN
candidates.

The HGPS sources have diverse characteristics (Sect.~\ref{sec:results}).  Apart
from the shell-like sources, most source morphologies are generally well-modeled
as symmetric two-dimensional Gaussians, but their sizes range from point-like
($\la 0.1\degr$) to $0.6\degr$ (Fig.~\ref{fig:hgps_sources_flux_extension}).
Their fluxes cover a wide range as well from 0.6\%~Crab to 103\%~Crab of which
the majority are in the range 1-20\%~Crab
(Fig.~\ref{fig:hgps_sources_flux_extension}). The cumulative $\log N$ -- $\log
S$ distribution above 10\%~Crab (containing 32 sources) is well described by a
power law of slope $-1.3~\pm~0.2$ (Fig.~\ref{fig:hgps_sources_log_n_log_s}),
matching the expectation of a power law of slope $-1$ from a population of
equal-luminosity sources homogeneously distributed in the Galactic disk. Below
10\%~Crab, the HGPS source catalog is incomplete and can only provide a lower
limit on the true number of fainter VHE sources (70 above 1\%~Crab). Spectral
indices range from hard ($\Gamma \approx 2.0$) to very soft ($\Gamma \approx
3.0$) in an approximately normal distribution centered at $2.4 \pm 0.3$
(Fig.~\ref{fig:sources_index}). The VHE sources cluster narrowly along the
Galactic plane (median $b = -0.20\degr$, with a spread of $0.51\degr$), in good
agreement with the distributions of SNRs, energetic pulsars, molecular gas, and
HE \gammaray\ sources (Fig.~\ref{fig:hgps_sources_glat}). Their distribution in
longitude (Fig.~\ref{fig:hgps_sources_glon}) shows a general correlation with
molecular gas.

To study the origin of the VHE \gammarays, we performed a systematic search to
associate the HGPS sources with known or suspected VHE source classes, based
largely on spatial compatibility with objects in the SNR and PWN catalog SNRcat,
the ATNF pulsar catalog, and the \fermi\ 3FGL and 2FHL catalogs
(Sect.~\ref{sec:results:assoc_id}). By comparing the HGPS catalog to plausible
MWL counterpart catalogs, we come to one of the main conclusions of the HGPS
program: the majority (\hgpsSourceCountAssoc, or 86\%) of the HGPS sources are
associated with at least one astronomical object that could potentially account
for the production of \gammarays\ at TeV energies. The unassociated sources
(\hgpsSourceCountNotAssoc, 14\%) are not necessarily dark, i.e. emitting
exclusively in the VHE domain; it is also possible that their counterparts were
missed by our association procedure. In short, most HGPS sources have either
firm associations, plausible or potential counterparts in other wavelength
regimes. Whether there remains a population of truly dark VHE sources in the
HGPS can only be figured out with deeper MWL studies.

We then used additional, stricter criteria, such as shell-like morphology or
variability, to establish firm identifications for \hgpsSourceCountID sources
(Fig.~\ref{fig:hgps_source_id}). We found the largest identified VHE source
class to be PWNe (\hgpsSourceCountPWN sources, or 39\% of identified sources),
followed by shell-type SNRs (\hgpsSourceCountSNR, 26\%); composite SNRs
(\hgpsSourceCountCOMP, 26\%), where both the interior PWN and SNR shell may
contribute to the emission; and high-energy binary systems (\hgpsSourceCountBin,
10\%). At present, only 40\% of the HGPS sources can be firmly identified.  This
is typically due to difficulties resolving ambiguity among competing scenarios
involving multiple associated objects in large part because of the large
intrinsic sizes of VHE \gammaray\ sources.

The HGPS data set allows for population studies of sources. An early study of 15
globular clusters was published before the HGPS was completed \citep{GCPop}. Two
further such studies, on the primary Galactic VHE source classes of PWNe and
SNRs, are published as companion articles to this paper
\citep[][respectively]{HESS:PWNPOP,HESS:SNRUL}, together with more specific
studies on a number of microquasars \citep{HESS:MQ} and bow shocks of runaway
stars~\citep{HESS:Bow}. With the public release of the HGPS catalog along with
sky maps, more comprehensive such population studies will become possible.

Further insights into the Galactic VHE source population and diffuse emission in
the coming years can be expected. \hess, \fermi\ and HAWC are surveying the
Milky Way; the analysis methods for the individual gamma-ray data sets, and
joint analysis methods combining multiple data sets are improving; and new
surveys at lower wavelengths (especially those detecting nonthermal emission in
the radio and X-ray bands) will be come available soon. The next major leap
forward will be achieved by the Galactic plane survey of the Cherenkov Telescope
Observatory (CTA), which will consist of two arrays in the northern and southern
hemisphere \citep{2017arXiv170907997C}. The Galactic plane survey is a key
science project of CTA, and is planned to cover the whole Galactic plane, over a
wider energy band and with better angular resolution and sensitivity compared to
HGPS \citep{2013APh....43..317D, 2017arXiv170907997C}.

In conclusion, the additional exposure obtained since 2006, plus significant
improvements in analysis and reconstruction methods, allowed us to probe much
more of the Galaxy, whether it be more distant sources, fainter nearby sources,
or regions never before observed at TeV energies. The HGPS program clearly
demonstrates that sources of VHE \gammaray\ emission are common in the Galaxy
and are linked to diverse sites of high-energy particle acceleration.

\section{Online material}
\label{sec:online}

In this section, we provide further information about the public data products
released in electronic format. We also provide some guidance and caveats
regarding the correct use of these products. The HGPS survey maps and source
catalog presented in this paper are available for download at:

\begin{center}
\hgpsurl
\end{center}

In addition to the figures and tables present in Sect.~\ref{sec:results}, there
are a series of HGPS maps and tables available online:

\begin{itemize}

\item Figure~\ref{fig:fluxmap}: HGPS flux map

\item Figures~\ref{fig:hgps_survey_mwl_1}-\ref{fig:hgps_survey_mwl_4}:
Four-panel HGPS significance maps with all VHE sources and MWL associations labeled

\item Table~\ref{tab:hgps_catalog}: HGPS catalog source morphology summary

\item Table~\ref{tab:hgps_spectra}: HGPS catalog source spectrum summary

\item Table~\ref{tab:hgps_associations}: HGPS catalog source associations

\end{itemize}

\subsection{Sky maps}
\label{sec:online:maps}

\subsubsection*{Description}

Survey maps are released in \textit{FITS} format \citep{Pence:2010}, using a
Cartesian (CAR) projection in Galactic coordinates
\citep{2002AandA...395.1077C}. The maps contain the whole HGPS region
(\hgpsregion), with a binning of 0.02\degr\ per pixel corresponding to a total
size of 9400~$\times$~500~pixels. Maps are available for the following
quantities:

\begin{itemize}

\item Statistical significance (described in Sect.~\ref{sec:significancemaps})

\item Flux (described in Sect.~\ref{sec:fluxmaps})

\item 1$\sigma$ flux error (described in Sect.~\ref{sec:errfluxmap})

\item Flux upper limit (described in Sect.~\ref{sec:errfluxmap})

\item Sensitivity (described in Sect.~\ref{sec:sensmaps})

\end{itemize}

We provide all flux and flux-like quantities as integral photon fluxes
above 1~TeV assuming a PL spectrum for the differential flux with an index
$\Gamma = \hgpsAssumedSpecIndex$. Each map is provided for two correlation radii,
$R_{\mathrm{c}} = 0.1\degr$ and $0.2\degr$.

A total of ten files are released (five quantities, each for two $R_{\mathrm{c}}$),
with file names \verb=hgps_map_<quantity>_<radius>deg_v<version>.fits.gz=, e.g.,
the significance map with $R_{\mathrm{c}} = 0.2\degr$ can be found in the file
\verb=hgps_map_significance_0.2deg_v1.fits.gz=.

\subsubsection*{Usage notes and caveats}

\begin{itemize}

\item Since none of the released flux-derived maps are computed for a point-like
source hypothesis, information extracted from these maps should always be used
in the context of full containment of the PSF. Otherwise, this could yield
incorrect information, for example, a flux upper limit that is too low
(optimistic). In particular, since the \hess\ PSF has a size comparable to
0.1\degr, the maps computed with this correlation radius do not fully contain
the PSF. In this case, one only gets roughly 80\% of the flux when reading a
pixel value at a given position of a point-like source. Those maps should
therefore be used with care when extracting a flux value (see also below).

\item The released maps are already spatially correlated (oversampled);
therefore, pixel values should be read at the corresponding position of interest
for a circular region of radius $R_{\mathrm{c}}$. In the case of a region size
between two of the provided $R_{\mathrm{c}}$ values, interpolation could be used
as a first approximation.  The oversampling also implies that maps should not be
used for morphology studies (e.g., production of radial profiles or fitting).

\item Some caution should be taken for values in the $0.2\degr$ correlation maps
where a gradient in exposure is present, since the background is estimated at
the center of the ROI and not averaged across it (see
Sect.~\ref{sec:background_estimation}).

\item The significance maps contain, at each position, the statistical
significance of the \gammaray\ excess. This value is not corrected for trials
and the large-scale emission component is not taken into account in its
computation.

\item We recommend assuming a systematic error of 30\% on the flux values (see
Sect.~\ref{sec:cc:discussion}).

\end{itemize}

\subsection{Source catalog}
\label{sec:online:catalog}

\subsubsection*{Description}

The HGPS source catalog (construction described in Sect.~\ref{sec:cc}) and
a number of other tables are available as BINTABLE \emph{FITS} extensions in
the \verb=hgps_catalog_v1.fits.gz= file.

An overview of the available tables (including links to the tables in this paper
describing the columns in detail) is given in Table~\ref{tab:hgps_fits_catalog}.
Here is some further information on the content of the tables:

\begin{itemize}

\item \verb=HGPS_Sources= : The HGPS catalog, one source per row, identified via
the \verb=Source_Name= column, which is in the format
\verb=HESS JHHMM=$\pm$\verb=DDd=.

\item \verb=HGPS_Gauss_Components= : The HGPS Gaussian component list, one
component per row.  Reference back to \verb=HGPS_Sources= catalog via the
\verb=Source_Name= column (if the component is part of a source).

\item \verb=HGPS_Associations= : The HGPS association list, one association per
row. Reference back to \verb=HGPS_Sources= catalog via the \verb=Source_Name=
column. A given HGPS catalog source can appear zero, one, or multiple times in
this table.

\end{itemize}

\subsubsection*{Usage notes and caveats}

For reasons of reproducibility, we decided to release the complete emission
model on which the analysis is based. This includes the full list of Gaussian
components and parameters of the large-scale emission model. When working
with this data, beware of following usage notes and caveats:

\begin{itemize}

\item Some of the components are unstable and are not confirmed by the
cross-check analysis. Use the source catalog, not the component list, for
studies based on the HGPS.

\item For the HGPS catalog, we did not perform detailed per-source systematic
error estimates. In general, when using spectra from the HGPS catalog, we
recommend assuming a systematic error of 30\% on the absolute flux and 0.2 on
the spectral index.

\item In Fig.~\ref{fig:hgps_sources_spectrum_map_flux_comparison}, there are a
few sources where the integral flux estimate differs by more than 30\% when
using the two methods discussed in this paper. As discussed in
Sect.~\ref{sec:cc:discussion}, the estimate of a source spectrum is affected by
the assumed source morphology, diffuse gamma and atmospheric hadronic background
model and uncertainties in the instrument response functions.  In particular,
the integral flux estimate may be uncertain by more than 30\% for sources with
relatively low significance that are not spatially isolated from other sources.
In those cases, one can assume the difference between \verb=Flux_Map= and
\verb=Flux_Spec_Int_1TeV= to be a lower limit on the systematic error.

\end{itemize}

\begin{table*}
\caption{
HGPS catalog FITS tables. See Sect.~\ref{sec:online:catalog}.
}
\label{tab:hgps_fits_catalog}
\centering
\begin{tabular}{llrl}
\hline \hline
HDU Extension name & Description & Rows & Column description \\
\hline
HGPS\_Sources & HGPS source catalog & \hgpsSourceCountTotal & see Table~\ref{tab:hgps_sources_columns} \\
HGPS\_Gauss\_Components & HGPS component list & \hgpsComponentCountTotal & see Table~\ref{tab:hgps_component_columns} \\
HGPS\_Associations & HGPS association list & \HGPSTotalAssociationCount & see Table~\ref{tab:hgps_associations_columns} \\
HGPS\_Identifications & HGPS identification list & \hgpsSourceCountID & see Table~\ref{tab:hgps_identifications_columns} \\
HGPS\_Large\_Scale\_Component & HGPS large-scale emission model parameters & 50 & see Table~\ref{tab:hgps_diffuse_columns} \\
SNRcat & Bundled version of SNRcat used for associations & 282  &  \\
\hline
\end{tabular}
\end{table*}

\onecolumn
\LTcapwidth=\textwidth

\begin{longtable}{llp{0.5\textwidth}}
\caption[HGPS FITS table columns for \texttt{HGPS\_Sources}]{
HGPS FITS table columns for \texttt{HGPS\_Sources}, the main source catalog.
The column descriptions link back to sections and equations in the main text where needed.
}
\label{tab:hgps_sources_columns}\\
\hline \hline
Column & Unit & Description \\
\hline
\endfirsthead

\caption{continued.}\\
\hline\hline
Column & Unit & Description \\
\hline
\endhead

\hline
\endfoot

Source\_Name                        &                                      & Source name (HESS~JHHmm$\pm$DDd identifier) \\ 
Analysis\_Reference                 &                                      & Source analysis reference ("HGPS" for most sources, "EXTERN" if source in Table~\ref{tab:hgps_external_sources}) \\ 
Source\_Class                       &                                      & Source class (only filled for identified sources, see Table~\ref{tab:hgps_identified_sources}) \\ 
Identified\_Object                  &                                      & Identified object (only filled for identified sources, see Table~\ref{tab:hgps_identified_sources}) \\ 
Gamma\_Cat\_Source\_ID              &                                      & Source ID in the gamma-cat open TeV catalog \\ 
\hline 
RAJ2000                             & $\mathrm{deg}$                       & Right Ascension (J2000) \\ 
DEJ2000                             & $\mathrm{deg}$                       & Declination (J2000) \\ 
GLON                                & $\mathrm{deg}$                       & Galactic longitude \\ 
GLON\_Err                           & $\mathrm{deg}$                       & Statistical error (1 sigma) on GLON \\ 
GLAT                                & $\mathrm{deg}$                       & Galactic latitude \\ 
GLAT\_Err                           & $\mathrm{deg}$                       & Statistical error (1 sigma) on GLAT \\ 
Pos\_Err\_68                        & $\mathrm{deg}$                       & Position error ($68\%$ CL, including systematics, see Eq.~\ref{eq:position_error}) \\ 
Pos\_Err\_95                        & $\mathrm{deg}$                       & Position error ($95\%$ CL, including systematics, see Eq.~\ref{eq:position_error}) \\ 
\hline 
ROI\_Number                         &                                      & ROI number, see Fig.~\ref{fig:catalog:rois} for details \\ 
Spatial\_Model                      &                                      & Spatial model (one of "Gaussian", "X-Gaussian" or "Shell") \\ 
Components                          &                                      & List of Gaussian components the source is composed of \\ 
Sqrt\_TS                            &                                      & Square root of the sum of the test statistics of the individual components (Eq.~\ref{eq:detection:alternative_threshold}) \\ 
Size                                & $\mathrm{deg}$                       & Source size (1 sigma for single-Gaussian sources, RMS equivalent for multi-Gaussian sources and outer radius for SNRs) \\ 
Size\_Err                           & $\mathrm{deg}$                       & Statistical error (1 sigma) on Size \\ 
Size\_UL                            & $\mathrm{deg}$                       & Upper limit (95\% CL) on Size (Eq.~\ref{eq:extension_ul}, NULL if source is extended) \\ 
R70                                 & $\mathrm{deg}$                       & $70\%$ containment radius, computed on the PSF-convolved excess model image \\ 
RSpec                               & $\mathrm{deg}$                       & $R_{spec}$, the radius of the spectral analysis circular region \\ 
Excess\_Model\_Total                &                                      & Total excess from spatial model (this source only) \\ 
Excess\_RSpec                       &                                      & Data excess in R\_Spec (measured on maps) \\ 
Excess\_RSpec\_Model                &                                      & Model excess in R\_Spec (this source, other sources, large scale emission component) \\ 
Background\_RSpec                   &                                      & Background in R\_Spec \\ 
Livetime                            & $\mathrm{hour}$                      & Livetime for map \\ 
Energy\_Threshold                   & $\mathrm{TeV}$                       & Energy threshold for map (minimum) \\ 
\hline 
Flux\_Map                           & $\mathrm{cm^{-2}\,s^{-1}}$           & Integral flux above 1 TeV from the morphology fit on the map (total) \\ 
Flux\_Map\_Err                      & $\mathrm{cm^{-2}\,s^{-1}}$           & Statistical error (1 sigma) on Flux\_Model\_Total \\ 
Flux\_Map\_RSpec\_Data              & $\mathrm{cm^{-2}\,s^{-1}}$           & Data flux in R\_Spec (measured on maps) \\ 
Flux\_Map\_RSpec\_Source            & $\mathrm{cm^{-2}\,s^{-1}}$           & Model flux in R\_Spec (this source only) \\ 
Flux\_Map\_RSpec\_Other             & $\mathrm{cm^{-2}\,s^{-1}}$           & Model flux in R\_Spec (other sources only) \\ 
Flux\_Map\_RSpec\_LS                & $\mathrm{cm^{-2}\,s^{-1}}$           & Model flux in R\_Spec (large scale emission component only) \\ 
Flux\_Map\_RSpec\_Total             & $\mathrm{cm^{-2}\,s^{-1}}$           & Model flux in R\_Spec (this source, other sources, large scale emission component) \\ 
Containment\_RSpec                  &                                      & Containment fraction (Eq.~\ref{eq:containment}) \\ 
Contamination\_RSpec                &                                      & Contamination fraction (Eq.~\ref{eq:contamination}) \\ 
Flux\_Correction\_RSpec\_To\_Total  &                                      & Total flux correction factor (Eq.~\ref{eq:correction_factor_1}) \\ 
\hline 
Livetime\_Spec                      & $\mathrm{hour}$                      & Livetime for spectrum \\ 
Energy\_Range\_Spec\_Min            & $\mathrm{TeV}$                       & Minimum energy of counts spectrum \\ 
Energy\_Range\_Spec\_Max            & $\mathrm{TeV}$                       & Maximum energy of counts spectrum \\ 
Background\_Spec                    &                                      & Background from spectral analysis \\ 
Excess\_Spec                        &                                      & Excess from spectral analysis \\ 
Spectral\_Model                     &                                      & Spectral model, either "PL" or "ECPL" (Eq.~\ref{eqn:pl}) and (Eq.~\ref{eqn:ecpl}) \\ 
TS\_ECPL\_over\_PL                  &                                      & Test statistic difference of ECPL and PL model \\ 
Flux\_Spec\_Int\_1TeV               & $\mathrm{cm^{-2}\,s^{-1}}$           & PL or ECPL integral flux above 1~TeV, depending on Spectral\_Model \\ 
Flux\_Spec\_Int\_1TeV\_Err          & $\mathrm{cm^{-2}\,s^{-1}}$           & Statistical error (1 sigma) on Flux\_Spec\_Int\_1TeV \\ 
Flux\_Spec\_Energy\_1\_10\_TeV      & $\mathrm{erg\,cm^{-2}\,s^{-1}}$      & PL or ECPL energy flux in the 1 to 10 TeV range, depending on Spectral\_Model \\ 
Flux\_Spec\_Energy\_1\_10\_TeV\_Err & $\mathrm{erg\,cm^{-2}\,s^{-1}}$      & Statistical error (1 sigma) on Flux\_Spec\_Energy\_1\_10\_TeV \\ 
Energy\_Spec\_PL\_Pivot             & $\mathrm{TeV}$                       & Reference energy $E_0$, see Eq.~\ref{eqn:pl} \\ 
Flux\_Spec\_PL\_Diff\_Pivot         & $\mathrm{cm^{-2}\,s^{-1}\,TeV^{-1}}$ & Differential flux at pivot energy \\ 
Flux\_Spec\_PL\_Diff\_Pivot\_Err    & $\mathrm{cm^{-2}\,s^{-1}\,TeV^{-1}}$ & Statistical error (1 sigma) on Flux\_Spec\_PL\_Diff\_Pivot \\ 
Flux\_Spec\_PL\_Diff\_1TeV          & $\mathrm{cm^{-2}\,s^{-1}\,TeV^{-1}}$ & Differential flux at 1 TeV \\ 
Flux\_Spec\_PL\_Diff\_1TeV\_Err     & $\mathrm{cm^{-2}\,s^{-1}\,TeV^{-1}}$ & Statistical error (1 sigma) on Flux\_Spec\_PL\_Diff\_1TeV \\ 
Flux\_Spec\_PL\_Int\_1TeV           & $\mathrm{cm^{-2}\,s^{-1}}$           & Integral flux above 1 TeV \\ 
Flux\_Spec\_PL\_Int\_1TeV\_Err      & $\mathrm{cm^{-2}\,s^{-1}}$           & Statistical error (1 sigma) on Flux\_Spec\_PL\_Int\_1TeV \\ 
Index\_Spec\_PL                     &                                      & Spectral index \\ 
Index\_Spec\_PL\_Err                &                                      & Statistical error (1 sigma) on Index\_Spec\_PL \\ 
Energy\_Spec\_ECPL\_Pivot           & $\mathrm{TeV}$                       & Reference energy $E_0$ (Eq.~\ref{eqn:ecpl}) \\ 
Flux\_Spec\_ECPL\_Diff\_Pivot       & $\mathrm{cm^{-2}\,s^{-1}\,TeV^{-1}}$ & Differential flux at pivot energy \\ 
Flux\_Spec\_ECPL\_Diff\_Pivot\_Err  & $\mathrm{cm^{-2}\,s^{-1}\,TeV^{-1}}$ & Statistical error (1 sigma) on Flux\_Spec\_ECPL\_Diff\_Pivot \\ 
Flux\_Spec\_ECPL\_Diff\_1TeV        & $\mathrm{cm^{-2}\,s^{-1}\,TeV^{-1}}$ & Differential flux at 1 TeV \\ 
Flux\_Spec\_ECPL\_Diff\_1TeV\_Err   & $\mathrm{cm^{-2}\,s^{-1}\,TeV^{-1}}$ & Statistical error (1 sigma) on Flux\_Spec\_ECPL\_Diff\_1TeV \\ 
Flux\_Spec\_ECPL\_Int\_1TeV         & $\mathrm{cm^{-2}\,s^{-1}}$           & Integral flux above 1 TeV \\ 
Flux\_Spec\_ECPL\_Int\_1TeV\_Err    & $\mathrm{cm^{-2}\,s^{-1}}$           & Statistical error (1 sigma) on Flux\_Spec\_ECPL\_Int\_1TeV \\ 
Index\_Spec\_ECPL                   &                                      & Spectral index \\ 
Index\_Spec\_ECPL\_Err              &                                      & Statistical error (1 sigma) on Index\_Spec\_ECPL \\ 
Lambda\_Spec\_ECPL                  & $\mathrm{TeV^{-1}}$                  & Spectral cutoff fit parameter (inverse cutoff energy) \\ 
Lambda\_Spec\_ECPL\_Err             & $\mathrm{TeV^{-1}}$                  & Statistical error (1 sigma) on Lambda\_Spec\_ECPL \\ 
\hline 
N\_Flux\_Points                     &                                      & Number of flux points \\ 
Flux\_Points\_Energy                & $\mathrm{TeV}$                       & Energy value \\ 
Flux\_Points\_Energy\_Min           & $\mathrm{TeV}$                       & Lower bound of energy bin \\ 
Flux\_Points\_Energy\_Max           & $\mathrm{TeV}$                       & Upper bound of energy bin \\ 
Flux\_Points\_Flux                  & $\mathrm{cm^{-2}\,s^{-1}\,TeV^{-1}}$ & Differential flux at given energy \\ 
Flux\_Points\_Flux\_Err\_Lo         & $\mathrm{cm^{-2}\,s^{-1}\,TeV^{-1}}$ & Lower error on Flux\_Points\_Flux \\ 
Flux\_Points\_Flux\_Err\_Hi         & $\mathrm{cm^{-2}\,s^{-1}\,TeV^{-1}}$ & Upper error on Flux\_Points\_Flux \\ 
Flux\_Points\_Flux\_UL              & $\mathrm{cm^{-2}\,s^{-1}\,TeV^{-1}}$ & Upper limit on Flux\_Points\_Flux \\ 
Flux\_Points\_Flux\_Is\_UL          &                                      & Boolean flag when to use Flux\_Points\_Flux\_UL \\ 
\hline 

\end{longtable}

\begin{table*}[h!]
\caption[HGPS FITS table columns for \texttt{HGPS\_Gauss\_Components}]{
HGPS FITS table columns for \texttt{HGPS\_Gauss\_Components}.
See Sects.~\ref{sec:cc:components} and \ref{sec:cc:source_characterization}.
}
\label{tab:hgps_component_columns}
\centering
\begin{tabular}{llll}
\hline\hline
Column          & Unit & Description\\
\hline
Component\_ID                       &                                      & Gauss component identifier (HGPSC NNN) \\ 
Source\_Name                        &                                      & Source name (HESS~JHHmm$\pm$DDd identifier) the component belongs to \\ 
Component\_Class                    &                                      & Component class (see Sect.~\ref{sec:cc:component_classification}) \\ 
GLON                                & $\mathrm{deg}$                       & Galactic longitude \\ 
GLON\_Err                           & $\mathrm{deg}$                       & Statistical error (1 sigma) on GLON \\ 
GLAT                                & $\mathrm{deg}$                       & Galactic latitude \\ 
GLAT\_Err                           & $\mathrm{deg}$                       & Statistical error (1 sigma) on GLAT \\ 
Sqrt\_TS                            &                                      & Square root of the the test statistics of the component (see Eq.~\ref{eq:detection:alternative_threshold}) \\ 
Size                                & $\mathrm{deg}$                       & Component size (1 $\sigma$ Gaussian width) \\ 
Size\_Err                           & $\mathrm{deg}$                       & Statistical error (1 sigma) on Size \\ 
Flux\_Map                           & $\mathrm{cm^{-2}\,s^{-1}}$           & Integral flux above 1 TeV from the morphology fit on the map (total) \\ 
Flux\_Map\_Err                      & $\mathrm{cm^{-2}\,s^{-1}}$           & Statistical error (1 sigma) on Flux\_Map \\ 
Excess                              &                                      & Total model excess contained in the component \\ 
\hline 

\end{tabular}
\end{table*}

\begin{table*}[h!]
\caption[HGPS FITS table columns for \texttt{HGPS\_Associations}]{
HGPS FITS table columns for \texttt{HGPS\_Associations}.
See Sect.~\ref{sec:results:assoc_id}.
}
\label{tab:hgps_associations_columns}
\centering
\begin{tabular}{llll}
\hline\hline
Column & Unit & Description\\
\hline
Source\_Name                        &                                      & Source name (HESS~JHHmm$\pm$DDd identifier) \\ 
Association\_Catalog                &                                      & Association catalog name (see Table~\ref{tab:hgps_associations_catalogs}) \\ 
Association\_Name                   &                                      & Association source name \\ 
Separation                          & $\mathrm{deg}$                       & Angular separation to HGPS source position \\ 
\hline 

\end{tabular}
\end{table*}

\begin{table*}[h!]
\caption[HGPS FITS table columns for \texttt{HGPS\_Identifications}]{
HGPS FITS table columns for \texttt{HGPS\_Identifications}.
See Table~\ref{sec:identifications}.
}
\label{tab:hgps_identifications_columns}
\centering
\begin{tabular}{llll}
\hline\hline
Column & Unit & Description\\
\hline
Source\_Name                        &                                      & Source name (HESS~JHHmm$\pm$DDd identifier) \\ 
Identified\_Object                  &                                      & Identified object name \\ 
Class                               &                                      & Class of the identified object \\ 
Evidence                            &                                      & Evidence for the identification \\ 
Reference                           &                                      & Reference for the identification \\ 
Distance                            & $\mathrm{kpc}$                       & Distance of the identified object \\ 
Distance\_Min                       & $\mathrm{kpc}$                       & Minimum distance of the identified object \\ 
Distance\_Max                       & $\mathrm{kpc}$                       & Maximum distance of the identified object \\ 
Distance\_Reference                 &                                      & Reference for the distance estimate \\ 
\hline 

\end{tabular}
\end{table*}

\begin{table*}[h!]
\caption[HGPS FITS table columns for \texttt{HGPS\_Large\_Scale\_Component}]{
HGPS FITS table columns for \texttt{HGPS\_Large\_Scale\_Component}
See Sect.~\ref{sec:cc:large-scale-emission}.
}
\label{tab:hgps_diffuse_columns}
\centering
\begin{tabular}{llll}
\hline\hline
Column             & Unit & Description\\
\hline
GLON                                & $\mathrm{deg}$                       & Galactic longitude of window center \\ 
Surface\_Brightness                 & $\mathrm{cm^{-2}\,s^{-1}\,sr^{-1}}$  & Peak brightness \\ 
Surface\_Brightness\_Err            & $\mathrm{cm^{-2}\,s^{-1}\,sr^{-1}}$  & Statistical error (1 sigma) on peak surface brightness \\ 
GLAT                                & $\mathrm{deg}$                       & Peak latitude \\ 
GLAT\_Err                           & $\mathrm{deg}$                       & Statistical error (1 sigma) on peak latitude\ \\ 
Width                               & $\mathrm{deg}$                       & Gaussian width \\ 
Width\_Err                          & $\mathrm{deg}$                       & Statistical error (1 sigma) on Gaussian width \\ 
\hline 

\end{tabular}
\end{table*}

\section*{Acknowledgements}

This work made extensive use of gamma-cat\footnote{\urlGammacat},
SNRcat\footnote{\urlSnrcat} \citep{SNRcat}, ATNF\footnote{\urlAtnf}
\citep{Manchester:2005}, SIMBAD\footnote{\urlSimbad}
\citep{2000AnAS..143....9W} and NASA's Astrophysics Data System Bibliographic
Services.
For data analysis, we made extensive use of the Python packages
Gammapy\footnote{\urlGammapy} \citep{Donath2015, 2017arXiv170901751D},
Astropy\footnote{\urlAstropy} \citep{2013AandA...558A..33A} and
Sherpa\footnote{\urlSherpa} \citep{Freeman:2001}, as well as Numpy
\citep{numpy}, Scipy \citep{scipy} and Matplotlib \citep{matplotlib}.

The support of the Namibian authorities and the University of Namibia in
facilitating the construction and operation of H.E.S.S. is gratefully
acknowledged, as is the support by the German Ministry for Education and
Research (BMBF), the Max Planck Society, the German Research Foundation (DFG),
the French Ministry for Research, the CNRS-IN2P3 and the Astroparticle
Interdisciplinary Programme of the CNRS, the U.K. Science and Technology
Facilities Council (STFC), the IPNP of the Charles University, the Czech Science
Foundation, the Polish Ministry of Science and Higher Education, the South
African Department of Science and Technology and National Research Foundation,
the University of Namibia, the Innsbruck University, the Austrian Science Fund
(FWF), and the Austrian Federal Ministry for Science, Research and Economy, and
by the University of Adelaide and the Australian Research Council. We appreciate
the excellent work of the technical support staff in Berlin, Durham, Hamburg,
Heidelberg, Palaiseau, Paris, Saclay, and in Namibia in the construction and
operation of the equipment. This work benefited from services provided by the
H.E.S.S. Virtual Organisation, supported by the national resource providers of
the EGI Federation.

\listofobjects

\bibliographystyle{aa}
\bibliography{hgps_paper}

\begin{thebibliography}{235}
\expandafter\ifx\csname natexlab\endcsname\relax\def\natexlab#1{#1}\fi

\bibitem[{{Abdalla} {et~al.}(2016){Abdalla}, {Abramowski}, {Aharonian}, {Ait
  Benkhali}, {Akhperjanian}, {Ang{\"u}ner}, {Arrieta}, {Aubert}, {Backes},
  {Balzer}, \& et~al.}]{HESS:1808}
{Abdalla}, H., {Abramowski}, A., {Aharonian}, F., {et~al.} 2016, ArXiv e-prints
  [\eprint[arXiv]{1606.05404}]

\bibitem[{{Abdo} {et~al.}(2009{\natexlab{a}}){Abdo}, {Ackermann}, {Ajello},
  {Anderson}, {Atwood}, {Axelsson}, {Baldini}, {Ballet}, {Barbiellini},
  {Baring}, {Bastieri}, {Baughman}, {Bechtol}, {Bellazzini}, {Berenji},
  {Bignami}, {Blandford}, {Bloom}, {Bonamente}, {Borgland}, {Bregeon}, {Brez},
  {Brigida}, {Bruel}, {Burnett}, {Caliandro}, {Cameron}, {Caraveo},
  {Casandjian}, {Cecchi}, {{\c C}elik}, {Chekhtman}, {Cheung}, {Chiang},
  {Ciprini}, {Claus}, {Cohen-Tanugi}, {Conrad}, {Cutini}, {Dermer}, {de
  Angelis}, {de Luca}, {de Palma}, {Digel}, {Dormody}, {do Couto e Silva},
  {Drell}, {Dubois}, {Dumora}, {Farnier}, {Favuzzi}, {Fegan}, {Fukazawa},
  {Funk}, {Fusco}, {Gargano}, {Gasparrini}, {Gehrels}, {Germani}, {Giebels},
  {Giglietto}, {Giommi}, {Giordano}, {Glanzman}, {Godfrey}, {Grenier},
  {Grondin}, {Grove}, {Guillemot}, {Guiriec}, {Gwon}, {Hanabata}, {Harding},
  {Hayashida}, {Hays}, {Hughes}, {J{\'o}hannesson}, {Johnson}, {Johnson},
  {Johnson}, {Kamae}, {Katagiri}, {Kataoka}, {Kawai}, {Kerr}, {Kn{\"o}dlseder},
  {Kocian}, {Kuss}, {Lande}, {Latronico}, {Lemoine-Goumard}, {Longo},
  {Loparco}, {Lott}, {Lovellette}, {Lubrano}, {Madejski}, {Makeev}, {Marelli},
  {Mazziotta}, {McConville}, {McEnery}, {Meurer}, {Michelson}, {Mitthumsiri},
  {Mizuno}, {Monte}, {Monzani}, {Morselli}, {Moskalenko}, {Murgia}, {Nolan},
  {Norris}, {Nuss}, {Ohsugi}, {Omodei}, {Orlando}, {Ormes}, {Paneque},
  {Parent}, {Pelassa}, {Pepe}, {Pesce-Rollins}, {Pierbattista}, {Piron},
  {Porter}, {Primack}, {Rain{\`o}}, {Rando}, {Ray}, {Razzano}, {Rea}, {Reimer},
  {Reimer}, {Reposeur}, {Ritz}, {Rochester}, {Rodriguez}, {Romani}, {Ryde},
  {Sadrozinski}, {Sanchez}, {Sander}, {Parkinson}, {Scargle}, {Sgr{\`o}},
  {Siskind}, {Smith}, {Smith}, {Spandre}, {Spinelli}, {Starck}, {Strickman},
  {Suson}, {Tajima}, {Takahashi}, {Takahashi}, {Tanaka}, {Thayer}, {Thompson},
  {Tibaldo}, {Tibolla}, {Torres}, {Tosti}, {Tramacere}, {Uchiyama}, {Usher},
  {Van Etten}, {Vasileiou}, {Vilchez}, {Vitale}, {Waite}, {Wang}, {Watters},
  {Winer}, {Wolff}, {Wood}, {Ylinen}, {Ziegler}, \& {Fermi LAT
  Collaboration}}]{Abdo09}
{Abdo}, A.~A., {Ackermann}, M., {Ajello}, M., {et~al.} 2009{\natexlab{a}},
  Science, 325, 840

\bibitem[{{Abdo} {et~al.}(2009{\natexlab{b}}){Abdo}, {Ackermann}, {Ajello},
  {Atwood}, {Axelsson}, {Baldini}, {Ballet}, {Band}, {Barbiellini}, {Bastieri},
  {Battelino}, {Baughman}, {Bechtol}, {Bellazzini}, {Berenji}, {Bignami},
  {Blandford}, {Bloom}, {Bonamente}, {Borgland}, {Bouvier}, {Bregeon}, {Brez},
  {Brigida}, {Bruel}, {Burnett}, {Caliandro}, {Cameron}, {Caraveo},
  {Casandjian}, {Cavazzuti}, {Cecchi}, {Charles}, {Chekhtman}, {Cheung},
  {Chiang}, {Ciprini}, {Claus}, {Cohen-Tanugi}, {Cominsky}, {Conrad}, {Corbet},
  {Costamante}, {Cutini}, {Davis}, {Dermer}, {de Angelis}, {de Luca}, {de
  Palma}, {Digel}, {Dormody}, {do Couto e Silva}, {Drell}, {Dubois}, {Dumora},
  {Farnier}, {Favuzzi}, {Fegan}, {Ferrara}, {Focke}, {Frailis}, {Fukazawa},
  {Funk}, {Fusco}, {Gargano}, {Gasparrini}, {Gehrels}, {Germani}, {Giebels},
  {Giglietto}, {Giommi}, {Giordano}, {Glanzman}, {Godfrey}, {Grenier},
  {Grondin}, {Grove}, {Guillemot}, {Guiriec}, {Hanabata}, {Harding}, {Hartman},
  {Hayashida}, {Hays}, {Healey}, {Horan}, {Hughes}, {J{\'o}hannesson},
  {Johnson}, {Johnson}, {Johnson}, {Johnson}, {Kamae}, {Katagiri}, {Kataoka},
  {Kawai}, {Kerr}, {Kn{\"o}dlseder}, {Kocevski}, {Kocian}, {Komin}, {Kuehn},
  {Kuss}, {Lande}, {Latronico}, {Lee}, {Lemoine-Goumard}, {Longo}, {Loparco},
  {Lott}, {Lovellette}, {Lubrano}, {Madejski}, {Makeev}, {Marelli},
  {Mazziotta}, {McConville}, {McEnery}, {McGlynn}, {Meurer}, {Michelson},
  {Mitthumsiri}, {Mizuno}, {Moiseev}, {Monte}, {Monzani}, {Moretti},
  {Morselli}, {Moskalenko}, {Murgia}, {Nakamori}, {Nolan}, {Norris}, {Nuss},
  {Ohno}, {Ohsugi}, {Omodei}, {Orlando}, {Ormes}, {Ozaki}, {Paneque},
  {Panetta}, {Parent}, {Pelassa}, {Pepe}, {Pesce-Rollins}, {Piron}, {Porter},
  {Poupard}, {Rain{\`o}}, {Rando}, {Ray}, {Razzano}, {Rea}, {Reimer}, {Reimer},
  {Reposeur}, {Ritz}, {Rochester}, {Rodriguez}, {Romani}, {Roth}, {Ryde},
  {Sadrozinski}, {Sanchez}, {Sander}, {Saz Parkinson}, {Scargle}, {Schalk},
  {Sellerholm}, {Sgr{\`o}}, {Shaw}, {Shrader}, {Sierpowska-Bartosik},
  {Siskind}, {Smith}, {Smith}, {Spandre}, {Spinelli}, {Starck}, {Stephens},
  {Strickman}, {Strong}, {Suson}, {Tajima}, {Takahashi}, {Takahashi}, {Tanaka},
  {Thayer}, {Thayer}, {Thompson}, {Tibaldo}, {Tibolla}, {Torres}, {Tosti},
  {Tramacere}, {Uchiyama}, {Usher}, {Van Etten}, {Vilchez}, {Vitale}, {Waite},
  {Wallace}, {Wang}, {Watters}, {Winer}, {Wood}, {Ylinen}, {Ziegler}, \& {The
  Fermi/LAT Collaboration}}]{Abdo:2009e}
{Abdo}, A.~A., {Ackermann}, M., {Ajello}, M., {et~al.} 2009{\natexlab{b}},
  \apjs, 183, 46

\bibitem[{{Abdo} {et~al.}(2013){Abdo}, {Ajello}, {Allafort}, {Baldini},
  {Ballet}, {Barbiellini}, {Baring}, {Bastieri}, {Belfiore}, {Bellazzini}, \&
  et~al.}]{Abdo13}
{Abdo}, A.~A., {Ajello}, M., {Allafort}, A., {et~al.} 2013, \apjs, 208, 17

\bibitem[{{Abeysekara} {et~al.}(2017){Abeysekara}, {Albert}, {Alfaro},
  {Alvarez}, {{\'A}lvarez}, {Arceo}, {Arteaga-Vel{\'a}zquez}, {Ayala Solares},
  {Barber}, {Baughman}, {Bautista-Elivar}, {Becerra Gonzalez}, {Becerril},
  {Belmont-Moreno}, {BenZvi}, {Berley}, {Bernal}, {Braun}, {Brisbois},
  {Caballero-Mora}, {Capistr{\'a}n}, {Carrami{\~n}ana}, {Casanova}, {Castillo},
  {Cotti}, {Cotzomi}, {Couti{\~n}o de Le{\'o}n}, {de la Fuente}, {De Le{\'o}n},
  {Diaz Hernandez}, {Dingus}, {DuVernois}, {D{\'{\i}}az-V{\'e}lez},
  {Ellsworth}, {Engel}, {Fiorino}, {Fraija}, {Garc{\'{\i}}a-Gonz{\'a}lez},
  {Garfias}, {Gerhardt}, {Gonz{\'a}lez Mu{\~n}oz}, {Gonz{\'a}lez}, {Goodman},
  {Hampel-Arias}, {Harding}, {Hernandez}, {Hernandez-Almada}, {Hinton}, {Hui},
  {H{\"u}ntemeyer}, {Iriarte}, {Jardin-Blicq}, {Joshi}, {Kaufmann}, {Kieda},
  {Lara}, {Lauer}, {Lee}, {Lennarz}, {Le{\'o}n Vargas}, {Linnemann},
  {Longinotti}, {Raya}, {Luna-Garc{\'{\i}}a}, {L{\'o}pez-Coto}, {Malone},
  {Marinelli}, {Martinez}, {Martinez-Castellanos}, {Mart{\'{\i}}nez-Castro},
  {Mart{\'{\i}}nez-Huerta}, {Matthews}, {Miranda-Romagnoli}, {Moreno},
  {Mostaf{\'a}}, {Nellen}, {Newbold}, {Nisa}, {Noriega-Papaqui}, {Pelayo},
  {Pretz}, {P{\'e}rez-P{\'e}rez}, {Ren}, {Rho}, {Rivi{\`e}re},
  {Rosa-Gonz{\'a}lez}, {Rosenberg}, {Ruiz-Velasco}, {Salazar}, {Salesa Greus},
  {Sandoval}, {Schneider}, {Schoorlemmer}, {Sinnis}, {Smith}, {Springer},
  {Surajbali}, {Taboada}, {Tibolla}, {Tollefson}, {Torres}, {Ukwatta},
  {Vianello}, {Villase{\~n}or}, {Weisgarber}, {Westerhoff}, {Wisher}, {Wood},
  {Yapici}, {Younk}, {Zepeda}, \& {Zhou}}]{2017ApJ...843...40A}
{Abeysekara}, A.~U., {Albert}, A., {Alfaro}, R., {et~al.} 2017, \apj, 843, 40

\bibitem[{{Abramowski} {et~al.}(2012{\natexlab{a}}){Abramowski}, {Acero},
  {Aharonian}, {Akhperjanian}, {Anton}, {Balenderan}, {Balzer}, {Barnacka},
  {Becherini}, {Becker Tjus}, {Bernl{\"o}hr}, {Birsin}, {Biteau}, {Bochow},
  {Boisson}, {Bolmont}, {Bordas}, {Brucker}, {Brun}, {Brun}, {Bulik},
  {Carrigan}, {Casanova}, {Cerruti}, {Chadwick}, {Charbonnier}, {Chaves},
  {Cheesebrough}, {Cologna}, {Conrad}, {Couturier}, {Dalton}, {Daniel},
  {Davids}, {Degrange}, {Deil}, {deWilt}, {Dickinson}, {Djannati-Ata{\"i}},
  {Domainko}, {Drury}, {Dubois}, {Dubus}, {Dutson}, {Dyks}, {Dyrda}, {Egberts},
  {Eger}, {Espigat}, {Fallon}, {Farnier}, {Fegan}, {Feinstein}, {Fernandes},
  {Fernandez}, {Fiasson}, {Fontaine}, {F{\"o}rster}, {F{\"u}{\ss}ling},
  {Gajdus}, {Gallant}, {Garrigoux}, {Gast}, {Giebels}, {Glicenstein},
  {Gl{\"u}ck}, {G{\"o}ring}, {Grondin}, {H{\"a}ffner}, {Hague}, {Hahn},
  {Hampf}, {Harris}, {Heinz}, {Heinzelmann}, {Henri}, {Hermann}, {Hillert},
  {Hinton}, {Hofmann}, {Hofverberg}, {Holler}, {Horns}, {Jacholkowska}, {Jahn},
  {Jamrozy}, {Jung}, {Kastendieck}, {Katarzy{\'n}ski}, {Katz}, {Kaufmann},
  {Kh{\'e}lifi}, {Klochkov}, {Klu{\'z}niak}, {Kneiske}, {Komin}, {Kosack},
  {Kossakowski}, {Krayzel}, {Kr{\"u}ger}, {Laffon}, {Lamanna}, {Lenain},
  {Lennarz}, {Lohse}, {Lopatin}, {Lu}, {Marandon}, {Marcowith}, {Masbou},
  {Maurin}, {Maxted}, {Mayer}, {McComb}, {Medina}, {M{\'e}hault}, {Menzler},
  {Moderski}, {Mohamed}, {Moulin}, {Naumann}, {Naumann-Godo}, {de Naurois},
  {Nedbal}, {Nguyen}, {Niemiec}, {Nolan}, {Ohm}, {de O{\~n}a Wilhelmi},
  {Opitz}, {Ostrowski}, {Oya}, {Panter}, {Parsons}, {Paz Arribas}, {Pekeur},
  {Pelletier}, {Perez}, {Petrucci}, {Peyaud}, {Pita}, {P{\"u}hlhofer}, {Punch},
  {Quirrenbach}, {Raue}, {Reimer}, {Reimer}, {Renaud}, {de los Reyes},
  {Rieger}, {Ripken}, {Rob}, {Rosier-Lees}, {Rowell}, {Rudak}, {Rulten},
  {Sahakian}, {Sanchez}, {Santangelo}, {Schlickeiser}, {Schulz}, {Schwanke},
  {Schwarzburg}, {Schwemmer}, {Sheidaei}, {Skilton}, {Sol}, {Spengler},
  {Stawarz}, {Steenkamp}, {Stegmann}, {Stinzing}, {Stycz}, {Sushch}, {Szostek},
  {Tavernet}, {Terrier}, {Tluczykont}, {Trichard}, {Valerius}, {van Eldik},
  {Vasileiadis}, {Venter}, {Viana}, {Vincent}, {V{\"o}lk}, {Volpe}, {Vorobiov},
  {Vorster}, {Wagner}, {Ward}, {White}, {Wierzcholska}, {Wouters}, {Zacharias},
  {Zajczyk}, {Zdziarski}, {Zech}, \& {Zechlin}}]{2012A&A...548A..38A}
{Abramowski}, A., {Acero}, F., {Aharonian}, F., {et~al.} 2012{\natexlab{a}},
  \aap, 548, A38

\bibitem[{{Abramowski} {et~al.}(2012{\natexlab{b}}){Abramowski}, {Acero},
  {Aharonian}, {Akhperjanian}, {Anton}, {Balzer}, {Barnacka}, {Barres de
  Almeida}, {Becherini}, {Becker}, {Behera}, {Bernl{\"o}hr}, {Birsin},
  {Biteau}, {Bochow}, {Boisson}, {Bolmont}, {Bordas}, {Brucker}, {Brun},
  {Brun}, {Bulik}, {B{\"u}sching}, {Carrigan}, {Casanova}, {Cerruti},
  {Chadwick}, {Charbonnier}, {Chaves}, {Cheesebrough}, {Chounet}, {Clapson},
  {Coignet}, {Cologna}, {Conrad}, {Dalton}, {Daniel}, {Davids}, {Degrange},
  {Deil}, {Dickinson}, {Djannati-Ata{\"i}}, {Domainko}, {O'C.~Drury}, {Dubois},
  {Dubus}, {Dutson}, {Dyks}, {Dyrda}, {Egberts}, {Eger}, {Espigat}, {Fallon},
  {Farnier}, {Fegan}, {Feinstein}, {Fernandes}, {Fiasson}, {Fontaine},
  {F{\"o}rster}, {F{\"u}{\ss}ling}, {Gallant}, {Gast}, {G{\'e}rard}, {Gerbig},
  {Giebels}, {Glicenstein}, {Gl{\"u}ck}, {Goret}, {G{\"o}ring}, {H{\"a}ffner},
  {Hague}, {Hampf}, {Hauser}, {Heinz}, {Heinzelmann}, {Henri}, {Hermann},
  {Hinton}, {Hoffmann}, {Hofmann}, {Hofverberg}, {Holler}, {Horns},
  {Jacholkowska}, {de Jager}, {Jahn}, {Jamrozy}, {Jung}, {Kastendieck},
  {Katarzy{\'n}ski}, {Katz}, {Kaufmann}, {Keogh}, {Khangulyan}, {Kh{\'e}lifi},
  {Klochkov}, {Klu{\.z}niak}, {Kneiske}, {Komin}, {Kosack}, {Kossakowski},
  {Laffon}, {Lamanna}, {Lennarz}, {Lohse}, {Lopatin}, {Lu}, {Marandon},
  {Marcowith}, {Masbou}, {Maurin}, {Maxted}, {Mayer}, {McComb}, {Medina},
  {M{\'e}hault}, {Moderski}, {Moulin}, {Naumann}, {Naumann-Godo}, {de Naurois},
  {Nedbal}, {Nekrassov}, {Nguyen}, {Nicholas}, {Niemiec}, {Nolan}, {Ohm}, {de
  O{\~n}a Wilhelmi}, {Opitz}, {Ostrowski}, {Oya}, {Panter}, {Paz Arribas},
  {Pedaletti}, {Pelletier}, {Petrucci}, {Pita}, {P{\"u}hlhofer}, {Punch},
  {Quirrenbach}, {Raue}, {Rayner}, {Reimer}, {Reimer}, {Renaud}, {de Los
  Reyes}, {Rieger}, {Ripken}, {Rob}, {Rosier-Lees}, {Rowell}, {Rudak},
  {Rulten}, {Ruppel}, {Sahakian}, {Sanchez}, {Santangelo}, {Schlickeiser},
  {Sch{\"o}ck}, {Schulz}, {Schwanke}, {Schwarzburg}, {Schwemmer}, {Sheidaei},
  {Sikora}, {Skilton}, {Sol}, {Spengler}, {Stawarz}, {Steenkamp}, {Stegmann},
  {Stinzing}, {Stycz}, {Sushch}, {Szostek}, {Tavernet}, {Terrier},
  {Tluczykont}, {Valerius}, {van Eldik}, {Vasileiadis}, {Venter}, {Vialle},
  {Viana}, {Vincent}, {V{\"o}lk}, {Volpe}, {Vorobiov}, {Vorster}, {Wagner},
  {Ward}, {White}, {Wierzcholska}, {Zacharias}, {Zajczyk}, {Zdziarski}, {Zech},
  \& {Zechlin}}]{2012A&A...537A.114A}
{Abramowski}, A., {Acero}, F., {Aharonian}, F., {et~al.} 2012{\natexlab{b}},
  \aap, 537, A114

\bibitem[{{Abramowski} {et~al.}(2014{\natexlab{a}}){Abramowski}, {Aharonian},
  {Ait Benkhali}, {Akhperjanian}, {Ang{\"u}ner}, {Backes}, {Balenderan},
  {Balzer}, {Barnacka}, {Becherini}, \& et~al.}]{2014PhRvD..90l2007A}
{Abramowski}, A., {Aharonian}, F., {Ait Benkhali}, F., {et~al.}
  2014{\natexlab{a}}, \prd, 90, 122007

\bibitem[{{Abramowski} {et~al.}(2014{\natexlab{b}}){Abramowski}, {Aharonian},
  {Ait Benkhali}, {Akhperjanian}, {Ang{\"u}ner}, {Anton}, {Balenderan},
  {Balzer}, {Barnacka}, {Becherini}, \& et~al.}]{2014MNRAS.439.2828A}
{Abramowski}, A., {Aharonian}, F., {Ait Benkhali}, F.~A., {et~al.}
  2014{\natexlab{b}}, \mnras, 439, 2828

\bibitem[{{Acciari} {et~al.}(2010){Acciari}, {Aliu}, {Arlen}, {Aune},
  {Bautista}, {Beilicke}, {Benbow}, {Boltuch}, {Bradbury}, {Buckley}, {Bugaev},
  {Butt}, {Byrum}, {Cesarini}, {Ciupik}, {Cui}, {Dickherber}, {Duke}, {Finley},
  {Finnegan}, {Fortson}, {Furniss}, {Galante}, {Gall}, {Gillanders}, {Godambe},
  {Gotthelf}, {Grube}, {Guenette}, {Gyuk}, {Hanna}, {Holder}, {Hui},
  {Humensky}, {Imran}, {Kaaret}, {Karlsson}, {Kertzman}, {Kieda}, {Konopelko},
  {Krawczynski}, {Krennrich}, {Lang}, {LeBohec}, {Maier}, {McArthur}, {McCann},
  {McCutcheon}, {Moriarty}, {Muhkerjee}, {Ong}, {Otte}, {Pandel}, {Perkins},
  {Pohl}, {Quinn}, {Ragan}, {Reyes}, {Reynolds}, {Roache}, {Rose},
  {Schroedter}, {Sembroski}, {Senturk}, {Slane}, {Smith}, {Steele}, {Swordy},
  {T{\v e}si{\'c}}, {Theiling}, {Thibadeau}, {Vassiliev}, {Vincent}, {Wakely},
  {Ward}, {Weekes}, {Weinstein}, {Weisgarber}, {Williams}, {Wissel}, {Wood}, \&
  {Zitzer}}]{2010ApJ...719L..69A}
{Acciari}, V.~A., {Aliu}, E., {Arlen}, T., {et~al.} 2010, \apjl, 719, L69

\bibitem[{{Acero} {et~al.}(2015{\natexlab{a}}){Acero}, {Ackermann}, {Ajello},
  {Albert}, {Atwood}, {Axelsson}, {Baldini}, {Ballet}, {Barbiellini},
  {Bastieri}, {Belfiore}, {Bellazzini}, {Bissaldi}, {Blandford}, {Bloom},
  {Bogart}, {Bonino}, {Bottacini}, {Bregeon}, {Britto}, {Bruel}, {Buehler},
  {Burnett}, {Buson}, {Caliandro}, {Cameron}, {Caputo}, {Caragiulo}, {Caraveo},
  {Casandjian}, {Cavazzuti}, {Charles}, {Chaves}, {Chekhtman}, {Cheung},
  {Chiang}, {Chiaro}, {Ciprini}, {Claus}, {Cohen-Tanugi}, {Cominsky}, {Conrad},
  {Cutini}, {D'Ammando}, {de Angelis}, {DeKlotz}, {de Palma}, {Desiante},
  {Digel}, {Di Venere}, {Drell}, {Dubois}, {Dumora}, {Favuzzi}, {Fegan},
  {Ferrara}, {Finke}, {Franckowiak}, {Fukazawa}, {Funk}, {Fusco}, {Gargano},
  {Gasparrini}, {Giebels}, {Giglietto}, {Giommi}, {Giordano}, {Giroletti},
  {Glanzman}, {Godfrey}, {Grenier}, {Grondin}, {Grove}, {Guillemot}, {Guiriec},
  {Hadasch}, {Harding}, {Hays}, {Hewitt}, {Hill}, {Horan}, {Iafrate}, {Jogler},
  {J{\'o}hannesson}, {Johnson}, {Johnson}, {Johnson}, {Johnson}, {Kamae},
  {Kataoka}, {Katsuta}, {Kuss}, {La Mura}, {Landriu}, {Larsson}, {Latronico},
  {Lemoine-Goumard}, {Li}, {Li}, {Longo}, {Loparco}, {Lott}, {Lovellette},
  {Lubrano}, {Madejski}, {Massaro}, {Mayer}, {Mazziotta}, {McEnery},
  {Michelson}, {Mirabal}, {Mizuno}, {Moiseev}, {Mongelli}, {Monzani},
  {Morselli}, {Moskalenko}, {Murgia}, {Nuss}, {Ohno}, {Ohsugi}, {Omodei},
  {Orienti}, {Orlando}, {Ormes}, {Paneque}, {Panetta}, {Perkins},
  {Pesce-Rollins}, {Piron}, {Pivato}, {Porter}, {Racusin}, {Rando}, {Razzano},
  {Razzaque}, {Reimer}, {Reimer}, {Reposeur}, {Rochester}, {Romani},
  {Salvetti}, {S{\'a}nchez-Conde}, {Saz Parkinson}, {Schulz}, {Siskind},
  {Smith}, {Spada}, {Spandre}, {Spinelli}, {Stephens}, {Strong}, {Suson},
  {Takahashi}, {Takahashi}, {Tanaka}, {Thayer}, {Thayer}, {Thompson},
  {Tibaldo}, {Tibolla}, {Torres}, {Torresi}, {Tosti}, {Troja}, {Van Klaveren},
  {Vianello}, {Winer}, {Wood}, {Wood}, {Zimmer}, \& {Fermi-LAT
  Collaboration}}]{3FGL}
{Acero}, F., {Ackermann}, M., {Ajello}, M., {et~al.} 2015{\natexlab{a}}, \apjs,
  218, 23

\bibitem[{{Acero} {et~al.}(2013){Acero}, {Ackermann}, {Ajello}, {Allafort},
  {Baldini}, {Ballet}, {Barbiellini}, {Bastieri}, {Bechtol}, {Bellazzini},
  {Blandford}, {Bloom}, {Bonamente}, {Bottacini}, {Brandt}, {Bregeon},
  {Brigida}, {Bruel}, {Buehler}, {Buson}, {Caliandro}, {Cameron}, {Caraveo},
  {Cecchi}, {Charles}, {Chaves}, {Chekhtman}, {Chiang}, {Chiaro}, {Ciprini},
  {Claus}, {Cohen-Tanugi}, {Conrad}, {Cutini}, {Dalton}, {D'Ammando}, {de
  Palma}, {Dermer}, {Di Venere}, {Silva}, {Drell}, {Drlica-Wagner}, {Falletti},
  {Favuzzi}, {Fegan}, {Ferrara}, {Focke}, {Franckowiak}, {Fukazawa}, {Funk},
  {Fusco}, {Gargano}, {Gasparrini}, {Giglietto}, {Giordano}, {Giroletti},
  {Glanzman}, {Godfrey}, {Gr{\'e}goire}, {Grenier}, {Grondin}, {Grove},
  {Guiriec}, {Hadasch}, {Hanabata}, {Harding}, {Hayashida}, {Hayashi}, {Hays},
  {Hewitt}, {Hill}, {Horan}, {Hou}, {Hughes}, {Inoue}, {Jackson}, {Jogler},
  {J{\'o}hannesson}, {Johnson}, {Kamae}, {Kawano}, {Kerr}, {Kn{\"o}dlseder},
  {Kuss}, {Lande}, {Larsson}, {Latronico}, {Lemoine-Goumard}, {Longo},
  {Loparco}, {Lovellette}, {Lubrano}, {Marelli}, {Massaro}, {Mayer},
  {Mazziotta}, {McEnery}, {Mehault}, {Michelson}, {Mitthumsiri}, {Mizuno},
  {Monte}, {Monzani}, {Morselli}, {Moskalenko}, {Murgia}, {Nakamori}, {Nemmen},
  {Nuss}, {Ohsugi}, {Okumura}, {Orienti}, {Orlando}, {Ormes}, {Paneque},
  {Panetta}, {Perkins}, {Pesce-Rollins}, {Piron}, {Pivato}, {Porter},
  {Rain{\`o}}, {Rando}, {Razzano}, {Reimer}, {Reimer}, {Reposeur}, {Ritz},
  {Roth}, {Rousseau}, {Saz Parkinson}, {Schulz}, {Sgr{\`o}}, {Siskind},
  {Smith}, {Spandre}, {Spinelli}, {Suson}, {Takahashi}, {Takeuchi}, {Thayer},
  {Thayer}, {Thompson}, {Tibaldo}, {Tibolla}, {Tinivella}, {Torres}, {Tosti},
  {Troja}, {Uchiyama}, {Vandenbroucke}, {Vasileiou}, {Vianello}, {Vitale},
  {Werner}, {Winer}, {Wood}, \& {Yang}}]{Acero13}
{Acero}, F., {Ackermann}, M., {Ajello}, M., {et~al.} 2013, \apj, 773, 77

\bibitem[{{Acero} {et~al.}(2015{\natexlab{b}}){Acero}, {Ackermann}, {Ajello},
  {Baldini}, {Ballet}, {Barbiellini}, {Bastieri}, {Bellazzini}, {Bissaldi},
  {Blandford}, {Bloom}, {Bonino}, {Bottacini}, {Bregeon}, {Bruel}, {Buehler},
  {Buson}, {Caliandro}, {Cameron}, {Caputo}, {Caragiulo}, {Caraveo},
  {Casandjian}, {Cavazzuti}, {Cecchi}, {Chekhtman}, {Chiang}, {Chiaro},
  {Ciprini}, {Claus}, {Cohen}, {Cohen-Tanugi}, {Cominsky}, {Condon}, {Conrad},
  {Cutini}, {D'Ammando}, {Angelis}, {Palma}, {Desiante}, {Digel}, {Venere},
  {Drell}, {Drlica-Wagner}, {Favuzzi}, {Ferrara}, {Franckowiak}, {Fukazawa},
  {Funk}, {Fusco}, {Gargano}, {Gasparrini}, {Giglietto}, {Giommi}, {Giordano},
  {Giroletti}, {Glanzman}, {Godfrey}, {Gomez-Vargas}, {Grenier}, {Grondin},
  {Guillemot}, {Guiriec}, {Gustafsson}, {Hadasch}, {Harding}, {Hayashida},
  {Hays}, {Hewitt}, {Hill}, {Horan}, {Hou}, {Iafrate}, {Jogler},
  {J'ohannesson}, {Johnson}, {Kamae}, {Katagiri}, {Kataoka}, {Katsuta}, {Kerr},
  {Knodlseder}, {Kocevski}, {Kuss}, {Laffon}, {Lande}, {Larsson}, {Latronico},
  {Lemoine-Goumard}, {Li}, {Li}, {Longo}, {Loparco}, {Lovellette}, {Lubrano},
  {Magill}, {Maldera}, {Marelli}, {Mayer}, {Mazziotta}, {Michelson},
  {Mitthumsiri}, {Mizuno}, {Moiseev}, {Monzani}, {Moretti}, {Morselli},
  {Moskalenko}, {Murgia}, {Nemmen}, {Nuss}, {Ohsugi}, {Omodei}, {Orienti},
  {Orlando}, {Ormes}, {Paneque}, {Perkins}, {Pesce-Rollins}, {Petrosian},
  {Piron}, {Pivato}, {Porter}, {Rain`o}, {Rando}, {Razzano}, {Razzaque},
  {Reimer}, {Reimer}, {Renaud}, {Reposeur}, {Romain Rousseau}, {Parkinson},
  {Schmid}, {Schulz}, {Sgr`o}, {Siskind}, {Spada}, {Spandre}, {Spinelli},
  {Strong}, {Suson}, {Tajima}, {Takahashi}, {Tanaka}, {Thayer}, {Thompson},
  {Tibaldo}, {Tibolla}, {Torres}, {Tosti}, {Troja}, {Uchiyama}, {Vianello},
  {Wells}, {Wood}, {Wood}, {Yassine}, \& {Zimmer}}]{Fermi_SNR_cat}
{Acero}, F., {Ackermann}, M., {Ajello}, M., {et~al.} 2015{\natexlab{b}}, ArXiv
  e-prints [\eprint[arXiv]{1511.06778}]

\bibitem[{{Acero} {et~al.}(2010{\natexlab{a}}){Acero}, {Aharonian},
  {Akhperjanian}, {Anton}, {Barres de Almeida}, {Bazer-Bachi}, {Becherini},
  {Behera}, {Beilicke}, {Bernl{\"o}hr}, {Bochow}, {Boisson}, {Bolmont},
  {Borrel}, {Brucker}, {Brun}, {Brun}, {B{\"u}hler}, {Bulik}, {B{\"u}sching},
  {Boutelier}, {Chadwick}, {Charbonnier}, {Chaves}, {Cheesebrough}, {Conrad},
  {Chounet}, {Clapson}, {Coignet}, {Dalton}, {Daniel}, {Davids}, {Degrange},
  {Deil}, {Dickinson}, {Djannati-Ata{\"i}}, {Domainko}, {O'C.~Drury}, {Dubois},
  {Dubus}, {Dyks}, {Dyrda}, {Egberts}, {Eger}, {Espigat}, {Fallon}, {Farnier},
  {Fegan}, {Feinstein}, {Fiasson}, {F{\"o}rster}, {Fontaine},
  {F{\"u}{\ss}ling}, {Gabici}, {Gallant}, {G{\'e}rard}, {Gerbig}, {Giebels},
  {Glicenstein}, {Gl{\"u}ck}, {Goret}, {G{\"o}ring}, {Hauser}, {Hauser},
  {Heinz}, {Heinzelmann}, {Henri}, {Hermann}, {Hinton}, {Hoffmann}, {Hofmann},
  {Hofverberg}, {Holleran}, {Hoppe}, {Horns}, {Jacholkowska}, {de Jager},
  {Jahn}, {Jung}, {Katarzy{\'n}ski}, {Katz}, {Kaufmann}, {Kerschhaggl},
  {Khangulyan}, {Kh{\'e}lifi}, {Keogh}, {Klochkov}, {Klu{\'z}niak}, {Kneiske},
  {Komin}, {Kosack}, {Kossakowski}, {Lamanna}, {Lemoine-Goumard}, {Lenain},
  {Lohse}, {Marandon}, {Marcowith}, {Masbou}, {Maurin}, {McComb}, {Medina},
  {M{\'e}hault}, {Moderski}, {Moulin}, {Naumann-Godo}, {de Naurois}, {Nedbal},
  {Nekrassov}, {Nicholas}, {Niemiec}, {Nolan}, {Ohm}, {Olive}, {de O{\~n}a
  Wilhelmi}, {Orford}, {Ostrowski}, {Panter}, {Paz Arribas}, {Pedaletti},
  {Pelletier}, {Petrucci}, {Pita}, {P{\"u}hlhofer}, {Punch}, {Quirrenbach},
  {Raubenheimer}, {Raue}, {Rayner}, {Reimer}, {Renaud}, {de Los Reyes},
  {Rieger}, {Ripken}, {Rob}, {Rosier-Lees}, {Rowell}, {Rudak}, {Rulten},
  {Ruppel}, {Ryde}, {Sahakian}, {Santangelo}, {Schlickeiser}, {Sch{\"o}ck},
  {Sch{\"o}nwald}, {Schwanke}, {Schwarzburg}, {Schwemmer}, {Shalchi}, {Sushch},
  {Sikora}, {Skilton}, {Sol}, {Stawarz}, {Steenkamp}, {Stegmann}, {Stinzing},
  {Superina}, {Szostek}, {Tam}, {Tavernet}, {Terrier}, {Tibolla}, {Tluczykont},
  {van Eldik}, {Vasileiadis}, {Venter}, {Venter}, {Vialle}, {Vincent}, {Vink},
  {Vivier}, {V{\"o}lk}, {Volpe}, {Vorobiov}, {Wagner}, {Ward}, {Zdziarski},
  {Zech}, \& {H.E.S.S.~Collaboration}}]{2010A&A...516A..62A}
{Acero}, F., {Aharonian}, F., {Akhperjanian}, A.~G., {et~al.}
  2010{\natexlab{a}}, \aap, 516, A62

\bibitem[{{Acero} {et~al.}(2010{\natexlab{b}}){Acero}, {Aharonian},
  {Akhperjanian}, {Anton}, {Barres de Almeida}, {Bazer-Bachi}, {Becherini},
  {Behera}, {Bernl{\"o}hr}, {Bochow}, {Boisson}, {Bolmont}, {Borrel}, {Braun},
  {Brucker}, {Brun}, {Brun}, {B{\"u}hler}, {Bulik}, {B{\"u}sching},
  {Boutelier}, {Chadwick}, {Charbonnier}, {Chaves}, {Cheesebrough}, {Conrad},
  {Chounet}, {Clapson}, {Coignet}, {Dalton}, {Daniel}, {Davids}, {Degrange},
  {Deil}, {Dickinson}, {Djannati-Ata{\"i}}, {Domainko}, {Drury}, {Dubois},
  {Dubus}, {Dyks}, {Dyrda}, {Egberts}, {Eger}, {Espigat}, {Fallon}, {Farnier},
  {Fegan}, {Feinstein}, {Fiasson}, {F{\"o}rster}, {Fontaine},
  {F{\"u}{\ss}ling}, {Gabici}, {Gallant}, {G{\'e}rard}, {Gerbig}, {Giebels},
  {Glicenstein}, {Gl{\"u}ck}, {Goret}, {G{\"o}ring}, {Hauser}, {Heinz},
  {Heinzelmann}, {Henri}, {Hermann}, {Hinton}, {Hoffmann}, {Hofmann},
  {Hofverberg}, {Holleran}, {Hoppe}, {Horns}, {Jacholkowska}, {de Jager},
  {Jahn}, {Jung}, {Katarzy{\'n}ski}, {Katz}, {Kaufmann}, {Kerschhaggl},
  {Khangulyan}, {Kh{\'e}lifi}, {Keogh}, {Klochkov}, {Klu{\'z}niak}, {Kneiske},
  {Komin}, {Kosack}, {Kossakowski}, {Lamanna}, {Lenain}, {Lohse}, {Marandon},
  {Martineau-Huynh}, {Marcowith}, {Masbou}, {Maurin}, {McComb}, {Medina},
  {M{\'e}hault}, {Moderski}, {Moulin}, {Naumann-Godo}, {de Naurois}, {Nedbal},
  {Nekrassov}, {Nicholas}, {Niemiec}, {Nolan}, {Ohm}, {Olive}, {de O{\~n}a
  Wilhelmi}, {Orford}, {Ostrowski}, {Panter}, {Arribas}, {Pedaletti},
  {Pelletier}, {Petrucci}, {Pita}, {P{\"u}hlhofer}, {Punch}, {Quirrenbach},
  {Raubenheimer}, {Raue}, {Rayner}, {Reimer}, {Renaud}, {Rieger}, {Ripken},
  {Rob}, {Rosier-Lees}, {Rowell}, {Rudak}, {Rulten}, {Ruppel}, {Ryde},
  {Sahakian}, {Santangelo}, {Schlickeiser}, {Sch{\"o}ck}, {Sch{\"o}nwald},
  {Schwanke}, {Schwarzburg}, {Schwemmer}, {Shalchi}, {Sikora}, {Skilton},
  {Sol}, {Stawarz}, {Steenkamp}, {Stegmann}, {Stinzing}, {Superina}, {Sushch},
  {Szostek}, {Tam}, {Tavernet}, {Terrier}, {Tibolla}, {Tluczykont}, {van
  Eldik}, {Vasileiadis}, {Venter}, {Venter}, {Vialle}, {Vincent}, {Vivier},
  {V{\"o}lk}, {Volpe}, {Wagner}, {Ward}, {Zdziarski}, {Zech}, \&
  {H.E.S.S.~Collaboration}}]{2010MNRAS.402.1877A}
{Acero}, F., {Aharonian}, F., {Akhperjanian}, A.~G., {et~al.}
  2010{\natexlab{b}}, \mnras, 402, 1877

\bibitem[{{Acero} {et~al.}(2012){Acero}, {Djannati-Ata{\"\i}}, {F{\"o}rster},
  {Gallant}, {Renaud}, \& {for the H.E.S.S.~collaboration}}]{Acero:2012}
{Acero}, F., {Djannati-Ata{\"\i}}, A., {F{\"o}rster}, A., {et~al.} 2012, ArXiv
  e-prints [\eprint[arXiv]{1201.0481}]

\bibitem[{Ackermann {et~al.}(2016)Ackermann, Ajello, Atwood, Baldini, Ballet,
  Barbiellini, Bastieri, Gonzalez, Bellazzini, Bissaldi, Blandford, Bloom,
  Bonino, Bottacini, Brandt, Bregeon, Bruel, Buehler, Buson, Caliandro,
  Cameron, Caputo, Caragiulo, Caraveo, Cavazzuti, Cecchi, Charles, Chekhtman,
  Cheung, Chiang, Chiaro, Ciprini, Cohen, Cohen-Tanugi, Cominsky, Conrad,
  Cuoco, Cutini, D’Ammando, de~Angelis, de~Palma, Desiante, Mauro, Venere,
  Domínguez, Drell, Favuzzi, Fegan, Ferrara, Focke, Fortin, Franckowiak,
  Fukazawa, Funk, Furniss, Fusco, Gargano, Gasparrini, Giglietto, Giommi,
  Giordano, Giroletti, Glanzman, Godfrey, Grenier, Grondin, Guillemot, Guiriec,
  Harding, Hays, Hewitt, Hill, Horan, Iafrate, Hartmann, Jogler, Jóhannesson,
  Johnson, Kamae, Kataoka, Knödlseder, Kuss, Mura, Larsson, Latronico,
  Lemoine-Goumard, Li, Li, Longo, Loparco, Lott, Lovellette, Lubrano, Madejski,
  Maldera, Manfreda, Mayer, Mazziotta, Michelson, Mirabal, Mitthumsiri, Mizuno,
  Moiseev, Monzani, Morselli, Moskalenko, Murgia, Nuss, Ohsugi, Omodei,
  Orienti, Orlando, Ormes, Paneque, Perkins, Pesce-Rollins, Petrosian, Piron,
  Pivato, Porter, Rainò, Rando, Razzano, Razzaque, Reimer, Reimer, Reposeur,
  Romani, Sánchez-Conde, Parkinson, Schmid, Schulz, Sgrò, Siskind, Spada,
  Spandre, Spinelli, Suson, Tajima, Takahashi, Takahashi, Takahashi, Thayer,
  Thompson, Tibaldo, Torres, Tosti, Troja, Vianello, Wood, Wood, Yassine,
  Zaharijas, \& Zimmer}]{2FHL}
Ackermann, M., Ajello, M., Atwood, W.~B., {et~al.} 2016, The Astrophysical
  Journal Supplement Series, 222, 5

\bibitem[{{Ackermann} {et~al.}(2011){Ackermann}, {Ajello}, {Baldini}, {Ballet},
  {Barbiellini}, {Bastieri}, {Bechtol}, {Bellazzini}, {Berenji}, {Bloom},
  {Bonamente}, {Borgland}, {Bouvier}, {Bregeon}, {Brez}, {Brigida}, {Bruel},
  {Buehler}, {Buson}, {Caliandro}, {Cameron}, {Camilo}, {Caraveo},
  {Casandjian}, {Cecchi}, {{\c C}elik}, {Charles}, {Chekhtman}, {Cheung},
  {Chiang}, {Ciprini}, {Claus}, {Cognard}, {Cohen-Tanugi}, {Conrad}, {Dermer},
  {de Angelis}, {de Luca}, {de Palma}, {Digel}, {Silva}, {Drell}, {Dubois},
  {Dumora}, {Favuzzi}, {Focke}, {Frailis}, {Fukazawa}, {Funk}, {Fusco},
  {Gargano}, {Germani}, {Giglietto}, {Giommi}, {Giordano}, {Giroletti},
  {Glanzman}, {Godfrey}, {Grenier}, {Grondin}, {Grove}, {Guillemot}, {Guiriec},
  {Hadasch}, {Hanabata}, {Harding}, {Hayashi}, {Hays}, {Hobbs}, {Hughes},
  {J{\'o}hannesson}, {Johnson}, {Johnson}, {Johnston}, {Kamae}, {Katagiri},
  {Kataoka}, {Keith}, {Kerr}, {Kn{\"o}dlseder}, {Kramer}, {Kuss}, {Lande},
  {Latronico}, {Lee}, {Lemoine-Goumard}, {Longo}, {Loparco}, {Lovellette},
  {Lubrano}, {Lyne}, {Makeev}, {Marelli}, {Mazziotta}, {McEnery}, {Mehault},
  {Michelson}, {Mizuno}, {Moiseev}, {Monte}, {Monzani}, {Morselli},
  {Moskalenko}, {Murgia}, {Nakamori}, {Naumann-Godo}, {Nolan}, {Noutsos},
  {Nuss}, {Ohsugi}, {Okumura}, {Ormes}, {Paneque}, {Panetta}, {Parent},
  {Pelassa}, {Pepe}, {Pesce-Rollins}, {Piron}, {Porter}, {Rain{\`o}}, {Rando},
  {Ransom}, {Ray}, {Razzano}, {Rea}, {Reimer}, {Reimer}, {Reposeur}, {Ripken},
  {Ritz}, {Romani}, {Sadrozinski}, {Sander}, {Saz Parkinson}, {Sgr{\`o}},
  {Siskind}, {Smith}, {Smith}, {Spandre}, {Spinelli}, {Strickman}, {Suson},
  {Takahashi}, {Takahashi}, {Tanaka}, {Thayer}, {Thayer}, {Theureau},
  {Thompson}, {Thorsett}, {Tibaldo}, {Torres}, {Tosti}, {Tramacere},
  {Uchiyama}, {Uehara}, {Usher}, {Vandenbroucke}, {Van Etten}, {Vasileiou},
  {Vilchez}, {Vitale}, {Waite}, {Wang}, {Weltevrede}, {Winer}, {Wood}, {Yang},
  {Ylinen}, \& {Ziegler}}]{Ackermann11}
{Ackermann}, M., {Ajello}, M., {Baldini}, L., {et~al.} 2011, \apj, 726, 35

\bibitem[{{Aharonian} {et~al.}(2009){Aharonian}, {Akhperjanian}, {Anton},
  {Barres de Almeida}, {Bazer-Bachi}, {Becherini}, {Behera}, {Benbow},
  {Bernl{\"o}hr}, {Boisson}, {Bochow}, {Borrel}, {Braun}, {Brion}, {Brucker},
  {Brun}, {B{\"u}hler}, {Bulik}, {B{\"u}sching}, {Boutelier}, {Carrigan},
  {Chadwick}, {Charbonnier}, {Chaves}, {Cheesebrough}, {Chounet}, {Clapson},
  {Coignet}, {Dalton}, {Daniel}, {Degrange}, {Deil}, {Dickinson},
  {Djannati-Ata{\"i}}, {Domainko}, {O'C.~Drury}, {Dubois}, {Dubus}, {Dyks},
  {Dyrda}, {Egberts}, {Emmanoulopoulos}, {Espigat}, {Farnier}, {Feinstein},
  {Fiasson}, {F{\"o}rster}, {Fontaine}, {F{\"u}{\ss}ling}, {Gabici}, {Gallant},
  {G{\'e}rard}, {Giebels}, {Glicenstein}, {Gl{\"u}ck}, {Goret}, {Hauser},
  {Hauser}, {Heinz}, {Heinzelmann}, {Henri}, {Hermann}, {Hinton}, {Hoffmann},
  {Hofmann}, {Holleran}, {Hoppe}, {Horns}, {Jacholkowska}, {de Jager}, {Jung},
  {Katarzy{\'n}ski}, {Katz}, {Kaufmann}, {Kendziorra}, {Kerschhaggl},
  {Khangulyan}, {Kh{\'e}lifi}, {Keogh}, {Komin}, {Kosack}, {Lamanna}, {Lenain},
  {Lohse}, {Marandon}, {Martin}, {Martineau-Huynh}, {Marcowith}, {Maurin},
  {McComb}, {Medina}, {Moderski}, {Moulin}, {Naumann-Godo}, {de Naurois},
  {Nedbal}, {Nekrassov}, {Niemiec}, {Nolan}, {Ohm}, {Olive}, {de O{\~n}a
  Wilhelmi}, {Orford}, {Ostrowski}, {Panter}, {Paz Arribas}, {Pedaletti},
  {Pelletier}, {Petrucci}, {Pita}, {P{\"u}hlhofer}, {Punch}, {Quirrenbach},
  {Raubenheimer}, {Raue}, {Rayner}, {Renaud}, {Reimer}, {Rieger}, {Ripken},
  {Rob}, {Rosier-Lees}, {Rowell}, {Rudak}, {Rulten}, {Ruppel}, {Sahakian},
  {Santangelo}, {Schlickeiser}, {Sch{\"o}ck}, {Schr{\"o}der}, {Schwanke},
  {Schwarzburg}, {Schwemmer}, {Shalchi}, {Skilton}, {Sol}, {Spangler},
  {Stawarz}, {Steenkamp}, {Stegmann}, {Superina}, {Tam}, {Tavernet}, {Terrier},
  {Tibolla}, {van Eldik}, {Vasileiadis}, {Venter}, {Venter}, {Vialle},
  {Vincent}, {Vivier}, {V{\"o}lk}, {Volpe}, {Wagner}, {Ward}, {Zdziarski}, \&
  {Zech}}]{2009A&A...499..723A}
{Aharonian}, F., {Akhperjanian}, A.~G., {Anton}, G., {et~al.} 2009, \aap, 499,
  723

\bibitem[{{Aharonian} {et~al.}(2005{\natexlab{a}}){Aharonian}, {Akhperjanian},
  {Aye}, {Bazer-Bachi}, {Beilicke}, {Benbow}, {Berge}, {Berghaus},
  {Bernl{\"o}hr}, {Boisson}, {Bolz}, {Borgmeier}, {Braun}, {Breitling},
  {Brown}, {Bussons Gordo}, {Chadwick}, {Chounet}, {Cornils}, {Costamante},
  {Degrange}, {Djannati-Ata{\"i}}, {O'C.~Drury}, {Dubus}, {Ergin}, {Espigat},
  {Feinstein}, {Fleury}, {Fontaine}, {Funk}, {Gallant}, {Giebels}, {Gillessen},
  {Goret}, {Hadjichristidis}, {Hauser}, {Heinzelmann}, {Henri}, {Hermann},
  {Hinton}, {Hofmann}, {Holleran}, {Horns}, {de Jager}, {Jung}, {Kh{\'e}lifi},
  {Komin}, {Konopelko}, {Latham}, {Le Gallou}, {Lemi{\`e}re}, {Lemoine},
  {Leroy}, {Lohse}, {Marcowith}, {Masterson}, {McComb}, {de Naurois}, {Nolan},
  {Noutsos}, {Orford}, {Osborne}, {Ouchrif}, {Panter}, {Pelletier}, {Pita},
  {P{\"u}hlhofer}, {Punch}, {Raubenheimer}, {Raue}, {Raux}, {Rayner},
  {Redondo}, {Reimer}, {Reimer}, {Ripken}, {Rob}, {Rolland}, {Rowell},
  {Sahakian}, {Saug{\'e}}, {Schlenker}, {Schlickeiser}, {Schuster}, {Schwanke},
  {Siewert}, {Sol}, {Steenkamp}, {Stegmann}, {Tavernet}, {Terrier},
  {Th{\'e}oret}, {Tluczykont}, {Vasileiadis}, {Venter}, {Vincent}, {Visser},
  {V{\"o}lk}, \& {Wagner}}]{Aharonian:2005d}
{Aharonian}, F., {Akhperjanian}, A.~G., {Aye}, K.-M., {et~al.}
  2005{\natexlab{a}}, \aap, 432, L25

\bibitem[{{Aharonian} {et~al.}(2005{\natexlab{b}}){Aharonian}, {Akhperjanian},
  {Aye}, {Bazer-Bachi}, {Beilicke}, {Benbow}, {Berge}, {Berghaus},
  {Bernl{\"o}hr}, {Boisson}, {Bolz}, {Borgmeier}, {Braun}, {Breitling},
  {Brown}, {Gordo}, {Chadwick}, {Chounet}, {Cornils}, {Costamante}, {Degrange},
  {Djannati-Ata{\"i}}, {Drury}, {Dubus}, {Ergin}, {Espigat}, {Feinstein},
  {Fleury}, {Fontaine}, {Funk}, {Gallant}, {Giebels}, {Gillessen}, {Goret},
  {Hadjichristidis}, {Hauser}, {Heinzelmann}, {Henri}, {Hermann}, {Hinton},
  {Hofmann}, {Holleran}, {Horns}, {de Jager}, {Jung}, {Kh{\'e}lifi}, {Komin},
  {Konopelko}, {Latham}, {Le Gallou}, {Lemi{\`e}re}, {Lemoine}, {Leroy},
  {Lohse}, {Marcowith}, {Masterson}, {McComb}, {de Naurois}, {Nolan},
  {Noutsos}, {Orford}, {Osborne}, {Ouchrif}, {Panter}, {Pelletier}, {Pita},
  {P{\"u}hlhofer}, {Punch}, {Raubenheimer}, {Raue}, {Raux}, {Rayner},
  {Redondo}, {Reimer}, {Reimer}, {Ripken}, {Rob}, {Rolland}, {Rowell},
  {Sahakian}, {Saug{\'e}}, {Schlenker}, {Schlickeiser}, {Schuster}, {Schwanke},
  {Siewert}, {Sol}, {Steenkamp}, {Stegmann}, {Tavernet}, {Terrier},
  {Th{\'e}oret}, {Tluczykont}, {van der Walt}, {Vasileiadis}, {Venter},
  {Vincent}, {Visser}, {V{\"o}lk}, \& {Wagner}}]{2005Sci...307.1938A}
{Aharonian}, F., {Akhperjanian}, A.~G., {Aye}, K.-M., {et~al.}
  2005{\natexlab{b}}, Science, 307, 1938

\bibitem[{{Aharonian} {et~al.}(2005{\natexlab{c}}){Aharonian}, {Akhperjanian},
  {Aye}, {Bazer-Bachi}, {Beilicke}, {Benbow}, {Berge}, {Berghaus},
  {Bernl{\"o}hr}, {Boisson}, {Bolz}, {Braun}, {Breitling}, {Brown}, {Bussons
  Gordo}, {Chadwick}, {Chounet}, {Cornils}, {Costamante}, {Degrange},
  {Djannati-Ata{\"i}}, {O'C.~Drury}, {Dubus}, {Emmanoulopoulos}, {Espigat},
  {Feinstein}, {Fleury}, {Fontaine}, {Fuchs}, {Funk}, {Gallant}, {Giebels},
  {Gillessen}, {Glicenstein}, {Goret}, {Hadjichristidis}, {Hauser},
  {Heinzelmann}, {Henri}, {Hermann}, {Hinton}, {Hofmann}, {Holleran}, {Horns},
  {de Jager}, {Johnston}, {Kh{\'e}lifi}, {Kirk}, {Komin}, {Konopelko},
  {Latham}, {Le Gallou}, {Lemi{\`e}re}, {Lemoine-Goumard}, {Leroy},
  {Martineau-Huynh}, {Lohse}, {Marcowith}, {Masterson}, {McComb}, {de Naurois},
  {Nolan}, {Noutsos}, {Orford}, {Osborne}, {Ouchrif}, {Panter}, {Pelletier},
  {Pita}, {P{\"u}hlhofer}, {Punch}, {Raubenheimer}, {Raue}, {Raux}, {Rayner},
  {Redondo}, {Reimer}, {Reimer}, {Ripken}, {Rob}, {Rolland}, {Rowell},
  {Sahakian}, {Saug{\'e}}, {Schlenker}, {Schlickeiser}, {Schuster}, {Schwanke},
  {Siewert}, {Skj{\ae}raasen}, {Sol}, {Steenkamp}, {Stegmann}, {Tavernet},
  {Terrier}, {Th{\'e}oret}, {Tluczykont}, {Vasileiadis}, {Venter}, {Vincent},
  {V{\"o}lk}, \& {Wagner}}]{2005AandA...442....1A}
{Aharonian}, F., {Akhperjanian}, A.~G., {Aye}, K.-M., {et~al.}
  2005{\natexlab{c}}, \aap, 442, 1

\bibitem[{{Aharonian} {et~al.}(2005{\natexlab{d}}){Aharonian}, {Akhperjanian},
  {Aye}, {Bazer-Bachi}, {Beilicke}, {Benbow}, {Berge}, {Berghaus},
  {Bernl{\"o}hr}, {Boisson}, {Bolz}, {Braun}, {Breitling}, {Brown}, {Bussons
  Gordo}, {Chadwick}, {Chounet}, {Cornils}, {Costamante}, {Degrange},
  {Djannati-Ata{\"i}}, {O'C.~Drury}, {Dubus}, {Emmanoulopoulos}, {Espigat},
  {Feinstein}, {Fleury}, {Fontaine}, {Fuchs}, {Funk}, {Gallant}, {Giebels},
  {Gillessen}, {Glicenstein}, {Goret}, {Hadjichristidis}, {Hauser},
  {Heinzelmann}, {Henri}, {Hermann}, {Hinton}, {Hofmann}, {Holleran}, {Horns},
  {de Jager}, {Kh{\'e}lifi}, {Komin}, {Konopelko}, {Latham}, {Le Gallou},
  {Lemi{\`e}re}, {Lemoine-Goumard}, {Leroy}, {Lohse}, {Martineau-Huynh},
  {Marcowith}, {Masterson}, {McComb}, {de Naurois}, {Nolan}, {Noutsos},
  {Orford}, {Osborne}, {Ouchrif}, {Panter}, {Pelletier}, {Pita},
  {P{\"u}hlhofer}, {Punch}, {Raubenheimer}, {Raue}, {Raux}, {Rayner},
  {Redondo}, {Reimer}, {Reimer}, {Ripken}, {Rob}, {Rolland}, {Rowell},
  {Sahakian}, {Saug{\'e}}, {Schlenker}, {Schlickeiser}, {Schuster}, {Schwanke},
  {Siewert}, {Sol}, {Steenkamp}, {Stegmann}, {Tavernet}, {Terrier},
  {Th{\'e}oret}, {Tluczykont}, {Vasileiadis}, {Venter}, {Vincent}, {V{\"o}lk},
  \& {Wagner}}]{Aharonian:2005c}
{Aharonian}, F., {Akhperjanian}, A.~G., {Aye}, K.-M., {et~al.}
  2005{\natexlab{d}}, \aap, 435, L17

\bibitem[{{Aharonian} {et~al.}(2004{\natexlab{a}}){Aharonian}, {Akhperjanian},
  {Aye}, {Bazer-Bachi}, {Beilicke}, {Benbow}, {Berge}, {Berghaus},
  {Bernl{\"o}hr}, {Bolz}, {Boisson}, {Borgmeier}, {Breitling}, {Brown},
  {Chadwick}, {Chitnis}, {Chounet}, {Cornils}, {Costamante}, {Degrange}, {de
  Jager}, {Djannati-Ata{\"i}}, {Drury}, {Ergin}, {Espigat}, {Feinstein},
  {Fleury}, {Fontaine}, {Funk}, {Gallant}, {Giebels}, {Gillessen}, {Goret},
  {Guy}, {Hadjichristidis}, {Hauser}, {Heinzelmann}, {Henri}, {Hermann},
  {Hinton}, {Hofmann}, {Holleran}, {Horns}, {Jung}, {Kh{\'e}lifi}, {Komin},
  {Konopelko}, {Latham}, {Le Gallou}, {Lemoine}, {Lemi{\`e}re}, {Leroy},
  {Lohse}, {Marcowith}, {Masterson}, {McComb}, {de Naurois}, {Nolan},
  {Noutsos}, {Orford}, {Osborne}, {Ouchrif}, {Panter}, {Pelletier}, {Pita},
  {Pohl}, {P{\"u}hlhofer}, {Punch}, {Raubenheimer}, {Raue}, {Raux}, {Rayner},
  {Redondo}, {Reimer}, {Reimer}, {Ripken}, {Rivoal}, {Rob}, {Rolland},
  {Rowell}, {Sahakian}, {Sauge}, {Schlenker}, {Schlickeiser}, {Schuster},
  {Schwanke}, {Siewert}, {Sol}, {Steenkamp}, {Stegmann}, {Tavernet},
  {Th{\'e}oret}, {Tluczykont}, {van der Walt}, {Vasileiadis}, {Vincent},
  {Visser}, {Volk}, \& {Wagner}}]{ref:hesscalib}
{Aharonian}, F., {Akhperjanian}, A.~G., {Aye}, K.-M., {et~al.}
  2004{\natexlab{a}}, Astroparticle Physics, 22, 109

\bibitem[{{Aharonian} {et~al.}(2008{\natexlab{a}}){Aharonian}, {Akhperjanian},
  {Barres de Almeida}, {Bazer-Bachi}, {Behera}, {Beilicke}, {Benbow},
  {Bernl{\"o}hr}, {Boisson}, {Bolz}, {Borrel}, {Braun}, {Brion}, {Brown},
  {B{\"u}hler}, {Bulik}, {B{\"u}sching}, {Boutelier}, {Carrigan}, {Chadwick},
  {Chounet}, {Clapson}, {Coignet}, {Cornils}, {Costamante}, {Dalton},
  {Degrange}, {Dickinson}, {Djannati-Ata{\"i}}, {Domainko}, {Drury}, {Dubois},
  {Dubus}, {Dyks}, {Egberts}, {Emmanoulopoulos}, {Espigat}, {Farnier},
  {Feinstein}, {Fiasson}, {F{\"o}rster}, {Fontaine}, {Funk}, {F{\"u}{\ss}ling},
  {Gallant}, {Giebels}, {Glicenstein}, {Gl{\"u}ck}, {Goret}, {Hadjichristidis},
  {Hauser}, {Hauser}, {Heinzelmann}, {Henri}, {Hermann}, {Hinton}, {Hoffmann},
  {Hofmann}, {Holleran}, {Hoppe}, {Horns}, {Jacholkowska}, {de Jager}, {Jung},
  {Katarzy{\'n}ski}, {Kendziorra}, {Kerschhaggl}, {Kh{\'e}lifi}, {Keogh},
  {Komin}, {Kosack}, {Lamanna}, {Latham}, {Lemi{\`e}re}, {Lemoine-Goumard},
  {Lenain}, {Lohse}, {Martin}, {Martineau-Huynh}, {Marcowith}, {Masterson},
  {Maurin}, {Maurin}, {McComb}, {Moderski}, {Moulin}, {de Naurois}, {Nedbal},
  {Nolan}, {Ohm}, {Olive}, {de O{\~n}a Wilhelmi}, {Orford}, {Osborne},
  {Ostrowski}, {Panter}, {Pedaletti}, {Pelletier}, {Petrucci}, {Pita},
  {P{\"u}hlhofer}, {Punch}, {Ranchon}, {Raubenheimer}, {Raue}, {Rayner},
  {Renaud}, {Ripken}, {Rob}, {Rolland}, {Rosier-Lees}, {Rowell}, {Rudak},
  {Ruppel}, {Sahakian}, {Santangelo}, {Schlickeiser}, {Sch{\"o}ck},
  {Schr{\"o}der}, {Schwanke}, {Schwarzburg}, {Schwemmer}, {Shalchi}, {Sol},
  {Spangler}, {Stawarz}, {Steenkamp}, {Stegmann}, {Superina}, {Tam},
  {Tavernet}, {Terrier}, {van Eldik}, {Vasileiadis}, {Venter}, {Vialle},
  {Vincent}, {Vivier}, {V{\"o}lk}, {Volpe}, {Wagner}, {Ward}, {Zdziarski}, \&
  {Zech}}]{ref_gps_unids2008}
{Aharonian}, F., {Akhperjanian}, A.~G., {Barres de Almeida}, U., {et~al.}
  2008{\natexlab{a}}, \aap, 477, 353

\bibitem[{{Aharonian} {et~al.}(2008{\natexlab{b}}){Aharonian}, {Akhperjanian},
  {Barres de Almeida}, {Bazer-Bachi}, {Behera}, {Beilicke}, {Benbow},
  {Bernl{\"o}hr}, {Boisson}, {Bolz}, {Borrel}, {Braun}, {Brion}, {Brown},
  {B{\"u}hler}, {Bulik}, {B{\"u}sching}, {Boutelier}, {Carrigan}, {Chadwick},
  {Chounet}, {Clapson}, {Coignet}, {Cornils}, {Costamante}, {Dalton},
  {Degrange}, {Dickinson}, {Djannati-Ata{\"i}}, {Domainko}, {O'C.~Drury},
  {Dubois}, {Dubus}, {Dyks}, {Egberts}, {Emmanoulopoulos}, {Espigat},
  {Farnier}, {Feinstein}, {Fiasson}, {F{\"o}rster}, {Fontaine}, {Funk},
  {F{\"u}{\ss}ling}, {Gallant}, {Giebels}, {Glicenstein}, {Gl{\"u}ck}, {Goret},
  {Hadjichristidis}, {Hauser}, {Hauser}, {Heinzelmann}, {Henri}, {Hermann},
  {Hinton}, {Hoffmann}, {Hofmann}, {Holleran}, {Hoppe}, {Horns},
  {Jacholkowska}, {de Jager}, {Jung}, {Katarzy{\'n}ski}, {Kendziorra},
  {Kerschhaggl}, {Kh{\'e}lifi}, {Keogh}, {Komin}, {Kosack}, {Lamanna},
  {Latham}, {Lemoine-Goumard}, {Lenain}, {Lohse}, {Martin}, {Martineau-Huynh},
  {Marcowith}, {Masterson}, {Maurin}, {McComb}, {Moderski}, {Moulin},
  {Naumann-Godo}, {de Naurois}, {Nedbal}, {Nekrassov}, {Nolan}, {Ohm}, {Olive},
  {de O{\~n}a Wilhelmi}, {Orford}, {Osborne}, {Ostrowski}, {Panter},
  {Pedaletti}, {Pelletier}, {Petrucci}, {Pita}, {P{\"u}hlhofer}, {Punch},
  {Raubenheimer}, {Raue}, {Rayner}, {Renaud}, {Ripken}, {Rob}, {Rosier-Lees},
  {Rowell}, {Rudak}, {Ruppel}, {Sahakian}, {Santangelo}, {Schlickeiser},
  {Sch{\"o}ck}, {Schr{\"o}der}, {Schwanke}, {Schwarzburg}, {Schwemmer},
  {Shalchi}, {Sol}, {Spangler}, {Stawarz}, {Steenkamp}, {Stegmann}, {Superina},
  {Tam}, {Tavernet}, {Terrier}, {van Eldik}, {Vasileiadis}, {Venter}, {Vialle},
  {Vincent}, {Vivier}, {V{\"o}lk}, {Volpe}, {Wagner}, {Ward}, {Zdziarski}, \&
  {Zech}}]{2008A&A...483..509A}
{Aharonian}, F., {Akhperjanian}, A.~G., {Barres de Almeida}, U., {et~al.}
  2008{\natexlab{b}}, \aap, 483, 509

\bibitem[{{Aharonian} {et~al.}(2008{\natexlab{c}}){Aharonian}, {Akhperjanian},
  {Barres de Almeida}, {Bazer-Bachi}, {Behera}, {Beilicke}, {Benbow},
  {Bernl{\"o}hr}, {Boisson}, {Borrel}, {Braun}, {Brion}, {Brucker},
  {B{\"u}hler}, {Bulik}, {B{\"u}sching}, {Boutelier}, {Carrigan}, {Chadwick},
  {Chaves}, {Chounet}, {Clapson}, {Coignet}, {Cornils}, {Costamante}, {Dalton},
  {Degrange}, {Dickinson}, {Djannati-Ata{\"i}}, {Domainko}, {O'C.~Drury},
  {Dubois}, {Dubus}, {Dyks}, {Egberts}, {Emmanoulopoulos}, {Espigat},
  {Farnier}, {Feinstein}, {Fiasson}, {F{\"o}rster}, {Fontaine}, {Funk},
  {F{\"u}{\ss}ling}, {Gabici}, {Gallant}, {Giebels}, {Glicenstein},
  {Gl{\"u}ck}, {Goret}, {Hadjichristidis}, {Hauser}, {Hauser}, {Heinzelmann},
  {Henri}, {Hermann}, {Hinton}, {Hoffmann}, {Hofmann}, {Holleran}, {Hoppe},
  {Horns}, {Jacholkowska}, {de Jager}, {Jung}, {Katarzy{\'n}ski}, {Kaufmann},
  {Kendziorra}, {Kerschhaggl}, {Khangulyan}, {Kh{\'e}lifi}, {Keogh}, {Komin},
  {Kosack}, {Lamanna}, {Latham}, {Lemoine-Goumard}, {Lenain}, {Lohse},
  {Martin}, {Martineau-Huynh}, {Marcowith}, {Masterson}, {Maurin}, {McComb},
  {Moderski}, {Moulin}, {Nakajima}, {Naumann-Godo}, {de Naurois}, {Nedbal},
  {Nekrassov}, {Nolan}, {Ohm}, {Olive}, {de O{\~n}a Wilhelmi}, {Orford},
  {Osborne}, {Ostrowski}, {Panter}, {Pedaletti}, {Pelletier}, {Petrucci},
  {Pita}, {P{\"u}hlhofer}, {Punch}, {Quirrenbach}, {Raubenheimer}, {Raue},
  {Rayner}, {Reimer}, {Renaud}, {Rieger}, {Ripken}, {Rob}, {Rosier-Lees},
  {Rowell}, {Rudak}, {Ruppel}, {Sahakian}, {Santangelo}, {Schlickeiser},
  {Sch{\"o}ck}, {Schr{\"o}der}, {Schwanke}, {Schwarzburg}, {Schwemmer},
  {Shalchi}, {Skilton}, {Sol}, {Spangler}, {Stawarz}, {Steenkamp}, {Stegmann},
  {Superina}, {Tam}, {Tavernet}, {Terrier}, {Tibolla}, {van Eldik},
  {Vasileiadis}, {Venter}, {Vialle}, {Vincent}, {Vivier}, {V{\"o}lk}, {Volpe},
  {Wagner}, {Ward}, {Zdziarski}, \& {Zech}}]{2008AA...490..685A}
{Aharonian}, F., {Akhperjanian}, A.~G., {Barres de Almeida}, U., {et~al.}
  2008{\natexlab{c}}, \aap, 490, 685

\bibitem[{{Aharonian} {et~al.}(2008{\natexlab{d}}){Aharonian}, {Akhperjanian},
  {Bazer-Bachi}, {Behera}, {Beilicke}, {Benbow}, {Berge}, {Bernl{\"o}hr},
  {Boisson}, {Bolz}, {Borrel}, {Braun}, {Brion}, {Brown}, {B{\"u}hler},
  {Bulik}, {B{\"u}sching}, {Boutelier}, {Carrigan}, {Chadwick}, {Chounet},
  {Clapson}, {Coignet}, {Cornils}, {Costamante}, {Degrange}, {Dickinson},
  {Djannati-Ata{\"i}}, {Domainko}, {O'C.~Drury}, {Dubus}, {Dyks}, {Egberts},
  {Emmanoulopoulos}, {Espigat}, {Farnier}, {Feinstein}, {Fiasson},
  {F{\"o}rster}, {Fontaine}, {Fukui}, {Funk}, {Funk}, {F{\"u}{\ss}ling},
  {Gallant}, {Giebels}, {Glicenstein}, {Gl{\"u}ck}, {Goret}, {Hadjichristidis},
  {Hauser}, {Hauser}, {Heinzelmann}, {Henri}, {Hermann}, {Hinton}, {Hoffmann},
  {Hofmann}, {Holleran}, {Hoppe}, {Horns}, {Jacholkowska}, {de Jager},
  {Kendziorra}, {Kerschhaggl}, {Kh{\'e}lifi}, {Komin}, {Kosack}, {Lamanna},
  {Latham}, {Le Gallou}, {Lemi{\`e}re}, {Lemoine-Goumard}, {Lenain}, {Lohse},
  {Martin}, {Martineau-Huynh}, {Marcowith}, {Masterson}, {Maurin}, {McComb},
  {Moderski}, {Moriguchi}, {Moulin}, {de Naurois}, {Nedbal}, {Nolan}, {Olive},
  {Orford}, {Osborne}, {Ostrowski}, {Panter}, {Pedaletti}, {Pelletier},
  {Petrucci}, {Pita}, {P{\"u}hlhofer}, {Punch}, {Ranchon}, {Raubenheimer},
  {Raue}, {Rayner}, {Reimer}, {Renaud}, {Ripken}, {Rob}, {Rolland},
  {Rosier-Lees}, {Rowell}, {Rudak}, {Ruppel}, {Sahakian}, {Santangelo},
  {Saug{\'e}}, {Schlenker}, {Schlickeiser}, {Schr{\"o}der}, {Schwanke},
  {Schwarzburg}, {Schwemmer}, {Shalchi}, {Sol}, {Spangler}, {Stawarz},
  {Steenkamp}, {Stegmann}, {Superina}, {Takeuchi}, {Tam}, {Tavernet},
  {Terrier}, {van Eldik}, {Vasileiadis}, {Venter}, {Vialle}, {Vincent},
  {Vivier}, {V{\"o}lk}, {Volpe}, {Wagner}, \& {Ward}}]{Aharonian:2008f}
{Aharonian}, F., {Akhperjanian}, A.~G., {Bazer-Bachi}, A.~R., {et~al.}
  2008{\natexlab{d}}, \aap, 481, 401

\bibitem[{{Aharonian} {et~al.}(2006{\natexlab{a}}){Aharonian}, {Akhperjanian},
  {Bazer-Bachi}, {Beilicke}, {Benbow}, {Berge}, {Bernl{\"o}hr}, {Boisson},
  {Bolz}, {Borrel}, {Braun}, {Breitling}, {Brown}, {B{\"u}hler},
  {B{\"u}sching}, {Carrigan}, {Chadwick}, {Chounet}, {Cornils}, {Costamante},
  {Degrange}, {Dickinson}, {Djannati-Ata{\"i}}, {O'C.~Drury}, {Dubus},
  {Egberts}, {Emmanoulopoulos}, {Espigat}, {Feinstein}, {Ferrero}, {Fiasson},
  {Fontaine}, {Funk}, {Funk}, {Gallant}, {Giebels}, {Glicenstein}, {Goret},
  {Hadjichristidis}, {Hauser}, {Hauser}, {Heinzelmann}, {Henri}, {Hermann},
  {Hinton}, {Hofmann}, {Holleran}, {Horns}, {Jacholkowska}, {de Jager},
  {Kh{\'e}lifi}, {Komin}, {Konopelko}, {Kosack}, {Latham}, {Le Gallou},
  {Lemi{\`e}re}, {Lemoine-Goumard}, {Lohse}, {Martin}, {Martineau-Huynh},
  {Marcowith}, {Masterson}, {McComb}, {de Naurois}, {Nedbal}, {Nolan},
  {Noutsos}, {Orford}, {Osborne}, {Ouchrif}, {Panter}, {Pelletier}, {Pita},
  {P{\"u}hlhofer}, {Punch}, {Raubenheimer}, {Raue}, {Rayner}, {Reimer},
  {Reimer}, {Ripken}, {Rob}, {Rolland}, {Rowell}, {Sahakian}, {Saug{\'e}},
  {Schlenker}, {Schlickeiser}, {Schwanke}, {Sol}, {Spangler}, {Spanier},
  {Steenkamp}, {Stegmann}, {Superina}, {Tavernet}, {Terrier}, {Th{\'e}oret},
  {Tluczykont}, {van Eldik}, {Vasileiadis}, {Venter}, {Vincent}, {V{\"o}lk},
  {Wagner}, \& {Ward}}]{2006A&A...457..899A}
{Aharonian}, F., {Akhperjanian}, A.~G., {Bazer-Bachi}, A.~R., {et~al.}
  2006{\natexlab{a}}, \aap, 457, 899

\bibitem[{{Aharonian} {et~al.}(2006{\natexlab{b}}){Aharonian}, {Akhperjanian},
  {Bazer-Bachi}, {Beilicke}, {Benbow}, {Berge}, {Bernl{\"o}hr}, {Boisson},
  {Bolz}, {Borrel}, {Braun}, {Breitling}, {Brown}, {B{\"u}hler},
  {B{\"u}sching}, {Carrigan}, {Chadwick}, {Chounet}, {Cornils}, {Costamante},
  {Degrange}, {Dickinson}, {Djannati-Ata{\"i}}, {O'C.~Drury}, {Dubus},
  {Egberts}, {Emmanoulopoulos}, {Espigat}, {Feinstein}, {Ferrero}, {Fiasson},
  {Fontaine}, {Funk}, {Funk}, {Gallant}, {Giebels}, {Glicenstein}, {Goret},
  {Hadjichristidis}, {Hauser}, {Hauser}, {Heinzelmann}, {Henri}, {Hermann},
  {Hinton}, {Hofmann}, {Holleran}, {Horns}, {Jacholkowska}, {de Jager},
  {Kh{\'e}lifi}, {Komin}, {Konopelko}, {Kosack}, {Latham}, {Le Gallou},
  {Lemi{\`e}re}, {Lemoine-Goumard}, {Lohse}, {Martin}, {Martineau-Huynh},
  {Marcowith}, {Masterson}, {McComb}, {de Naurois}, {Nedbal}, {Nolan},
  {Noutsos}, {Orford}, {Osborne}, {Ouchrif}, {Panter}, {Pelletier}, {Pita},
  {P{\"u}hlhofer}, {Punch}, {Raubenheimer}, {Raue}, {Rayner}, {Reimer},
  {Reimer}, {Ripken}, {Rob}, {Rolland}, {Rowell}, {Sahakian}, {Saug{\'e}},
  {Schlenker}, {Schlickeiser}, {Schwanke}, {Sol}, {Spangler}, {Spanier},
  {Steenkamp}, {Stegmann}, {Superina}, {Tavernet}, {Terrier}, {Th{\'e}oret},
  {Tluczykont}, {van Eldik}, {Vasileiadis}, {Venter}, {Vincent}, {V{\"o}lk},
  {Wagner}, \& {Ward}}]{ref:hesscrab}
{Aharonian}, F., {Akhperjanian}, A.~G., {Bazer-Bachi}, A.~R., {et~al.}
  2006{\natexlab{b}}, \aap, 457, 899

\bibitem[{{Aharonian} {et~al.}(2006{\natexlab{c}}){Aharonian}, {Akhperjanian},
  {Bazer-Bachi}, {Beilicke}, {Benbow}, {Berge}, {Bernl{\"o}hr}, {Boisson},
  {Bolz}, {Borrel}, {Braun}, {Breitling}, {Brown}, {B{\"u}hler},
  {B{\"u}sching}, {Carrigan}, {Chadwick}, {Chounet}, {Cornils}, {Costamante},
  {Degrange}, {Dickinson}, {Djannati-Ata{\"i}}, {O'C.~Drury}, {Dubus},
  {Egberts}, {Emmanoulopoulos}, {Epinat}, {Espigat}, {Feinstein}, {Ferrero},
  {Fontaine}, {Funk}, {Funk}, {Gallant}, {Giebels}, {Glicenstein}, {Goret},
  {Hadjichristidis}, {Hauser}, {Hauser}, {Heinzelmann}, {Henri}, {Hermann},
  {Hinton}, {Hofmann}, {Holleran}, {Horns}, {Jacholkowska}, {de Jager},
  {Kh{\'e}lifi}, {Komin}, {Konopelko}, {Latham}, {Le Gallou}, {Lemi{\`e}re},
  {Lemoine-Goumard}, {Lohse}, {Martin}, {Martineau-Huynh}, {Marcowith},
  {Masterson}, {McComb}, {de Naurois}, {Nedbal}, {Nolan}, {Noutsos}, {Orford},
  {Osborne}, {Ouchrif}, {Panter}, {Pelletier}, {Pita}, {P{\"u}hlhofer},
  {Punch}, {Raubenheimer}, {Raue}, {Rayner}, {Reimer}, {Reimer}, {Ripken},
  {Rob}, {Rolland}, {Rowell}, {Sahakian}, {Saug{\'e}}, {Schlenker},
  {Schlickeiser}, {Schwanke}, {Sol}, {Spangler}, {Spanier}, {Steenkamp},
  {Stegmann}, {Superina}, {Tavernet}, {Terrier}, {Th{\'e}oret}, {Tluczykont},
  {van Eldik}, {Vasileiadis}, {Venter}, {Vincent}, {V{\"o}lk}, {Wagner}, \&
  {Ward}}]{Aharonian:2006d}
{Aharonian}, F., {Akhperjanian}, A.~G., {Bazer-Bachi}, A.~R., {et~al.}
  2006{\natexlab{c}}, \aap, 448, L43

\bibitem[{{Aharonian} {et~al.}(2006{\natexlab{d}}){Aharonian}, {Akhperjanian},
  {Bazer-Bachi}, {Beilicke}, {Benbow}, {Berge}, {Bernl{\"o}hr}, {Boisson},
  {Bolz}, {Borrel}, {Braun}, {Breitling}, {Brown}, {Chadwick}, {Chounet},
  {Cornils}, {Costamante}, {Degrange}, {Dickinson}, {Djannati-Ata{\"i}},
  {Drury}, {Dubus}, {Emmanoulopoulos}, {Espigat}, {Feinstein}, {Fontaine},
  {Fuchs}, {Funk}, {Gallant}, {Giebels}, {Gillessen}, {Glicenstein}, {Goret},
  {Hadjichristidis}, {Hauser}, {Heinzelmann}, {Henri}, {Hermann}, {Hinton},
  {Hofmann}, {Holleran}, {Horns}, {Jacholkowska}, {de Jager}, {Kh{\'e}lifi},
  {Komin}, {Konopelko}, {Latham}, {Le Gallou}, {Lemi{\`e}re},
  {Lemoine-Goumard}, {Leroy}, {Lohse}, {Martin}, {Martineau-Huynh},
  {Marcowith}, {Masterson}, {McComb}, {de Naurois}, {Nolan}, {Noutsos},
  {Orford}, {Osborne}, {Ouchrif}, {Panter}, {Pelletier}, {Pita},
  {P{\"u}hlhofer}, {Punch}, {Raubenheimer}, {Raue}, {Raux}, {Rayner}, {Reimer},
  {Reimer}, {Ripken}, {Rob}, {Rolland}, {Rowell}, {Sahakian}, {Saug{\'e}},
  {Schlenker}, {Schlickeiser}, {Schuster}, {Schwanke}, {Siewert}, {Sol},
  {Spangler}, {Steenkamp}, {Stegmann}, {Tavernet}, {Terrier}, {Th{\'e}oret},
  {Tluczykont}, {Vasileiadis}, {Venter}, {Vincent}, {V{\"o}lk}, \&
  {Wagner}}]{ref:gps2006}
{Aharonian}, F., {Akhperjanian}, A.~G., {Bazer-Bachi}, A.~R., {et~al.}
  2006{\natexlab{d}}, \apj, 636, 777

\bibitem[{{Aharonian} {et~al.}(2005{\natexlab{e}}){Aharonian}, {Akhperjanian},
  {Bazer-Bachi}, {Beilicke}, {Benbow}, {Berge}, {Bernl{\"o}hr}, {Boisson},
  {Bolz}, {Borrel}, {Braun}, {Breitling}, {Brown}, {Chadwick}, {Chounet},
  {Cornils}, {Costamante}, {Degrange}, {Dickinson}, {Djannati-Ata{\"i}},
  {O'C.~Drury}, {Dubus}, {Emmanoulopoulos}, {Espigat}, {Feinstein}, {Fontaine},
  {Fuchs}, {Funk}, {Gallant}, {Giebels}, {Gillessen}, {Glicenstein}, {Goret},
  {Hadjichristidis}, {Hauser}, {Heinzelmann}, {Henri}, {Hermann}, {Hinton},
  {Hofmann}, {Holleran}, {Horns}, {Jacholkowska}, {de Jager}, {Kh{\'e}lifi},
  {Komin}, {Konopelko}, {Latham}, {Le Gallou}, {Lemi{\`e}re},
  {Lemoine-Goumard}, {Leroy}, {Lohse}, {Martin}, {Martineau-Huynh},
  {Marcowith}, {Masterson}, {McComb}, {de Naurois}, {Nolan}, {Noutsos},
  {Orford}, {Osborne}, {Ouchrif}, {Panter}, {Pelletier}, {Pita},
  {P{\"u}hlhofer}, {Punch}, {Raubenheimer}, {Raue}, {Raux}, {Rayner}, {Reimer},
  {Reimer}, {Ripken}, {Rob}, {Rolland}, {Rowell}, {Sahakian}, {Saug{\'e}},
  {Schlenker}, {Schlickeiser}, {Schuster}, {Schwanke}, {Siewert}, {Sol},
  {Spangler}, {Steenkamp}, {Stegmann}, {Tavernet}, {Terrier}, {Th{\'e}oret},
  {Tluczykont}, {Vasileiadis}, {Venter}, {Vincent}, {V{\"o}lk}, \&
  {Wagner}}]{Aharonian:2005b}
{Aharonian}, F., {Akhperjanian}, A.~G., {Bazer-Bachi}, A.~R., {et~al.}
  2005{\natexlab{e}}, \aap, 437, L7

\bibitem[{{Aharonian} {et~al.}(2006{\natexlab{e}}){Aharonian}, {Akhperjanian},
  {Bazer-Bachi}, {Beilicke}, {Benbow}, {Berge}, {Bernl{\"o}hr}, {Boisson},
  {Bolz}, {Borrel}, {Braun}, {Brown}, {B{\"u}hler}, {B{\"u}sching}, {Carrigan},
  {Chadwick}, {Chounet}, {Cornils}, {Costamante}, {Degrange}, {Dickinson},
  {Djannati-Ata{\"i}}, {O'C.~Drury}, {Dubus}, {Egberts}, {Emmanoulopoulos},
  {Espigat}, {Feinstein}, {Ferrero}, {Fiasson}, {Fontaine}, {Funk}, {Funk},
  {F{\"u}{\ss}ling}, {Gallant}, {Giebels}, {Glicenstein}, {Goret},
  {Hadjichristidis}, {Hauser}, {Hauser}, {Heinzelmann}, {Henri}, {Hermann},
  {Hinton}, {Hoffmann}, {Hofmann}, {Holleran}, {Horns}, {Jacholkowska}, {de
  Jager}, {Kendziorra}, {Kh{\'e}lifi}, {Komin}, {Konopelko}, {Kosack},
  {Latham}, {Le Gallou}, {Lemi{\`e}re}, {Lemoine-Goumard}, {Lohse}, {Martin},
  {Martineau-Huynh}, {Marcowith}, {Masterson}, {Maurin}, {McComb}, {Moulin},
  {de Naurois}, {Nedbal}, {Nolan}, {Noutsos}, {Orford}, {Osborne}, {Ouchrif},
  {Panter}, {Pelletier}, {Pita}, {P{\"u}hlhofer}, {Punch}, {Raubenheimer},
  {Raue}, {Rayner}, {Reimer}, {Reimer}, {Ripken}, {Rob}, {Rolland}, {Rowell},
  {Sahakian}, {Santangelo}, {Saug{\'e}}, {Schlenker}, {Schlickeiser},
  {Schr{\"o}der}, {Schwanke}, {Schwarzburg}, {Shalchi}, {Sol}, {Spangler},
  {Spanier}, {Steenkamp}, {Stegmann}, {Superina}, {Tavernet}, {Terrier},
  {Tluczykont}, {van Eldik}, {Vasileiadis}, {Venter}, {Vincent}, {V{\"o}lk},
  {Wagner}, \& {Ward}}]{Aharonian:2006a}
{Aharonian}, F., {Akhperjanian}, A.~G., {Bazer-Bachi}, A.~R., {et~al.}
  2006{\natexlab{e}}, \aap, 460, 743

\bibitem[{{Aharonian} {et~al.}(2006{\natexlab{f}}){Aharonian}, {Akhperjanian},
  {Bazer-Bachi}, {Beilicke}, {Benbow}, {Berge}, {Bernl{\"o}hr}, {Boisson},
  {Bolz}, {Borrel}, {Braun}, {Brown}, {B{\"u}hler}, {B{\"u}sching}, {Carrigan},
  {Chadwick}, {Chounet}, {Cornils}, {Costamante}, {Degrange}, {Dickinson},
  {Djannati-Ata{\"i}}, {O'C.~Drury}, {Dubus}, {Egberts}, {Emmanoulopoulos},
  {Espigat}, {Feinstein}, {Ferrero}, {Fiasson}, {Fontaine}, {Funk}, {Funk},
  {F{\"u}{\ss}ling}, {Gallant}, {Giebels}, {Glicenstein}, {Goret},
  {Hadjichristidis}, {Hauser}, {Hauser}, {Heinzelmann}, {Henri}, {Hermann},
  {Hinton}, {Hoffmann}, {Hofmann}, {Holleran}, {Horns}, {Jacholkowska}, {de
  Jager}, {Kendziorra}, {Kh{\'e}lifi}, {Komin}, {Konopelko}, {Kosack},
  {Latham}, {Le Gallou}, {Lemi{\`e}re}, {Lemoine-Goumard}, {Lohse}, {Martin},
  {Martineau-Huynh}, {Marcowith}, {Masterson}, {Maurin}, {McComb}, {de
  Naurois}, {Nedbal}, {Nolan}, {Noutsos}, {Orford}, {Osborne}, {Ouchrif},
  {Panter}, {Pelletier}, {Pita}, {P{\"u}hlhofer}, {Punch}, {Raubenheimer},
  {Raue}, {Rayner}, {Reimer}, {Reimer}, {Ripken}, {Rob}, {Rolland}, {Rowell},
  {Sahakian}, {Santangelo}, {Saug{\'e}}, {Schlenker}, {Schlickeiser},
  {Schr{\"o}der}, {Schwanke}, {Schwarzburg}, {Shalchi}, {Sol}, {Spangler},
  {Spanier}, {Steenkamp}, {Stegmann}, {Superina}, {Tavernet}, {Terrier},
  {Th{\'e}oret}, {Tluczykont}, {van Eldik}, {Vasileiadis}, {Venter}, {Vincent},
  {V{\"o}lk}, {Wagner}, \& {Ward}}]{Aharonian:2006b}
{Aharonian}, F., {Akhperjanian}, A.~G., {Bazer-Bachi}, A.~R., {et~al.}
  2006{\natexlab{f}}, \aap, 456, 245

\bibitem[{{Aharonian} {et~al.}(2006{\natexlab{g}}){Aharonian}, {Akhperjanian},
  {Bazer-Bachi}, {Beilicke}, {Benbow}, {Berge}, {Bernl{\"o}hr}, {Boisson},
  {Bolz}, {Borrel}, {Braun}, {Brown}, {B{\"u}hler}, {B{\"u}sching}, {Carrigan},
  {Chadwick}, {Chounet}, {Cornils}, {Costamante}, {Degrange}, {Dickinson},
  {Djannati-Ata{\"i}}, {O'C.~Drury}, {Dubus}, {Egberts}, {Emmanoulopoulos},
  {Espigat}, {Feinstein}, {Ferrero}, {Fiasson}, {Fontaine}, {Funk}, {Funk},
  {F{\"u}{\ss}ling}, {Gallant}, {Giebels}, {Glicenstein}, {Goret},
  {Hadjichristidis}, {Hauser}, {Hauser}, {Heinzelmann}, {Henri}, {Hermann},
  {Hinton}, {Hoffmann}, {Hofmann}, {Holleran}, {Horns}, {Jacholkowska}, {de
  Jager}, {Kendziorra}, {Kh{\'e}lifi}, {Komin}, {Konopelko}, {Kosack},
  {Latham}, {Le Gallou}, {Lemi{\`e}re}, {Lemoine-Goumard}, {Lohse}, {Martin},
  {Martineau-Huynh}, {Marcowith}, {Masterson}, {Maurin}, {McComb}, {Moulin},
  {de Naurois}, {Nedbal}, {Nolan}, {Noutsos}, {Orford}, {Osborne}, {Ouchrif},
  {Panter}, {Pelletier}, {Pita}, {P{\"u}hlhofer}, {Punch}, {Raubenheimer},
  {Raue}, {Rayner}, {Reimer}, {Reimer}, {Ripken}, {Rob}, {Rolland}, {Rowell},
  {Sahakian}, {Santangelo}, {Saug{\'e}}, {Schlenker}, {Schlickeiser},
  {Schr{\"o}der}, {Schwanke}, {Schwarzburg}, {Shalchi}, {Sol}, {Spangler},
  {Spanier}, {Steenkamp}, {Stegmann}, {Superina}, {Tavernet}, {Terrier},
  {Th{\'e}oret}, {Tluczykont}, {van Eldik}, {Vasileiadis}, {Venter}, {Vincent},
  {V{\"o}lk}, {Wagner}, \& {Ward}}]{Aharonian:2006g}
{Aharonian}, F., {Akhperjanian}, A.~G., {Bazer-Bachi}, A.~R., {et~al.}
  2006{\natexlab{g}}, \aap, 460, 365

\bibitem[{{Aharonian}(1991)}]{Aharonian91}
{Aharonian}, F.~A. 1991, \apss, 180, 305

\bibitem[{{Aharonian} {et~al.}(2004{\natexlab{b}}){Aharonian}, {Akhperjanian},
  {Aye}, {Bazer-Bachi}, {Beilicke}, {Benbow}, {Berge}, {Berghaus},
  {Bernl{\"o}hr}, {Bolz}, {Boisson}, {Borgmeier}, {Breitling}, {Brown},
  {Bussons Gordo}, {Chadwick}, {Chitnis}, {Chounet}, {Cornils}, {Costamante},
  {Degrange}, {Djannati-Ata{\"i}}, {Drury}, {Ergin}, {Espigat}, {Feinstein},
  {Fleury}, {Fontaine}, {Funk}, {Gallant}, {Giebels}, {Gillessen}, {Goret},
  {Guy}, {Hadjichristidis}, {Hauser}, {Heinzelmann}, {Henri}, {Hermann},
  {Hinton}, {Hofmann}, {Holleran}, {Horns}, {de Jager}, {Jung}, {Kh{\'e}lifi},
  {Komin}, {Konopelko}, {Latham}, {Le Gallou}, {Lemoine}, {Lemi{\`e}re},
  {Leroy}, {Lohse}, {Marcowith}, {Masterson}, {McComb}, {de Naurois}, {Nolan},
  {Noutsos}, {Orford}, {Osborne}, {Ouchrif}, {Panter}, {Pelletier}, {Pita},
  {Pohl}, {P{\"u}hlhofer}, {Punch}, {Raubenheimer}, {Raue}, {Raux}, {Rayner},
  {Redondo}, {Reimer}, {Reimer}, {Ripken}, {Rivoal}, {Rob}, {Rolland},
  {Rowell}, {Sahakian}, {Saug{\'e}}, {Schlenker}, {Schlickeiser}, {Schuster},
  {Schwanke}, {Siewert}, {Sol}, {Steenkamp}, {Stegmann}, {Tavernet},
  {Th{\'e}oret}, {Tluczykont}, {van der Walt}, {Vasileiadis}, {Vincent},
  {Visser}, {V{\"o}lk}, \& {Wagner}}]{2004Natur.432...75A}
{Aharonian}, F.~A., {Akhperjanian}, A.~G., {Aye}, K.-M., {et~al.}
  2004{\natexlab{b}}, \nat, 432, 75

\bibitem[{{Aharonian} {et~al.}(2007){Aharonian}, {Akhperjanian}, {Bazer-Bachi},
  {Behera}, {Beilicke}, {Benbow}, {Berge}, {Bernl{\"o}hr}, {Boisson}, {Bolz},
  {Borrel}, {Braun}, {Brion}, {Brown}, {B{\"u}hler}, {B{\"u}sching},
  {Boutelier}, {Carrigan}, {Chadwick}, {Chounet}, {Coignet}, {Cornils},
  {Costamante}, {Degrange}, {Dickinson}, {Djannati-Ata{\"i}}, {Domainko},
  {O'C.~Drury}, {Dubus}, {Egberts}, {Emmanoulopoulos}, {Espigat}, {Farnier},
  {Feinstein}, {Fiasson}, {F{\"o}rster}, {Fontaine}, {Funk}, {Funk},
  {F{\"u}{\ss}ling}, {Gallant}, {Giebels}, {Glicenstein}, {Gl{\"u}ck}, {Goret},
  {Hadjichristidis}, {Hauser}, {Hauser}, {Heinzelmann}, {Henri}, {Hermann},
  {Hinton}, {Hoffmann}, {Hofmann}, {Holleran}, {Hoppe}, {Horns},
  {Jacholkowska}, {de Jager}, {Kendziorra}, {Kerschhaggl}, {Kh{\'e}lifi},
  {Komin}, {Kosack}, {Lamanna}, {Latham}, {Le Gallou}, {Lemi{\`e}re},
  {Lemoine-Goumard}, {Lohse}, {Martin}, {Martineau-Huynh}, {Marcowith},
  {Masterson}, {Maurin}, {McComb}, {Moulin}, {de Naurois}, {Nedbal}, {Nolan},
  {Noutsos}, {Olive}, {Orford}, {Osborne}, {Panter}, {Pedaletti}, {Pelletier},
  {Petrucci}, {Pita}, {P{\"u}hlhofer}, {Punch}, {Ranchon}, {Raubenheimer},
  {Raue}, {Rayner}, {Reimer}, {Ripken}, {Rob}, {Rolland}, {Rosier-Lees},
  {Rowell}, {Ruppel}, {Sahakian}, {Santangelo}, {Saug{\'e}}, {Schlenker},
  {Schlickeiser}, {Schr{\"o}der}, {Schwanke}, {Schwarzburg}, {Schwemmer},
  {Shalchi}, {Sol}, {Spangler}, {Steenkamp}, {Stegmann}, {Superina}, {Tam},
  {Tavernet}, {Terrier}, {Tluczykont}, {van Eldik}, {Vasileiadis}, {Venter},
  {Vialle}, {Vincent}, {V{\"o}lk}, {Wagner}, {Ward}, {Moriguchi}, \&
  {Fukui}}]{2007A&A...469L...1A}
{Aharonian}, F.~A., {Akhperjanian}, A.~G., {Bazer-Bachi}, A.~R., {et~al.} 2007,
  \aap, 469, L1

\bibitem[{{Aharonian} {et~al.}(2002){Aharonian}, {Akhperjanian}, {Beilicke},
  {Bernloehr}, {Bojahr}, {Bolz}, {Boerst}, {Coarasa}, {Contreras}, {Cortina},
  {Denninghoff}, {Fonseca}, {Girma}, {Goetting}, {Heinzelmann}, {Hermann},
  {Heusler}, {Hofmann}, {Horns}, {Jung}, {Kankanyan}, {Kestel}, {Kettler},
  {Kohnle}, {Konopelko}, {Kornmeyer}, {Kranich}, {Krawczynski}, {Lampeitl},
  {Lopez}, {Lorenz}, {Lucarelli}, {Mang}, {Meyer}, {Mirzoyan}, {Moralejo},
  {Ona}, {Panter}, {Plyasheshnikov}, {Puehlhofer}, {Rauterberg}, {Reyes},
  {Rhode}, {Ripken}, {Roehring}, {Rowell}, {Sahakian}, {Samorski}, {Schilling},
  {Siems}, {Sobzynska}, {Stamm}, {Tluczykont}, {Voelk}, {Wiedner}, \&
  {Wittek}}]{ref:hegrasurvey}
{Aharonian}, F.~A., {Akhperjanian}, A.~G., {Beilicke}, M., {et~al.} 2002, \aap,
  395, 803

\bibitem[{{Aharonian} \& {Atoyan}(1996)}]{Aharonian1996}
{Aharonian}, F.~A. \& {Atoyan}, A.~M. 1996, \aap, 309, 917

\bibitem[{{Aharonian} {et~al.}(1997){Aharonian}, {Atoyan}, \&
  {Kifune}}]{1997MNRAS.291..162A}
{Aharonian}, F.~A., {Atoyan}, A.~M., \& {Kifune}, T. 1997, \mnras, 291, 162

\bibitem[{{Akiyama} {et~al.}(2016){Akiyama}, {Stawarz}, {Tanaka}, {Nagai},
  {Giroletti}, \& {Honma}}]{2016ApJ...823L..26A}
{Akiyama}, K., {Stawarz}, {\L}., {Tanaka}, Y.~T., {et~al.} 2016, \apjl, 823,
  L26

\bibitem[{{Aleksi{\'c}} {et~al.}(2012){Aleksi{\'c}}, {Alvarez}, {Antonelli},
  {Antoranz}, {Asensio}, {Backes}, {Barres de Almeida}, {Barrio}, {Bastieri},
  {Becerra Gonz{\'a}lez}, {Bednarek}, {Berger}, {Bernardini}, {Biland},
  {Blanch}, {Bock}, {Boller}, {Bonnoli}, {Borla Tridon}, {Bretz},
  {Ca{\~n}ellas}, {Carmona}, {Carosi}, {Colin}, {Colombo}, {Contreras},
  {Cortina}, {Cossio}, {Covino}, {Da Vela}, {Dazzi}, {De Angelis}, {De Caneva},
  {De Cea del Pozo}, {De Lotto}, {Delgado Mendez}, {Diago Ortega}, {Doert},
  {Dom{\'{\i}}nguez}, {Dominis Prester}, {Dorner}, {Doro}, {Eisenacher},
  {Elsaesser}, {Ferenc}, {Fonseca}, {Font}, {Fruck}, {Garc{\'{\i}}a L{\'o}pez},
  {Garczarczyk}, {Garrido}, {Giavitto}, {Godinovi{\'c}}, {Gonz{\'a}lez
  Mu{\~n}oz}, {Gozzini}, {Hadasch}, {H{\"a}fner}, {Herrero}, {Hildebrand},
  {Hose}, {Hrupec}, {Huber}, {Jankowski}, {Jogler}, {Kadenius}, {Kellermann},
  {Klepser}, {Kr{\"a}henb{\"u}hl}, {Krause}, {La Barbera}, {Lelas}, {Leonardo},
  {Lewandowska}, {Lindfors}, {Lombardi}, {L{\'o}pez}, {L{\'o}pez-Coto},
  {L{\'o}pez-Oramas}, {Lorenz}, {Makariev}, {Maneva}, {Mankuzhiyil},
  {Mannheim}, {Maraschi}, {Mariotti}, {Mart{\'{\i}}nez}, {Mazin}, {Meucci},
  {Miranda}, {Mirzoyan}, {Mold{\'o}n}, {Moralejo}, {Munar-Adrover},
  {Niedzwiecki}, {Nieto}, {Nilsson}, {Nowak}, {Orito}, {Paiano}, {Paneque},
  {Paoletti}, {Pardo}, {Paredes}, {Partini}, {Perez-Torres}, {Persic}, {Pilia},
  {Pochon}, {Prada}, {Prada Moroni}, {Prandini}, {Puerto Gimenez}, {Puljak},
  {Reichardt}, {Reinthal}, {Rhode}, {Rib{\'o}}, {Rico}, {R{\"u}gamer},
  {Saggion}, {Saito}, {Saito}, {Salvati}, {Satalecka}, {Scalzotto}, {Scapin},
  {Schultz}, {Schweizer}, {Shore}, {Sillanp{\"a}{\"a}}, {Sitarek}, {Snidaric},
  {Sobczynska}, {Spanier}, {Spiro}, {Stamatescu}, {Stamerra}, {Steinke},
  {Storz}, {Strah}, {Sun}, {Suri{\'c}}, {Takalo}, {Takami}, {Tavecchio},
  {Temnikov}, {Terzi{\'c}}, {Tescaro}, {Teshima}, {Tibolla}, {Torres},
  {Treves}, {Uellenbeck}, {Vogler}, {Wagner}, {Weitzel}, {Zabalza}, {Zandanel},
  \& {Zanin}}]{MAGIC:W51}
{Aleksi{\'c}}, J., {Alvarez}, E.~A., {Antonelli}, L.~A., {et~al.} 2012, \aap,
  541, A13

\bibitem[{{Aliu} {et~al.}(2014{\natexlab{a}}){Aliu}, {Archambault}, {Aune},
  {Behera}, {Beilicke}, {Benbow}, {Berger}, {Bird}, {Bouvier}, {Buckley}, \&
  et~al.}]{2014ApJ...780..168A}
{Aliu}, E., {Archambault}, S., {Aune}, T., {et~al.} 2014{\natexlab{a}}, \apj,
  780, 168

\bibitem[{{Aliu} {et~al.}(2014{\natexlab{b}}){Aliu}, {Archambault}, {Aune},
  {Behera}, {Beilicke}, {Benbow}, {Berger}, {Bird}, {Buckley}, {Bugaev},
  {Cardenzana}, {Cerruti}, {Chen}, {Ciupik}, {Collins-Hughes}, {Connolly},
  {Cui}, {Dumm}, {Dwarkadas}, {Errando}, {Falcone}, {Federici}, {Feng},
  {Finley}, {Fleischhack}, {Fortin}, {Fortson}, {Furniss}, {Galante}, {Gall},
  {Gillanders}, {Griffin}, {Griffiths}, {Grube}, {Gyuk}, {Hanna}, {Holder},
  {Hughes}, {Humensky}, {Kaaret}, {Kertzman}, {Khassen}, {Kieda}, {Krennrich},
  {Kumar}, {Lang}, {Madhavan}, {Maier}, {McCann}, {Meagher}, {Millis},
  {Moriarty}, {Mukherjee}, {Nieto}, {O'Faol{\'a}in de Bhr{\'o}ithe}, {Ong},
  {Otte}, {Pandel}, {Park}, {Pohl}, {Popkow}, {Prokoph}, {Quinn}, {Ragan},
  {Rajotte}, {Ratliff}, {Reyes}, {Reynolds}, {Richards}, {Roache}, {Rousselle},
  {Sembroski}, {Shahinyan}, {Sheidaei}, {Smith}, {Staszak}, {Telezhinsky},
  {Tsurusaki}, {Tucci}, {Tyler}, {Varlotta}, {Vassiliev}, {Vincent}, {Wakely},
  {Ward}, {Weinstein}, {Welsing}, \& {Wilhelm}}]{2014ApJ...787..166A}
{Aliu}, E., {Archambault}, S., {Aune}, T., {et~al.} 2014{\natexlab{b}}, \apj,
  787, 166

\bibitem[{{Ang{\"u}ner} {et~al.}(2017){Ang{\"u}ner}, {Aharonian}, {Bordas},
  {Casanova}, {Hoischen}, {Oya}, {Ziegler}, \& {for the
  H.E.S.S.~Collaboration}}]{2017arXiv170107002A}
{Ang{\"u}ner}, E.~O., {Aharonian}, F., {Bordas}, P., {et~al.} 2017, ArXiv
  e-prints [\eprint[arXiv]{1701.07002}]

\bibitem[{{Antonopoulou} {et~al.}(2016){Antonopoulou}, {Vasilopoulos}, \&
  {Espinoza}}]{2016ATel.9282....1A}
{Antonopoulou}, D., {Vasilopoulos}, G., \& {Espinoza}, C.~M. 2016, The
  Astronomer's Telegram, 9282

\bibitem[{{Astropy Collaboration} {et~al.}(2013){Astropy Collaboration},
  {Robitaille}, {Tollerud}, {Greenfield}, {Droettboom}, {Bray}, {Aldcroft},
  {Davis}, {Ginsburg}, {Price-Whelan}, {Kerzendorf}, {Conley}, {Crighton},
  {Barbary}, {Muna}, {Ferguson}, {Grollier}, {Parikh}, {Nair}, {Unther},
  {Deil}, {Woillez}, {Conseil}, {Kramer}, {Turner}, {Singer}, {Fox}, {Weaver},
  {Zabalza}, {Edwards}, {Azalee Bostroem}, {Burke}, {Casey}, {Crawford},
  {Dencheva}, {Ely}, {Jenness}, {Labrie}, {Lim}, {Pierfederici}, {Pontzen},
  {Ptak}, {Refsdal}, {Servillat}, \& {Streicher}}]{2013AandA...558A..33A}
{Astropy Collaboration}, {Robitaille}, T.~P., {Tollerud}, E.~J., {et~al.} 2013,
  \aap, 558, A33

\bibitem[{{Bamba} {et~al.}(2016){Bamba}, {Terada}, {Hewitt}, {Petre},
  {Angelini}, {Safi-Harb}, {Zhou}, {Bocchino}, \& {Sawada}}]{Bamba16}
{Bamba}, A., {Terada}, Y., {Hewitt}, J., {et~al.} 2016, \apj, 818, 63

\bibitem[{{Bamba} {et~al.}(2009){Bamba}, {Yamazaki}, {Kohri}, {Matsumoto},
  {Wagner}, {P{\"u}hlhofer}, \& {Kosack}}]{2009ApJ...691.1854B}
{Bamba}, A., {Yamazaki}, R., {Kohri}, K., {et~al.} 2009, \apj, 691, 1854

\bibitem[{{Berge} {et~al.}(2007){Berge}, {Funk}, \& {Hinton}}]{ref:bgmodeling}
{Berge}, D., {Funk}, S., \& {Hinton}, J. 2007, \aap, 466, 1219

\bibitem[{{Bernl{\"o}hr}(2008)}]{ref:simulations}
{Bernl{\"o}hr}, K. 2008, Astroparticle Physics, 30, 149

\bibitem[{{Bietenholz} \& {Bartel}(2008)}]{2008MNRAS.386.1411B}
{Bietenholz}, M.~F. \& {Bartel}, N. 2008, \mnras, 386, 1411

\bibitem[{{Blum} {et~al.}(1999){Blum}, {Damineli}, \& {Conti}}]{Blum99}
{Blum}, R.~D., {Damineli}, A., \& {Conti}, P.~S. 1999, \aj, 117, 1392

\bibitem[{{Bocchino} {et~al.}(2001){Bocchino}, {Parmar}, {Mereghetti},
  {Orlandini}, {Santangelo}, \& {Angelini}}]{Bocchino01}
{Bocchino}, F., {Parmar}, A.~N., {Mereghetti}, S., {et~al.} 2001, \aap, 367,
  629

\bibitem[{{Bocchino} {et~al.}(2005){Bocchino}, {van der Swaluw}, {Chevalier},
  \& {Bandiera}}]{Bocchino05}
{Bocchino}, F., {van der Swaluw}, E., {Chevalier}, R., \& {Bandiera}, R. 2005,
  \aap, 442, 539

\bibitem[{{Bolz}(2004)}]{ref:muonsbolz}
{Bolz}, O. 2004, PhD thesis, Ruprecht-Karls-Universit\"{a}t Heidelberg

\bibitem[{{Brogan} {et~al.}(2006){Brogan}, {Gelfand}, {Gaensler}, {Kassim}, \&
  {Lazio}}]{Brogan06}
{Brogan}, C.~L., {Gelfand}, J.~D., {Gaensler}, B.~M., {Kassim}, N.~E., \&
  {Lazio}, T.~J.~W. 2006, \apjl, 639, L25

\bibitem[{{Burgay} {et~al.}(2013){Burgay}, {Bailes}, {Bates}, {Bhat},
  {Burke-Spolaor}, {Champion}, {Coster}, {D'Amico}, {Johnston}, {Keith},
  {Kramer}, {Levin}, {Lyne}, {Milia}, {Ng}, {Possenti}, {Stappers}, {Thornton},
  {Tiburzi}, {van Straten}, \& {Bassa}}]{Burgay2013}
{Burgay}, M., {Bailes}, M., {Bates}, S.~D., {et~al.} 2013, \mnras, 433, 259

\bibitem[{{Calabretta} \& {Greisen}(2002)}]{2002AandA...395.1077C}
{Calabretta}, M.~R. \& {Greisen}, E.~W. 2002, \aap, 395, 1077

\bibitem[{{Camilo} {et~al.}(2000){Camilo}, {Kaspi}, {Lyne}, {Manchester},
  {Bell}, {D'Amico}, {McKay}, \& {Crawford}}]{Camilo00}
{Camilo}, F., {Kaspi}, V.~M., {Lyne}, A.~G., {et~al.} 2000, \apj, 541, 367

\bibitem[{{Camilo} {et~al.}(2006){Camilo}, {Ransom}, {Gaensler}, {Slane},
  {Lorimer}, {Reynolds}, {Manchester}, \& {Murray}}]{Camilo06}
{Camilo}, F., {Ransom}, S.~M., {Gaensler}, B.~M., {et~al.} 2006, \apj, 637, 456

\bibitem[{{Carrigan} {et~al.}(2013{\natexlab{a}}){Carrigan}, {Brun}, {Chaves},
  {Deil}, {Donath}, {Gast}, {Marandon}, {Renaud}, \& {for the
  H.E.S.S.~Collaboration}}]{Carrigan13a_GPS}
{Carrigan}, S., {Brun}, F., {Chaves}, R.~C.~G., {et~al.} 2013{\natexlab{a}},
  ArXiv e-prints [\eprint[arXiv]{1307.4690}]

\bibitem[{{Carrigan} {et~al.}(2013{\natexlab{b}}){Carrigan}, {Brun}, {Chaves},
  {Deil}, {Gast}, {Marandon}, \& {for the
  H.E.S.S.~Collaboration}}]{Carrigan13b_GPS}
{Carrigan}, S., {Brun}, F., {Chaves}, R.~C.~G., {et~al.} 2013{\natexlab{b}},
  ArXiv e-prints [\eprint[arXiv]{1307.4868}]

\bibitem[{{Cash}(1979)}]{1979ApJ...228..939C}
{Cash}, W. 1979, \apj, 228, 939

\bibitem[{{Caswell} {et~al.}(2004){Caswell}, {McClure-Griffiths}, \&
  {Cheung}}]{Caswell04}
{Caswell}, J.~L., {McClure-Griffiths}, N.~M., \& {Cheung}, M.~C.~M. 2004,
  \mnras, 352, 1405

\bibitem[{{Chaves}(2009)}]{ref:icrc09}
{Chaves}, R.~C.~G. 2009, ArXiv e-prints [\eprint[arXiv]{0907.0768}]

\bibitem[{{Chaves} {et~al.}(2008{\natexlab{a}}){Chaves}, {de O{\~n}a Wilhemi},
  \& {Hoppe}}]{Chaves08_GPS}
{Chaves}, R.~C.~G., {de O{\~n}a Wilhemi}, E., \& {Hoppe}, S.
  2008{\natexlab{a}}, in American Institute of Physics Conference Series, Vol.
  1085, American Institute of Physics Conference Series, ed. F.~A. {Aharonian},
  W.~{Hofmann}, \& F.~{Rieger}, 219--222

\bibitem[{{Chaves} {et~al.}(2008{\natexlab{b}}){Chaves}, {Renaud},
  {Lemoine-Goumard}, \& {Goret}}]{Chaves:2008}
{Chaves}, R.~C.~G., {Renaud}, M., {Lemoine-Goumard}, M., \& {Goret}, P.
  2008{\natexlab{b}}, in American Institute of Physics Conference Series, Vol.
  1085, American Institute of Physics Conference Series, ed. F.~A. {Aharonian},
  W.~{Hofmann}, \& F.~{Rieger}, 372--375

\bibitem[{{Clifton} \& {Lyne}(1986)}]{Clifton1986}
{Clifton}, T.~R. \& {Lyne}, A.~G. 1986, \nat, 320, 43

\bibitem[{{Cordes} \& {Lazio}(2002)}]{CordesLazio:2002}
{Cordes}, J.~M. \& {Lazio}, T.~J.~W. 2002, ArXiv Astrophysics e-prints
  [\eprint{astro-ph/0207156}]

\bibitem[{{Cui} {et~al.}(2016){Cui}, {P{\"u}hlhofer}, \&
  {Santangelo}}]{Cui2016}
{Cui}, Y., {P{\"u}hlhofer}, G., \& {Santangelo}, A. 2016, \aap, 591, A68

\bibitem[{{Dame} {et~al.}(2001){Dame}, {Hartmann}, \& {Thaddeus}}]{Dame01}
{Dame}, T.~M., {Hartmann}, D., \& {Thaddeus}, P. 2001, \apj, 547, 792

\bibitem[{{Davies} {et~al.}(2009){Davies}, {Figer}, {Kudritzki}, {Trombley},
  {Kouveliotou}, \& {Wachter}}]{Davies:2009}
{Davies}, B., {Figer}, D.~F., {Kudritzki}, R.-P., {et~al.} 2009, \apj, 707, 844

\bibitem[{{de Jager} \& {Djannati-Ata{\"i}}(2009)}]{deJager09}
{de Jager}, O.~C. \& {Djannati-Ata{\"i}}, A. 2009, in Astrophysics and Space
  Science Library, Vol. 357, Astrophysics and Space Science Library, ed.
  W.~{Becker}, 451

\bibitem[{{de Jager} {et~al.}(1995){de Jager}, {Harding}, {Baring}, \&
  {Mastichiadis}}]{deJager95}
{de Jager}, O.~C., {Harding}, A.~K., {Baring}, M.~G., \& {Mastichiadis}, A.
  1995, International Cosmic Ray Conference, 2, 528

\bibitem[{{de los Reyes} {et~al.}(2012){de los Reyes}, {Zajczyk}, {Chaves}, \&
  {for the H.E.S.S.~Collaboration}}]{delosReyes12}
{de los Reyes}, R., {Zajczyk}, A., {Chaves}, R.~C.~G., \& {for the
  H.E.S.S.~Collaboration}. 2012, ArXiv e-prints [\eprint[arXiv]{1205.0719}]

\bibitem[{{de Naurois} \& {Rolland}(2009)}]{2009APh....32..231D}
{de Naurois}, M. \& {Rolland}, L. 2009, Astroparticle Physics, 32, 231

\bibitem[{{Deil}(2012)}]{Deil12_GPS}
{Deil}, C. 2012, in IAU Symposium, Vol. 284, The Spectral Energy Distribution
  of Galaxies - SED 2011, ed. R.~J. {Tuffs} \& C.~C. {Popescu}, 365--370

\bibitem[{{Deil} {et~al.}(2017){Deil}, {Zanin}, {Lefaucheur}, {Boisson},
  {Kh{\'e}lifi}, {Terrier}, {Wood}, {Mohrmann}, {Chakraborty}, {Watson},
  {L{\'o}pez Coto}, {Klepser}, {Cerruti}, {Lenain}, {Acero},
  {Djannati-Ata{\"i}}, {Pita}, {Bosnjak}, {Ruiz}, {Trichard}, {Vuillaume},
  {Donath}, {King}, {Jouvin}, {Owen}, {Arribas}, {Sipocz}, {Lennarz},
  {Voruganti}, \& {CTA Consortium}}]{2017arXiv170901751D}
{Deil}, C., {Zanin}, R., {Lefaucheur}, J., {et~al.} 2017, ArXiv e-prints
  [\eprint[arXiv]{1709.01751}]

\bibitem[{{Djannati-Ata{\"i}} {et~al.}(2008){Djannati-Ata{\"i}}, {de Jager},
  {Terrier}, {Gallant}, \& {Hoppe}}]{Djannati-Atai:2008a}
{Djannati-Ata{\"i}}, A., {de Jager}, O.~C., {Terrier}, R., {Gallant}, Y.~A., \&
  {Hoppe}, S. 2008, International Cosmic Ray Conference, 2, 823

\bibitem[{{Djannati-Ata{\"\i}} {et~al.}(2009){Djannati-Ata{\"\i}}, {Marandon},
  {Chaves}, {Terrier}, {Komin}, \& {H.E.S.S.~Collaboration}}]{Djannati-Atai09}
{Djannati-Ata{\"\i}}, A., {Marandon}, V., {Chaves}, R.~C.~G., {et~al.} 2009,
  {HESS discovery of VHE gamma-ray emission of a remarkable young composite
  SNR}, Online presentations from the Workshop on Supernova Remnants and Pulsar
  Wind Nebulae in the Chandra Era,
  http://cxc.harvard.edu/cdo/snr09/program.html

\bibitem[{{Donath} {et~al.}(2015){Donath}, {Deil}, {Arribas}, {King}, {Owen},
  {Terrier}, {Reichardt}, {Harris}, {B{\"u}hler}, \& {Klepser}}]{Donath2015}
{Donath}, A., {Deil}, C., {Arribas}, M.~P., {et~al.} 2015, ArXiv e-prints
  [\eprint[arXiv]{1509.07408}]

\bibitem[{{Drury} {et~al.}(1994){Drury}, {Aharonian}, \& {Voelk}}]{Drury94}
{Drury}, L.~O., {Aharonian}, F.~A., \& {Voelk}, H.~J. 1994, \aap, 287, 959

\bibitem[{{Dubner} {et~al.}(1998){Dubner}, {Holdaway}, {Goss}, \&
  {Mirabel}}]{1998AJ....116.1842D}
{Dubner}, G.~M., {Holdaway}, M., {Goss}, W.~M., \& {Mirabel}, I.~F. 1998, \aj,
  116, 1842

\bibitem[{Dubois {et~al.}(1996)Dubois, Hinsen, \& Hugunin}]{numpy}
Dubois, P.~F., Hinsen, K., \& Hugunin, J. 1996, Computers in Physics, 10

\bibitem[{{Dubus} {et~al.}(2013){Dubus}, {Contreras}, {Funk}, {Gallant},
  {Hassan}, {Hinton}, {Inoue}, {Kn{\"o}dlseder}, {Martin}, {Mirabal}, {de
  Naurois}, {Renaud}, \& {CTA Consortium}}]{2013APh....43..317D}
{Dubus}, G., {Contreras}, J.~L., {Funk}, S., {et~al.} 2013, Astroparticle
  Physics, 43, 317

\bibitem[{{Eger} {et~al.}(2011){Eger}, {Rowell}, {Kawamura}, {Fukui},
  {Rolland}, \& {Stegmann}}]{2011A&A...526A..82E}
{Eger}, P., {Rowell}, G., {Kawamura}, A., {et~al.} 2011, \aap, 526, A82

\bibitem[{{Ferrand} \& {Safi-Harb}(2012)}]{SNRcat}
{Ferrand}, G. \& {Safi-Harb}, S. 2012, Advances in Space Research, 49, 1313

\bibitem[{{Fiasson} {et~al.}(2009){Fiasson}, {Marandon}, {Chaves}, {Tibolla},
  \& {for the H.E.S.S.~Collaboration}}]{W51:HESS_ICRC}
{Fiasson}, A., {Marandon}, V., {Chaves}, R.~C.~G., {Tibolla}, O., \& {for the
  H.E.S.S.~Collaboration}. 2009, International Cosmic Ray Conference

\bibitem[{{Fomin} {et~al.}(1994){Fomin}, {Stepanian}, {Lamb}, {Lewis}, {Punch},
  \& {Weekes}}]{1994APh.....2..137F}
{Fomin}, V.~P., {Stepanian}, A.~A., {Lamb}, R.~C., {et~al.} 1994, Astroparticle
  Physics, 2, 137

\bibitem[{{Frail} {et~al.}(1991){Frail}, {Cordes}, {Hankins}, \&
  {Weisberg}}]{Frail91}
{Frail}, D.~A., {Cordes}, J.~M., {Hankins}, T.~H., \& {Weisberg}, J.~M. 1991,
  \apj, 382, 168

\bibitem[{{Frail} {et~al.}(1999){Frail}, {Kulkarni}, \& {Bloom}}]{Frail99}
{Frail}, D.~A., {Kulkarni}, S.~R., \& {Bloom}, J.~S. 1999, \nat, 398, 127

\bibitem[{{Freeman} {et~al.}(2001){Freeman}, {Doe}, \&
  {Siemiginowska}}]{Freeman:2001}
{Freeman}, P., {Doe}, S., \& {Siemiginowska}, A. 2001, in Society of
  Photo-Optical Instrumentation Engineers (SPIE) Conference Series, Vol. 4477,
  Society of Photo-Optical Instrumentation Engineers (SPIE) Conference Series,
  ed. {J.-L.~Starck \& F.~D.~Murtagh}, 76--87

\bibitem[{{Fuerst} {et~al.}(1987){Fuerst}, {Reich}, {Reich}, {Handa}, \&
  {Sofue}}]{Fuerst:1987}
{Fuerst}, E., {Reich}, W., {Reich}, P., {Handa}, T., \& {Sofue}, Y. 1987,
  \aaps, 69, 403

\bibitem[{{Funk} {et~al.}(2007){Funk}, {Hinton}, {Moriguchi}, {Aharonian},
  {Fukui}, {Hofmann}, {Horns}, {P{\"u}hlhofer}, {Reimer}, {Rowell}, {Terrier},
  {Vink}, \& {Wagner}}]{2007AA...470..249F}
{Funk}, S., {Hinton}, J.~A., {Moriguchi}, Y., {et~al.} 2007, \aap, 470, 249

\bibitem[{{Gabici} {et~al.}(2007){Gabici}, {Aharonian}, \& {Blasi}}]{Gabici07}
{Gabici}, S., {Aharonian}, F.~A., \& {Blasi}, P. 2007, \apss, 309, 365

\bibitem[{{Gabici} \& {Montmerle}(2015)}]{2015arXiv151002102G}
{Gabici}, S. \& {Montmerle}, T. 2015, International Cosmic Ray Conference, 29

\bibitem[{{Gaensler} {et~al.}(2002){Gaensler}, {Arons}, {Kaspi}, {Pivovaroff},
  {Kawai}, \& {Tamura}}]{2002ApJ...569..878G}
{Gaensler}, B.~M., {Arons}, J., {Kaspi}, V.~M., {et~al.} 2002, \apj, 569, 878

\bibitem[{{Gaensler} \& {Slane}(2006)}]{Gaensler06}
{Gaensler}, B.~M. \& {Slane}, P.~O. 2006, \araa, 44, 17

\bibitem[{{Gast} {et~al.}(2011){Gast}, {Brun}, {Carrigan}, {Chaves}, {Deil},
  {Djannati-Ata{\"i}}, {Gallant}, {Marandon}, {de Naurois}, \& {de los
  Reyes}}]{ref:icrc11}
{Gast}, H., {Brun}, F., {Carrigan}, S., {et~al.} 2011, International Cosmic Ray
  Conference, 7, 158

\bibitem[{{Gavriil} {et~al.}(2008){Gavriil}, {Gonzalez}, {Gotthelf}, {Kaspi},
  {Livingstone}, \& {Woods}}]{Gavriil08}
{Gavriil}, F.~P., {Gonzalez}, M.~E., {Gotthelf}, E.~V., {et~al.} 2008, Science,
  319, 1802

\bibitem[{{Gonzalez} \& {Safi-Harb}(2003)}]{Gonzalez03}
{Gonzalez}, M. \& {Safi-Harb}, S. 2003, \apjl, 591, L143

\bibitem[{{Gotthelf} \& {Halpern}(2009)}]{2009ApJ...700L.158G}
{Gotthelf}, E.~V. \& {Halpern}, J.~P. 2009, \apjl, 700, L158

\bibitem[{{Gotthelf} {et~al.}(2011){Gotthelf}, {Halpern}, {Terrier}, \&
  {Mattana}}]{2011ApJ...729L..16G}
{Gotthelf}, E.~V., {Halpern}, J.~P., {Terrier}, R., \& {Mattana}, F. 2011,
  \apjl, 729, L16

\bibitem[{{Gotthelf} {et~al.}(2014){Gotthelf}, {Tomsick}, {Halpern}, {Gelfand},
  {Harrison}, {Boggs}, {Christensen}, {Craig}, {Hailey}, {Kaspi}, {Stern}, \&
  {Zhang}}]{2014ApJ...788..155G}
{Gotthelf}, E.~V., {Tomsick}, J.~A., {Halpern}, J.~P., {et~al.} 2014, \apj,
  788, 155

\bibitem[{{G{\"o}{\u g}{\"u}{\c s}} {et~al.}(2016){G{\"o}{\u g}{\"u}{\c s}},
  {Lin}, {Kaneko}, {Kouveliotou}, {Watts}, {Chakraborty}, {Alpar},
  {Huppenkothen}, {Roberts}, {Younes}, \& {van der
  Horst}}]{2016ApJ...829L..25G}
{G{\"o}{\u g}{\"u}{\c s}}, E., {Lin}, L., {Kaneko}, Y., {et~al.} 2016, \apjl,
  829, L25

\bibitem[{{Green}(2014)}]{2014BASI...42...47G}
{Green}, D.~A. 2014, Bulletin of the Astronomical Society of India, 42, 47

\bibitem[{{Gupta} {et~al.}(2005){Gupta}, {Mitra}, {Green}, \&
  {Acharyya}}]{Gupta05}
{Gupta}, Y., {Mitra}, D., {Green}, D.~A., \& {Acharyya}, A. 2005, Current
  Science, 89, 853

\bibitem[{{Hahn} {et~al.}(2014){Hahn}, {de los Reyes}, {Bernl{\"o}hr},
  {Kr{\"u}ger}, {Lo}, {Chadwick}, {Daniel}, {Deil}, {Gast}, {Kosack}, \&
  {Marandon}}]{2014APh....54...25H}
{Hahn}, J., {de los Reyes}, R., {Bernl{\"o}hr}, K., {et~al.} 2014,
  Astroparticle Physics, 54, 25

\bibitem[{{Helfand} {et~al.}(2006){Helfand}, {Becker}, {White}, {Fallon}, \&
  {Tuttle}}]{Helfand06}
{Helfand}, D.~J., {Becker}, R.~H., {White}, R.~L., {Fallon}, A., \& {Tuttle},
  S. 2006, \aj, 131, 2525

\bibitem[{{Helfand} {et~al.}(1989){Helfand}, {Velusamy}, {Becker}, \&
  {Lockman}}]{Helfand89}
{Helfand}, D.~J., {Velusamy}, T., {Becker}, R.~H., \& {Lockman}, F.~J. 1989,
  \apj, 341, 151

\bibitem[{{H.E.S.S.~Collaboration}
  {et~al.}(2017{\natexlab{a}}){H.E.S.S.~Collaboration}, {Abdalla},
  {Abramowski}, {Aharonian}, {Ait Benkhali}, \& {Akhperjanian}}]{HESS:Shells}
{H.E.S.S.~Collaboration}, {Abdalla}, H., {Abramowski}, A., {et~al.}
  2017{\natexlab{a}}, {A search for new supernova remnant shells in the
  Galactic plane with H.E.S.S.}, \aap\ forthcoming

\bibitem[{{H.E.S.S.~Collaboration}
  {et~al.}(2017{\natexlab{b}}){H.E.S.S.~Collaboration}, {Abdalla},
  {Abramowski}, {Aharonian}, {Ait Benkhali}, \& {Akhperjanian}}]{HESS:SNRUL}
{H.E.S.S.~Collaboration}, {Abdalla}, H., {Abramowski}, A., {et~al.}
  2017{\natexlab{b}}, {Galactic Supernova Remnants Population Study at Very
  High Gamma-Ray Energies with H.E.S.S.}, \aap\ forthcoming

\bibitem[{{H.E.S.S.~Collaboration}
  {et~al.}(2017{\natexlab{c}}){H.E.S.S.~Collaboration}, {Abdalla},
  {Abramowski}, {Aharonian}, {Ait Benkhali}, \& {Akhperjanian}}]{HESS:1741}
{H.E.S.S.~Collaboration}, {Abdalla}, H., {Abramowski}, A., {et~al.}
  2017{\natexlab{c}}, {HESS J1741-302: a hidden accelerator in the Galactic
  plane}, \aap\ forthcoming

\bibitem[{{H.E.S.S.~Collaboration}
  {et~al.}(2017{\natexlab{d}}){H.E.S.S.~Collaboration}, {Abdalla},
  {Abramowski}, {Aharonian}, {Ait Benkhali}, \& {Akhperjanian}}]{HESS:SS433}
{H.E.S.S.~Collaboration}, {Abdalla}, H., {Abramowski}, A., {et~al.}
  2017{\natexlab{d}}, {Observations of SS433 with the MAGIC and H.E.S.S.
  telescopes}, \aap\ forthcoming

\bibitem[{{H.E.S.S.~Collaboration}
  {et~al.}(2016{\natexlab{a}}){H.E.S.S.~Collaboration}, {Abdalla},
  {Abramowski}, {Aharonian}, {Ait Benkhali}, {Akhperjanian}, {Andersson},
  {Ang{\"u}ner}, {Arakawa}, {Arrieta}, \& et~al.}]{HESS:VelaJnr}
{H.E.S.S.~Collaboration}, {Abdalla}, H., {Abramowski}, A., {et~al.}
  2016{\natexlab{a}}, ArXiv e-prints [\eprint[arXiv]{1611.01863}]

\bibitem[{{H.E.S.S.~Collaboration}
  {et~al.}(2017{\natexlab{e}}){H.E.S.S.~Collaboration}, {Abdalla},
  {Abramowski}, {Aharonian}, {Ait Benkhali}, {Akhperjanian}, {Andersson},
  {Ang{\"u}ner}, {Arakawa}, \& et~al.}]{HESS:Bow}
{H.E.S.S.~Collaboration}, {Abdalla}, H., {Abramowski}, A., {et~al.}
  2017{\natexlab{e}}, ArXiv e-prints [\eprint[arXiv]{1705.02263}]

\bibitem[{{H.E.S.S.~Collaboration}
  {et~al.}(2016{\natexlab{b}}){H.E.S.S.~Collaboration}, {Abdalla},
  {Abramowski}, {Aharonian}, {Ait Benkhali}, {Akhperjanian}, {Andersson},
  {Ang{\"u}ner}, {Arrieta}, {Aubert}, \& et~al.}]{HESS:W49}
{H.E.S.S.~Collaboration}, {Abdalla}, H., {Abramowski}, A., {et~al.}
  2016{\natexlab{b}}, ArXiv e-prints [\eprint[arXiv]{1609.00600}]

\bibitem[{{H.E.S.S.~Collaboration}
  {et~al.}(2017{\natexlab{f}}){H.E.S.S.~Collaboration}, {Abdalla},
  {Abramowski}, {Aharonian}, {Ait Benkhali}, {Akhperjanian}, {Andersson},
  {Ang{\"u}ner}, {Arrieta}, \& et~al.}]{HESS:PWNPOP}
{H.E.S.S.~Collaboration}, {Abdalla}, H., {Abramowski}, A., {et~al.}
  2017{\natexlab{f}}, ArXiv e-prints [\eprint[arXiv]{1702.08280}]

\bibitem[{{H.E.S.S.~Collaboration}
  {et~al.}(2016{\natexlab{c}}){H.E.S.S.~Collaboration}, {Abdalla},
  {Abramowski}, {Aharonian}, {Ait Benkhali}, {Akhperjanian}, {Andersson},
  {Ang{\"u}ner}, \& et~al.}]{HESS:RXJ1713}
{H.E.S.S.~Collaboration}, {Abdalla}, H., {Abramowski}, A., {et~al.}
  2016{\natexlab{c}}, ArXiv e-prints [\eprint[arXiv]{1609.08671}]

\bibitem[{{H.E.S.S.~Collaboration}
  {et~al.}(2016{\natexlab{d}}){H.E.S.S.~Collaboration}, {Abdalla},
  {Abramowski}, {Aharonian}, {Ait Benkhali}, {Akhperjanian}, {Ang{\"u}ner},
  {Arrieta}, {Aubert}, {Backes}, \& et~al.}]{HESS:MQ}
{H.E.S.S.~Collaboration}, {Abdalla}, H., {Abramowski}, A., {et~al.}
  2016{\natexlab{d}}, ArXiv e-prints [\eprint[arXiv]{1607.04613}]

\bibitem[{{H.E.S.S.~Collaboration}
  {et~al.}(2017{\natexlab{g}}){H.E.S.S.~Collaboration}, {Abdalla},
  {Abramowski}, {Aharonian}, {Ait Benkhali}, {Akhperjaniany}, {Andersson},
  {Ang{\"u}ner}, {Arakawa}, \& et~al.}]{HESS:Arc}
{H.E.S.S.~Collaboration}, {Abdalla}, H., {Abramowski}, A., {et~al.}
  2017{\natexlab{g}}, ArXiv e-prints [\eprint[arXiv]{1706.04535}]

\bibitem[{{H.E.S.S.~Collaboration} {et~al.}(2012){H.E.S.S.~Collaboration},
  {Abramowski}, {Acero}, {Aharonian}, {Akhperjanian}, {Anton}, {Balenderan},
  {Balzer}, {Barnacka}, {Becherini}, {Becker}, {Bernl{\"o}hr}, {Birsin},
  {Biteau}, {Bochow}, {Boisson}, {Bolmont}, {Bordas}, {Brucker}, {Brun},
  {Brun}, {Bulik}, {B{\"u}sching}, {Carrigan}, {Casanova}, {Cerruti},
  {Chadwick}, {Charbonnier}, {Chaves}, {Cheesebrough}, {Cologna}, {Conrad},
  {Couturier}, {Dalton}, {Daniel}, {Davids}, {Degrange}, {Deil}, {Dickinson},
  {Djannati-Ata{\"i}}, {Domainko}, {Drury}, {Dubus}, {Dutson}, {Dyks}, {Dyrda},
  {Egberts}, {Eger}, {Espigat}, {Fallon}, {Farnier}, {Fegan}, {Feinstein},
  {Fernandes}, {Fiasson}, {Fontaine}, {F{\"o}rster}, {F{\"u}{\ss}ling},
  {Gajdus}, {Gallant}, {Garrigoux}, {Gast}, {G{\'e}rard}, {Giebels},
  {Glicenstein}, {Gl{\"u}ck}, {G{\"o}ring}, {Grondin}, {H{\"a}ffner}, {Hague},
  {Hahn}, {Hampf}, {Harris}, {Hauser}, {Heinz}, {Heinzelmann}, {Henri},
  {Hermann}, {Hillert}, {Hinton}, {Hofmann}, {Hofverberg}, {Holler}, {Horns},
  {Jacholkowska}, {Jahn}, {Jamrozy}, {Jung}, {Kastendieck}, {Katarzy{\'n}ski},
  {Katz}, {Kaufmann}, {Kh{\'e}lifi}, {Klochkov}, {Klu{\'z}niak}, {Kneiske},
  {Komin}, {Kosack}, {Kossakowski}, {Krayzel}, {Laffon}, {Lamanna}, {Lenain},
  {Lennarz}, {Lohse}, {Lopatin}, {Lu}, {Marandon}, {Marcowith}, {Masbou},
  {Maurin}, {Maxted}, {Mayer}, {McComb}, {Medina}, {M{\'e}hault}, {Menzler},
  {Moderski}, {Mohamed}, {Moulin}, {Naumann}, {Naumann-Godo}, {de Naurois},
  {Nedbal}, {Nekrassov}, {Nguyen}, {Nicholas}, {Niemiec}, {Nolan}, {Ohm}, {de
  O{\~n}a Wilhelmi}, {Opitz}, {Ostrowski}, {Oya}, {Panter}, {Paz Arribas},
  {Pekeur}, {Pelletier}, {Perez}, {Petrucci}, {Peyaud}, {Pita},
  {P{\"u}hlhofer}, {Punch}, {Quirrenbach}, {Raue}, {Reimer}, {Reimer},
  {Renaud}, {de los Reyes}, {Rieger}, {Ripken}, {Rob}, {Rosier-Lees}, {Rowell},
  {Rudak}, {Rulten}, {Sahakian}, {Sanchez}, {Santangelo}, {Schlickeiser},
  {Schulz}, {Schwanke}, {Schwarzburg}, {Schwemmer}, {Sheidaei}, {Skilton},
  {Sol}, {Spengler}, {Stawarz}, {Steenkamp}, {Stegmann}, {Stinzing}, {Stycz},
  {Sushch}, {Szostek}, {Tavernet}, {Terrier}, {Tluczykont}, {Valerius}, {van
  Eldik}, {Vasileiadis}, {Venter}, {Viana}, {Vincent}, {V{\"o}lk}, {Volpe},
  {Vorobiov}, {Vorster}, {Wagner}, {Ward}, {White}, {Wierzcholska},
  {Zacharias}, {Zajczyk}, {Zdziarski}, {Zech}, \&
  {Zechlin}}]{2012AandA...548A..46H}
{H.E.S.S.~Collaboration}, {Abramowski}, A., {Acero}, F., {et~al.} 2012, \aap,
  548, A46

\bibitem[{{H.E.S.S.~Collaboration} {et~al.}(2013){H.E.S.S.~Collaboration},
  {Abramowski}, {Acero}, {Aharonian}, {Akhperjanian}, {Anton}, {Balenderan},
  {Balzer}, {Barnacka}, {Becherini}, {Becker Tjus}, {Bernl{\"o}hr}, {Birsin},
  {Biteau}, {Bochow}, {Boisson}, {Bolmont}, {Bordas}, {Brucker}, {Brun},
  {Brun}, {Bulik}, {Carrigan}, {Casanova}, {Cerruti}, {Chadwick}, {Chaves},
  {Cheesebrough}, {Colafrancesco}, {Cologna}, {Conrad}, {Couturier}, {Dalton},
  {Daniel}, {Davids}, {Degrange}, {Deil}, {deWilt}, {Dickinson},
  {Djannati-Ata{\"i}}, {Domainko}, {O'C.~Drury}, {Dubus}, {Dutson}, {Dyks},
  {Dyrda}, {Egberts}, {Eger}, {Espigat}, {Fallon}, {Farnier}, {Fegan},
  {Feinstein}, {Fernandes}, {Fernandez}, {Fiasson}, {Fontaine}, {F{\"o}rster},
  {F{\"u}{\ss}ling}, {Gajdus}, {Gallant}, {Garrigoux}, {Gast}, {Giebels},
  {Glicenstein}, {Gl{\"u}ck}, {G{\"o}ring}, {Grondin}, {Grudzi{\'n}ska},
  {H{\"a}ffner}, {Hague}, {Hahn}, {Hampf}, {Harris}, {Heinz}, {Heinzelmann},
  {Henri}, {Hermann}, {Hillert}, {Hinton}, {Hofmann}, {Hofverberg}, {Holler},
  {Horns}, {Jacholkowska}, {Jahn}, {Jamrozy}, {Jung}, {Kastendieck},
  {Katarzy{\'n}ski}, {Katz}, {Kaufmann}, {Kh{\'e}lifi}, {Klepser}, {Klochkov},
  {Klu{\'z}niak}, {Kneiske}, {Kolitzus}, {Komin}, {Kosack}, {Kossakowski},
  {Krayzel}, {Kr{\"u}ger}, {Laffon}, {Lamanna}, {Lefaucheur},
  {Lemoine-Goumard}, {Lenain}, {Lennarz}, {Lohse}, {Lopatin}, {Lu}, {Marandon},
  {Marcowith}, {Masbou}, {Maurin}, {Maxted}, {Mayer}, {McComb}, {Medina},
  {M{\'e}hault}, {Menzler}, {Moderski}, {Mohamed}, {Moulin}, {Naumann},
  {Naumann-Godo}, {de Naurois}, {Nedbal}, {Nguyen}, {Niemiec}, {Nolan}, {Ohm},
  {de O{\~n}a Wilhelmi}, {Opitz}, {Ostrowski}, {Oya}, {Panter}, {Parsons}, {Paz
  Arribas}, {Pekeur}, {Pelletier}, {Perez}, {Petrucci}, {Peyaud}, {Pita},
  {P{\"u}hlhofer}, {Punch}, {Quirrenbach}, {Raab}, {Raue}, {Reimer}, {Reimer},
  {Renaud}, {de los Reyes}, {Rieger}, {Ripken}, {Rob}, {Rosier-Lees}, {Rowell},
  {Rudak}, {Rulten}, {Sahakian}, {Sanchez}, {Santangelo}, {Schlickeiser},
  {Schulz}, {Schwanke}, {Schwarzburg}, {Schwemmer}, {Sheidaei}, {Skilton},
  {Sol}, {Spengler}, {Stawarz}, {Steenkamp}, {Stegmann}, {Stinzing}, {Stycz},
  {Sushch}, {Szostek}, {Tavernet}, {Terrier}, {Tluczykont}, {Trichard},
  {Valerius}, {van Eldik}, {Vasileiadis}, {Venter}, {Viana}, {Vincent},
  {V{\"o}lk}, {Volpe}, {Vorobiov}, {Vorster}, {Wagner}, {Ward}, {White},
  {Wierzcholska}, {Wouters}, {Zacharias}, {Zajczyk}, {Zdziarski}, {Zech}, \&
  {Zechlin}}]{GCPop}
{H.E.S.S.~Collaboration}, {Abramowski}, A., {Acero}, F., {et~al.} 2013, \aap,
  551, A26

\bibitem[{{H.E.S.S.~Collaboration}
  {et~al.}(2011{\natexlab{a}}){H.E.S.S.~Collaboration}, {Abramowski}, {Acero},
  {Aharonian}, {Akhperjanian}, {Anton}, {Balzer}, {Barnacka}, {Barres de
  Almeida}, {Becherini}, {Becker}, {Behera}, {Bernl{\"o}hr}, {Bochow},
  {Boisson}, {Bolmont}, {Bordas}, {Brucker}, {Brun}, {Brun}, {Bulik},
  {B{\"u}sching}, {Carrigan}, {Casanova}, {Cerruti}, {Chadwick}, {Charbonnier},
  {Chaves}, {Cheesebrough}, {Chounet}, {Clapson}, {Coignet}, {Cologna},
  {Conrad}, {Dalton}, {Daniel}, {Davids}, {Degrange}, {Deil}, {Dickinson},
  {Djannati-Ata{\"i}}, {Domainko}, {Drury}, {Dubois}, {Dubus}, {Dutson},
  {Dyks}, {Dyrda}, {Egberts}, {Eger}, {Espigat}, {Fallon}, {Farnier}, {Fegan},
  {Feinstein}, {Fernandes}, {Fiasson}, {Fontaine}, {F{\"o}rster},
  {F{\"u}{\ss}ling}, {Gallant}, {Gast}, {G{\'e}rard}, {Gerbig}, {Giebels},
  {Glicenstein}, {Gl{\"u}ck}, {Goret}, {G{\"o}ring}, {H{\"a}ffner}, {Hague},
  {Hampf}, {Hauser}, {Heinz}, {Heinzelmann}, {Henri}, {Hermann}, {Hinton},
  {Hoffmann}, {Hofmann}, {Hofverberg}, {Holler}, {Horns}, {Jacholkowska}, {de
  Jager}, {Jahn}, {Jamrozy}, {Jung}, {Kastendieck}, {Katarzy{\'n}ski}, {Katz},
  {Kaufmann}, {Keogh}, {Khangulyan}, {Kh{\'e}lifi}, {Klochkov}, {Klu{\'z}niak},
  {Kneiske}, {Komin}, {Kosack}, {Kossakowski}, {Laffon}, {Lamanna}, {Lennarz},
  {Lohse}, {Lopatin}, {Lu}, {Marandon}, {Marcowith}, {Masbou}, {Maurin},
  {Maxted}, {McComb}, {Medina}, {M{\'e}hault}, {Moderski}, {Moulin}, {Naumann},
  {Naumann-Godo}, {de Naurois}, {Nedbal}, {Nekrassov}, {Nguyen}, {Nicholas},
  {Niemiec}, {Nolan}, {Ohm}, {de O{\~n}a Wilhelmi}, {Opitz}, {Ostrowski},
  {Oya}, {Panter}, {Paz Arribas}, {Pedaletti}, {Pelletier}, {Petrucci}, {Pita},
  {P{\"u}hlhofer}, {Punch}, {Quirrenbach}, {Raue}, {Rayner}, {Reimer},
  {Reimer}, {Renaud}, {de los Reyes}, {Rieger}, {Ripken}, {Rob}, {Rosier-Lees},
  {Rowell}, {Rudak}, {Rulten}, {Ruppel}, {Ryde}, {Sahakian}, {Santangelo},
  {Schlickeiser}, {Sch{\"o}ck}, {Schulz}, {Schwanke}, {Schwarzburg},
  {Schwemmer}, {Sikora}, {Skilton}, {Sol}, {Spengler}, {Stawarz}, {Steenkamp},
  {Stegmann}, {Stinzing}, {Stycz}, {Sushch}, {Szostek}, {Tavernet}, {Terrier},
  {Tluczykont}, {Valerius}, {van Eldik}, {Vasileiadis}, {Venter}, {Vialle},
  {Viana}, {Vincent}, {V{\"o}lk}, {Volpe}, {Vorobiov}, {Vorster}, {Wagner},
  {Ward}, {White}, {Wierzcholska}, {Zacharias}, {Zajczyk}, {Zdziarski}, {Zech},
  \& {Zechlin}}]{2011AandA...531A..81H}
{H.E.S.S.~Collaboration}, {Abramowski}, A., {Acero}, F., {et~al.}
  2011{\natexlab{a}}, \aap, 531, A81

\bibitem[{{H.E.S.S.~Collaboration}
  {et~al.}(2011{\natexlab{b}}){H.E.S.S.~Collaboration}, {Abramowski}, {Acero},
  {Aharonian}, {Akhperjanian}, {Anton}, {Balzer}, {Barnacka}, {Barres de
  Almeida}, {Becherini}, {Becker}, {Behera}, {Bernl{\"o}hr}, {Bochow},
  {Boisson}, {Bolmont}, {Bordas}, {Brucker}, {Brun}, {Brun}, {Bulik},
  {B{\"u}sching}, {Carrigan}, {Casanova}, {Cerruti}, {Chadwick}, {Charbonnier},
  {Chaves}, {Cheesebrough}, {Chounet}, {Clapson}, {Coignet}, {Cologna},
  {Conrad}, {Dalton}, {Daniel}, {Davids}, {Degrange}, {Deil}, {Dickinson},
  {Djannati-Ata{\"i}}, {Domainko}, {O'C.~Drury}, {Dubois}, {Dubus}, {Dutson},
  {Dyks}, {Dyrda}, {Egberts}, {Eger}, {Espigat}, {Fallon}, {Farnier}, {Fegan},
  {Feinstein}, {Fernandes}, {Fiasson}, {Fontaine}, {F{\"o}rster},
  {F{\"u}ssling}, {Gallant}, {Gast}, {G{\'e}rard}, {Gerbig}, {Giebels},
  {Glicenstein}, {Gl{\"u}ck}, {Goret}, {G{\"o}ring}, {H{\"a}ffner}, {Hague},
  {Hampf}, {Hauser}, {Heinz}, {Heinzelmann}, {Henri}, {Hermann}, {Hinton},
  {Hoffmann}, {Hofmann}, {Hofverberg}, {Holler}, {Horns}, {Jacholkowska}, {de
  Jager}, {Jahn}, {Jamrozy}, {Jung}, {Kastendieck}, {Katarzynski}, {Katz},
  {Kaufmann}, {Keogh}, {Khangulyan}, {Kh{\'e}lifi}, {Klochkov}, {Kluzniak},
  {Kneiske}, {Komin}, {Kosack}, {Kossakowski}, {Laffon}, {Lamanna}, {Lennarz},
  {Lohse}, {Lopatin}, {Lu}, {Marandon}, {Marcowith}, {Masbou}, {Maurin},
  {Maxted}, {Mayer}, {McComb}, {Medina}, {M{\'e}hault}, {Moderski}, {Moulin},
  {Naumann}, {Naumann-Godo}, {de Naurois}, {Nedbal}, {Nekrassov}, {Nguyen},
  {Nicholas}, {Niemiec}, {Nolan}, {Ohm}, {de Ona Wilhelmi}, {Opitz},
  {Ostrowski}, {Oya}, {Panter}, {Paz Arribas}, {Pedaletti}, {Pelletier},
  {Petrucci}, {Pita}, {P{\"u}hlhofer}, {Punch}, {Quirrenbach}, {Raue},
  {Rayner}, {Reimer}, {Reimer}, {Renaud}, {de Los Reyes}, {Rieger}, {Ripken},
  {Rob}, {Rosier-Lees}, {Rowell}, {Rudak}, {Rulten}, {Ruppel}, {Sahakian},
  {Sanchez}, {Santangelo}, {Schlickeiser}, {Sch{\"o}ck}, {Schulz}, {Schwanke},
  {Schwarzburg}, {Schwemmer}, {Sikora}, {Skilton}, {Sol}, {Spengler},
  {Stawarz}, {Steenkamp}, {Stegmann}, {Stinzing}, {Stycz}, {Sushch}, {Szostek},
  {Tavernet}, {Terrier}, {Tluczykont}, {Valerius}, {van Eldik}, {Vasileiadis},
  {Venter}, {Vialle}, {Viana}, {Vincent}, {V{\"o}lk}, {Volpe}, {Vorobiov},
  {Vorster}, {Wagner}, {Ward}, {White}, {Wierzcholska}, {Zacharias}, {Zajczyk},
  {Zdziarski}, {Zech}, \& {Zechlin}}]{2011AandA...533A.103H}
{H.E.S.S.~Collaboration}, {Abramowski}, A., {Acero}, F., {et~al.}
  2011{\natexlab{b}}, \aap, 533, A103

\bibitem[{{H.E.S.S.~Collaboration}
  {et~al.}(2015{\natexlab{a}}){H.E.S.S.~Collaboration}, {Abramowski},
  {Aharonian}, {Ait Benkhali}, {Akhperjanian}, {Ang{\"u}ner}, {Anton},
  {Backes}, {Balenderan}, {Balzer}, \& et~al.}]{2015AA...574A..27H}
{H.E.S.S.~Collaboration}, {Abramowski}, A., {Aharonian}, F., {et~al.}
  2015{\natexlab{a}}, \aap, 574, A27

\bibitem[{{H.E.S.S.~Collaboration}
  {et~al.}(2014{\natexlab{a}}){H.E.S.S.~Collaboration}, {Abramowski},
  {Aharonian}, {Ait Benkhali}, {Akhperjanian}, {Ang{\"u}ner}, {Anton},
  {Balenderan}, {Balzer}, {Barnacka}, \& et~al.}]{HESS2014_J1818}
{H.E.S.S.~Collaboration}, {Abramowski}, A., {Aharonian}, F., {et~al.}
  2014{\natexlab{a}}, \aap, 562, A40

\bibitem[{{H.E.S.S.~Collaboration}
  {et~al.}(2014{\natexlab{b}}){H.E.S.S.~Collaboration}, {Abramowski},
  {Aharonian}, {Ait Benkhali}, {Akhperjanian}, {Ang{\"u}ner}, {Anton},
  {Balenderan}, {Balzer}, {Barnacka}, \& et~al.}]{2014A&A...562L...4H}
{H.E.S.S.~Collaboration}, {Abramowski}, A., {Aharonian}, F., {et~al.}
  2014{\natexlab{b}}, \aap, 562, L4

\bibitem[{{H.E.S.S.~Collaboration}
  {et~al.}(2015{\natexlab{b}}){H.E.S.S.~Collaboration}, {Abramowski},
  {Aharonian}, {Ait Benkhali}, {Akhperjanian}, {Ang{\"u}ner}, {Backes},
  {Balenderan}, {Balzer}, {Barnacka}, \& et~al.}]{2015AandA...574A.100H}
{H.E.S.S.~Collaboration}, {Abramowski}, A., {Aharonian}, F., {et~al.}
  2015{\natexlab{b}}, \aap, 574, A100

\bibitem[{{H.E.S.S.~Collaboration}
  {et~al.}(2015{\natexlab{c}}){H.E.S.S.~Collaboration}, {Abramowski},
  {Aharonian}, {Ait Benkhali}, {Akhperjanian}, {Ang{\"u}ner}, {Backes},
  {Balenderan}, {Balzer}, {Barnacka}, \& et~al.}]{2015AA...575A..81H}
{H.E.S.S.~Collaboration}, {Abramowski}, A., {Aharonian}, F., {et~al.}
  2015{\natexlab{c}}, \aap, 575, A81

\bibitem[{{H.E.S.S.~Collaboration}
  {et~al.}(2015{\natexlab{d}}){H.E.S.S.~Collaboration}, {Abramowski},
  {Aharonian}, {Ait Benkhali}, {Akhperjanian}, {Ang{\"u}ner}, {Backes},
  {Balenderan}, {Balzer}, {Barnacka}, \& et~al.}]{HESS15_LMC}
{H.E.S.S.~Collaboration}, {Abramowski}, A., {Aharonian}, F., {et~al.}
  2015{\natexlab{d}}, Science, 347, 406

\bibitem[{{H.E.S.S.~Collaboration}
  {et~al.}(2015{\natexlab{e}}){H.E.S.S.~Collaboration}, {Abramowski},
  {Aharonian}, {Ait Benkhali}, {Akhperjanian}, {Ang{\"u}ner}, {Backes},
  {Balzer}, {Becherini}, {Becker Tjus}, \& et~al.}]{J1018}
{H.E.S.S.~Collaboration}, {Abramowski}, A., {Aharonian}, F., {et~al.}
  2015{\natexlab{e}}, \aap, 577, A131

\bibitem[{{H.E.S.S.~Collaboration}
  {et~al.}(2016{\natexlab{e}}){H.E.S.S.~Collaboration}, {Abramowski},
  {Aharonian}, {Ait Benkhali}, {Akhperjanian}, {Ang{\"u}ner}, {Backes},
  {Balzer}, {Becherini}, {Becker Tjus}, \& et~al.}]{2016arXiv160104461H}
{H.E.S.S.~Collaboration}, {Abramowski}, A., {Aharonian}, F., {et~al.}
  2016{\natexlab{e}}, ArXiv e-prints [\eprint[arXiv]{1601.04461}]

\bibitem[{{H.E.S.S.~Collaboration}
  {et~al.}(2016{\natexlab{f}}){H.E.S.S.~Collaboration}, {Abramowski},
  {Aharonian}, {Ait Benkhali}, {Akhperjanian}, {Ang{\"u}ner}, {Backes},
  {Balzer}, {Becherini}, {Tjus}, \& et~al.}]{GCPevatron}
{H.E.S.S.~Collaboration}, {Abramowski}, A., {Aharonian}, F., {et~al.}
  2016{\natexlab{f}}, \nat, 531, 476

\bibitem[{{Hinton} \& {Hofmann}(2009)}]{ref:hintonhofmann}
{Hinton}, J.~A. \& {Hofmann}, W. 2009, \araa, 47, 523

\bibitem[{{Hobbs} {et~al.}(2004{\natexlab{a}}){Hobbs}, {Faulkner}, {Stairs},
  {Camilo}, {Manchester}, {Lyne}, {Kramer}, {D'Amico}, {Kaspi}, {Possenti},
  {McLaughlin}, {Lorimer}, {Burgay}, {Joshi}, \& {Crawford}}]{Hobbs04}
{Hobbs}, G., {Faulkner}, A., {Stairs}, I.~H., {et~al.} 2004{\natexlab{a}},
  \mnras, 352, 1439

\bibitem[{{Hobbs} {et~al.}(2004{\natexlab{b}}){Hobbs}, {Lyne}, {Kramer},
  {Martin}, \& {Jordan}}]{Hobbs2004}
{Hobbs}, G., {Lyne}, A.~G., {Kramer}, M., {Martin}, C.~E., \& {Jordan}, C.
  2004{\natexlab{b}}, \mnras, 353, 1311

\bibitem[{{Hofverberg} {et~al.}(2010){Hofverberg}, {Chaves}, {Fiasson},
  {Kosack}, {M{\'e}hault}, {de On{\~a} Wilhelmi}, \&
  {H.E.S.S.~Collaboration}}]{Hofverberg10}
{Hofverberg}, P., {Chaves}, R.~C.~G., {Fiasson}, A., {et~al.} 2010, in 25th
  Texas Symposium on Relativistic Astrophysics, 196

\bibitem[{{Holder}(2009)}]{Holder09}
{Holder}, J. 2009, ArXiv e-prints [\eprint[arXiv]{0907.3918}]

\bibitem[{{Hoppe}(2008{\natexlab{a}})}]{Hoppe:2008c}
{Hoppe}, S. 2008{\natexlab{a}}, PhD thesis, Universit\"at Heidelberg,
  Max-Planck-Institut f\"ur Kernphysik

\bibitem[{{Hoppe}(2008{\natexlab{b}})}]{2008ICRC....2..579H}
{Hoppe}, S. 2008{\natexlab{b}}, International Cosmic Ray Conference, 2, 579

\bibitem[{Hunter(2007)}]{matplotlib}
Hunter, J.~D. 2007, Computing In Science \& Engineering, 9, 90

\bibitem[{{Hurley} {et~al.}(1999){Hurley}, {Cline}, {Mazets}, {Barthelmy},
  {Butterworth}, {Marshall}, {Palmer}, {Aptekar}, {Golenetskii}, {Il'Inskii},
  {Frederiks}, {McTiernan}, {Gold}, \& {Trombka}}]{Hurley:1999}
{Hurley}, K., {Cline}, T., {Mazets}, E., {et~al.} 1999, \nat, 397, 41

\bibitem[{{Johanson} \& {Kerton}(2009)}]{Johanson09}
{Johanson}, A.~K. \& {Kerton}, C.~R. 2009, \aj, 138, 1615

\bibitem[{{Johnston} \& {Galloway}(1999)}]{Johnston1999}
{Johnston}, S. \& {Galloway}, D. 1999, \mnras, 306, L50

\bibitem[{Jones {et~al.}(2001)Jones, Oliphant, Peterson, {et~al.}}]{scipy}
Jones, E., Oliphant, T., Peterson, P., {et~al.} 2001, {SciPy}: Open source
  scientific tools for {Python}

\bibitem[{{Kang} \& {Koo}(2007)}]{Kang07}
{Kang}, J.-H. \& {Koo}, B.-C. 2007, \apjs, 173, 85

\bibitem[{{Kang} {et~al.}(2012){Kang}, {Koo}, \& {Salter}}]{Kang12}
{Kang}, J.-H., {Koo}, B.-C., \& {Salter}, C. 2012, \aj, 143, 75

\bibitem[{{Kargaltsev} \& {Pavlov}(2008)}]{Kargaltsev08}
{Kargaltsev}, O. \& {Pavlov}, G.~G. 2008, in American Institute of Physics
  Conference Series, Vol. 983, 40 Years of Pulsars: Millisecond Pulsars,
  Magnetars and More, ed. C.~{Bassa}, Z.~{Wang}, A.~{Cumming}, \& V.~M.
  {Kaspi}, 171--185

\bibitem[{{Kargaltsev} {et~al.}(2013){Kargaltsev}, {Rangelov}, \&
  {Pavlov}}]{Kargaltsev13}
{Kargaltsev}, O., {Rangelov}, B., \& {Pavlov}, G.~G. 2013, ArXiv e-prints
  [\eprint[arXiv]{1305.2552}]

\bibitem[{{Kesteven}(1968)}]{Kesteven68}
{Kesteven}, M.~J.~L. 1968, Australian Journal of Physics, 21, 369

\bibitem[{{Koo} {et~al.}(2006){Koo}, {Kang}, \& {Salter}}]{Koo06}
{Koo}, B.-C., {Kang}, J.-H., \& {Salter}, C.~J. 2006, \apjl, 643, L49

\bibitem[{{Koo} {et~al.}(2005){Koo}, {Lee}, {Seward}, \& {Moon}}]{Koo2005}
{Koo}, B.-C., {Lee}, J.-J., {Seward}, F.~D., \& {Moon}, D.-S. 2005, \apj, 633,
  946

\bibitem[{{Koralesky} {et~al.}(1998){Koralesky}, {Frail}, {Goss}, {Claussen},
  \& {Green}}]{Koralesky98}
{Koralesky}, B., {Frail}, D.~A., {Goss}, W.~M., {Claussen}, M.~J., \& {Green},
  A.~J. 1998, \aj, 116, 1323

\bibitem[{{Kosack} {et~al.}(2011){Kosack}, {Chaves}, \& {Acero}}]{Kosack11}
{Kosack}, K., {Chaves}, R.~C.~G., \& {Acero}, F. 2011, International Cosmic Ray
  Conference, 7, 76

\bibitem[{{Krivonos} {et~al.}(2012){Krivonos}, {Tsygankov}, {Lutovinov},
  {Revnivtsev}, {Churazov}, \& {Sunyaev}}]{2012AandA...545A..27K}
{Krivonos}, R., {Tsygankov}, S., {Lutovinov}, A., {et~al.} 2012, \aap, 545, A27

\bibitem[{{Kuiper} \& {Hermsen}(2015)}]{2015MNRAS.449.3827K}
{Kuiper}, L. \& {Hermsen}, W. 2015, \mnras, 449, 3827

\bibitem[{{Kumar} \& {Safi-Harb}(2008)}]{Kumar08}
{Kumar}, H.~S. \& {Safi-Harb}, S. 2008, \apjl, 678, L43

\bibitem[{{Kumar} {et~al.}(2012){Kumar}, {Safi-Harb}, \& {Gonzalez}}]{Kumar12}
{Kumar}, H.~S., {Safi-Harb}, S., \& {Gonzalez}, M.~E. 2012, \apj, 754, 96

\bibitem[{{Leahy} \& {Tian}(2008)}]{2008A&A...480L..25L}
{Leahy}, D.~A. \& {Tian}, W.~W. 2008, \aap, 480, L25

\bibitem[{{Lemoine-Goumard} {et~al.}(2011){Lemoine-Goumard}, {Ferrara},
  {Grondin}, {Martin}, \& {Renaud}}]{Lemoine-Goumard11}
{Lemoine-Goumard}, M., {Ferrara}, E., {Grondin}, M.-H., {Martin}, P., \&
  {Renaud}, M. 2011, \memsai, 82, 739

\bibitem[{{Leroy}(2004)}]{ref:muonsleroy}
{Leroy}, N. 2004, PhD thesis, Ecole Polytechnique

\bibitem[{{Li} \& {Ma}(1983)}]{ref:lima}
{Li}, T.-P. \& {Ma}, Y.-Q. 1983, \apj, 272, 317

\bibitem[{{Livingstone} {et~al.}(2006){Livingstone}, {Kaspi}, {Gotthelf}, \&
  {Kuiper}}]{Livingstone06}
{Livingstone}, M.~A., {Kaspi}, V.~M., {Gotthelf}, E.~V., \& {Kuiper}, L. 2006,
  \apj, 647, 1286

\bibitem[{{Manchester} {et~al.}(2005){Manchester}, {Hobbs}, {Teoh}, \&
  {Hobbs}}]{Manchester:2005}
{Manchester}, R.~N., {Hobbs}, G.~B., {Teoh}, A., \& {Hobbs}, M. 2005, \aj, 129,
  1993

\bibitem[{{Manchester} {et~al.}(2001){Manchester}, {Lyne}, {Camilo}, {Bell},
  {Kaspi}, {D'Amico}, {McKay}, {Crawford}, {Stairs}, {Possenti}, {Kramer}, \&
  {Sheppard}}]{Manchester:2001}
{Manchester}, R.~N., {Lyne}, A.~G., {Camilo}, F., {et~al.} 2001, \mnras, 328,
  17

\bibitem[{{Marandon} {et~al.}(2008){Marandon}, {Djannati-Atai}, {Terrier},
  {Puehlhofer}, {Hauser}, {Schwarzburg}, \& {Horns}}]{2008AIPC.1085..320M}
{Marandon}, V., {Djannati-Atai}, A., {Terrier}, R., {et~al.} 2008, in American
  Institute of Physics Conference Series, Vol. 1085, American Institute of
  Physics Conference Series, ed. F.~A. {Aharonian}, W.~{Hofmann}, \&
  F.~{Rieger}, 320--323

\bibitem[{{Marelli} {et~al.}(2014){Marelli}, {Harding}, {Pizzocaro}, {De Luca},
  {Wood}, {Caraveo}, {Salvetti}, {Saz Parkinson}, \& {Acero}}]{Marelli14}
{Marelli}, M., {Harding}, A., {Pizzocaro}, D., {et~al.} 2014, \apj, 795, 168

\bibitem[{{Mazets} {et~al.}(1979){Mazets}, {Golenetskij}, \&
  {Guryan}}]{Mazets:1979}
{Mazets}, E.~P., {Golenetskij}, S.~V., \& {Guryan}, Y.~A. 1979, Soviet
  Astronomy Letters, 5, 343

\bibitem[{{McClure-Griffiths} {et~al.}(2002){McClure-Griffiths}, {Dickey},
  {Gaensler}, \& {Green}}]{McClure-Griffiths02}
{McClure-Griffiths}, N.~M., {Dickey}, J.~M., {Gaensler}, B.~M., \& {Green},
  A.~J. 2002, \apj, 578, 176

\bibitem[{{Morton} {et~al.}(2007){Morton}, {Slane}, {Borkowski}, {Reynolds},
  {Helfand}, {Gaensler}, \& {Hughes}}]{Morton07}
{Morton}, T.~D., {Slane}, P., {Borkowski}, K.~J., {et~al.} 2007, \apj, 667, 219

\bibitem[{{Motte} {et~al.}(2003){Motte}, {Schilke}, \& {Lis}}]{Motte03}
{Motte}, F., {Schilke}, P., \& {Lis}, D.~C. 2003, \apj, 582, 277

\bibitem[{{Neronov} \& {Semikoz}(2010)}]{Neronov10}
{Neronov}, A. \& {Semikoz}, D.~V. 2010, ArXiv e-prints
  [\eprint[arXiv]{1011.0210}]

\bibitem[{{Ng} {et~al.}(2008){Ng}, {Slane}, {Gaensler}, \& {Hughes}}]{Ng08}
{Ng}, C.-Y., {Slane}, P.~O., {Gaensler}, B.~M., \& {Hughes}, J.~P. 2008, \apj,
  686, 508

\bibitem[{{Nguyen Luong} {et~al.}(2011){Nguyen Luong}, {Motte}, {Schuller},
  {Schneider}, {Bontemps}, {Schilke}, {Menten}, {Heitsch}, {Wyrowski},
  {Carlhoff}, {Bronfman}, \& {Henning}}]{Nguyen11}
{Nguyen Luong}, Q., {Motte}, F., {Schuller}, F., {et~al.} 2011, \aap, 529, A41

\bibitem[{{Ohm} {et~al.}(2009){Ohm}, {van Eldik}, \& {Egberts}}]{ref:tmva}
{Ohm}, S., {van Eldik}, C., \& {Egberts}, K. 2009, Astroparticle Physics, 31,
  383

\bibitem[{{Paredes} {et~al.}(2014){Paredes}, {Ishwara-Chandra}, {Bosch-Ramon},
  {Zabalza}, {Iwasawa}, \& {Rib{\'o}}}]{2014A&A...561A..56P}
{Paredes}, J.~M., {Ishwara-Chandra}, C.~H., {Bosch-Ramon}, V., {et~al.} 2014,
  \aap, 561, A56

\bibitem[{{Parent} {et~al.}(2011){Parent}, {Kerr}, {den Hartog}, {Baring},
  {DeCesar}, {Espinoza}, {Gotthelf}, {Harding}, {Johnston}, {Kaspi},
  {Livingstone}, {Romani}, {Stappers}, {Watters}, {Weltevrede}, {Abdo},
  {Burgay}, {Camilo}, {Craig}, {Freire}, {Giordano}, {Guillemot}, {Hobbs},
  {Keith}, {Kramer}, {Lyne}, {Manchester}, {Noutsos}, {Possenti}, \&
  {Smith}}]{Parent11}
{Parent}, D., {Kerr}, M., {den Hartog}, P.~R., {et~al.} 2011, \apj, 743, 170

\bibitem[{{Pence} {et~al.}(2010){Pence}, {Chiappetti}, {Page}, {Shaw}, \&
  {Stobie}}]{Pence:2010}
{Pence}, W.~D., {Chiappetti}, L., {Page}, C.~G., {Shaw}, R.~A., \& {Stobie}, E.
  2010, \aap, 524, A42+

\bibitem[{{Peter} {et~al.}(2014){Peter}, {Domainko}, {Sanchez}, {van der Wel},
  \& {G{\"a}ssler}}]{2014A&A...571A..41P}
{Peter}, D., {Domainko}, W., {Sanchez}, D.~A., {van der Wel}, A., \&
  {G{\"a}ssler}, W. 2014, \aap, 571, A41

\bibitem[{{Piron} {et~al.}(2001){Piron}, {Djannati-Atai}, {Punch}, {Tavernet},
  {Barrau}, {Bazer-Bachi}, {Chounet}, {Debiais}, {Degrange}, {Dezalay},
  {Espigat}, {Fabre}, {Fleury}, {Fontaine}, {Goret}, {Gouiffes}, {Khelifi},
  {Malet}, {Masterson}, {Mohanty}, {Nuss}, {Renault}, {Rivoal}, {Rob}, \&
  {Vorobiov}}]{Piron:2001}
{Piron}, F., {Djannati-Atai}, A., {Punch}, M., {et~al.} 2001, \aap, 374, 895

\bibitem[{{Planck Collaboration} {et~al.}(2016){Planck Collaboration}, {Adam},
  {Ade}, {Aghanim}, {Alves}, {Arnaud}, {Ashdown}, {Aumont}, {Baccigalupi},
  {Banday}, \& et~al.}]{Planck15}
{Planck Collaboration}, {Adam}, R., {Ade}, P.~A.~R., {et~al.} 2016, \aap, 594,
  A10

\bibitem[{{Ptuskin} \& {Zirakashvili}(2005)}]{Ptuskin05}
{Ptuskin}, V.~S. \& {Zirakashvili}, V.~N. 2005, \aap, 429, 755

\bibitem[{{Ray} {et~al.}(2011){Ray}, {Kerr}, {Parent}, {Abdo}, {Guillemot},
  {Ransom}, {Rea}, {Wolff}, {Makeev}, {Roberts}, {Camilo}, {Dormody}, {Freire},
  {Grove}, {Gwon}, {Harding}, {Johnston}, {Keith}, {Kramer}, {Michelson},
  {Romani}, {Saz Parkinson}, {Thompson}, {Weltevrede}, {Wood}, \&
  {Ziegler}}]{Ray11}
{Ray}, P.~S., {Kerr}, M., {Parent}, D., {et~al.} 2011, \apjs, 194, 17

\bibitem[{{Reich}(1982)}]{Reich82}
{Reich}, W. 1982, \aap, 106, 314

\bibitem[{{Reimer} {et~al.}(2006){Reimer}, {Pohl}, \& {Reimer}}]{Reimer06}
{Reimer}, A., {Pohl}, M., \& {Reimer}, O. 2006, \apj, 644, 1118

\bibitem[{{Renaud} {et~al.}(2008){Renaud}, {Goret}, \& {Chaves}}]{Renaud08}
{Renaud}, M., {Goret}, P., \& {Chaves}, R.~C.~G. 2008, in American Institute of
  Physics Conference Series, Vol. 1085, American Institute of Physics
  Conference Series, ed. F.~A. {Aharonian}, W.~{Hofmann}, \& F.~{Rieger},
  281--284

\bibitem[{{Roberts} {et~al.}(2007){Roberts}, {Gotthelf}, {Halpern}, {Brogan},
  \& {Ransom}}]{Roberts07}
{Roberts}, M.~S.~E., {Gotthelf}, E.~V., {Halpern}, J.~P., {Brogan}, C.~L., \&
  {Ransom}, S.~M. 2007, in WE-Heraeus Seminar on Neutron Stars and Pulsars 40
  years after the Discovery, ed. W.~{Becker} \& H.~H. {Huang}, 24

\bibitem[{Rolke {et~al.}(2005)Rolke, L\'opez, \& Conrad}]{ref:rolke}
Rolke, W.~A., L\'opez, A.~M., \& Conrad, J. 2005, Nuclear Instruments and
  Methods in Physics Research Section A: Accelerators, Spectrometers, Detectors
  and Associated Equipment, 551, 493

\bibitem[{{Romero}(2010)}]{Romero10}
{Romero}, G.~E. 2010, \memsai, 81, 181

\bibitem[{{Russeil}(2003)}]{Russeil03}
{Russeil}, D. 2003, \aap, 397, 133

\bibitem[{{Safi-Harb} {et~al.}(2001){Safi-Harb}, {Harrus}, {Petre}, {Pavlov},
  {Koptsevich}, \& {Sanwal}}]{Safi-Harb01}
{Safi-Harb}, S., {Harrus}, I.~M., {Petre}, R., {et~al.} 2001, \apj, 561, 308

\bibitem[{{Safi-Harb} \& {Kumar}(2008)}]{Safi-Harb08}
{Safi-Harb}, S. \& {Kumar}, H.~S. 2008, \apj, 684, 532

\bibitem[{{Shahinyan} \& {the VERITAS
  Collaboration}(2016)}]{2016arXiv161005799S}
{Shahinyan}, K. \& {the VERITAS Collaboration}. 2016, ArXiv e-prints
  [\eprint[arXiv]{1610.05799}]

\bibitem[{{Sheidaei}(2011)}]{2011ICRC....7..244S}
{Sheidaei}, F. 2011, International Cosmic Ray Conference, 7, 244

\bibitem[{{Slane} {et~al.}(2015){Slane}, {Bykov}, {Ellison}, {Dubner}, \&
  {Castro}}]{2015SSRv..188..187S}
{Slane}, P., {Bykov}, A., {Ellison}, D.~C., {Dubner}, G., \& {Castro}, D. 2015,
  \ssr, 188, 187

\bibitem[{{Stewart}(2009)}]{Stewart:2009}
{Stewart}, I.~M. 2009, \aap, 495, 989

\bibitem[{{Stil} {et~al.}(2006){Stil}, {Taylor}, {Dickey}, {Kavars}, {Martin},
  {Rothwell}, {Boothroyd}, {Lockman}, \&
  {McClure-Griffiths}}]{2006AJ....132.1158S}
{Stil}, J.~M., {Taylor}, A.~R., {Dickey}, J.~M., {et~al.} 2006, \aj, 132, 1158

\bibitem[{{Straal} {et~al.}(2016){Straal}, {Gab{\'a}nyi}, {van Leeuwen},
  {Clarke}, {Dubner}, {Frey}, {Giacani}, \& {Paragi}}]{2016ApJ...822..117S}
{Straal}, S.~M., {Gab{\'a}nyi}, K.~{\'E}., {van Leeuwen}, J., {et~al.} 2016,
  \apj, 822, 117

\bibitem[{Stycz(2016)}]{Stycz16}
Stycz, K. 2016, PhD thesis, Humboldt-Universit\"{a}t Berlin

\bibitem[{{Su} {et~al.}(2009){Su}, {Chen}, {Yang}, {Koo}, {Zhou}, {Jeong}, \&
  {Zhang}}]{Su09}
{Su}, Y., {Chen}, Y., {Yang}, J., {et~al.} 2009, \apj, 694, 376

\bibitem[{{Sugizaki} {et~al.}(2001){Sugizaki}, {Mitsuda}, {Kaneda},
  {Matsuzaki}, {Yamauchi}, \& {Koyama}}]{Sugizaki01}
{Sugizaki}, M., {Mitsuda}, K., {Kaneda}, H., {et~al.} 2001, \apjs, 134, 77

\bibitem[{{Sun} {et~al.}(1999){Sun}, {Wang}, \& {Chen}}]{Sun99}
{Sun}, M., {Wang}, Z.-r., \& {Chen}, Y. 1999, \apj, 511, 274

\bibitem[{{Tam} {et~al.}(2010){Tam}, {Wagner}, {Tibolla}, \& {Chaves}}]{Tam10}
{Tam}, P.~H.~T., {Wagner}, S.~J., {Tibolla}, O., \& {Chaves}, R.~C.~G. 2010,
  \aap, 518, A8

\bibitem[{{Tanaka} \& {Takahara}(2011)}]{2011ApJ...741...40T}
{Tanaka}, S.~J. \& {Takahara}, F. 2011, \apj, 741, 40

\bibitem[{{Temim} {et~al.}(2012){Temim}, {Slane}, {Arendt}, \&
  {Dwek}}]{Temim12}
{Temim}, T., {Slane}, P., {Arendt}, R.~G., \& {Dwek}, E. 2012, \apj, 745, 46

\bibitem[{{Temim} {et~al.}(2013){Temim}, {Slane}, {Castro}, {Plucinsky},
  {Gelfand}, \& {Dickel}}]{2013ApJ...768...61T}
{Temim}, T., {Slane}, P., {Castro}, D., {et~al.} 2013, \apj, 768, 61

\bibitem[{{Temim} {et~al.}(2009){Temim}, {Slane}, {Gaensler}, {Hughes}, \& {Van
  Der Swaluw}}]{Temim09}
{Temim}, T., {Slane}, P., {Gaensler}, B.~M., {Hughes}, J.~P., \& {Van Der
  Swaluw}, E. 2009, \apj, 691, 895

\bibitem[{{Temim} {et~al.}(2015){Temim}, {Slane}, {Kolb}, {Blondin}, {Hughes},
  \& {Bucciantini}}]{Temim15}
{Temim}, T., {Slane}, P., {Kolb}, C., {et~al.} 2015, \apj, 808, 100

\bibitem[{{Terrier} {et~al.}(2008{\natexlab{a}}){Terrier}, {Djannati-Atai},
  {Hoppe}, {Marandon}, {Renaud}, \& {de Jager}}]{2008AIPC.1085..316T}
{Terrier}, R., {Djannati-Atai}, A., {Hoppe}, S., {et~al.} 2008{\natexlab{a}},
  in American Institute of Physics Conference Series, Vol. 1085, American
  Institute of Physics Conference Series, ed. F.~A. {Aharonian}, W.~{Hofmann},
  \& F.~{Rieger}, 316--319

\bibitem[{{Terrier} {et~al.}(2008{\natexlab{b}}){Terrier}, {Mattana},
  {Djannati-Atai}, {Marandon}, {Renaud}, \& {Dubois}}]{2008AIPC.1085..312T}
{Terrier}, R., {Mattana}, F., {Djannati-Atai}, A., {et~al.} 2008{\natexlab{b}},
  in American Institute of Physics Conference Series, Vol. 1085, American
  Institute of Physics Conference Series, ed. F.~A. {Aharonian}, W.~{Hofmann},
  \& F.~{Rieger}, 312--315

\bibitem[{{The Cherenkov Telescope Array Consortium} {et~al.}(2017){The
  Cherenkov Telescope Array Consortium}, {Acharya}, {Agudo}, {Samarai},
  {Alfaro}, {Alfaro}, {Alispach}, {Alves Batista}, {Amans}, \&
  et~al.}]{2017arXiv170907997C}
{The Cherenkov Telescope Array Consortium}, {Acharya}, B.~S., {Agudo}, I.,
  {et~al.} 2017, ArXiv e-prints [\eprint[arXiv]{1709.07997}]

\bibitem[{{The Fermi-LAT Collaboration}(2017)}]{2017arXiv170200664T}
{The Fermi-LAT Collaboration}. 2017, ArXiv e-prints
  [\eprint[arXiv]{1702.00664}]

\bibitem[{{Tian} \& {Leahy}(2008)}]{Tian08}
{Tian}, W.~W. \& {Leahy}, D.~A. 2008, \mnras, 391, L54

\bibitem[{{Torres} {et~al.}(2014){Torres}, {Cillis}, {Mart{\'{\i}}n}, \& {de
  O{\~n}a Wilhelmi}}]{2014JHEAp...1...31T}
{Torres}, D.~F., {Cillis}, A., {Mart{\'{\i}}n}, J., \& {de O{\~n}a Wilhelmi},
  E. 2014, Journal of High Energy Astrophysics, 1, 31

\bibitem[{{Ueno} {et~al.}(2003){Ueno}, {Bamba}, {Koyama}, \&
  {Ebisawa}}]{Ueno03}
{Ueno}, M., {Bamba}, A., {Koyama}, K., \& {Ebisawa}, K. 2003, \apj, 588, 338

\bibitem[{{Vall{\'e}e}(2014)}]{2014ApJS..215....1V}
{Vall{\'e}e}, J.~P. 2014, \apjs, 215, 1

\bibitem[{{Van Etten} {et~al.}(2009){Van Etten}, {Funk}, \&
  {Hinton}}]{2009ApJ...707.1717V}
{Van Etten}, A., {Funk}, S., \& {Hinton}, J. 2009, \apj, 707, 1717

\bibitem[{{Vasisht} {et~al.}(2000){Vasisht}, {Gotthelf}, {Torii}, \&
  {Gaensler}}]{Vasisht00}
{Vasisht}, G., {Gotthelf}, E.~V., {Torii}, K., \& {Gaensler}, B.~M. 2000,
  \apjl, 542, L49

\bibitem[{{Velusamy} \& {Kundu}(1974)}]{Velusamy74}
{Velusamy}, T. \& {Kundu}, M.~R. 1974, \aap, 32, 375

\bibitem[{{Voisin} {et~al.}(2016){Voisin}, {Rowell}, {Burton}, {Walsh},
  {Fukui}, \& {Aharonian}}]{2016MNRAS.458.2813V}
{Voisin}, F., {Rowell}, G., {Burton}, M.~G., {et~al.} 2016, \mnras, 458, 2813

\bibitem[{{Wachter} {et~al.}(2008){Wachter}, {Ramirez-Ruiz}, {Dwarkadas},
  {Kouveliotou}, {Granot}, {Patel}, \& {Figer}}]{Wachter:2008}
{Wachter}, S., {Ramirez-Ruiz}, E., {Dwarkadas}, V.~V., {et~al.} 2008, \nat,
  453, 626

\bibitem[{{Weinstein}(2009)}]{Weinstein:2009}
{Weinstein}, A. 2009, ArXiv e-prints [\eprint[arXiv]{0912.4492}]

\bibitem[{{Weltevrede} {et~al.}(2011){Weltevrede}, {Johnston}, \&
  {Espinoza}}]{Weltevrede11}
{Weltevrede}, P., {Johnston}, S., \& {Espinoza}, C.~M. 2011, \mnras, 411, 1917

\bibitem[{{Wenger} {et~al.}(2000){Wenger}, {Ochsenbein}, {Egret}, {Dubois},
  {Bonnarel}, {Borde}, {Genova}, {Jasniewicz}, {Lalo{\"e}}, {Lesteven}, \&
  {Monier}}]{2000AnAS..143....9W}
{Wenger}, M., {Ochsenbein}, F., {Egret}, D., {et~al.} 2000, \aaps, 143, 9

\bibitem[{{Whiteoak} \& {Green}(1996)}]{Whiteoak96}
{Whiteoak}, J.~B.~Z. \& {Green}, A.~J. 1996, \aaps, 118, 329

\bibitem[{Wilks(1938)}]{Wilks38}
Wilks, S.~S. 1938, The Annals of Mathematical Statistics, 9, 60

\bibitem[{{Wilson} \& {Weiler}(1976)}]{Wilson76}
{Wilson}, A.~S. \& {Weiler}, K.~W. 1976, \aap, 53, 89

\bibitem[{{Xu} \& {Zhang}(2009)}]{Xu09}
{Xu}, J. \& {Zhang}, H. 2009, ArXiv e-prints [\eprint[arXiv]{0909.0394}]

\bibitem[{{Younes} {et~al.}(2016){Younes}, {Archibald}, {Kouveliotou}, {Kaspi},
  {Ray}, {McEnery}, \& {Fermi LAT Collaboration}}]{2016ATel.9378....1Y}
{Younes}, G., {Archibald}, R., {Kouveliotou}, C., {et~al.} 2016, The
  Astronomer's Telegram, 9378

\bibitem[{{Zajczyk} {et~al.}(2012){Zajczyk}, {Gallant}, {Slane}, {Reynolds},
  {Bandiera}, {Gouiff{\`e}s}, {Le Floc'h}, {Comer{\'o}n}, \& {Koch
  Miramond}}]{Zajczyk12}
{Zajczyk}, A., {Gallant}, Y.~A., {Slane}, P., {et~al.} 2012, \aap, 542, A12

\bibitem[{{Zhou} \& {Chen}(2011)}]{Zhou11}
{Zhou}, P. \& {Chen}, Y. 2011, \apj, 743, 4

\end{thebibliography}


\begin{figure*}
\centering
\includegraphics[angle=-90, width=18cm]{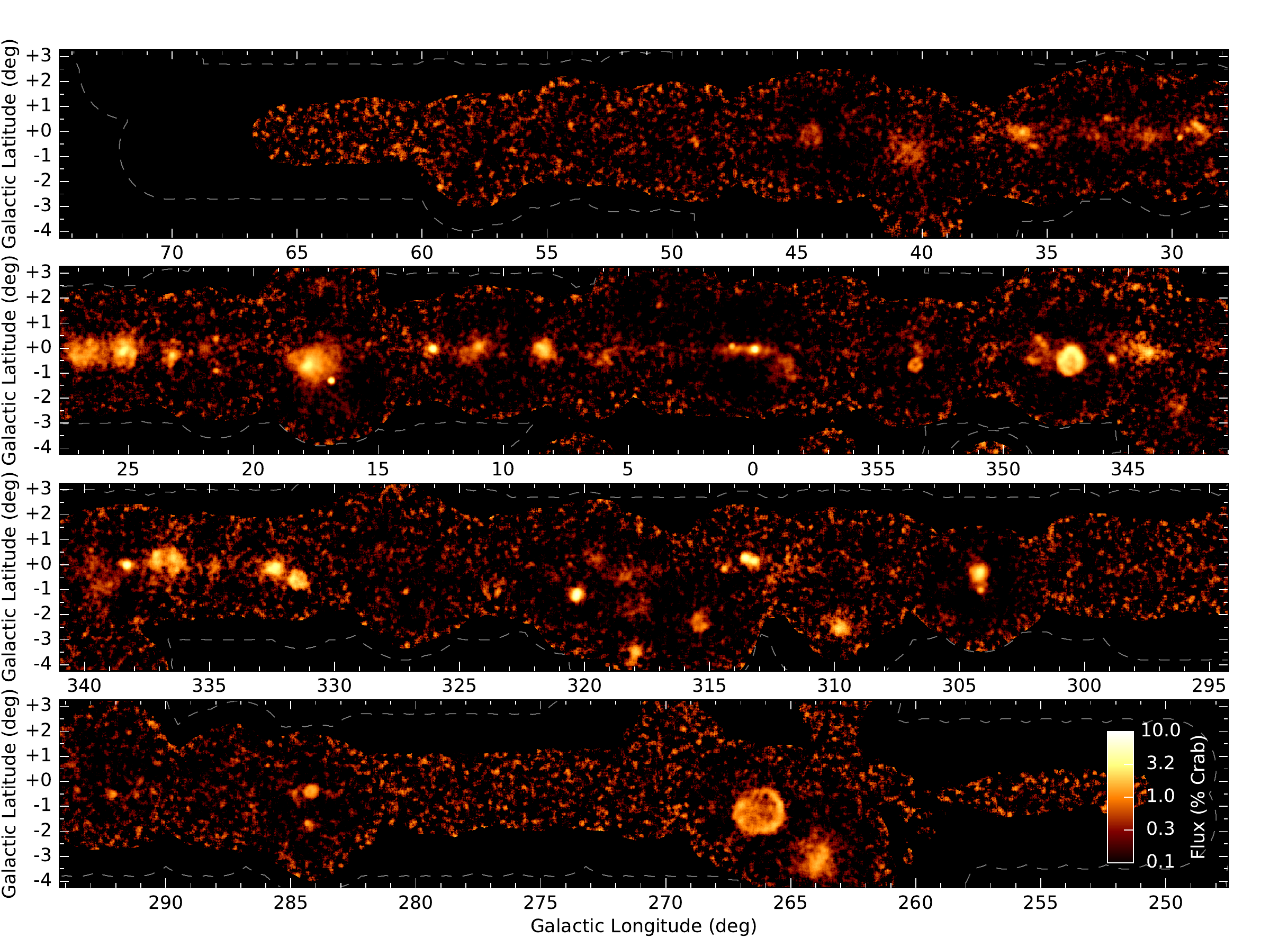}
\caption[HGPS flux map]{
Integral flux above 1~TeV using a correlation radius $R_{\mathrm{c}} = 0.1\degr$
and assuming spectral index $\Gamma = \hgpsAssumedSpecIndex$, in units of
\%~Crab. The map is only filled where the point-source sensitivity for a
5-$\sigma$ detection (c.f. Fig.~\ref{fig:hgps_sensitivity}) is better (lower)
than 2.5\% Crab.
}
\label{fig:fluxmap}
\end{figure*}

\begin{figure*}
\centering
\includegraphics[angle=90, width=18cm]{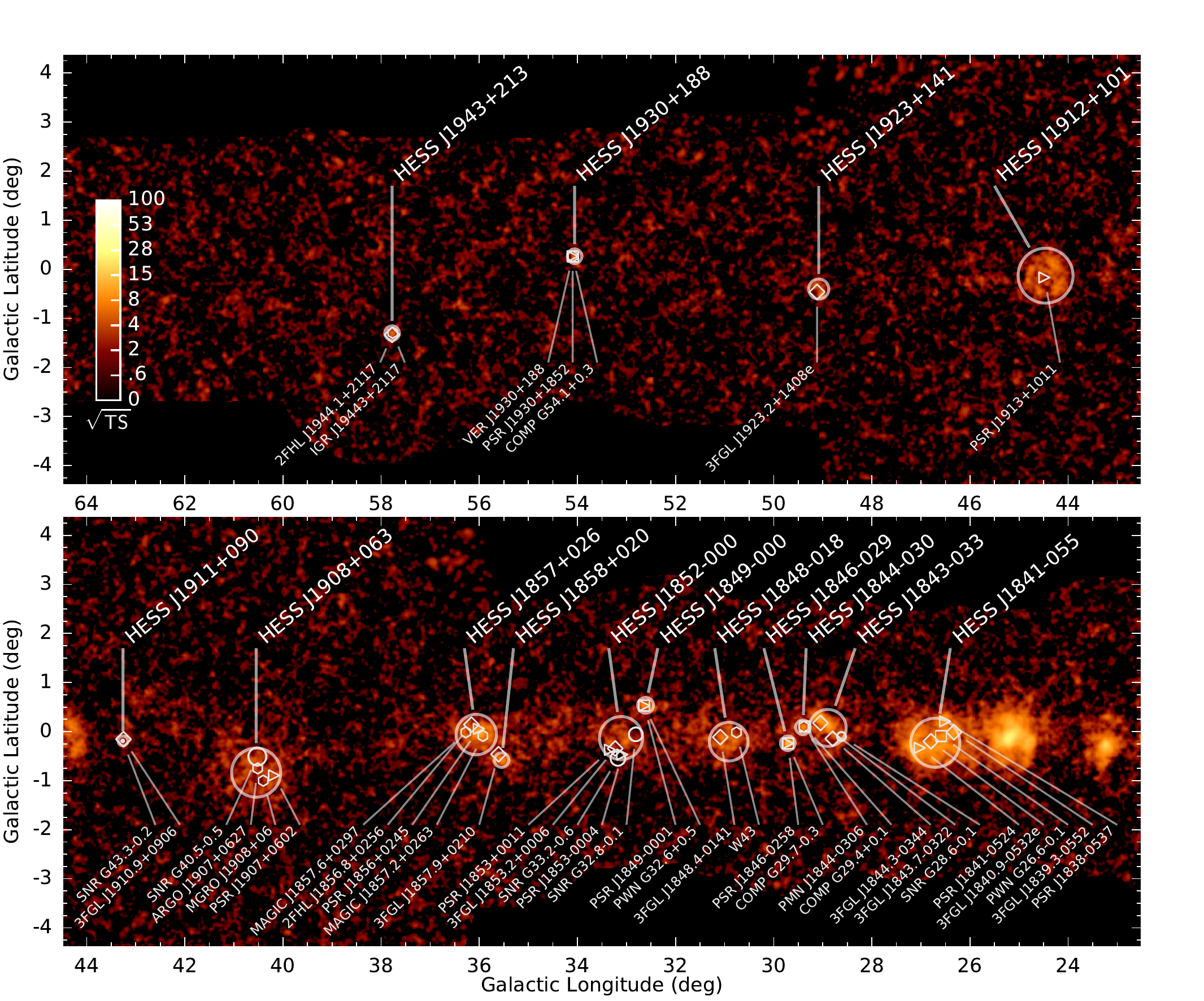}
\caption[Maps: HGPS in a MWL context (1 of 4)]{
HGPS sources and associations in context (1 of 4). The background image shows
$\sqrt{\TSinFormula}$ of the VHE \gammaray\ excess in the Galactic plane
assuming a point source morphology. All HGPS catalog sources are shown on top
with transparent circles that correspond to the measured size of the source.
Source names are labeled above the Galactic plane. Associations for the sources
are shown with markers in white and corresponding labels. Pulsar (PSR)
associations from the ATNF catalog are shown with triangles; SNRs as white
circles with the radius representing the size, PWN and Composite (COMP)
associations from the SNRcat catalog are marked with squares; associations with
\fermi\ 3FGL and 2FHL sources are shown with diamonds; extra associations are
shown with hexagonal markers. 3FGL sources are not labeled when they are
identical to the pulsar. 2FHL sources are not labeled, when they are identical
to the 3FGL source. The various object categories and the association criteria
applied are detailed in Sect.~\ref{sec:results:assoc_id}.
}
\label{fig:hgps_survey_mwl_1}
\end{figure*}

\begin{figure*}
\centering
\includegraphics[angle=-90, width=18cm]{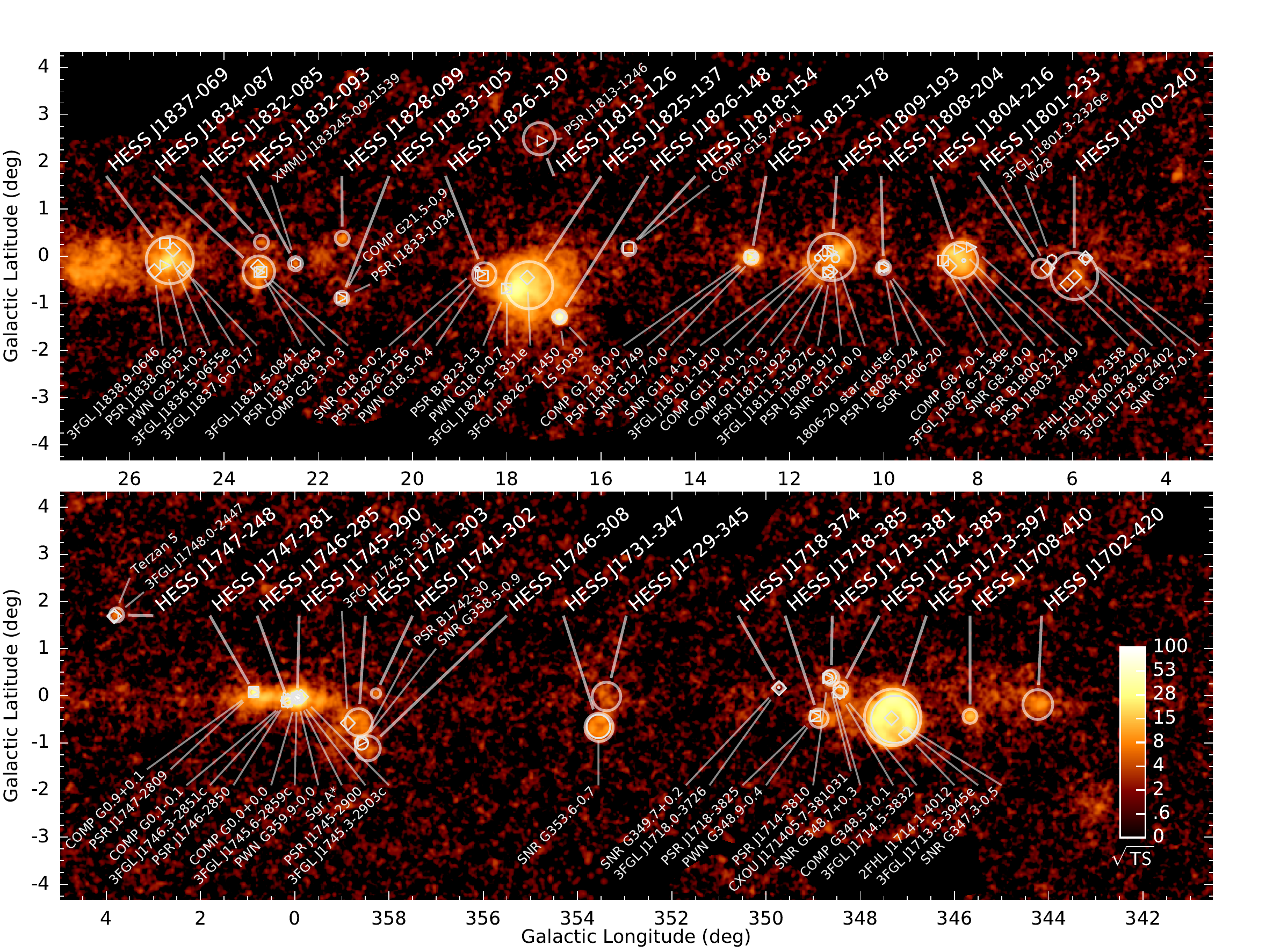}
\caption[Maps: HGPS in a MWL context (2 of 4)]{
HGPS in a MWL context (2 of 4). Fig.~\ref{fig:hgps_survey_mwl_1} continued.
}
\label{fig:hgps_survey_mwl_2}
\end{figure*}

\begin{figure*}
\centering
\includegraphics[angle=90, width=18cm]{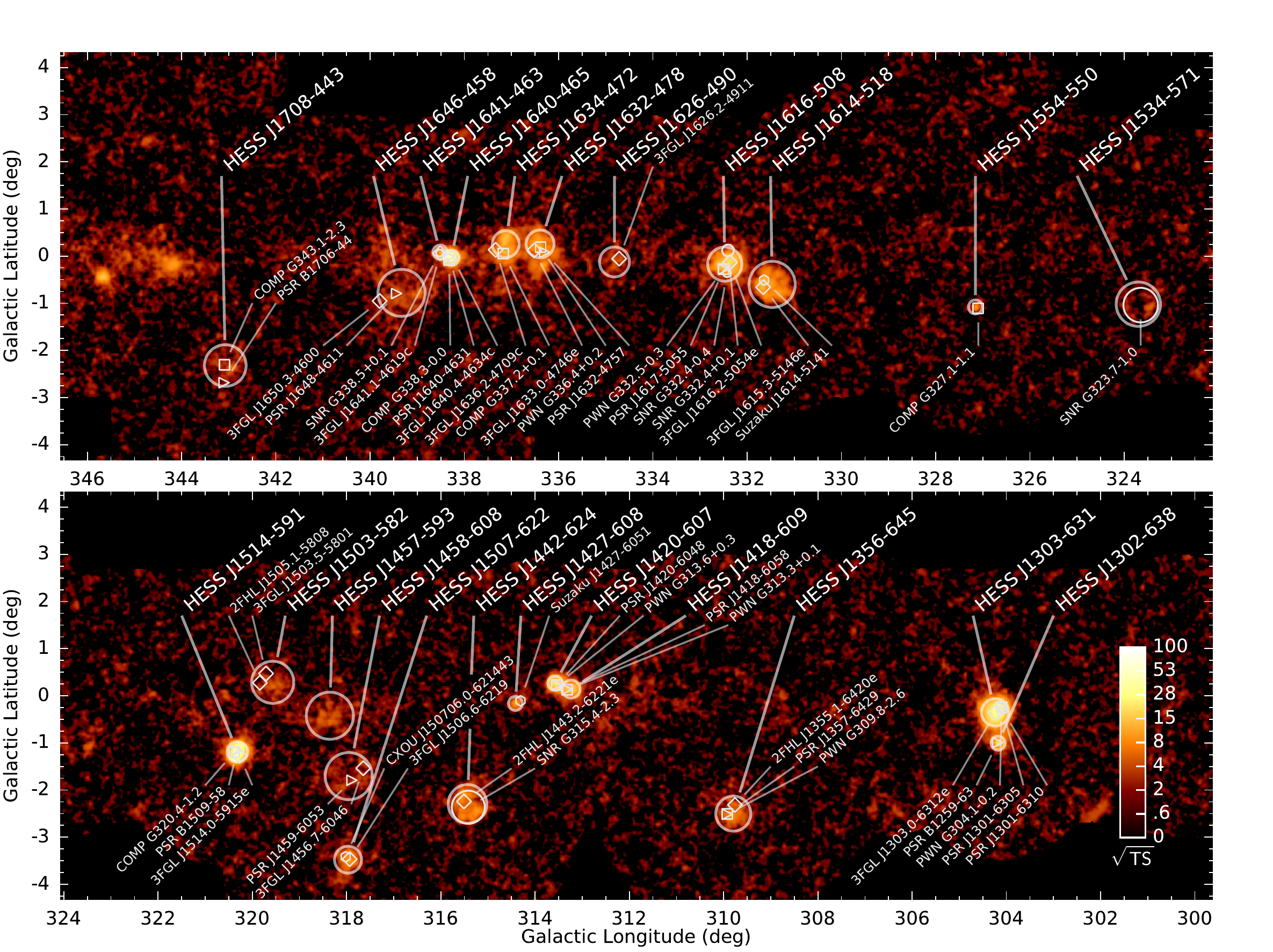}
\caption[Maps: HGPS in a MWL context (3 of 4)]{
HGPS in a MWL context (3 of 4). Fig.~\ref{fig:hgps_survey_mwl_1} continued.
}
\label{fig:hgps_survey_mwl_3}
\end{figure*}

\begin{figure*}
\centering
\includegraphics[angle=-90, width=18cm]{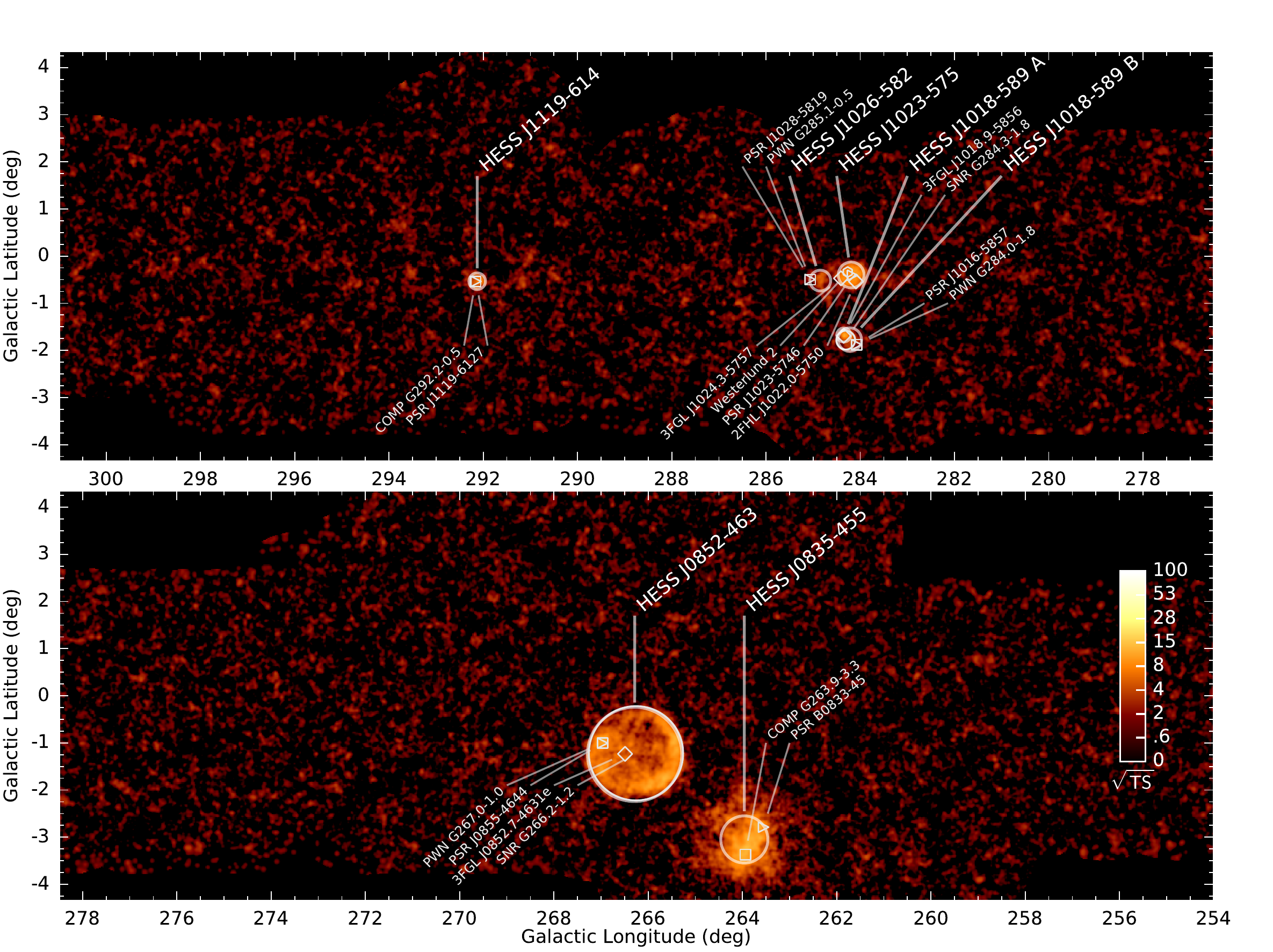}
\caption[Maps: HGPS in a MWL context (4 of 4)]{
HGPS in a MWL context (4 of 4). Fig.~\ref{fig:hgps_survey_mwl_1} continued.
}
\label{fig:hgps_survey_mwl_4}
\end{figure*}

\begin{figure*}
\includegraphics[angle=90, width=18cm]{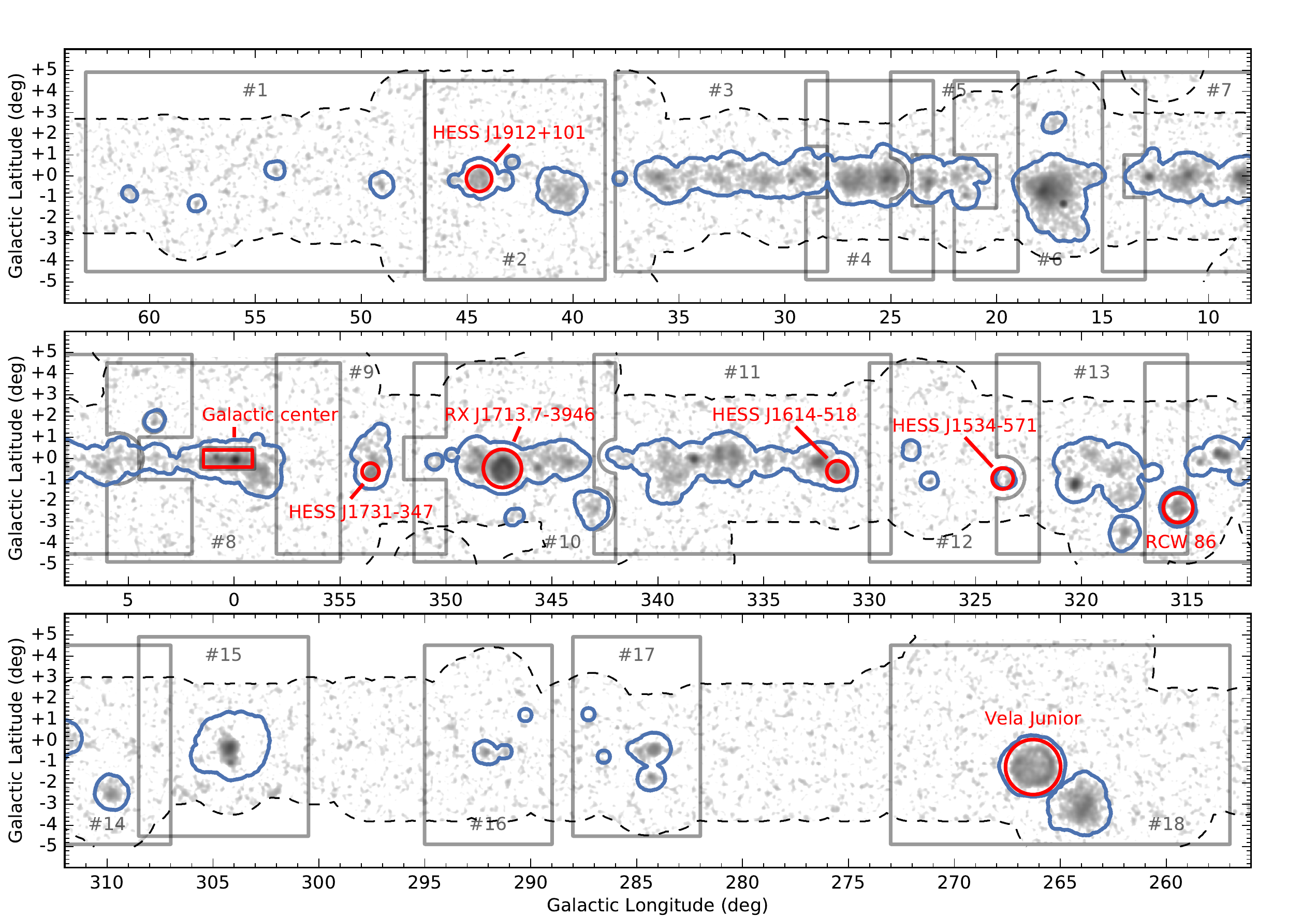}
\caption[Illustration of cut-out regions, exclusion regions and modeling regions (ROIs) in the HGPS map]{
Illustration of cut-out regions (red), exclusion regions (blue) and modeling
regions (ROIs; gray, numbered) in the HGPS map. The background image displays
significance ($R_{\mathrm{c}} = 0.1\degr$) in inverse grayscale.
}
\label{fig:catalog:rois}
\end{figure*}

\clearpage
\onecolumn
\LTcapwidth=\textwidth
\begin{longtable}{lrrrrrrrr}
\caption[HGPS source catalog morphology results]{
HGPS source catalog -- summary of map-based measurements. This is a small
excerpt of the information available in the FITS catalog. The data shown here
correspond to the following catalog columns (described in
Table~\ref{tab:hgps_sources_columns}): \texttt{Source\_Name},
\texttt{Spatial\_Model}, \texttt{GLON}, \texttt{GLAT}, \texttt{Pos\_Err\_95},
\texttt{Size}, \texttt{Size\_Err}, \texttt{Size\_UL}, \texttt{Flux\_Map},
\texttt{Flux\_Map\_Err}, and \texttt{Sqrt\_TS}. The values for sources indicated
with an asterisk are taken from external references; see
Table~\ref{tab:hgps_external_sources} for details.
}
\label{tab:hgps_catalog}\\
\hline\hline
\aligncell{c}{Name} & \aligncell{c}{Spatial Model} & \aligncell{c}{GLON} & \aligncell{c}{GLAT} & \aligncell{c}{$R_{95}$} &
\aligncell{c}{Size} & \aligncell{c}{$F(>1~\mathrm{TeV})$} &  \aligncell{c}{$F(>1~\mathrm{TeV})$} & \aligncell{c}{$\sqrt{TS}$} \\
\hline
& & \aligncell{c}{deg} & \aligncell{c}{deg} & \aligncell{c}{deg} & \aligncell{c}{deg} &
\aligncell{c}{$10^{-12}$ cm$^{-2}$ s$^{-1}$} & \aligncell{c}{\%~Crab} & \\
\hline
\endfirsthead

\caption{continued.}\\
\hline\hline
\aligncell{c}{Name} & \aligncell{c}{Spatial Model} & \aligncell{c}{GLON} & \aligncell{c}{GLAT} & \aligncell{c}{$R_{95}$} &
\aligncell{c}{Size} & \aligncell{c}{$F(>1~\mathrm{TeV})$} &  \aligncell{c}{$F(>1~\mathrm{TeV})$} & \aligncell{c}{$\sqrt{TS}$} \\
\hline
& & \aligncell{c}{deg} & \aligncell{c}{deg} & \aligncell{c}{deg} & \aligncell{c}{deg} &
\aligncell{c}{$10^{-12}$ cm$^{-2}$ s$^{-1}$} & \aligncell{c}{\%~Crab} & \\
\hline
\endhead

\hline
\endfoot
HESS~J0835$-$455 & 3-Gaussian & 263.96 & $-$3.05 & 0.09 & 0.58 $\pm$ 0.052 & 15.36 $\pm$ 0.53 & 67.7 $\pm$ 2.4 & 39.4\\ 
HESS~J0852$-$463$^{*}$ & Shell & 266.29 & $-$1.24 & -- & 1.00  & 23.39 $\pm$ 2.35 & 103.2 $\pm$ 10.3 & -- \\ 
HESS~J1018$-$589~A & Gaussian & 284.35 & $-$1.67 & 0.03 & 0.00 $\pm$ 0.012 & 0.30 $\pm$ 0.05 & 1.3 $\pm$ 0.2 & 8.7\\ 
HESS~J1018$-$589~B & Gaussian & 284.22 & $-$1.77 & 0.12 & 0.15 $\pm$ 0.026 & 0.83 $\pm$ 0.17 & 3.7 $\pm$ 0.8 & 7.6\\ 
HESS~J1023$-$575 & Gaussian & 284.19 & $-$0.40 & 0.05 & 0.17 $\pm$ 0.009 & 2.56 $\pm$ 0.17 & 11.3 $\pm$ 0.8 & 21.4\\ 
HESS~J1026$-$582 & Gaussian & 284.85 & $-$0.52 & 0.12 & 0.13 $\pm$ 0.039 & 0.69 $\pm$ 0.19 & 3.0 $\pm$ 0.9 & 7.3\\ 
HESS~J1119$-$614 & Gaussian & 292.13 & $-$0.53 & 0.06 & 0.10 $\pm$ 0.014 & 0.87 $\pm$ 0.13 & 3.8 $\pm$ 0.6 & 10.2\\ 
HESS~J1302$-$638 & Gaussian & 304.18 & $-$1.00 & 0.02 & 0.01 $\pm$ 0.009 & 0.40 $\pm$ 0.05 & 1.7 $\pm$ 0.2 & 16.6\\ 
HESS~J1303$-$631 & 2-Gaussian & 304.24 & $-$0.35 & 0.04 & 0.18 $\pm$ 0.015 & 5.26 $\pm$ 0.27 & 23.2 $\pm$ 1.2 & 54.5\\ 
HESS~J1356$-$645 & Gaussian & 309.79 & $-$2.50 & 0.08 & 0.23 $\pm$ 0.020 & 5.53 $\pm$ 0.53 & 24.4 $\pm$ 2.3 & 17.3\\ 
HESS~J1418$-$609 & Gaussian & 313.24 & 0.14 & 0.04 & 0.11 $\pm$ 0.011 & 3.01 $\pm$ 0.31 & 13.3 $\pm$ 1.4 & 21.9\\ 
HESS~J1420$-$607 & Gaussian & 313.58 & 0.27 & 0.03 & 0.08 $\pm$ 0.006 & 3.28 $\pm$ 0.24 & 14.5 $\pm$ 1.1 & 27.6\\ 
HESS~J1427$-$608 & Gaussian & 314.42 & $-$0.16 & 0.04 & 0.05 $\pm$ 0.009 & 0.74 $\pm$ 0.10 & 3.3 $\pm$ 0.5 & 10.5\\ 
HESS~J1442$-$624$^{*}$ & Shell & 315.43 & $-$2.29 & -- & 0.30 $\pm$ 0.020 & 2.44 $\pm$ 0.67 & 10.8 $\pm$ 3.0 & -- \\ 
HESS~J1457$-$593 & Gaussian & 318.35 & $-$0.42 & 0.15 & 0.33 $\pm$ 0.045 & 2.50 $\pm$ 0.40 & 11.0 $\pm$ 1.8 & 12.5\\ 
HESS~J1458$-$608 & Gaussian & 317.95 & $-$1.70 & 0.17 & 0.37 $\pm$ 0.031 & 2.44 $\pm$ 0.30 & 10.8 $\pm$ 1.3 & 11.5\\ 
HESS~J1503$-$582 & Gaussian & 319.57 & 0.29 & 0.14 & 0.28 $\pm$ 0.033 & 1.89 $\pm$ 0.28 & 8.3 $\pm$ 1.2 & 10.8\\ 
HESS~J1507$-$622 & Gaussian & 317.97 & $-$3.48 & 0.06 & 0.18 $\pm$ 0.017 & 2.99 $\pm$ 0.31 & 13.2 $\pm$ 1.4 & 17.0\\ 
HESS~J1514$-$591 & 3-Gaussian & 320.32 & $-$1.19 & 0.03 & 0.14 $\pm$ 0.026 & 6.43 $\pm$ 0.21 & 28.4 $\pm$ 0.9 & 42.0\\ 
HESS~J1534$-$571$^{*}$ & Shell & 323.70 & $-$1.02 & -- & 0.40 $\pm$ 0.040 & 1.98 $\pm$ 0.23 & 8.7 $\pm$ 1.0 & -- \\ 
HESS~J1554$-$550 & Gaussian & 327.16 & $-$1.08 & 0.03 & 0.02 $\pm$ 0.009 & 0.36 $\pm$ 0.06 & 1.6 $\pm$ 0.3 & 9.1\\ 
HESS~J1614$-$518$^{*}$ & Shell & 331.47 & $-$0.60 & -- & 0.42 $\pm$ 0.010 & 5.87 $\pm$ 0.42 & 25.9 $\pm$ 1.9 & -- \\ 
HESS~J1616$-$508 & 2-Gaussian & 332.48 & $-$0.17 & 0.12 & 0.23 $\pm$ 0.035 & 8.48 $\pm$ 0.44 & 37.4 $\pm$ 1.9 & 34.3\\ 
HESS~J1626$-$490 & Gaussian & 334.82 & $-$0.12 & 0.14 & 0.20 $\pm$ 0.035 & 1.65 $\pm$ 0.33 & 7.3 $\pm$ 1.5 & 8.4\\ 
HESS~J1632$-$478 & Gaussian & 336.39 & 0.26 & 0.08 & 0.18 $\pm$ 0.020 & 2.93 $\pm$ 0.51 & 12.9 $\pm$ 2.3 & 14.8\\ 
HESS~J1634$-$472 & Gaussian & 337.12 & 0.26 & 0.06 & 0.17 $\pm$ 0.013 & 2.90 $\pm$ 0.37 & 12.8 $\pm$ 1.6 & 17.8\\ 
HESS~J1640$-$465 & 2-Gaussian & 338.28 & $-$0.04 & 0.05 & 0.11 $\pm$ 0.034 & 3.33 $\pm$ 0.19 & 14.7 $\pm$ 0.8 & 41.1\\ 
HESS~J1641$-$463 & Gaussian & 338.52 & 0.08 & 0.05 & 0.04 $\pm$ 0.013 & 0.27 $\pm$ 0.06 & 1.2 $\pm$ 0.3 & 6.9\\ 
HESS~J1646$-$458 & Gaussian & 339.33 & $-$0.78 & 0.15 & 0.50 $\pm$ 0.030 & 5.48 $\pm$ 0.46 & 24.2 $\pm$ 2.0 & 18.6\\ 
HESS~J1702$-$420 & Gaussian & 344.23 & $-$0.19 & 0.08 & 0.20 $\pm$ 0.025 & 3.91 $\pm$ 0.65 & 17.3 $\pm$ 2.9 & 15.0\\ 
HESS~J1708$-$410 & Gaussian & 345.67 & $-$0.44 & 0.03 & 0.06 $\pm$ 0.006 & 0.88 $\pm$ 0.09 & 3.9 $\pm$ 0.4 & 17.0\\ 
HESS~J1708$-$443 & Gaussian & 343.07 & $-$2.32 & 0.14 & 0.28 $\pm$ 0.031 & 2.28 $\pm$ 0.32 & 10.1 $\pm$ 1.4 & 11.0\\ 
HESS~J1713$-$381 & Gaussian & 348.62 & 0.38 & 0.05 & 0.09 $\pm$ 0.017 & 0.65 $\pm$ 0.13 & 2.9 $\pm$ 0.6 & 11.6\\ 
HESS~J1713$-$397$^{*}$ & Shell & 347.31 & $-$0.46 & -- & 0.50  & 16.88 $\pm$ 0.82 & 74.4 $\pm$ 3.6 & -- \\ 
HESS~J1714$-$385 & Gaussian & 348.42 & 0.14 & 0.04 & 0.03 $\pm$ 0.011 & 0.25 $\pm$ 0.05 & 1.1 $\pm$ 0.2 & 8.6\\ 
HESS~J1718$-$374$^{*}$ & Point-Like & 349.72 & 0.17 & -- & -- &0.12 $\pm$ 0.04 & 0.6 $\pm$ 0.2 & -- \\ 
HESS~J1718$-$385 & Gaussian & 348.88 & $-$0.48 & 0.06 & 0.12 $\pm$ 0.015 & 0.80 $\pm$ 0.14 & 3.5 $\pm$ 0.6 & 11.6\\ 
HESS~J1729$-$345 & Gaussian & 353.39 & $-$0.02 & 0.13 & 0.19 $\pm$ 0.031 & 0.86 $\pm$ 0.17 & 3.8 $\pm$ 0.8 & 8.4\\ 
HESS~J1731$-$347$^{*}$ & Shell & 353.54 & $-$0.67 & -- & 0.27 $\pm$ 0.020 & 2.01 $\pm$ 0.15 & 8.8 $\pm$ 0.6 & -- \\ 
HESS~J1741$-$302$^{*}$ & Point-Like & 358.28 & 0.05 & -- & -- &0.16 $\pm$ 0.04 & 0.7 $\pm$ 0.2 & -- \\ 
HESS~J1745$-$290$^{*}$ & Point-Like & 359.94 & $-$0.04 & -- & -- &1.70 $\pm$ 0.08 & 7.5 $\pm$ 0.3 & -- \\ 
HESS~J1745$-$303 & Gaussian & 358.64 & $-$0.56 & 0.11 & 0.18 $\pm$ 0.020 & 0.94 $\pm$ 0.21 & 4.1 $\pm$ 0.9 & 13.7\\ 
HESS~J1746$-$285$^{*}$ & Point-Like & 0.14 & $-$0.11 & -- & -- &0.15 $\pm$ 0.05 & 0.7 $\pm$ 0.2 & -- \\ 
HESS~J1746$-$308 & Gaussian & 358.45 & $-$1.11 & 0.15 & 0.16 $\pm$ 0.036 & 0.68 $\pm$ 0.22 & 3.0 $\pm$ 1.0 & 8.7\\ 
HESS~J1747$-$248 & Gaussian & 3.78 & 1.71 & 0.06 & 0.06 $\pm$ 0.012 & 0.29 $\pm$ 0.05 & 1.3 $\pm$ 0.2 & 8.1\\ 
HESS~J1747$-$281$^{*}$ & Point-Like & 0.87 & 0.08 & -- & -- &0.60 $\pm$ 0.13 & 2.6 $\pm$ 0.6 & -- \\ 
HESS~J1800$-$240 & Gaussian & 5.96 & $-$0.42 & 0.13 & 0.32 $\pm$ 0.039 & 2.44 $\pm$ 0.35 & 10.8 $\pm$ 1.5 & 12.6\\ 
HESS~J1801$-$233$^{*}$ & Gaussian & 6.66 & $-$0.27 & -- & 0.17 $\pm$ 0.030 & 0.45 $\pm$ 0.10 & 2.0 $\pm$ 0.4 & -- \\ 
HESS~J1804$-$216 & 2-Gaussian & 8.38 & $-$0.09 & 0.15 & 0.24 $\pm$ 0.034 & 5.88 $\pm$ 0.27 & 25.9 $\pm$ 1.2 & 34.2\\ 
HESS~J1808$-$204 & Gaussian & 10.01 & $-$0.24 & 0.07 & 0.06 $\pm$ 0.014 & 0.19 $\pm$ 0.04 & 0.8 $\pm$ 0.2 & 6.4\\ 
HESS~J1809$-$193 & 3-Gaussian & 11.11 & $-$0.02 & 0.21 & 0.40 $\pm$ 0.048 & 5.27 $\pm$ 0.29 & 23.2 $\pm$ 1.3 & 26.6\\ 
HESS~J1813$-$126 & Gaussian & 17.31 & 2.49 & 0.19 & 0.21 $\pm$ 0.032 & 1.08 $\pm$ 0.24 & 4.8 $\pm$ 1.1 & 6.1\\ 
HESS~J1813$-$178 & Gaussian & 12.82 & $-$0.03 & 0.02 & 0.05 $\pm$ 0.004 & 1.98 $\pm$ 0.15 & 8.7 $\pm$ 0.7 & 26.4\\ 
HESS~J1818$-$154 & Gaussian & 15.41 & 0.16 & 0.04 & 0.00 $\pm$ 0.046 & 0.17 $\pm$ 0.04 & 0.7 $\pm$ 0.2 & 5.6\\ 
HESS~J1825$-$137 & 3-Gaussian & 17.53 & $-$0.62 & 0.20 & 0.46 $\pm$ 0.032 & 18.41 $\pm$ 0.56 & 81.2 $\pm$ 2.5 & 76.5\\ 
HESS~J1826$-$130 & Gaussian & 18.48 & $-$0.39 & 0.10 & 0.15 $\pm$ 0.021 & 0.86 $\pm$ 0.17 & 3.8 $\pm$ 0.7 & 9.4\\ 
HESS~J1826$-$148 & Gaussian & 16.88 & $-$1.29 & 0.02 & 0.01 $\pm$ 0.004 & 1.28 $\pm$ 0.04 & 5.7 $\pm$ 0.2 & 58.1\\ 
HESS~J1828$-$099 & Gaussian & 21.49 & 0.38 & 0.05 & 0.05 $\pm$ 0.011 & 0.43 $\pm$ 0.07 & 1.9 $\pm$ 0.3 & 8.9\\ 
HESS~J1832$-$085 & Gaussian & 23.21 & 0.29 & 0.05 & 0.02 $\pm$ 0.012 & 0.21 $\pm$ 0.05 & 0.9 $\pm$ 0.2 & 5.9\\ 
HESS~J1832$-$093 & Gaussian & 22.48 & $-$0.16 & 0.03 & 0.00 $\pm$ 0.012 & 0.17 $\pm$ 0.03 & 0.8 $\pm$ 0.1 & 6.8\\ 
HESS~J1833$-$105 & Gaussian & 21.50 & $-$0.90 & 0.03 & 0.02 $\pm$ 0.017 & 0.39 $\pm$ 0.07 & 1.7 $\pm$ 0.3 & 11.4\\ 
HESS~J1834$-$087 & 2-Gaussian & 23.26 & $-$0.33 & 0.06 & 0.21 $\pm$ 0.037 & 3.34 $\pm$ 0.24 & 14.7 $\pm$ 1.1 & 21.0\\ 
HESS~J1837$-$069 & 3-Gaussian & 25.15 & $-$0.09 & 0.05 & 0.36 $\pm$ 0.031 & 12.05 $\pm$ 0.45 & 53.1 $\pm$ 2.0 & 41.5\\ 
HESS~J1841$-$055 & 2-Gaussian & 26.71 & $-$0.23 & 0.17 & 0.41 $\pm$ 0.033 & 10.16 $\pm$ 0.42 & 44.8 $\pm$ 1.9 & 33.9\\ 
HESS~J1843$-$033 & 2-Gaussian & 28.90 & 0.07 & 0.20 & 0.24 $\pm$ 0.063 & 2.88 $\pm$ 0.30 & 12.7 $\pm$ 1.3 & 16.0\\ 
HESS~J1844$-$030 & Gaussian & 29.41 & 0.09 & 0.04 & 0.02 $\pm$ 0.013 & 0.26 $\pm$ 0.05 & 1.1 $\pm$ 0.2 & 7.3\\ 
HESS~J1846$-$029 & Gaussian & 29.71 & $-$0.24 & 0.03 & 0.01 $\pm$ 0.013 & 0.45 $\pm$ 0.05 & 2.0 $\pm$ 0.2 & 13.8\\ 
HESS~J1848$-$018 & Gaussian & 30.92 & $-$0.21 & 0.12 & 0.25 $\pm$ 0.032 & 1.74 $\pm$ 0.35 & 7.7 $\pm$ 1.6 & 12.0\\ 
HESS~J1849$-$000 & Gaussian & 32.61 & 0.53 & 0.06 & 0.09 $\pm$ 0.015 & 0.53 $\pm$ 0.09 & 2.3 $\pm$ 0.4 & 9.1\\ 
HESS~J1852$-$000 & Gaussian & 33.11 & $-$0.13 & 0.18 & 0.28 $\pm$ 0.042 & 1.30 $\pm$ 0.25 & 5.7 $\pm$ 1.1 & 9.0\\ 
HESS~J1857+026 & 2-Gaussian & 36.06 & $-$0.06 & 0.10 & 0.26 $\pm$ 0.056 & 3.77 $\pm$ 0.40 & 16.6 $\pm$ 1.8 & 16.8\\ 
HESS~J1858+020 & Gaussian & 35.54 & $-$0.58 & 0.07 & 0.08 $\pm$ 0.016 & 0.53 $\pm$ 0.11 & 2.3 $\pm$ 0.5 & 8.4\\ 
HESS~J1908+063 & Gaussian & 40.55 & $-$0.84 & 0.13 & 0.49 $\pm$ 0.027 & 6.53 $\pm$ 0.50 & 28.8 $\pm$ 2.2 & 19.0\\ 
HESS~J1911+090$^{*}$ & Point-Like & 43.26 & $-$0.19 & -- & -- &0.15 $\pm$ 0.03 & 0.6 $\pm$ 0.1 & -- \\ 
HESS~J1912+101$^{*}$ & Shell & 44.46 & $-$0.13 & -- & 0.49 $\pm$ 0.040 & 2.49 $\pm$ 0.35 & 11.0 $\pm$ 1.5 & -- \\ 
HESS~J1923+141 & Gaussian & 49.08 & $-$0.40 & 0.10 & 0.12 $\pm$ 0.019 & 0.78 $\pm$ 0.15 & 3.5 $\pm$ 0.7 & 7.3\\ 
HESS~J1930+188 & Gaussian & 54.06 & 0.27 & 0.05 & 0.02 $\pm$ 0.025 & 0.29 $\pm$ 0.09 & 1.3 $\pm$ 0.4 & 5.8\\ 
HESS~J1943+213 & Gaussian & 57.78 & $-$1.30 & 0.05 & 0.03 $\pm$ 0.022 & 0.32 $\pm$ 0.10 & 1.4 $\pm$ 0.4 & 5.9\\ 

 \end{longtable}
\twocolumn
{
\clearpage
\onecolumn
\LTcapwidth=\textwidth
\begin{longtable}{lrrrrrrrr}
\caption[HGPS source catalog spectral results] {
HGPS source catalog -- summary of spectral measurements. This is a small excerpt
of the information available in the FITS catalog. The data shown here correspond
to the following catalog columns (described in
Table~\ref{tab:hgps_sources_columns}): \texttt{Source\_Name}, \texttt{RSpec},
\texttt{Energy\_Range\_Spec\_Lo}, \texttt{Flux\_Spec\_Int\_1TeV}
(\texttt{Flux\_Map} for extern sources), \texttt{Index\_Spec\_PL},
\texttt{Index\_Spec\_PL\_Err} (for PL spectrum sources),
\texttt{Index\_Spec\_ECPL}, \texttt{Index\_Spec\_ECPL\_Err},
\texttt{Lambda\_Spec\_ECPL}, \texttt{Lambda\_Spec\_ECPL\_Err} (for ECPL spectrum
sources), \texttt{Containment\_RSpec}, \texttt{Contamination\_RSpec} and
\texttt{Flux\_Correction\_RSpec\_To\_Total}. The values for sources indicated with
an asterisk are taken from external references, see
Table~\ref{tab:hgps_external_sources} for details.
}
\label{tab:hgps_spectra}\\
\hline\hline
\aligncell{c}{Name} &
\aligncell{c}{$R_{\mathrm{spec}}$} &
\aligncell{c}{$E_{\mathrm{min}}$} &
\aligncell{c}{$F(>1~\mathrm{TeV})$} &
\aligncell{c}{$\Gamma$} &
\aligncell{c}{$\lambda$} &
\aligncell{c}{Contain.} &
\aligncell{c}{Contam.} &
\aligncell{c}{CF} \\
\hline &
\aligncell{c}{deg} &
\aligncell{c}{TeV} &
\aligncell{c}{$10^{-12}$ cm$^{-2}$ s$^{-1}$} &
&
\aligncell{c}{$\mathrm{TeV}^{-1}$}&
\% &
\% &
\% \\
\hline
\endfirsthead

\caption{continued.}\\
\hline\hline
\aligncell{c}{Name} &
\aligncell{c}{$R_{\mathrm{spec}}$} &
\aligncell{c}{$E_{\mathrm{min}}$} &
\aligncell{c}{$F(>1~\mathrm{TeV})$} &
\aligncell{c}{$\Gamma$} &
\aligncell{c}{$\lambda$} &
\aligncell{c}{Contain.} &
\aligncell{c}{Contam.} &
\aligncell{c}{CF} \\
\hline &
\aligncell{c}{deg} &
\aligncell{c}{TeV} &
\aligncell{c}{$10^{-12}$ cm$^{-2}$ s$^{-1}$} &
&
\aligncell{c}{$\mathrm{TeV}^{-1}$}&
\% &
\% &
\% \\
\hline
\endhead

\hline
\endfoot
HESS~J0835$-$455 & 0.50 & 0.3 & 17.43 $\pm$ 1.40 & 1.35 $\pm$ 0.08 & 0.08 $\pm$ 0.01 & 37 & 0 & 271.4   \\ 
HESS J0852-463$^{*}$ & 1.00 & 0.3 & 23.39 $\pm$ 2.35 & 1.81 $\pm$ 0.08 & 0.15 $\pm$ 0.03 & -- & -- & --   \\ 
HESS~J1018$-$589~A & 0.15 & 0.4 & 0.21 $\pm$ 0.03 & 2.24 $\pm$ 0.13 & -- &92 & 42 & 63.5   \\ 
HESS~J1018$-$589~B & 0.25 & 0.5 & 0.70 $\pm$ 0.09 & 2.20 $\pm$ 0.09 & -- &70 & 32 & 96.8   \\ 
HESS~J1023$-$575 & 0.27 & 0.4 & 2.41 $\pm$ 0.13 & 2.36 $\pm$ 0.05 & -- &70 & 5 & 135.5   \\ 
HESS~J1026$-$582 & 0.22 & 0.5 & 0.66 $\pm$ 0.09 & 1.81 $\pm$ 0.10 & -- &70 & 11 & 126.9   \\ 
HESS~J1119$-$614 & 0.18 & 0.4 & 0.92 $\pm$ 0.09 & 2.64 $\pm$ 0.12 & -- &70 & 4 & 137.9   \\ 
HESS~J1302$-$638 & 0.15 & 0.4 & 0.39 $\pm$ 0.03 & 2.59 $\pm$ 0.09 & -- &90 & 40 & 67.4   \\ 
HESS~J1303$-$631 & 0.29 & 0.4 & 5.21 $\pm$ 0.35 & 2.04 $\pm$ 0.06 & 0.07 $\pm$ 0.01 & 70 & 5 & 136.2   \\ 
HESS~J1356$-$645 & 0.37 & 0.5 & 4.39 $\pm$ 0.39 & 2.20 $\pm$ 0.08 & -- &70 & 0 & 142.8   \\ 
HESS~J1418$-$609 & 0.19 & 0.4 & 2.69 $\pm$ 0.15 & 2.26 $\pm$ 0.05 & -- &70 & 6 & 134.7   \\ 
HESS~J1420$-$607 & 0.15 & 0.4 & 2.77 $\pm$ 0.15 & 2.20 $\pm$ 0.05 & -- &70 & 4 & 138.5   \\ 
HESS~J1427$-$608 & 0.15 & 0.4 & 0.48 $\pm$ 0.09 & 2.85 $\pm$ 0.22 & -- &84 & 5 & 113.3   \\ 
HESS J1442-624$^{*}$ & 0.41 & 0.4 & 2.44 $\pm$ 0.67 & 1.59 $\pm$ 0.22 & 0.29 $\pm$ 0.10 & -- & -- & --   \\ 
HESS~J1457$-$593 & 0.50 & 0.5 & 4.31 $\pm$ 0.56 & 2.52 $\pm$ 0.14 & -- &67 & 10 & 135.1   \\ 
HESS~J1458$-$608 & 0.50 & 0.5 & 1.40 $\pm$ 0.35 & 1.81 $\pm$ 0.14 & -- &58 & 1 & 170.2   \\ 
HESS~J1503$-$582 & 0.45 & 0.4 & 3.07 $\pm$ 0.24 & 2.68 $\pm$ 0.08 & -- &70 & 8 & 131.1   \\ 
HESS~J1507$-$622 & 0.29 & 0.5 & 2.60 $\pm$ 0.21 & 2.22 $\pm$ 0.07 & -- &70 & 0 & 142.9   \\ 
HESS~J1514$-$591 & 0.22 & 0.4 & 5.72 $\pm$ 0.42 & 2.05 $\pm$ 0.06 & 0.05 $\pm$ 0.01 & 70 & 0 & 142.8   \\ 
HESS J1534-571$^{*}$ & 0.47 & 0.4 & 1.98 $\pm$ 0.23 & 2.51 $\pm$ 0.09 & -- &-- & -- & --   \\ 
HESS~J1554$-$550 & 0.15 & 0.4 & 0.29 $\pm$ 0.06 & 2.19 $\pm$ 0.17 & -- &92 & 0 & 108.6   \\ 
HESS J1614-518$^{*}$ & 0.49 & 0.3 & 5.87 $\pm$ 0.42 & 2.42 $\pm$ 0.06 & -- &-- & -- & --   \\ 
HESS~J1616$-$508 & 0.36 & 0.3 & 7.99 $\pm$ 0.55 & 2.32 $\pm$ 0.06 & -- &70 & 2 & 139.9   \\ 
HESS~J1626$-$490 & 0.32 & 0.3 & 2.13 $\pm$ 0.26 & 2.47 $\pm$ 0.11 & -- &70 & 11 & 126.9   \\ 
HESS~J1632$-$478 & 0.30 & 0.3 & 2.32 $\pm$ 0.16 & 2.52 $\pm$ 0.06 & -- &70 & 34 & 93.9   \\ 
HESS~J1634$-$472 & 0.28 & 0.3 & 2.87 $\pm$ 0.15 & 2.31 $\pm$ 0.05 & -- &70 & 31 & 98.8   \\ 
HESS~J1640$-$465 & 0.16 & 0.3 & 2.84 $\pm$ 0.73 & 2.12 $\pm$ 0.13 & 0.24 $\pm$ 0.09 & 70 & 4 & 137.6   \\ 
HESS~J1641$-$463 & 0.15 & 0.3 & 0.22 $\pm$ 0.03 & 2.47 $\pm$ 0.11 & -- &90 & 58 & 47.4   \\ 
HESS~J1646$-$458 & 0.50 & 0.3 & 5.81 $\pm$ 0.73 & 2.54 $\pm$ 0.13 & -- &39 & 2 & 254.4   \\ 
HESS~J1702$-$420 & 0.32 & 0.2 & 4.45 $\pm$ 0.36 & 2.09 $\pm$ 0.07 & -- &70 & 14 & 122.9   \\ 
HESS~J1708$-$410 & 0.15 & 0.2 & 0.65 $\pm$ 0.05 & 2.54 $\pm$ 0.07 & -- &81 & 9 & 112.8   \\ 
HESS~J1708$-$443 & 0.44 & 0.2 & 3.32 $\pm$ 0.37 & 2.17 $\pm$ 0.08 & -- &70 & 0 & 142.9   \\ 
HESS~J1713$-$381 & 0.16 & 0.2 & 0.52 $\pm$ 0.07 & 2.74 $\pm$ 0.12 & -- &70 & 15 & 121.4   \\ 
HESS J1713-397$^{*}$ & 0.60 & 0.2 & 16.88 $\pm$ 0.82 & 2.06 $\pm$ 0.02 & 0.08 $\pm$ 0.01 & -- & -- & --   \\ 
HESS~J1714$-$385 & 0.15 & 0.2 & 0.21 $\pm$ 0.03 & 2.52 $\pm$ 0.12 & -- &91 & 47 & 57.9   \\ 
HESS J1718-374$^{*}$ & 0.10 & 0.2 & 0.12 $\pm$ 0.04 & 2.80 $\pm$ 0.27 & -- &-- & -- & --   \\ 
HESS~J1718$-$385 & 0.20 & 0.2 & 0.62 $\pm$ 0.14 & 0.98 $\pm$ 0.22 & 0.09 $\pm$ 0.03 & 70 & 23 & 110.4   \\ 
HESS~J1729$-$345 & 0.30 & 0.2 & 0.82 $\pm$ 0.09 & 2.43 $\pm$ 0.09 & -- &70 & 25 & 108.0   \\ 
HESS J1731-347$^{*}$ & 0.30 & 0.2 & 2.01 $\pm$ 0.15 & 2.32 $\pm$ 0.06 & -- &-- & -- & --   \\ 
HESS J1741-302$^{*}$ & 0.10 & 0.4 & 0.16 $\pm$ 0.04 & 2.30 $\pm$ 0.20 & -- &-- & -- & --   \\ 
HESS J1745-290$^{*}$ & 0.10 & -- & 1.70 $\pm$ 0.08 & 2.14 $\pm$ 0.02 & 0.09 $\pm$ 0.02 & -- & -- & --   \\ 
HESS~J1745$-$303 & 0.29 & 0.2 & 1.09 $\pm$ 0.08 & 2.57 $\pm$ 0.06 & -- &70 & 30 & 99.6   \\ 
HESS J1746-285$^{*}$ & 0.09 & 0.3 & 0.15 $\pm$ 0.05 & 2.17 $\pm$ 0.24 & -- &-- & -- & --   \\ 
HESS~J1746$-$308 & 0.26 & 0.2 & 0.30 $\pm$ 0.09 & 3.27 $\pm$ 0.22 & -- &70 & 23 & 110.2   \\ 
HESS~J1747$-$248 & 0.15 & 0.2 & 0.27 $\pm$ 0.04 & 2.36 $\pm$ 0.14 & -- &83 & 0 & 120.5   \\ 
HESS J1747-281$^{*}$ & 0.10 & 0.3 & 0.60 $\pm$ 0.13 & 2.40 $\pm$ 0.11 & -- &-- & -- & --   \\ 
HESS~J1800$-$240 & 0.50 & 0.2 & 2.90 $\pm$ 0.31 & 2.47 $\pm$ 0.09 & -- &70 & 17 & 118.9   \\ 
HESS J1801-233$^{*}$ & 0.20 & 0.3 & 0.45 $\pm$ 0.10 & 2.66 $\pm$ 0.27 & -- &-- & -- & --   \\ 
HESS~J1804$-$216 & 0.38 & 0.2 & 5.12 $\pm$ 0.23 & 2.69 $\pm$ 0.04 & -- &70 & 8 & 131.6   \\ 
HESS~J1808$-$204 & 0.15 & 0.2 & 0.19 $\pm$ 0.03 & 2.19 $\pm$ 0.14 & -- &85 & 27 & 86.1   \\ 
HESS~J1809$-$193 & 0.50 & 0.2 & 5.37 $\pm$ 0.45 & 2.38 $\pm$ 0.07 & -- &54 & 16 & 154.2   \\ 
HESS~J1813$-$126 & 0.34 & 0.2 & 1.04 $\pm$ 0.21 & 1.99 $\pm$ 0.14 & -- &70 & 0 & 143.6   \\ 
HESS~J1813$-$178 & 0.15 & 0.2 & 2.12 $\pm$ 0.40 & 1.64 $\pm$ 0.12 & 0.14 $\pm$ 0.04 & 89 & 14 & 96.6   \\ 
HESS~J1818$-$154 & 0.15 & 0.2 & 0.23 $\pm$ 0.05 & 2.21 $\pm$ 0.15 & -- &95 & 29 & 74.6   \\ 
HESS~J1825$-$137 & 0.50 & 0.2 & 19.15 $\pm$ 1.85 & 2.15 $\pm$ 0.06 & 0.07 $\pm$ 0.02 & 47 & 3 & 203.6   \\ 
HESS~J1826$-$130 & 0.25 & 0.2 & 1.14 $\pm$ 0.16 & 2.04 $\pm$ 0.10 & -- &70 & 41 & 84.4   \\ 
HESS~J1826$-$148 & 0.15 & 0.2 & 0.84 $\pm$ 0.08 & 2.32 $\pm$ 0.07 & -- &95 & 10 & 94.5   \\ 
HESS~J1828$-$099 & 0.15 & 0.2 & 0.38 $\pm$ 0.05 & 2.25 $\pm$ 0.12 & -- &89 & 11 & 100.3   \\ 
HESS~J1832$-$085 & 0.15 & 0.2 & 0.23 $\pm$ 0.04 & 2.38 $\pm$ 0.14 & -- &94 & 27 & 77.7   \\ 
HESS~J1832$-$093 & 0.15 & 0.2 & 0.16 $\pm$ 0.04 & 2.54 $\pm$ 0.22 & -- &95 & 25 & 78.9   \\ 
HESS~J1833$-$105 & 0.15 & 0.2 & 0.26 $\pm$ 0.06 & 2.42 $\pm$ 0.19 & -- &94 & 2 & 104.6   \\ 
HESS~J1834$-$087 & 0.34 & 0.2 & 2.47 $\pm$ 0.22 & 2.61 $\pm$ 0.07 & -- &70 & 8 & 131.2   \\ 
HESS~J1837$-$069 & 0.50 & 0.2 & 11.55 $\pm$ 0.49 & 2.54 $\pm$ 0.04 & -- &63 & 8 & 145.6   \\ 
HESS~J1841$-$055 & 0.50 & 0.2 & 11.58 $\pm$ 1.36 & 2.21 $\pm$ 0.07 & 0.09 $\pm$ 0.03 & 51 & 10 & 178.3   \\ 
HESS~J1843$-$033 & 0.38 & 0.2 & 3.04 $\pm$ 0.20 & 2.15 $\pm$ 0.05 & -- &70 & 15 & 121.0   \\ 
HESS~J1844$-$030 & 0.15 & 0.2 & 0.28 $\pm$ 0.04 & 2.48 $\pm$ 0.12 & -- &94 & 28 & 77.0   \\ 
HESS~J1846$-$029 & 0.15 & 0.2 & 0.48 $\pm$ 0.05 & 2.41 $\pm$ 0.09 & -- &94 & 10 & 95.8   \\ 
HESS~J1848$-$018 & 0.39 & 0.3 & 1.11 $\pm$ 0.15 & 2.57 $\pm$ 0.11 & -- &70 & 26 & 105.9   \\ 
HESS~J1849$-$000 & 0.16 & 0.3 & 0.58 $\pm$ 0.07 & 1.97 $\pm$ 0.09 & -- &70 & 9 & 129.9   \\ 
HESS~J1852$-$000 & 0.44 & 0.3 & 1.21 $\pm$ 0.15 & 2.17 $\pm$ 0.10 & -- &70 & 25 & 106.8   \\ 
HESS~J1857+026 & 0.41 & 0.3 & 4.00 $\pm$ 0.29 & 2.57 $\pm$ 0.06 & -- &70 & 11 & 127.6   \\ 
HESS~J1858+020 & 0.15 & 0.3 & 0.47 $\pm$ 0.06 & 2.39 $\pm$ 0.12 & -- &72 & 14 & 120.6   \\ 
HESS~J1908+063 & 0.50 & 0.3 & 8.35 $\pm$ 0.57 & 2.26 $\pm$ 0.06 & -- &41 & 2 & 240.9   \\ 
HESS J1911+090$^{*}$ & 0.10 & 0.3 & 0.15 $\pm$ 0.03 & 3.14 $\pm$ 0.24 & -- &-- & -- & --   \\ 
HESS J1912+101$^{*}$ & 0.56 & 0.7 & 2.49 $\pm$ 0.35 & 2.56 $\pm$ 0.09 & -- &-- & -- & --   \\ 
HESS~J1923+141 & 0.21 & 0.4 & 0.69 $\pm$ 0.11 & 2.55 $\pm$ 0.17 & -- &70 & 3 & 138.7   \\ 
HESS~J1930+188 & 0.15 & 0.5 & 0.32 $\pm$ 0.07 & 2.59 $\pm$ 0.26 & -- &92 & 8 & 100.3   \\ 
HESS~J1943+213 & 0.15 & 0.6 & 0.39 $\pm$ 0.08 & 2.83 $\pm$ 0.22 & -- &91 & 1 & 109.0   \\ 

 \end{longtable}
\twocolumn
}
{
\onecolumn
\LTcapwidth=\textwidth
\begin{longtable}{llll}
\caption
[HGPS associations]{
HGPS source associations (see Sect.~\ref{sec:results:assoc_id}).
}
\label{tab:hgps_associations}\\
\hline\hline
H.E.S.S. Source & Association \\
\hline
\endfirsthead
\caption{continued.}\\
\hline\hline
H.E.S.S. Source & Association \\
\hline
\endhead
\hline
\endfoot
HESS~J0835$-$455     & G263.9$-$3.3         (COMP) \\ 
                     & 2FHL~J0835.3$-$4511  (2FHL) \\ 
                     & 3FGL~J0835.3$-$4510  (3FGL) \\ 
                     & B0833$-$45           (PSR) \\ 
\hline
HESS~J0852$-$463     & G266.2$-$1.2         (SNR) \\ 
                     & 2FHL~J0852.8$-$4631e (2FHL) \\ 
                     & 3FGL~J0852.7$-$4631e (3FGL) \\ 
                     & J0855$-$4644         (PSR) \\ 
                     & G267.0$-$1.0         (PWN) \\ 
\hline
HESS~J1018$-$589~A   & 3FGL~J1018.9$-$5856  (3FGL) \\ 
                     & G284.3$-$1.8         (SNR) \\ 
\hline
HESS~J1018$-$589~B   & G284.3$-$1.8         (SNR) \\ 
                     & 3FGL~J1018.9$-$5856  (3FGL) \\ 
                     & J1016$-$5857         (PSR) \\ 
                     & G284.0$-$1.8         (PWN) \\ 
                     & 3FGL~J1016.3$-$5858  (3FGL) \\ 
\hline
HESS~J1023$-$575     & 3FGL~J1023.1$-$5745  (3FGL) \\ 
                     & J1023$-$5746         (PSR) \\ 
                     & Westerlund~2         (EXTRA) \\ 
                     & 2FHL~J1022.0$-$5750  (2FHL) \\ 
                     & 3FGL~J1024.3$-$5757  (3FGL) \\ 
\hline
HESS~J1026$-$582     & G285.1$-$0.5         (PWN) \\ 
                     & J1028$-$5819         (PSR) \\ 
\hline
HESS~J1119$-$614     & 3FGL~J1119.1$-$6127  (3FGL) \\ 
                     & J1119$-$6127         (PSR) \\ 
                     & G292.2$-$0.5         (COMP) \\ 
\hline
HESS~J1302$-$638     & B1259$-$63           (PSR) \\ 
\hline
HESS~J1303$-$631     & 2FHL~J1303.4$-$6312e (2FHL) \\ 
                     & 3FGL~J1303.0$-$6312e (3FGL) \\ 
                     & J1301$-$6305         (PSR) \\ 
                     & J1301$-$6310         (PSR) \\ 
                     & G304.1$-$0.2         (PWN) \\ 
\hline
HESS~J1356$-$645     & 3FGL~J1356.6$-$6428  (3FGL) \\ 
                     & G309.8$-$2.6         (PWN) \\ 
                     & J1357$-$6429         (PSR) \\ 
                     & 2FHL~J1355.1$-$6420e (2FHL) \\ 
\hline
HESS~J1418$-$609     & 3FGL~J1418.6$-$6058  (3FGL) \\ 
                     & G313.3+0.1           (PWN) \\ 
                     & J1418$-$6058         (PSR) \\ 
\hline
HESS~J1420$-$607     & G313.6+0.3           (PWN) \\ 
                     & J1420$-$6048         (PSR) \\ 
                     & 3FGL~J1420.0$-$6048  (3FGL) \\ 
                     & 2FHL~J1419.3$-$6047e (2FHL) \\ 
\hline
HESS~J1427$-$608     & Suzaku~J1427$-$6051  (EXTRA) \\ 
\hline
HESS~J1442$-$624     & G315.4$-$2.3         (SNR) \\ 
                     & 2FHL~J1443.2$-$6221e (2FHL) \\ 
\hline
HESS~J1457$-$593     & -- \\ 
\hline
HESS~J1458$-$608     & J1459$-$6053         (PSR) \\ 
                     & 3FGL~J1459.4$-$6053  (3FGL) \\ 
                     & 3FGL~J1456.7$-$6046  (3FGL) \\ 
\hline
HESS~J1503$-$582     & 3FGL~J1503.5$-$5801  (3FGL) \\ 
                     & 2FHL~J1505.1$-$5808  (2FHL) \\ 
\hline
HESS~J1507$-$622     & 3FGL~J1506.6$-$6219  (3FGL) \\ 
                     & CXOU~J150706.0$-$621443 (EXTRA) \\ 
                     & 2FHL~J1507.4$-$6213  (2FHL) \\ 
\hline
HESS~J1514$-$591     & B1509$-$58           (PSR) \\ 
                     & 3FGL~J1513.9$-$5908  (3FGL) \\ 
                     & G320.4$-$1.2         (COMP) \\ 
                     & 3FGL~J1514.0$-$5915e (3FGL) \\ 
                     & 2FHL~J1514.0$-$5915e (2FHL) \\ 
\hline
HESS~J1534$-$571     & G323.7$-$1.0         (SNR) \\ 
\hline
HESS~J1554$-$550     & G327.1$-$1.1         (COMP) \\ 
\hline
HESS~J1614$-$518     & Suzaku~J1614$-$5141  (EXTRA) \\ 
                     & 2FHL~J1615.3$-$5146e (2FHL) \\ 
                     & 3FGL~J1615.3$-$5146e (3FGL) \\ 
\hline
HESS~J1616$-$508     & J1617$-$5055         (PSR) \\ 
                     & G332.5$-$0.3         (PWN) \\ 
                     & 2FHL~J1616.2$-$5054e (2FHL) \\ 
                     & 3FGL~J1616.2$-$5054e (3FGL) \\ 
                     & G332.4$-$0.4         (SNR) \\ 
                     & G332.4+0.1           (SNR) \\ 
\hline
HESS~J1626$-$490     & 3FGL~J1626.2$-$4911  (3FGL) \\ 
\hline
HESS~J1632$-$478     & G336.4+0.2           (PWN) \\ 
                     & 2FHL~J1633.5$-$4746e (2FHL) \\ 
                     & 3FGL~J1633.0$-$4746e (3FGL) \\ 
                     & J1632$-$4757         (PSR) \\ 
\hline
HESS~J1634$-$472     & G337.2+0.1           (COMP) \\ 
                     & 3FGL~J1636.2$-$4709c (3FGL) \\ 
\hline
HESS~J1640$-$465     & 2FHL~J1640.6$-$4632  (2FHL) \\ 
                     & 3FGL~J1640.4$-$4634c (3FGL) \\ 
                     & J1640$-$4631         (PSR) \\ 
                     & G338.3$-$0.0         (COMP) \\ 
\hline
HESS~J1641$-$463     & G338.5+0.1           (SNR) \\ 
                     & 3FGL~J1641.1$-$4619c (3FGL) \\ 
\hline
HESS~J1646$-$458     & 3FGL~J1648.3$-$4611  (3FGL) \\ 
                     & J1648$-$4611         (PSR) \\ 
                     & 3FGL~J1650.3$-$4600  (3FGL) \\ 
\hline
HESS~J1702$-$420     & -- \\ 
\hline
HESS~J1708$-$410     & -- \\ 
\hline
HESS~J1708$-$443     & G343.1$-$2.3         (COMP) \\ 
                     & B1706$-$44           (PSR) \\ 
                     & 3FGL~J1709.7$-$4429  (3FGL) \\ 
\hline
HESS~J1713$-$381     & G348.7+0.3           (SNR) \\ 
                     & CXOU~J171405.7$-$381031 (EXTRA) \\ 
                     & J1714$-$3810         (PSR) \\ 
\hline
HESS~J1713$-$397     & 2FHL~J1713.5$-$3945e (2FHL) \\ 
                     & 3FGL~J1713.5$-$3945e (3FGL) \\ 
                     & G347.3$-$0.5         (SNR) \\ 
                     & 2FHL~J1714.1$-$4012  (2FHL) \\ 
\hline
HESS~J1714$-$385     & 3FGL~J1714.5$-$3832  (3FGL) \\ 
                     & G348.5+0.1           (COMP) \\ 
\hline
HESS~J1718$-$374     & G349.7+0.2           (SNR) \\ 
                     & 3FGL~J1718.0$-$3726  (3FGL) \\ 
\hline
HESS~J1718$-$385     & 3FGL~J1718.1$-$3825  (3FGL) \\ 
                     & G348.9$-$0.4         (PWN) \\ 
                     & J1718$-$3825         (PSR) \\ 
\hline
HESS~J1729$-$345     & -- \\ 
\hline
HESS~J1731$-$347     & G353.6$-$0.7         (SNR) \\ 
\hline
HESS~J1741$-$302     & -- \\ 
\hline
HESS~J1745$-$290     & Sgr~A*               (EXTRA) \\ 
                     & J1745$-$2900         (PSR) \\ 
                     & G359.9$-$0.0         (PWN) \\ 
                     & 2FHL~J1745.7$-$2900  (2FHL) \\ 
                     & 3FGL~J1745.6$-$2859c (3FGL) \\ 
                     & G0.0+0.0             (COMP) \\ 
                     & 3FGL~J1745.3$-$2903c (3FGL) \\ 
\hline
HESS~J1745$-$303     & 3FGL~J1745.1$-$3011  (3FGL) \\ 
                     & 2FHL~J1745.1$-$3035  (2FHL) \\ 
\hline
HESS~J1746$-$285     & 3FGL~J1746.3$-$2851c (3FGL) \\ 
                     & G0.1$-$0.1           (COMP) \\ 
                     & J1746$-$2850         (PSR) \\ 
\hline
HESS~J1746$-$308     & G358.5$-$0.9         (SNR) \\ 
                     & B1742$-$30           (PSR) \\ 
\hline
HESS~J1747$-$248     & 3FGL~J1748.0$-$2447  (3FGL) \\ 
                     & Terzan~5             (EXTRA) \\ 
\hline
HESS~J1747$-$281     & J1747$-$2809         (PSR) \\ 
                     & G0.9+0.1             (COMP) \\ 
\hline
HESS~J1800$-$240     & 3FGL~J1800.8$-$2402  (3FGL) \\ 
                     & 2FHL~J1801.7$-$2358  (2FHL) \\ 
                     & 3FGL~J1758.8$-$2402  (3FGL) \\ 
                     & G5.7$-$0.1           (SNR) \\ 
\hline
HESS~J1801$-$233     & 2FHL~J1801.3$-$2326e (2FHL) \\ 
                     & 3FGL~J1801.3$-$2326e (3FGL) \\ 
                     & W28                  (EXTRA) \\ 
\hline
HESS~J1804$-$216     & G8.3$-$0.0           (SNR) \\ 
                     & B1800$-$21           (PSR) \\ 
                     & 3FGL~J1805.6$-$2136e (3FGL) \\ 
                     & 2FHL~J1805.6$-$2136e (2FHL) \\ 
                     & G8.7$-$0.1           (COMP) \\ 
                     & J1803$-$2149         (PSR) \\ 
                     & 3FGL~J1803.1$-$2147  (3FGL) \\ 
\hline
HESS~J1808$-$204     & J1808$-$2024         (PSR) \\ 
                     & SGR~1806$-$20        (EXTRA) \\ 
                     & 1806$-$20~star~cluster (EXTRA) \\ 
\hline
HESS~J1809$-$193     & G11.0$-$0.0          (SNR) \\ 
                     & J1809$-$1917         (PSR) \\ 
                     & G11.1+0.1            (COMP) \\ 
                     & 3FGL~J1810.1$-$1910  (3FGL) \\ 
                     & G11.4$-$0.1          (SNR) \\ 
                     & 3FGL~J1811.3$-$1927c (3FGL) \\ 
                     & G11.2$-$0.3          (COMP) \\ 
                     & J1811$-$1925         (PSR) \\ 
\hline
HESS~J1813$-$126     & J1813$-$1246         (PSR) \\ 
                     & 3FGL~J1813.4$-$1246  (3FGL) \\ 
\hline
HESS~J1813$-$178     & J1813$-$1749         (PSR) \\ 
                     & G12.8$-$0.0          (COMP) \\ 
                     & G12.7$-$0.0          (SNR) \\ 
\hline
HESS~J1818$-$154     & G15.4+0.1            (COMP) \\ 
\hline
HESS~J1825$-$137     & 3FGL~J1824.5$-$1351e (3FGL) \\ 
                     & 2FHL~J1824.5$-$1350e (2FHL) \\ 
                     & G18.0$-$0.7          (PWN) \\ 
                     & B1823$-$13           (PSR) \\ 
\hline
HESS~J1826$-$130     & G18.5$-$0.4          (PWN) \\ 
                     & J1826$-$1256         (PSR) \\ 
                     & 3FGL~J1826.1$-$1256  (3FGL) \\ 
                     & G18.6$-$0.2          (SNR) \\ 
\hline
HESS~J1826$-$148     & LS~5039              (EXTRA) \\ 
                     & 3FGL~J1826.2$-$1450  (3FGL) \\ 
                     & 2FHL~J1826.3$-$1450  (2FHL) \\ 
\hline
HESS~J1828$-$099     & -- \\ 
\hline
HESS~J1832$-$085     & -- \\ 
\hline
HESS~J1832$-$093     & XMMU~J183245$-$0921539 (EXTRA) \\ 
\hline
HESS~J1833$-$105     & G21.5$-$0.9          (COMP) \\ 
                     & J1833$-$1034         (PSR) \\ 
                     & 3FGL~J1833.5$-$1033  (3FGL) \\ 
\hline
HESS~J1834$-$087     & J1834$-$0845         (PSR) \\ 
                     & G23.3$-$0.3          (COMP) \\ 
                     & 2FHL~J1834.5$-$0846e (2FHL) \\ 
                     & 3FGL~J1834.5$-$0841  (3FGL) \\ 
\hline
HESS~J1837$-$069     & J1838$-$0655         (PSR) \\ 
                     & 2FHL~J1836.5$-$0655e (2FHL) \\ 
                     & 3FGL~J1836.5$-$0655e (3FGL) \\ 
                     & 2FHL~J1837.4$-$0717  (2FHL) \\ 
                     & 3FGL~J1837.6$-$0717  (3FGL) \\ 
                     & G25.2+0.3            (PWN) \\ 
                     & 3FGL~J1838.9$-$0646  (3FGL) \\ 
\hline
HESS~J1841$-$055     & 2FHL~J1840.9$-$0532e (2FHL) \\ 
                     & 3FGL~J1840.9$-$0532e (3FGL) \\ 
                     & G26.6$-$0.1          (PWN) \\ 
                     & J1841$-$0524         (PSR) \\ 
                     & 3FGL~J1839.3$-$0552  (3FGL) \\ 
                     & J1838$-$0537         (PSR) \\ 
                     & 3FGL~J1838.9$-$0537  (3FGL) \\ 
\hline
HESS~J1843$-$033     & 3FGL~J1843.7$-$0322  (3FGL) \\ 
                     & 3FGL~J1844.3$-$0344  (3FGL) \\ 
                     & G28.6$-$0.1          (SNR) \\ 
\hline
HESS~J1844$-$030     & PMN~J1844$-$0306     (EXTRA) \\ 
                     & G29.4+0.1            (COMP) \\ 
\hline
HESS~J1846$-$029     & J1846$-$0258         (PSR) \\ 
                     & G29.7$-$0.3          (COMP) \\ 
\hline
HESS~J1848$-$018     & 3FGL~J1848.4$-$0141  (3FGL) \\ 
                     & W43                  (EXTRA) \\ 
\hline
HESS~J1849$-$000     & G32.6+0.5            (PWN) \\ 
                     & J1849$-$0001         (PSR) \\ 
\hline
HESS~J1852$-$000     & 3FGL~J1853.2+0006    (3FGL) \\ 
                     & G32.8$-$0.1          (SNR) \\ 
                     & J1853+0011           (PSR) \\ 
                     & J1853$-$0004         (PSR) \\ 
                     & G33.2$-$0.6          (SNR) \\ 
\hline
HESS~J1857+026       & J1856+0245           (PSR) \\ 
                     & MAGIC~J1857.2+0263   (EXTRA) \\ 
                     & MAGIC~J1857.6+0297   (EXTRA) \\ 
                     & 2FHL~J1856.8+0256    (2FHL) \\ 
\hline
HESS~J1858+020       & 3FGL~J1857.9+0210    (3FGL) \\ 
\hline
HESS~J1908+063       & ARGO~J1907+0627      (EXTRA) \\ 
                     & MGRO~J1908+06        (EXTRA) \\ 
                     & G40.5$-$0.5          (SNR) \\ 
                     & 3FGL~J1907.9+0602    (3FGL) \\ 
                     & J1907+0602           (PSR) \\ 
\hline
HESS~J1911+090       & G43.3$-$0.2          (SNR) \\ 
                     & 2FHL~J1911.0+0905    (2FHL) \\ 
                     & 3FGL~J1910.9+0906    (3FGL) \\ 
\hline
HESS~J1912+101       & J1913+1011           (PSR) \\ 
\hline
HESS~J1923+141       & 3FGL~J1923.2+1408e   (3FGL) \\ 
                     & 2FHL~J1923.2+1408e   (2FHL) \\ 
\hline
HESS~J1930+188       & G54.1+0.3            (COMP) \\ 
                     & J1930+1852           (PSR) \\ 
                     & VER~J1930+188        (EXTRA) \\ 
\hline
HESS~J1943+213       & IGR~J19443+2117      (EXTRA) \\ 
                     & 2FHL~J1944.1+2117    (2FHL) \\ 
\hline

\end{longtable}
\twocolumn
}

\end{document}